\documentclass[a4paper,11pt]{article}
\pdfoutput=1 

\usepackage{jheppub} 

\usepackage[T1]{fontenc} 
\usepackage[usenames,dvipsnames]{xcolor}
\usepackage{feynmp-auto}

\newcommand{\TT}[1]{\textcolor{Blue}{\textbf{TT : #1}}}

\newcommand{\half}{\frac{1}{2}}
\newcommand{\la}[1] {\left\langle #1 \right\rvert}
\newcommand{\ls}[1] {\left\lbrack #1 \bf \right\rvert}
\newcommand{\ra}[1] {\left\lvert #1 \right\rangle}
\newcommand{\rs}[1] {\left\lvert #1 \bf \right\rbrack}
\newcommand{\da}[1] {\left\langle #1 \right\rangle}
\newcommand{\ds}[1] {\left\lbrack #1 \bf \right\rbrack}

\newcommand{\bigzero}{\mbox{\normalfont\Large\bfseries 0}}
\newcommand{\bigI}{\mbox{\normalfont\Large\bfseries I}}
\newcommand{\rvline}{\hspace*{-\arraycolsep}\vline\hspace*{-\arraycolsep}}

\usepackage{verbatim}
\usepackage{amssymb}
\usepackage{amstext} 
\usepackage{array}   
\newcolumntype{L}{>{$}l<{$}} 

\allowdisplaybreaks[1]

\title{Supersymmetry, Supergravity and the Consistency of On-Shell Massive Superamplitudes}

\author[a,1]{Timothy Trott%
\note{Currently at private address.}}
\affiliation[a]{Department of Physics, University of California, \\
Santa Barbara, CA 93106, U.S.A.}

\emailAdd{drtimothytrott@gmail.com}

\abstract{I study constraints from consistent complex factorisation of $2\rightarrow 2$ scattering amplitudes in $4d$ and their relationship with supersymmetry. I complete the argument demonstrating that a massless helicity-$3/2$ particle must be a gravitino in a theory of supergravity and derive the structure and couplings of massive supermultiplets from first principles. For massive BPS particles in theories with extended supersymmetry, further constraints on the couplings are derived from consistent factorisation of massive superamplitudes. Among these is the requirement that BPS vector bosons must have couplings conforming to Lie algebra structure constants or generalised Chern-Simons terms. The gravitational coupling is shown to participate and draws the graviphotons into the Lie algebra. All $2\rightarrow 2$ tree-level amplitudes of massive particles with spin $s\leq 1$ are calculated in (super)gravity using on-shell methods and the double copy. These are assembled and decomposed into (super)amplitudes with varying amounts of supersymmetry. Finally, I study scattering of BPS gravitinos with unbroken $\mathcal{N}\geq 4$ supergravity and show that consistent factorisation leads to the full reconstruction of the super-Higgs mechanism. This argument completely determines the perturbative structure of all $\mathcal{N}=4$ Minkowski vacua of gauged maximal supergravity. Gravitinos in theories not admitting a super-Higgs mechanism are ruled-out.} 

\begin{document} 
\maketitle
\flushbottom

\section{Introduction}\label{sec:intro}

In the modern field of scattering amplitudes, direct recourse to fundamental principles of relativity and quantum mechanics has led to the development of efficient computational techniques that bypass obstructive complications of field theoretic formalism (see reviews \cite{elvang2015scattering,Dixon:2013uaa} and numerous references therein). These fundamental principles are incarnate through assumed analytic properties of the $S$-matrix - see \cite{Mizera:2023tfe,Caron-Huot:2025ymc} for recent reviews of these ideas and their connection to microcausality in field theory. Free of much of the convoluted redundancies in field theory, on-shell methods and the $S$-matrix have yielded numerous insights into fundamental physics, in particular colour-kinematics and the double copy \cite{Bern:2022wqg,Bern:2023zkg,Bern:2008qj}, non-renormalisation theorems \cite{Bern:2023zkg,Bern:2019wie,Jiang:2020rwz,Jones:2019nev}, the landscape of effective field theories (EFTs) \cite{DeAngelis:2022qco,Dong:2022mcv,Low:2019ynd,Cheung:2014dqa,Cheung:2016drk,Balkin:2021dko}, imprints of integrable structures \cite{Drummond:2009fd,Arkani-Hamed:2013jha,Arkani-Hamed:2012zlh,Caron-Huot:2018ape} and classical gravity \cite{Ema:2025qgd,Guevara:2018wpp,Chung:2018kqs,Arkani-Hamed:2019ymq,Bern:2020buy,Chiodaroli:2021eug,Bjerrum-Bohr:2013bxa,Correia:2024yfx} (a tiny handful of numerous examples). 

The ability to reconstruct scattering processes from elementary on-shell building blocks has also illuminated the structural rigidity of the known rules of fundamental physics and their origins. Older constraints arising from consistency of soft limits \cite{weinberg1964photons,Weinberg:1980kq,weinberg1965photons} (egs. soft photons must couple to conserved charges, soft gravitons are unique and couple universally to everything, higher spin massless particles are inconsistent with gravity) have been identified as special instances of a deeper argument traceable back to the bootstrapping of massless on-shell $3$- and $4$-particle amplitudes from the fundamental principles 
\cite{Benincasa:2007xk,McGady:2013sga,Arkani-Hamed:2017jhn}. Consistency restrictions on the hypothetical couplings appearing in the $3$-particle amplitudes provide a foundation upon which rest the known theories of particle physics: electrodynamics (QED), Yang-Mills (YM) gauge theory, general relativity (GR) and supergravity (SUGRA). The origin of these constraints will be elaborated upon further below. See \cite{Trott:2026cjj} for further introduction to the topic of the emergence of fundamental physics from perturbative $S$-matrix consistency, complex factorisation and high energy unitarity. 

Beyond these theories, the nature of the broader landscape of viable low energy EFTs and its boundaries remains an ongoing focus of contemporary research \cite{Adams:2006sv,Bellazzini:2020cot,Bellazzini:2021oaj,Bellazzini:2023nqj,Bellazzini:2025shd,Caron-Huot:2020cmc,Caron-Huot:2022ugt,Caron-Huot:2024lbf,Tolley:2020gtv,Hillman:2024ouy,Haring:2022sdp,Calisto:2025tjo} (and many more too numerous to cite). However, even the implications of the analytic structure of the $S$-matrix for perturbative theories at tree-level remain to be fully elucidated, as this study (and its prequel \cite{Trott:2026cjj}) investigates. The specific theme of this study is the relationship between supersymmetry (SUSY), SUGRA and the $S$-matrix as both an output (``emergence'') and an input (``imprints''), in particular for theories of massive particles. 

Scattering amplitudes of massive particles are less explored than their massless counterparts. This is partly because on-shell methods have been less developed for massive theories \cite{Wu:2021nmq} (although see \cite{Ema:2024vww,Ema:2024rss,Gherghetta:2024tob} for some interesting recent developments) and partly because of the much larger space of independent Lorentz structures and particle spectra that are permitted to exist. The implications and imprints of SUSY on the massive $S$-matrix have also received relatively little attention \cite{Herderschee:2019dmc,Herderschee:2019ofc,Caron-Huot:2018ape,Chen:2021hjl,Chen:2021huj,Abhishek:2022nqv,Chiodaroli:2015rdg,Chiodaroli:2018dbu,Johansson:2023ymb}, but provide an idealised arena of exploration in which it can be hoped that patterns or clues toward the structure of more general theories can be elucidated. Of course, the $S$-matrix of massless SUSY theories has revealed many unexpected surprises \cite{ArkaniHamed:2008gz,Drummond:2008cr,Drummond:2009fd,Arkani-Hamed:2013jha,Beisert:2010jr,Bern:2023zkg,elvang2015scattering} that could be hopefully mirrored in special massive theories. However, a necessary preliminary step pursued here is to identify, from first principles, the space of massive SUSY theories generated by $3$-particle amplitudes and their properties. The construction of $4$-particle superamplitudes from factorised $3$-particle superamplitudes then provides a simple test case for understanding the technical ingredients required for developing systematic unitarity-based methods.

Superamplitudes have been utilised (e.g. \cite{ArkaniHamed:2008gz,Drummond:2008cr}, see \cite{elvang2015scattering} for review) as a technique for representing scattering amplitudes in a form that exhibits manifest compliance with the supersymmetric Ward identities (SWIs) - relations between amplitudes involving supermultiplet members that must hold as a consequence of SUSY. In this formalism, supermultiplets of single particle states are grouped into fermionic coherent state representations of the SUSY algebra (``on-shell superfields''). Superamplitudes are Grassmann polynomials describing the transition amplitudes between these fermionic coherent states. They are effectively generating functions for the underlying amplitudes between component particles in the supermultiplets and they also inherit many of the algebraic properties of regular scattering amplitudes. This includes on-shell factorisation, once the intermediate state sum is extended to include an integration over intermediate Grassmann variables. For this reason, superamplitudes are an ideal tool for automating the management of SUSY and accounting for its constraints on the structure of particle interactions. 

Lorentz invariance places two restrictions upon the structure of scattering amplitudes. Firstly, scattering amplitudes must be decomposable into sums of tensors that represent each external particles' little groups. Secondly, scattering amplitudes must otherwise be spacetime scalars. The tensors are naturally constructed out of products and contractions of each external particles' helicity spinors (the natural variables that the $S$-matrix should be regarded as a function of) that obey these requirements \cite{Arkani-Hamed:2017jhn}. There are typically multiple linearly independent combination. These are the independent ``Lorentz structures''. Likewise, SUSY invariance places restrictions on the structure of the permissible Grassmann polynomials that can appear in superamplitudes. Systematic construction of superamplitudes requires a linearly independent basis of Grassmann polynomials (independent SUSY structures), analogous to the construction of a basis of independent Lorentz structures for non-SUSY amplitudes \cite{Arkani-Hamed:2017jhn,DeAngelis:2022qco,Dong:2022mcv}. Once constructed however, the SWIs between component amplitudes are automatically obeyed. Superamplitudes for massive supermultiplets have received attention previously in \cite{Herderschee:2019ofc,KNBalasubramanian:2022sae}. I will follow the framework and conventions of \cite{Herderschee:2019ofc} throughout this study. The reader seeking further background is referred to the sources cited in this paragraph and the one above. 

It will be assumed throughout the entirety of this study that the $S$-matrix has a simplified ``tree-level'' analytic structure (which is presumed to represent the leading order of some weak-coupling expansion). Following \cite{McGady:2013sga}, when viewed as functions of the external particles' helicity spinors, unitarity and locality materialise as the following necessary properties of a theory's scattering amplitudes: 
\begin{itemize}
    \item The only permitted singularities are simple poles of Mandelstam variables with residues that factorise into on-shell subamplitudes as
    \begin{align}\label{GenFact}
    A_n(1,\dots n)\sim\frac{-1}{s_{1\dots m}-m_P^2}\sum_PA_{m+1}(P, 1\dots m)A_{n-m+1}(P\rightarrow m+1\dots n)\end{align} 
    when $s_{1\dots m}\rightarrow m_P^2$, where $s_{1\dots m}=-(p_1+\dots p_m)^2$ for some $1<m<n-1$. Here $P$ indexes the possible intermediate particle species of mass $m_P$. I will always assume that all external particles are outgoing (unless otherwise specified) and that the amplitudes related to each other by crossed particles can be analytically continued into each other (see \cite{Trott:2026cjj} and \cite{Herderschee:2019ofc} for further explanation of the treatment of crossing and the spacetime related conventions adopted here). 
    \item Any factorisation channel consistent with the spectrum of particles and their interactions necessarily occurs.
\end{itemize}
The structure of a $4$-leg amplitude is thus assumed to consist of two sets of components:
\begin{enumerate}
    \item Terms containing simple poles of Mandelstam invariants. These must altogether have residues consistent with the factorisation rules stated above i.e. for an exchange of a mass $m$ particle $P$ in the $s$-channel, unitarity implies that, as $s\rightarrow m^2$, the amplitude has the form $A(12\rightarrow34)\sim\frac{-A(12\rightarrow P)A(P\rightarrow 34)}{s-m^2}$ (I drop the subscript on the amplitude denoting the number of legs henceforth as it will always be obvious from context). 
    \item Contact terms given by polynomials of Mandelstam invariants. These must be dressed by factors of spinor bilinears (Lorentz structures) that account for the spin information of the external particles.
\end{enumerate}
The separation of terms into these two classes is basis dependent. See \cite{Arkani-Hamed:2017jhn} for further commentary. The combination of subamplitudes appearing in the factorisation residues can itself contain cross-channel Mandlestam poles, even at $4$-legs. These poles can potentially introduce tension with locality: the poles appearing in a residue of a hypothetical amplitude's factorisation channel must themselves be consistent with residues of the other channels. This is the ``$4$-particle test'' discovered by \cite{Benincasa:2007xk} and requiring that it be obeyed imposes stringent constraints upon the couplings appearing in the $3$-particle amplitudes \cite{McGady:2013sga}. 

I will codify the $4$-particle test in several precise ways below in Sections \ref{SoftLimits} and \ref{sec:N=2}. However, it ultimately always amounts to bootstrapping hypothetical $4$-particle amplitudes from Lorentz invariance, little group covariance and the singularity structure mandated above. Candidate $4$-particle amplitudes are then required to factorise into elementary $3$-particle amplitudes in complete accordance with the unitarity and locality requirements. The elementary $3$-particle amplitudes are themselves fully determined by Lorentz invariance, little group covariance and possibly SUSY \cite{McGady:2013sga,Arkani-Hamed:2017jhn}. Imposing all possible factorisation requirements on the candidate $4$-particle amplitude can then lead to further consistency conditions on the permissible $3$-particle amplitudes and couplings. This reasoning can be leveraged to extract constraints on the space of viable theories consistent with fundamental principles while avoiding potential field theoretic convolution and complication. For massless particles, this is responsible for the consistency conditions cited above that pinpoint QED, YM, GR and SUGRA as unique special classes of theories of particles with helicities $\geq 1$. For massive particles however, the $4$-particle test is usually trivial, since the fully massive $3$-particle amplitudes do not contain singularities. A special exception to this occurs in the case of two equal mass particles with one massless leg \cite{Arkani-Hamed:2017jhn}. As these typify the couplings of gravitons, photons and gluons to matter, it is unsurprising that they would be responsible for extending the implications of the consistency constraints to include the coupling of massless particles to massive particles (the equivalence principle, conservation of charge and YM Lie algebra representations). 

The first objective of this study is to establish the emergence of unbroken SUSY from consistent factorisation of the $S$-matrix. It has been a longstanding expectation that a massless helicity-$3/2$ Rarita-Schwinger (RS) particle can only be consistent with causality if it is a gravitino in a theory of supergravity. A demonstration of this from consistency of soft-limits was first presented in \cite{GRISARU1977323}, the conclusions of which are uncontroversial. However, this analysis makes numerous assumptions about the hypothetical RS particle's couplings that drastically shortcut most of the necessary work. A significantly more thorough analysis from consistent factorisation of on-shell $3$-particle amplitudes was conducted in \cite{McGady:2013sga}, which established the structure of most of the RS particle's permitted $3$-particle couplings to massless particles. Subsequently, \cite{Elvang:2016qvq} presented a formulation of complex on-shell soft limits that allowed for the conclusions of consistent factorisation to be extended to amplitudes of arbitrary legs (and hence interactions of arbitrary valency). This demonstrates that, given the structure of the gravitino's $3$-particle couplings (which must have similar universality to that of the graviton), all higher-point interactions must obey the SWIs and the theory must be supersymmetric. 

Nevertheless, two holes remained in the analysis of \cite{McGady:2013sga}. Firstly, a couple of hypothetical $3$-particle amplitudes involving RS particles (that is, amplitudes with correct little group covariance, Lorentz invariance and that pass the pole counting criterion of \cite{McGady:2013sga}) but inconsistent with SUGRA remained unexcluded on factorisation grounds. Secondly, massive particles were not included in the analysis. In Section \ref{SoftLimits}, I address both of these gaps. The remaining spurious massless RS amplitudes are excluded through inconsistency with the all-channel pole, identified in \cite{Trott:2026cjj} where it was used to elucidate the complete $4$-particle test for gluons and photons. The inclusion of couplings to massive particles is more demanding. Consistent factorisation of Compton scattering of RS particles off massive spinning matter leads to a system of quadratic equations in the RS-matter couplings (again a consequence of the special properties of the two equal mass, one massless $3$-particle amplitudes). These can be solved for in order to determine the SUSY structure of the massive multiplets and their (almost) universal coupling to the gravitinos. The inconsistency of deviations to the gravitational dipole away from minimal coupling, identified in \cite{Chung:2018kqs}, also plays a critical role. Given then the structure of the $3$-particle couplings of massless gravitinos to massive matter (and the subsequent structure of the massive supermultiplets that it implies), the enforcement of SUSY follows from extending the SWIs to higher-leg amplitudes, paralleling the argument sketched-out in \cite{Elvang:2016qvq}. The whole argument is then extended to multiple flavours of gravitinos, corresponding to multiple supersymmetries. Of particular note is the on-shell derivation of the BPS bound. 

The two equal mass, one massless $3$-particle amplitude is a special case of a more general class of on-shell $3$-particle massive amplitudes in which the external particles have special complex kinematic configurations analogous to those underpinning the massless on-shell $3$-particle amplitudes. This special $3$-particle massive kinematics and its implications are described in detail in Section \ref{3PSK}. It generally arises in circumstances where the particle masses can be linked to a conserved charge (so that the mass is ``conserved''). The typical examples of this are the central charges of the SUSY algebra for BPS particles. Without SUSY, this special kinematics is otherwise only essential for describing the two equal mass, one massless particle amplitude. 

With enough SUSY however, massive amplitudes between supermultiplet members can only be combined into a supersymmetric Grassmann generating function at the expense of manifest locality. The $3$-particle superamplitudes of BPS particles in theories of extended SUSY necessarily contain factors with inverse dependence upon the external kinematic variables, analogous to those present in massless on-shell $3$-particle amplitudes. This is in spite of the fact that each of the constituent component amplitudes can be expressed in manifestly local forms (that is, a polynomial in helicity spinors). The assemblage of the component amplitudes into the on-shell $3$-particle SUSY invariant Grassmann polynomial is underpinned by the special kinematics of the BPS particles. The on-shell superfields of BPS particles, the structure of their SUSY invariants and superamplitudes are the subjects of Section \ref{SUSYAlgebra}. The unusual kinematic dependence of the $3$-particle superamplitudes of BPS particles gives them a similar tension with locality exhibited by the massless on-shell $3$-particle amplitudes. The superresidues produced by combining SUSY invariants across a factorisation channel contain cross-channel Mandelstam poles. The interactions among BPS particles are therefore also subjected to the $4$-particle test. The simplest example of this is provided by the self-interactions of BPS vector multiplets with $\mathcal{N}=4$ SUSY, where consistent complex factorisation of the $4$-leg superamplitude requires that the couplings be Lie algebra structure constants. Combined with the restrictions imposed directly on the couplings by SUSY, this demonstrates that the only $\mathcal{N}=4$ theory of BPS vector multiplets must be ``spontaneously broken'' super-Yang-Mills (SYM).

Before continuing to explore the implications of consistent factorisation for massive SUSY theories, Section \ref{Sec:LowSpinAmp} is devoted to the construction of $2\rightarrow 2$ (super)amplitudes with external particles of spin $s\leq 1/2$. This serves two purposes. Firstly, these low spin examples provide a useful warm-up for the calculations involving higher spin particles. The simplicity of these examples allows for an examination of several issues regarding the construction of $4$-leg superamplitudes. Of particular note is the basis construction for chiral SUSY invariants in the $\mathcal{N}=1$ on-shell superspace and also the role of electric-magnetic duality and the interpretation of the central charges in $\mathcal{N}=2$ SUGRA. The second purpose is to compile together the low spin on-shell residues for $2\rightarrow 2$ scattering in electrodynamics and gravity. These provide the seeds for the construction of higher spin gravitational amplitudes through double copy-like relations. One of the goals of this study is to extend the construction of the $2\rightarrow 2$ amplitudes of spin $\leq 1$ massive particles presented in \cite{Trott:2026cjj} to include gravity. I further demonstrate how the amplitudes of this Section assemble into the especially elegant superamplitudes of BPS hypermultiplets for $\mathcal{N}=2$ super-electrodynamics (SQED) and SUGRA. 

Section \ref{sec:N=2} advances to theories of BPS vector multiplets with $\mathcal{N}=2$ SUSY, the scattering amplitudes of which have received little previous attention. After establishing the structure of the permissible $3$-particle superamplitudes from first principles (which make critical use of the special kinematics of BPS particles), the $4$-leg superamplitudes are tested for consistent factorisation. Unlike the $4$-particle amplitudes of massless particles, the residues of the factorisation channels here contain multiple independent Lorentz structures and basis construction becomes an issue requiring careful attention. Once accounted for, consistent factorisation is shown to demand that the parity ($P$) symmetric couplings are Lie algebra structure constants, possibly non-semisimple or non-compact, while $P$ violating couplings are permitted to exist and must be identified as ``generalised Chern-Simons terms'' (GCSs). These are the same criteria identified in \cite{Trott:2026cjj} for the partial unitarisation of vector boson scattering amplitudes at high energies in general theories. However, in the $\mathcal{N}=2$ case, these conditions are obligatory consequences of causality. 

Section \ref{Sec:VecGrav} is devoted to the $2\rightarrow 2$ gravitational scattering amplitudes of massive vector bosons. These are easily constructed in theories with various amounts of SUSY by double copying the QED residues constructed in Section \ref{Sec:LowSpinAmp}. When the vector bosons are part of BPS vector multiplets in extended SUGRA, the graviton multiplet exchange residue itself contains a cross-channel Mandelstam pole. This draws the graviton exchange into the $4$-particle test, the outcome of which is to extend the Lie algebra to include the central charges as generators corresponding to the graviphotons. This Lie algebra is typically non-compact or non-semisimple. I further disassemble the superamplitudes into components produced by the exchanges of different states in order to show how their elegant SUSY expressions emerge. 

Finally, Section \ref{sec:ExtendedSSSB} combines both themes of the ``imprints'' and ``emergence'' of SUSY by studying the superamplitudes and consistency of BPS gravitino multiplets in the presence of unbroken extended SUGRA. The simplest theory of broken SUSY (in the sense of \cite{ArkaniHamed:2008gz}) is $\mathcal{N}=6$ SUGRA with a pair of BPS gravitinos. In this theory, the super-Higgs mechanism \cite{deser1977broken} is trivial: the gravitinos and the graviton multiplets alone comprise the entire particle content of the $\mathcal{N}=8$ graviton. It is easy to show that the gravitational scattering of the gravitinos factorises consistently and the resulting superamplitude has the same high energy growth as the regular graviton amplitude in the UV. However, problems immediately begin to arise when the theory is truncated to $\mathcal{N}=5$ SUGRA. In this case, the gravitational scattering superresidues (which may be constructed with the double copy) of the gravitinos fail the $4$-particle test, demonstrating that $\mathcal{N}=5$ SUGRA cannot arise from spontaneous SUSY breaking (at least in the absence of massive higher spin particles). The case with unbroken $\mathcal{N}=4$ SUGRA is more intricate, yet nevertheless still highly constrained by consistent factorisation. Pure gravitational scattering of $\frac{1}{2}$BPS gravitino multiplets is again inconsistent, but can be saved by the introduction of a pair of massless vector ``super-Higgs'' multiplets. The super-Higgs mechanism in this theory is therefore a necessary requirement of causality and not merely a property of a special subclass of EFTs with an extended range of validity into the UV. Consistent factorisation allows also for the introduction of a second species of gravitino multiplet while preserving a mass gap with the UV, provided that further vector super-Higgs multiplets are introduced with a coupling structure that is fully constrained up to a free phase. These arguments fully determine the perturbative sector of a theory of $\mathcal{N}=8\rightarrow 4$ spontaneous SUSY breaking on a flat background and without higher spin particles, verifying the gaugings derived in \cite{Catino:2013ppa} and establishing their uniqueness. Off-shell geometric complexity is entirely bypassed with on-shell simplicity. I additionally present a factorisation argument ruling-out $\frac{1}{4}$BPS gravitino multiplets. Finally, Section \ref{sec:ExtendedSSSB} also presents a derivation of Kaluza-Klein (KK) gravity from consistency a $\frac{1}{2}$BPS spin-2 supermultiplet in $\mathcal{N}=8$ SUGRA and the its relationship to the $\mathcal{N}=6$ broken SUSY theory through the Scherk-Schwarz (SS) mechanism.

Appendix \ref{sec:MasslessSUGRA} recapitulates and makes some further comments upon the $S$-matrix derivation of SUGRA from consistency of massless RS particles as argued in \cite{GRISARU1977323} and \cite{McGady:2013sga} (mostly the latter). Appendix \ref{sec:OSsuperfields} presents the on-shell superfields (or fermionic coherent state particle representations of the SUSY algebra) that underpin the superamplitudes constructed and analysed in this study. Appendix \ref{Sec:3legAmp} provides a catalogue of all of the elementary on-shell $3$-particle superamplitudes that can be constructed from first principles involving consistent massless multiplets and massive multiplets with particles of spin $s\leq 1$. These are presented for theories with $1\leq \mathcal{N}\leq 4$. Appendix \ref{Dyons} sketches some cursory ideas regarding supersymmetrisations of the dyon $S$-matrix proposed by \cite{Csaki:2020inw}. Finally, Appendix \ref{Quadrupoles} presents contributions to gravitational scattering amplitudes of vector bosons from insertions of quadrupole moments, which are of tangential relevance to the main narrative of this study. 

I generally follow the conventions of \cite{Srednicki:2007qs}. The more specialised conventions regarding massive amplitudes and superamplitudes follow \cite{elvang2015scattering} and \cite{Herderschee:2019ofc}. An exposition of the more rudimentary properties and conventions of the $S$-matrix adopted here (such as the relations imposed by unitarity and the way that crossing is managed) can be found in \cite{Trott:2026cjj}. However, unlike for \cite{Trott:2026cjj}, this time I do not promise perfect accuracy with negative signs, in particular overall signs of residues of superamplitudes. It's too hard. All of the important ones (and most of the unimportant ones) are correct though.

\section{Self-Consistency, Soft Limits and Supergravity}\label{SoftLimits}

This Section is dedicated to deriving SUSY and SUGRA from the existence of massless helicity-$3/2$ particles and the fundamental rules of relativity and quantum mechanics as imprinted upon the $S$-matrix. The subsequent Sections of the paper can be read largely independently of the content presented here and focus on the amplitudes of theories with pre-established SUSY. The reader may prefer to skip ahead if their interest lies elsewhere.

\subsection{SUGRA with massless particles}

\subsubsection{Addendum on 3-particle amplitudes in SUGRA}\label{sec:WrongGravitino}

One of the goals of this Section is to review and patch the critical parts of the argument that $S$-matrix self-consistency of a massless helicity-$3/2$ (RS) particle requires it to be a gravitino in a theory of supergravity \cite{GRISARU1977323}. For massless particles, the kinematic parts of the permissible elementary $3$-particle amplitudes, which determine the structure of the supermultiplets, were mostly deduced in \cite{McGady:2013sga}. However, a class of hypothetical amplitudes consisting of
\begin{align}
A\left(\widetilde{\psi}^+,\gamma^+,\chi^+\right)&=\ds{12}^2\ds{13}\label{MasslessGrInoDipole1}\\  A\left(\widetilde{\psi}^+,\widetilde{\psi}^+,\varphi\right)&=\ds{12}^3\label{MasslessGrInoDipole2}
\end{align}
(omitting overall dimensionful coupling constants and internal labels for brevity) and their parity conjugates were not able to be eliminated. I denote the RS particle or candidate gravitino by $\widetilde{\psi}$, a photon by $\gamma$, a (helicity-$1/2$) fermion by $\chi$ and scalar by $\varphi$. These amplitudes are not consistent with supergravity. It was argued that the first of these is not consistent with factorisation if the vector boson is a gluon or if the fermion is charged under the vector, but no further conclusion was drawn otherwise. These interactions are also too soft to modify the tree-level soft factors.

An obvious test of the consistency of these amplitudes with gravity (which should couple universally to each particle) is to consider the factorisation of an amplitude in which an additional graviton $h^-$ is radiated. However, correctly factorising amplitudes may nevertheless be constructed. For example, 
\begin{align}
A\left(\chi^+,\widetilde{\psi}^+,h^-,\gamma^+\right)=\frac{1}{M_{Pl}}\frac{1}{stu}\ds{12}^3\da{13}^3\da{32}\ds{42}\ds{41}
\end{align}
satisfies all of the requirements for each channel, despite the insertion of (\ref{MasslessGrInoDipole1}). An obvious next test is of the analogous process with a positive helicity graviton instead. This amplitude has a pole in which $\da{ij}\rightarrow 0$ for all legs $i$ and $j$. This is a single simple pole because of the relation
\begin{align}\label{AllChanPole}
    \frac{\ds{12}\ds{34}}{s}=\frac{\ds{14}\ds{23}}{u}=\frac{\ds{13}\ds{42}}{t}.
\end{align}
Poles of this form were used in \cite{Trott:2026cjj} to elucidate the nature of the Yang-Mills Lie algebra and derive parity and time-reversal symmetries of various massless $3$-particle amplitudes. 
\begin{figure}[h]
\begin{fmffile}{WeirdRSAmp}

\begin{center}
\begin{tabular}{ c c c c c }
& & & & \\
 \begin{fmfgraph*}(100,67)
   \fmfleft{i1,i2}
   \fmfright{o1,o2}
   \fmf{plain}{i1,v1}
   \fmf{plain}{i1,v1}
   \fmf{plain}{i2,v1}
   \fmf{boson}{i2,v1}
   \fmf{zigzag}{v2,o1}
   \fmf{zigzag}{v2,o1}
   \fmf{dbl_wiggly}{v2,o2}
   \fmf{dbl_wiggly}{v2,o2}
   \fmf{zigzag,label=$+\quad-$}{v1,v2}
   \fmf{zigzag}{v1,v2}
   \fmfv{decor.shape=circle,decor.filled=empty,decor.size=0.15w}{v1,v2}
   \fmflabel{$\chi^+$}{i1}
   \fmflabel{$\widetilde{\psi}^+$}{i2}
   \fmflabel{$h^+$}{o2}
   \fmflabel{$\gamma^+$}{o1}
 \end{fmfgraph*} & \, &
 \begin{fmfgraph*}(100,67)
   \fmfleft{i1,i2}
   \fmfright{o1,o2}
   \fmf{plain}{i1,v1}
   \fmf{plain}{i1,v1}
   \fmf{zigzag}{v2,o1}
   \fmf{zigzag}{v2,o1}
   \fmf{phantom}{v1,i2}
   \fmf{phantom}{v1,i2}
   \fmf{phantom}{v2,o2}
   \fmf{phantom}{v2,o2}
   \fmf{plain,tension=-0.25}{v2,i2}
   \fmf{boson,tension=-0.25}{v2,i2}
   \fmf{dbl_wiggly,tension=-0.25}{v1,o2}
   \fmf{dbl_wiggly,tension=-0.25}{v1,o2}
   \fmf{plain,label=$-\quad+$}{v1,v2}
   \fmf{plain}{v1,v2} \fmfv{decor.shape=circle,decor.filled=empty,decor.size=0.15w}{v1,v2}
   \fmflabel{$\chi^+$}{i1}
   \fmflabel{$\widetilde{\psi}^+$}{i2}
   \fmflabel{$h^+$}{o2}
   \fmflabel{$\gamma^+$}{o1}
 \end{fmfgraph*} & \, &
 \begin{fmfgraph*}(100,67)
   \fmfleft{i1,i2}
   \fmfright{o1,o2}
   \fmf{plain}{v1,v2}
   \fmf{boson}{v1,v2}
   \fmf{plain}{i1,v1}
   \fmf{plain}{i1,v1}
   \fmf{plain}{i2,v2}
   \fmf{boson}{i2,v2}
   \fmf{zigzag}{o1,v1}
   \fmf{zigzag}{o1,v1}
   \fmf{dbl_wiggly}{o2,v2}
   \fmf{dbl_wiggly}{o2,v2} \fmfv{decor.shape=circle,decor.filled=empty,decor.size=0.15w}{v1,v2}
   \fmflabel{$\chi^+$}{i1}
   \fmflabel{$\widetilde{\psi}^+$}{i2}
   \fmflabel{$h^+$}{o2}
   \fmflabel{$\gamma^+$}{o1}
   \fmfv{label=$-$,label.angle=-45,label.dist=0.1w}{v2}
   \fmfv{label=$+$,label.angle=45,label.dist=0.1w}{v1}
 \end{fmfgraph*}\nonumber\\
  $(s)$ & \, &  $(t)$ & \, &  $(u)$
\end{tabular}
\end{center}
\end{fmffile}
\caption{On-shell diagrams for $A\left(\chi^+,\widetilde{\psi}^+,h^+,\gamma^+\right)$ mediated by $A\left(\widetilde{\psi}^+,\gamma^+,\chi^+\right)$ insertions.}
\end{figure}
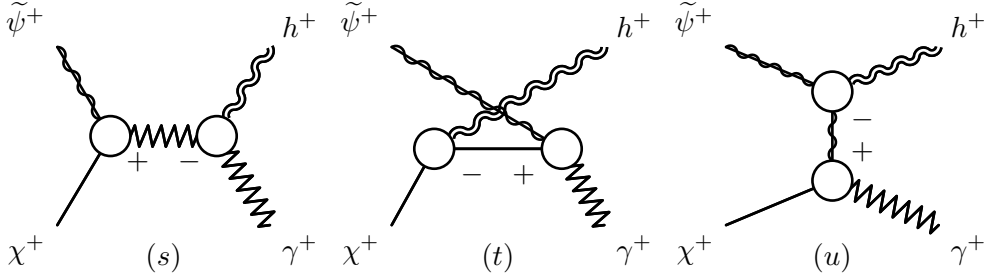
In this case, the candidate amplitude determined from factorisation is 
\begin{align}\label{All+GravitinoRes}
A\left(\chi^+,\widetilde{\psi}^+,h^+,\gamma^+\right)&=\frac{-1}{M_{Pl}}\frac{\ds{12}^5}{\ds{1P_s}^2\ds{2P_s}}\frac{1}{s}\ds{3P_s}\ds{4P_s}^2-\frac{1}{M_{Pl}}\frac{\ds{23}^3\ds{2P_u}}{\ds{3P_u}^2}\frac{1}{u}\ds{14}^2\ds{P_u1}\nonumber\\
&\qquad-\frac{1}{M_{Pl}}\frac{\ds{24}^4}{\ds{4P_t}^2}\frac{1}{t}\ds{1P_t}^2\ds{31}\nonumber\\
&=\frac{1}{M_{Pl}}\frac{\da{3q}}{\da{1q}}\frac{\ds{12}\ds{34}}{s}\ds{12}\ds{23}\ds{24},
\end{align}
where $P_s$, $P_t$ and $P_u$ are the on-shell internal momenta for each respective channel (more precisely, I define $P_s=p_3+p_4$, $P_t=p_2+p_4$ and $P_u=p_1+p_4$) and $\ra{q}$ is some spinor that has been introduced satisfying $\da{qi}\neq 0$ for any $\ra{i}$ on the residue. I have used the standard $3$-particle on-shell amplitudes between the graviton and each other massless species. However, lifting (\ref{All+GravitinoRes}) to a Lorentz invariant expression off-shell is not possible because it explicitly depends upon the spurious reference spinor $\ra{q}$. This demonstrates that no Lorentz invariant expression exists that correctly factorises on the all-channel pole. The interaction (\ref{MasslessGrInoDipole1}) is therefore inconsistent (at least in the presence of gravity). The computation with (\ref{MasslessGrInoDipole2}) is very similar with analogous results. 

This completes the argument of \cite{McGady:2013sga} that all massless $3$-particle amplitudes involving a RS particle must be consistent with a theory of supergravity in which the RS particle is a gravitino. The subsequent steps required to establish SUGRA consist of showing that the RS coupling is universally $1/M_{Pl}$, that the particles correctly arrange into supermultiplets (the structure of which is determined by their couplings to the gravitinos) and finally that the amplitudes obey the SWIs. The former two requirements are encapsulated in the coupling constants of the $3$-particle gravitino amplitudes. In Appendix \ref{sec:MasslessSUGRA}, I will review the derivation of these two requirements and give an additional illustrative example. The derivation mostly follows \cite{McGady:2013sga}. The results of this analysis are contained in the catalogue of SUGRA $3$-particle superamplitudes presented in Appendix \ref{Sec:3legAmp}. Given the $3$-particle couplings of the gravitinos, soft limits lead to the enforcement of the SWIs, as will be explained next.

\subsubsection{Consistent factorisation from soft limits}\label{sec:MasslessSoft}

Soft limits can be used to systematically derive constraints from consistent factorisation. The on-shell formulation presented in \cite{Elvang:2016qvq} makes the  tension with locality direct, rather than through convoluted issues of off-shell gauge invariance through which this was historically formulated. Choosing a soft particle with right-handed helicity, the soft limit can be taken in a chiral way so that only its right-handed spinor polarisation shrinks (in the all outgoing convention, outgoing right-handed particles have left-handed chiral polarisations). With this complexified soft momentum, soft factors can be exactly factorised from the hard amplitude and are described by the on-shell $3$-particle amplitudes. In \cite{Elvang:2016qvq}, the complex recoil of the soft leg is absorbed in the hard amplitude by a shift to two of the hard legs' momenta. This is additionally supplemented by a BCFW shift, which separates the kinematical singularities describing each of the distinct factorisations triggered by the soft limit into different complex configurations. Precisely, the shifted momenta are
\begin{align}\label{MasslessSoftShift}
   \hat{p}_s&=-\ra{\hat{s}}\ls{s}=-\left(\epsilon\ra{s}-z\ra{X}\right)\ls{s}\nonumber\\
   \hat{p}_i&=-\rs{\hat{i}}\la{i}=-\left(\rs{i}+\frac{1}{\da{ij}}\left(\epsilon\da{js}-z\da{jX}\right)\rs{s}\right)\la{i}\nonumber\\
   \hat{p}_j&=-\rs{\hat{j}}\la{j}=-\left(\rs{j}-\frac{1}{\da{ij}}\left(\epsilon\da{is}-z\da{iX}\right)\rs{s}\right)\la{j}
\end{align}
where $s$ denotes the soft leg and $i$ and $j$ are arbitrarily chosen hard legs, here assumed massless (the massive adaptation is presented further below). The spurious $\ra{X}$ spinor is arbitrary but chosen so that $\da{Xs},\da{Xk}\neq 0$ for any hard leg $k$. The spinor-level shifts can be easily read-off. Setting $z=0$, the parameter $\epsilon\rightarrow 0$ determines the soft limit. The BCFW shift parameter $z$ then separates the location of the poles describing factorisation of the soft leg with each hard leg $k$ to 
\begin{align}\label{softpole}
    \epsilon\rightarrow\epsilon^{(k)}_*=\frac{\la{X}p_k\rs{s}}{2p_s\cdot p_k}z=\frac{\da{Xk}}{\da{sk}}z,
\end{align}
where the last equality applies if $k$ is massless, but the first does not. The limit $z\ll\epsilon$ should be taken to ensure that all of the poles are still soft. The divergent terms in the soft limit can then be expressed as 
\begin{align}\label{MasterSoft}
A_{n+1}(1,2,\dots)=\sum_k\sum_P A_3(\hat{P}_{sk}\rightarrow\hat{k},\hat{s}^+)\frac{1}{P_{sk}^2} \frac{1}{\epsilon}\frac{1}{1-\frac{z}{\epsilon}\frac{\la{X}p_k\rs{s}}{2p_s\cdot p_k}}A_n^{(k)}(1,2,\ldots,\hat{P}_{sk},\ldots)
\end{align}
where $\hat{P}_{sk}=\left(\hat{p}_k+\hat{p}_s\right)$ (and $P_{sk}=\left(p_k+p_s\right)$). The sum over $P$ indicates the possible intermediate particle quantum numbers (internal or spin related). The hard amplitudes $A_n^{(k)}$ in the sum are evaluated on the poles (\ref{softpole}) and involve the hard leg $k$ being replaced with the intermediate particle with momentum $\hat{P}_{sk}$. For massless particles, the amplitudes $A_3(\hat{P}_{sk}\rightarrow \hat{k},\hat{s}^+)$ are built entirely out of the shifted left-handed spinor bilinears 
\begin{align}
    \ds{\hat{s}\hat{P}_{sk}}=\ds{s k}\qquad
    \ds{\hat{s}\hat{k}}=\ds{sk}\qquad
    \ds{\hat{k}\hat{P}_{sk}}=z\frac{\da{Xs}\ds{ks}}{\da{sk}}
\end{align}
for any hard leg $k$ (shifted or not - $\hat{k}=k$ in most cases).

The shift parameter $z$ only appears in the hard amplitudes through the momentum and polarisation of the leg  $\hat{P}_{sk}$. In the limit $z\rightarrow 0$, no singularity exists in the hard amplitude. However, spurious $z\rightarrow 0$ singularities can appear through the $A_3$ factors. Demanding that the amplitude be free of these singularities (and therefore factorises correctly) places constraints upon the theory. 

The formula (\ref{MasterSoft}) can be applied directly to $4$-leg amplitudes with a soft leg to systematically re-derive the ``$4$-particle test'' \cite{Benincasa:2007xk,McGady:2013sga}. For example, applying this to MHV scattering of four coloured vectors and assuming that the three vector amplitudes have couplings constants $f_{ABC}$, then taking e.g. the gluon $2^+_B$ soft, the spurious part of (\ref{MasterSoft}) becomes
\begin{align}
    A_4(1^+_A,2^+_B,3^-_C,4^-_D)&=f_{ABE}\frac{\ds{\hat{2}\hat{P}_s}^3}{\ds{\hat{1}\hat{2}}\ds{\hat{1}\hat{P}_s}}\frac{1}{s}f_{ECD}\frac{\da{4P_s}^3}{\da{43}\da{3P_s}}+f_{DBE}\frac{\ds{\hat{2}\hat{P}_t}^3}{\ds{\hat{2}\hat{4}}\ds{\hat{4}\hat{P}_t}}\frac{1}{t}f_{ECA}\frac{\da{3P_t}^3}{\da{13}\da{1P_t}}\nonumber\\
    &\qquad\qquad+f_{BCE}\frac{\ds{\hat{2}\hat{3}}^3}{\ds{\hat{2}\hat{P}_u}\ds{\hat{3}\hat{P}_u}}\frac{1}{u}f_{EDA}\frac{\da{41}^3}{\da{1P_u}\da{4P_u}}\nonumber\\
    &=\frac{\da{14}^3}{z\da{2X}\da{43}\da{13}}\left(f_{ABE}f_{ECD}+f_{DBE}f_{ECA}+f_{BCE}f_{EDA}\right).
\end{align}
(I am already assuming for simplicity that the couplings are real and fully antisymmetric, the derivation of the complete argument is given in \cite{Trott:2026cjj,Fonseca:2025mzj}). The coefficient of the spurious $z$-pole must vanish and this therefore necessitates the Jacobi identity. Another example given in \cite{Elvang:2016qvq} is that of soft gravitons. I will not bother to reproduce this here, but will assume the well-known conclusion that all particles universally couple to the graviton. 

The soft limits formulation of \cite{Elvang:2016qvq} encompasses most of the analysis of consistent factorisation for $4$-particle amplitudes and extends the conclusions to all higher leg amplitudes. For massless RS particles, this allows for the demonstration that all amplitudes must obey the SWIs, given the RS-matter $3$-particle amplitudes (derived in Appendix \ref{sec:MasslessSUGRA} following \cite{McGady:2013sga}). With the allowed $3$-leg amplitudes involving the gravitino identified, they can be used in (\ref{MasterSoft}) to derive constraints from soft gravitino limits. The $(+)$ helicity soft gravitinos effectively decrement the helicity of the hard legs that they are radiated off, while $(-)$ helicity gravitinos increment it. If the particle is the lowest helicity state in the multiplet, then the $(+)$ helicity gravitinos ignore it (and similarly the $(-)$ helicity gravitinos ignore the highest helicity states). 

Substituting the soft gravitino factors in (\ref{MasterSoft}), then the spurious terms all reduce to 
\begin{align}\label{Softino}
A_3(\hat{P}_{sk}\rightarrow \hat{k},\hat{\widetilde{\psi}}^+_a)\frac{1}{P_{sk}^2} \frac{1}{\epsilon}=\frac{1}{M_{Pl}}\frac{\ds{sk}}{\epsilon z\da{Xs}},
\end{align}
regardless of the identity of the hard leg but provided that it does couple to the gravitino $\widetilde{\psi}^+_a$. The consistency constraint in (\ref{MasterSoft}) then becomes 
\begin{align}\label{MasslessSWI}
\frac{1}{M_{Pl}}\frac{\ls{s}}{\epsilon z\da{Xs}}\sum_{k}\rs{k}A_n^{(k)}(1,2,\ldots,k^{h-1/2}_a,\ldots)=0,
\end{align}
where the sum over $k$ only includes particles that the $\widetilde{\psi}^+_a$ gravitino couples to. The amplitudes appearing in each term in the sum, $A_n^{(k)}(1,2,\ldots,k^{h-1/2}_a,\ldots)$, are intended to denote amplitudes of a list of particles $(1,2\ldots,n)$ with particle $k$ replaced by its partner to which it couples in the $3$-particle amplitude with the soft gravitino. This corresponds to the state in the supermultiplet with helicity decremented by $\frac{1}{2}$ and with $R$-structure modified to account for the gravitino's possible flavour index. The spinor $\ls{s}$ is arbitrary, along with everything else appearing in front of the sum in (\ref{MasslessSWI}). The equations that remain upon eliminating these junk factors are the SWIs and enforce supersymmetry on the theory. The RS particles are specifically always gravitinos - that is, superpartners of the graviton - as a consequence of the universality of the gravitational coupling \cite{Elvang:2016qvq}. 


I will adapt these arguments to include massive matter (as was set-up in \cite{Falkowski:2020aso}) in the next Section, again entirely from single soft RS limits. The soft limits approach systematises the derivation of the gravitino coupling and the massive SWIs which, at $3$-legs, determines the supermultiplet structure of the theory. This all represents a modernisation and completion of the original argument presented in \cite{GRISARU1977323}. Ultimately, the formulation of \cite{Elvang:2016qvq} and \cite{Falkowski:2020aso} also provides a convenient automation of the $4$-particle test for the specific case in which at least one particle is massless. It is always possible to bootstrap the same constraints instead from writing down a general expression for a candidate amplitudes (guided by the necessary little group properties, Lorentz invariance and the restrictions on singularities as being only simple Mandelstam poles) and then demanding that it correctly factorise in every possible on-shell limit. While sufficiently comprehensive for the derivations that I present below, this method can occasionally miss constraints, in particular those arising from comparing multiple possible limits of the same channel (e.g. consistency of $\ds{12},\da{34}\rightarrow 0$ with $\da{12},\ds{34}\rightarrow 0$ for a massless $4$-leg amplitude \cite{Trott:2026cjj}).

\subsection{Massive matter}\label{sec:MassiveMatter}

Massive particle interactions admit a much larger space of possible independent Lorentz structures than massless particles and the fact that higher spin particles are also permitted significantly exacerbates this. On-shell massive $3$-particle (super)amplitudes are discussed extensively in the remainder of this work, building on \cite{Arkani-Hamed:2017jhn} and \cite{Herderschee:2019ofc}. In this Section, I will focus on some remarks concerning the extensions of the massless soft theorems described above to include the presence of massive matter.

\subsubsection{Consistent soft limits with mass}

The extension of \cite{Elvang:2016qvq} to massive matter has been presented in \cite{Falkowski:2020aso} and applied to soft photons and gravitons. I will recapitulate the formulation of the soft-limits present above for massive hard legs, following the conventions of \cite{Elvang:2016qvq}.

The soft leg is shifted as in (\ref{MasslessSoftShift}) above, but now the hard leg BFCW shifts are:
\begin{align}
    \hat{p}_i&=p_i-\frac{1}{\ls{s}p_ip_j\rs{s}}\left(\epsilon\ls{s}p_j\ra{s}-z\ls{s}p_j\ra{X}\right)p_i\rs{s}\ls{s}\\
    \hat{p}_j&=p_j+\frac{1}{\ls{s}p_ip_j\rs{s}}\left(\epsilon\ls{s}p_i\ra{s}-z\ls{s}p_i\ra{X}\right)p_j\rs{s}\ls{s}.
\end{align}
These expressions are uniquely determined by placing the usual requirements on multi-line shifts: that the shift vectors be light-like and orthogonal to both each other and their corresponding unshifted momenta (and that momentum remains conserved). The momentum shifts can be adapted to the spinor level:
\begin{align}\label{MassSpinShift}
\ls{\hat{\mathbf{i}}}&=\ls{\mathbf{i}}-\frac{1}{\ls{s}p_ip_j\rs{s}}\left(\epsilon\ls{s}p_j\ra{s}-z\ls{s}p_j\ra{X}\right)\ds{\mathbf{i}s}\ls{s}\nonumber\\
\ls{\hat{\mathbf{j}}}&=\ls{\mathbf{j}}+\frac{1}{\ls{s}p_ip_j\rs{s}}\left(\epsilon\ls{s}p_i\ra{s}-z\ls{s}p_i\ra{X}\right)\ds{\mathbf{i}s}\ls{s}.
\end{align}
The spinor expressions that explicitly appear in the soft factors are
\begin{align}\label{MassiveSoftBis}
    \ds{\hat{s}\hat{P}_{sk}^K}=\ds{sk^K}\qquad
    \ds{\hat{s}\hat{k}^K}=\ds{sk^K}\qquad
    \da{\hat{k}^K\hat{P}^M_{sk}}=m_k\epsilon^{KM}\qquad
    \frac{1}{\hat{x}}=\frac{\ls{s}p_k\ra{s}}{m_k\da{Xs}z}.
\end{align}
The factor $x$ is defined just below in (\ref{Prelimx}). In practice, I have never found it necessary to actually keep track of the shifts to the hard legs in (\ref{MassSpinShift}).

Ordinarily, a test of consistent factorisation requires first the construction of a candidate expression for a $4$-particle amplitude obeying little group covariance, Lorentz invariance and the assumed analytic structure. This expression must then be matched onto the expected behaviour in each of the possible factorisation limits. The soft limit procedure described here instead conveniently bypasses the construction of the candidate amplitude. Taking the soft limit directly of a $4$-leg amplitude isolates all of the required physical singularities without requiring an expression for general momenta. The master formula (\ref{MasterSoft}) can then be used to systematically determine consistency requirements from factorisation. Note that, with mass and spin, (\ref{MasterSoft}) is modified by changing the intermediate propagator factor $P_{sk}^2$ to $P_{sk}^2+m_{P_{sk}}^2$, where $m_{P_{sk}}$ is the mass of the intermediate particle, while the sum over $P$ must also be interpreted as including spin.

\subsubsection{Multipoles and gravitational couplings}\label{sec:Multi}

A general spin $s$ massive particle $S$ has interactions involving a massless boson $b$ with helicity $h$ given by 
\begin{align}\label{Multipole}
A(S_i,S_j,b^+)&=\sum_{n=0}^{2s}{g_{ij}}_n^+ x^{n-h}\ds{3\bf{1}}^n\ds{3\bf{2}}^n\ds{\bf{12}}^{2s-n}\nonumber\\
A(S_i,S_j,b^-)&=\sum_{n=0}^{2s}\frac{{g_{ij}}_n^-}{x^{n-h}}\da{3\bf{1}}^n\da{3\bf{2}}^n\da{\bf{12}}^{2s-n}.
\end{align}
I allow for a possible species index denoted by $i$ and $j$ that distinguishes between otherwise degenerate particles. The factor $x$ is defined most precisely below in Section \ref{3PSK}, but for the purposes here I present its historical definition \cite{Arkani-Hamed:2017jhn}
\begin{align}\label{Prelimx}
x\rs{3}=\frac{p_1}{m}\ra{3}\qquad\Leftrightarrow\qquad x=\frac{\ls{q}p_1\ra{3}}{m\ds{q3}},
\end{align}
where $m$ is the mass of $S$. The reference spinor $\rs{q}$ satisfies $\ds{3q}\neq 0$ but is otherwise a placeholder that the expression is independent of. Coupling to photons is given by $h=1$, while gravitons are $h=2$. In this form, each term indexed by $n$ is identified as the $n$th order in the multipole expansion \cite{Arkani-Hamed:2017jhn} (the coupling constants ${g_{ij}}_n^\pm$ are related to the multipole moments). This is at the expense of manifest locality because of the presence of the factors of $x$ or $1/x$. However, using (\ref{Prelimx}), these factoris can be absorbed into the spinor bilinears involving the massless spinors leaving only bilinears of opposite chirality. It is only in the cases of the monopole (for $h=1$) and the gravitational dipole terms (for $h=2$) where there are insufficient appearances of the massless spinors for all factors of $x$ to be removed. When taking a soft vector or graviton limit, the $x$ factors source the spurious poles in (\ref{MasterSoft}) that are responsible for extending the conclusions about consistency of massless particles' couplings to the couplings involving massive particles. The special cases of minimal coupling to photons
\begin{align}\label{MinGluon}
A(S,\overline{S},\gamma^+)&=-e\frac{1}{x}\frac{\da{\bf{12}}^{2s}}{m^{2s-1}}\nonumber\\
A(S,\overline{S},\gamma^-)&=-ex\frac{\ds{\bf{12}}^{2s}}{m^{2s-1}}.
\end{align}
(for electric charge $e$) and gravitons
\begin{align}\label{MinGraviton}
A(S_i,S_j,h^+)&=\frac{\delta_{ij}}{M_{Pl}}\frac{1}{x^2}\frac{\da{\bf{12}}^{2s}}{m^{2(s-1)}}\nonumber\\
A(S_i,S_j,h^-)&=\frac{\delta_{ij}}{M_{Pl}}x^2\frac{\ds{\bf{12}}^{2s}}{m^{2(s-1)}}.
\end{align}
can be converted to the multipole form through use of 
\begin{align}\label{SimpleId}
\ds{\bf{12}}=\da{\bf{12}}+\frac{\da{3\bf{2}}\ds{3\bf{1}}}{m}.
\end{align}
I automatically pair each particle with a charge conjugate antiparticle in (\ref{MinGluon}). Exchanging the monopole term in (\ref{Multipole}) for minimal coupling, the remaining terms can be identified with ``anomalous'' multipole moments. Throughout this study, I will misuse ``minimal coupling'' to mean the specific Lorentz structures of the form stated above in (\ref{MinGluon}) and (\ref{Multipole}) (and a fermionic analogue below in (\ref{SUSYmin})). In this sense, a general amplitude such as those in (\ref{Multipole}) can be decomposed into a basis consisting of a ``minimal coupling'' term plus independent deviations to the multipole moments (beyond monopole).

Parenthetically, I will take the opportunity to comment upon the properties of graviton couplings (\ref{MinGraviton}) under crossing, analogous to the analysis presented for massless vectors in \cite{Trott:2026cjj}, to which I refer the reader for an explanation of the treatment of unitarity and crossing adopted in this study. Under crossing, the graviton polarisations do not flip sign (unlike for vectors), while the squaring of the $x$-factors means that the sign of the interior factor of the massive momentum in (\ref{Prelimx}) does not matter either. For this reason, I will assume that the minimal graviton coupling amplitudes to massive matter transform under crossing in the same way as the analogous gluon coupling amplitudes in \cite{Trott:2026cjj}, but with an extra factor of $(-1)^{n_1+n_3}$ to account for the differences just described ($(-1)^{n_i}$ indicates that the amplitude acquires an extra sign if it is to be reinterpreted with particle $i$ incoming instead of outgoing). For similar reasons, external gravitons typically do not flip sign under crossing. 

The multipolar Lorentz structure of the massless boson couplings in (\ref{Multipole}) generalises to arbitrary massive matter coupled to a massless particle. For two equal mass particles $S_1$ and $S_2$ of spins $s_1$ and $s_2$ with $s_1\leq s_2$, calling $\Delta s=s_2-s_1$, the couplings in (\ref{Multipole}) generalise to 
\begin{align}\label{GenMultipole}
A(S_1,\bar{S}_2,l^+)&=\sum_{n=\Delta s}^{s_1+s_2}g_n^+ x^{n-h}\ds{3\bf{1}}^{n-\Delta s}\ds{3\bf{2}}^{n+\Delta s}\ds{\bf{12}}^{s_1+s_2-n}\nonumber\\
A(S_1,\bar{S}_2,l^-)&=\sum_{n=\Delta s}^{s_1+s_2}\frac{g_n^-}{x^{n-h}}\da{3\bf{1}}^{n-\Delta s}\da{3\bf{2}}^{n+\Delta s}\da{\bf{12}}^{s_1+s_2-n}.
\end{align}
As is clear in (\ref{GenMultipole}), the permitted multipoles range from order $\Delta s$ to $s_1+s_2$. If $l$ is a boson, then it is required that $\Delta s\in \mathbb{N}\cup\{0\}$ and the multipole order represented by each term is $2^{n}$, while if $l$ is a fermion, then $\Delta s$ is a positive half-integer and the (analogue of a) multipole order of each term is $2^{n-1/2}$.

If the masses are distinct, then the $x$ factors are no longer meaningfully defined. However, the multipole expansion remains valid if each term in (\ref{GenMultipole}) is reinterpreted as their corresponding expressions after the $x$-factors have been absorbed through application of (\ref{Prelimx}). As this is not possible for the monopole and gravitational dipole, these interactions are simply precluded from existing in these inelastic cases. The multipole expansion is simply forced to begin at higher order. 

A further self-consistency condition on massive $3$-particle amplitudes was given in \cite{Chung:2018kqs}, where the (hypothetical) anomalous gravitational dipole moment was ruled-out. I here review that argument by directly applying (\ref{MasterSoft}) to the gravitational Compton amplitude $A(S_i,h^+,h^-,\overline{S}_i)$ with massive matter of arbitrary spin. Here, $i$ is some internal ``flavour'' index. Like the minimal coupling, the hypothetical gravitational dipole term also acquires spurious $z$ dependence when substituted into (\ref{MasterSoft}), but the higher order multipoles do not. However, this appears as a simple pole, whereas the minimal coupling contains a double $z$ pole. The cancellation of this double pole in (\ref{MasterSoft}) demands the universality of the minimal gravitational coupling independently of the dipole, and this is also sufficient to eliminate single poles that also appear associated to minimal coupling. So taking $h^+$ soft and ignoring the purely minimal coupling terms, the remaining spurious terms are given by
\begin{align}\label{spurdipole}
    \frac{1}{z\epsilon\da{X2}}\sum_k\left({g_{ik}}_1\ds{2\bf{1}}\ds{21_M}A(S^{(M\ldots)}_k,h^-,\overline{S}_i)+{g_{ki}}_1\ds{2\bf{4}}\ds{24_M}A(S_i,h^-,\overline{S}^{(M\ldots)}_k)\right).
\end{align}
These terms are produced by the $s$- and $t$-channels respectively. The superscript on $S^{(M\ldots)}_k$ just means that one of the corresponding implicit $SU(2)$ spin indices on the amplitude in which this particle appears is contracted as part of a spin sum. Substituting in the general expressions for the $3$-particle amplitudes in terms of minimal coupling plus anomalous multipoles, the minimal coupling terms cancel between the two channels, but the terms proportional to the anomalous dipole moments do not and leave linearly independent terms each proportional to $\sum_k |{g_{ik}}_1|^2$ and $\sum_k |{g_{ki}}_1|^2$. The only way for (\ref{spurdipole}) to vanish for any $i$ is for ${g_{ik}}_1=0$ for each $i,k$. Hence the anomalous gravitational dipole moments are prohibited. 

Note that this argument applies to dipole interactions in which the spin number of the massive particle changes by one unit (the intermediate particle in the $s$ and $t$-channels just have a different spin than the external legs, but this does not change the conclusions). Thus, for example, the hypothetical interaction
\begin{align}\label{MixedSpinGravDi}
A(\varphi,W,h^+)\propto\frac{1}{x}\ds{3\bf{2}}\ds{3\bf{2}},
\end{align}
in which $\varphi$ and $W$ are an equal mass scalar and vector respectively, is ruled-out. In general, larger changes in spin do not admit anomalous dipole interactions, while potentially inelastic interactions in which the massive particle also changes mass are never consistent with a dipolar structure and begin at quadrupole order (just by Lorentz invariance). 



\subsubsection{Supergravity with massive spinning particles}

In this Section, 
I will extend the derivation of SUSY from soft helicity-$3/2$ limits to include massive particles. This will involve showing that the massive particles arrange into supermultiplets defined by (near-)universal gravitational $3$-particle coupling to the gravitinos. 
Only single soft limits of RS particles in $4$-particle amplitudes are needed to completely derive these results. Having established the consistent $3$-particle gravitino amplitudes, the fact that higher leg amplitudes obey the SWIs, and hence that the theory is supersymmetric, follows immediately from application of (\ref{MasterSoft}), analogously to the demonstration above for massless amplitudes. 

To begin with, I will assume that there is only a single flavour of gravitino (so $\mathcal{N}=1$). Assume then that there exists a massive, spin $s$ particle $S_s$. Consistent soft graviton limits already establish that $S_s$ has universal gravitional coupling (\ref{MinGraviton}). It can therefore gravitationally scatter off the gravitino to give the $u$-channel of the amplitude $A_4(S_s,\widetilde{\psi}^+,\widetilde{\psi}^-,S_s)$. However, the soft gravitino limit of this channel contains a spurious $z$ pole when substituted into (\ref{MasterSoft}), in addition to a further spurious Lorentz violation in the form of a factor of $x$. Since there are no other available massless particles with consistent $3$-particle amplitudes involving this helicity configuration of gravitinos, consistency can only be restored by introducing a new exchanged particle in the $s$ or $t$-channels. The only possible $3$-particle amplitudes in these channels that can contain spurious $z$-dependence to cancel the $u$-channel pole are the monopole terms in (\ref{GenMultipole}) with $h=3/2$. Therefore, new partner particles of spin $s\pm\frac{1}{2}$ must be introduced. These massive spinning particles must have a non-zero minimal coupling to the gravitino, which necessitates that they have equal mass. I will define minimal coupling constants $C^{s}_{ij}$ through 
\begin{align}\label{SUSYmin}
A(S_{s,i},S_{s+\frac{1}{2},j},\widetilde{\psi}^+)&=C^{s}_{ij}\frac{1}{m^{2s}x}\ds{3\bf{2}}\da{\bf{12}}^{2s}\nonumber\\
A(S_{s,i},S_{s+\frac{1}{2},j},\widetilde{\psi}^-)&=(C^{s}_{ij})^*\frac{x}{m^{2s}}\da{3\bf{2}}\ds{\bf{12}}^{2s}.
\end{align}
I assume self-conjugate massive particles and include possible internal indices $i,j$ on the massive matter. Since it was implicitly established in \cite{McGady:2013sga} (and the review in Appendix \ref{sec:MasslessSUGRA}) that the gravitinos are singlets under a possible (unbroken) YM gauge algebra, then invariance of the coupling implies that all dependence on the gauge algebra be factorised from the $C^{s}_{ij}$ and is of the form $\delta_{RR'}\delta_{AA'}$, where $R$ ($R'$) and $A$ ($A'$) denote the irrep and component of the particle $S_s$ ($S_{s+\frac{1}{2}}$). It is therefore automatic that the massive spinning particles in (\ref{SUSYmin}) must have the same gauge quantum numbers. I will leave this trivial dependence implicit. Finally, I admit that I do not fully understand the crossing signs for these amplitudes, but, accepting that a consistent theory likely exists, it is possible to easily infer the required signs in the construction of $4$-particle amplitudes from exchange symmetries and unitarity (where they matter). 

Having now introduced new spinning particles with potentially spurious yet Lorentz invariant couplings into the theory, a set of $4$-particle amplitudes with non-trivial soft gravitino limits emerges that must be simultaneously made self-consistent. This leads to a system of quadratic equations in the couplings $C^{s}_{ij}$ that must be solved for. Taking a soft right-handed gravitino (chosen as particle $2$ in the amplitudes below), then applying (\ref{MasterSoft}) and (\ref{MassiveSoftBis}) to various $4$-particle amplitudes and demanding the vanishing of terms with spurious $z$ poles produces the conditions:
\begin{figure}[h]
\begin{fmffile}{Gravitino1}

\begin{center}
\begin{tabular}{ c c c c c }
& & & & \\
 \begin{fmfgraph*}(100,67)
   \fmfleft{i1,i2}
   \fmfright{o1,o2}
   \fmf{dbl_plain}{i1,v1}
   \fmf{dbl_plain}{i1,v1}
   \fmf{dbl_plain}{v1,v2}
   \fmf{dbl_plain}{v1,v2}
   \fmf{dbl_plain}{v2,o1}
   \fmf{dbl_plain}{v2,o1}
   \fmf{plain}{i2,v1}
   \fmf{boson}{i2,v1}
   \fmf{plain}{o2,v2}
   \fmf{boson}{o2,v2} \fmfv{decor.shape=circle,decor.filled=gray50,decor.size=0.15w}{v1,v2}
   \fmflabel{$\widetilde{\psi}^+$}{i2}
   \fmflabel{$\widetilde{\psi}^-$}{o2}
   \fmflabel{$s_i$}{i1}
   \fmflabel{$s_j$}{o1}
   \fmfv{label=$(s\pm\frac{1}{2})_m$,label.angle=-65,label.dist=0.065w}{v1}
 \end{fmfgraph*} 
 &\,& \begin{fmfgraph*}(100,67)
   \fmfleft{i1,i2}
   \fmfright{o1,o2}
   \fmf{dbl_plain}{i1,v1}
   \fmf{dbl_plain}{i1,v1}
   \fmf{dbl_plain}{v1,v2}
   \fmf{dbl_plain}{v1,v2}
   \fmf{dbl_plain}{v2,o1}
   \fmf{dbl_plain}{v2,o1}
   \fmf{phantom}{v1,i2}
   \fmf{phantom}{v1,i2}
   \fmf{phantom}{v2,o2}
   \fmf{phantom}{v2,o2}
   \fmf{plain,tension=-0.25}{v2,i2}
   \fmf{plain,tension=-0.25}{v1,o2}
   \fmf{boson,tension=-0.25}{v2,i2}
   \fmf{boson,tension=-0.25}{v1,o2}
   \fmfv{decor.shape=circle,decor.filled=gray50,decor.size=0.15w}{v1,v2}
   \fmflabel{$\widetilde{\psi}^+$}{i2}
   \fmflabel{$\widetilde{\psi}^-$}{o2}
   \fmflabel{$s_i$}{i1}
   \fmflabel{$s_j$}{o1}
   \fmfv{label=$(s\pm\frac{1}{2})_m$,label.angle=-60,label.dist=0.1w}{v1}
 \end{fmfgraph*}
&\,& \begin{fmfgraph*}(100,67)
   \fmfleft{i1,i2}
   \fmfright{o1,o2}
   \fmf{dbl_plain}{i1,v2}
   \fmf{dbl_plain}{i1,v2}
   \fmf{dbl_wiggly}{v1,v2}
   \fmf{dbl_wiggly}{v1,v2}
   \fmf{dbl_plain}{v2,o1}
   \fmf{dbl_plain}{v2,o1}
   \fmf{plain}{i2,v1,o2}
   \fmf{boson}{i2,v1,o2}
   \fmfv{decor.shape=circle,decor.filled=gray50,decor.size=0.15w}{v1,v2}
   \fmflabel{$\widetilde{\psi}^+$}{i2}
   \fmflabel{$\widetilde{\psi}-$}{o2}
   \fmflabel{$s_i$}{i1}
   \fmflabel{$s_j$}{o1}
   \fmfv{label=$+$,label.angle=-45,label.dist=0.1w}{v1}
   \fmfv{label=$-$,label.angle=45,label.dist=0.1w}{v2}
 \end{fmfgraph*}\nonumber\\
 $(s)$ & \, &  $(t)$ & \, &  $(u)$
\end{tabular}
\end{center}
\end{fmffile}
\caption{On-shell diagrams for $A(S_{s,i},\widetilde{\psi}^+,\widetilde{\psi}^-,S_{s,j})$.}
\end{figure}
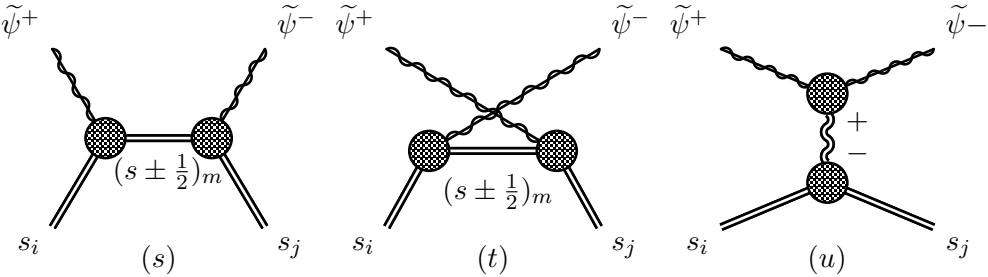
\begin{figure}[h]
\begin{fmffile}{Gravitino2}

\begin{center}
\begin{tabular}{ c c c }
& & \\
 \begin{fmfgraph*}(100,67)
   \fmfleft{i1,i2}
   \fmfright{o1,o2}
   \fmf{dbl_plain}{i1,v1}
   \fmf{dbl_plain}{i1,v1}
   \fmf{dbl_plain}{v1,v2}
   \fmf{dbl_plain}{v1,v2}
   \fmf{dbl_plain}{v2,o1}
   \fmf{dbl_plain}{v2,o1}
   \fmf{plain}{i2,v1}
   \fmf{boson}{i2,v1}
   \fmf{plain}{o2,v2}
   \fmf{boson}{o2,v2} \fmfv{decor.shape=circle,decor.filled=gray50,decor.size=0.15w}{v1,v2}
   \fmflabel{$\widetilde{\psi}^+$}{i2}
   \fmflabel{$\widetilde{\psi}^+$}{o2}
   \fmflabel{$s_i$}{i1}
   \fmflabel{$s_j$}{o1}
   \fmfv{label=$(s\pm\frac{1}{2})_m$,label.angle=-65,label.dist=0.065w}{v1}
 \end{fmfgraph*} 
 &\,& \begin{fmfgraph*}(100,67)
   \fmfleft{i1,i2}
   \fmfright{o1,o2}
   \fmf{dbl_plain}{i1,v1}
   \fmf{dbl_plain}{i1,v1}
   \fmf{dbl_plain}{v1,v2}
   \fmf{dbl_plain}{v1,v2}
   \fmf{dbl_plain}{v2,o1}
   \fmf{dbl_plain}{v2,o1}
   \fmf{phantom}{v1,i2}
   \fmf{phantom}{v1,i2}
   \fmf{phantom}{v2,o2}
   \fmf{phantom}{v2,o2}
   \fmf{plain,tension=-0.25}{v2,i2}
   \fmf{plain,tension=-0.25}{v1,o2}
   \fmf{boson,tension=-0.25}{v2,i2}
   \fmf{boson,tension=-0.25}{v1,o2}
   \fmfv{decor.shape=circle,decor.filled=gray50,decor.size=0.15w}{v1,v2}
   \fmflabel{$\widetilde{\psi}^+$}{i2}
   \fmflabel{$\widetilde{\psi}^+$}{o2}
   \fmflabel{$s_i$}{i1}
   \fmflabel{$s_j$}{o1}
   \fmfv{label=$(s\pm\frac{1}{2})_m$,label.angle=-60,label.dist=0.1w}{v1}
 \end{fmfgraph*}\nonumber\\
 $(s)$ & \, &  $(t)$ 
\end{tabular}
\end{center}
\end{fmffile}
\caption{On-shell diagrams for $A(S_{s,i},\widetilde{\psi}^+,\widetilde{\psi}^+,S_{s,j})$.}
\end{figure}
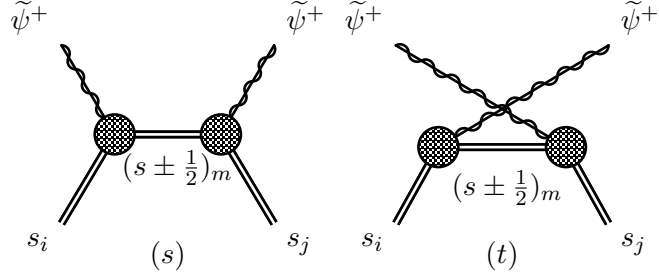
\begin{figure}[h]
\begin{fmffile}{Gravitino3}

\begin{center}
\begin{tabular}{ c c c }
& &  \\
 \begin{fmfgraph*}(100,67)
   \fmfleft{i1,i2}
   \fmfright{o1,o2}
   \fmf{dbl_plain}{i1,v1}
   \fmf{dbl_plain}{i1,v1}
   \fmf{dbl_plain}{v1,v2}
   \fmf{dbl_plain}{v1,v2}
   \fmf{dbl_plain}{v2,o1}
   \fmf{dbl_plain}{v2,o1}
   \fmf{plain}{i2,v1}
   \fmf{boson}{i2,v1}
   \fmf{plain}{o2,v2}
   \fmf{boson}{o2,v2} \fmfv{decor.shape=circle,decor.filled=gray50,decor.size=0.15w}{v1,v2}
   \fmflabel{$\widetilde{\psi}^+$}{i2}
   \fmflabel{$\widetilde{\psi}^\pm$}{o2}
   \fmflabel{$(s-\frac{1}{2})_i$}{i1}
   \fmflabel{$(s+\frac{1}{2})_j$}{o1}
   \fmfv{label=$s_m$,label.angle=-25,label.dist=0.15w}{v1}
 \end{fmfgraph*} 
 &\,\,\,\,\,\qquad\qquad\qquad & \begin{fmfgraph*}(100,67)
   \fmfleft{i1,i2}
   \fmfright{o1,o2}
   \fmf{dbl_plain}{i1,v1}
   \fmf{dbl_plain}{i1,v1}
   \fmf{dbl_plain}{v1,v2}
   \fmf{dbl_plain}{v1,v2}
   \fmf{dbl_plain}{v2,o1}
   \fmf{dbl_plain}{v2,o1}
   \fmf{phantom}{v1,i2}
   \fmf{phantom}{v1,i2}
   \fmf{phantom}{v2,o2}
   \fmf{phantom}{v2,o2}
   \fmf{plain,tension=-0.25}{v2,i2}
   \fmf{plain,tension=-0.25}{v1,o2}
   \fmf{boson,tension=-0.25}{v2,i2}
   \fmf{boson,tension=-0.25}{v1,o2}
   \fmfv{decor.shape=circle,decor.filled=gray50,decor.size=0.15w}{v1,v2}
   \fmflabel{$\widetilde{\psi}^+$}{i2}
   \fmflabel{$\widetilde{\psi}^\pm$}{o2}
   \fmflabel{$(s-\frac{1}{2})_i$}{i1}
   \fmflabel{$(s+\frac{1}{2})_j$}{o1}
   \fmfv{label=$s_m$,label.angle=-25,label.dist=0.2w}{v1}
 \end{fmfgraph*}\nonumber\\
 $(s)$ & \,\,\,\,\,\qquad\qquad\qquad &  $(t)$ 
\end{tabular}
\end{center}
\end{fmffile}
\caption{On-shell diagrams for $A(S_{s-\frac{1}{2},i},\widetilde{\psi}^+,\widetilde{\psi}^\pm,S_{s+\frac{1}{2},j})$.}
\end{figure}
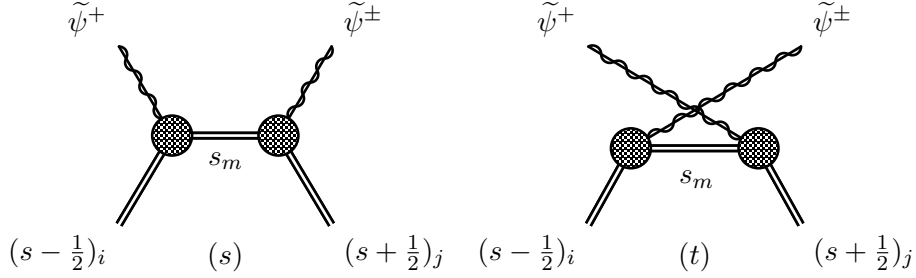
\begin{align}\label{GravitinoSoftCons}
&A(S_{s,i},\widetilde{\psi}^+,\widetilde{\psi}^-,S_{s,j}):&\left(\frac{m}{M_{Pl}}\right)^2\delta_{ij}&=C^{s}_{im}(C^{s}_{jm})^*+\frac{1}{2s+1}C^{s}_{jm}(C^{s}_{im})^*+C^{s-\frac{1}{2}}_{mj}(C^{s-\frac{1}{2}}_{mi})^*\nonumber\\
&&&=C^{s}_{jm}(C^{s}_{im})^*+\frac{1}{2s+1}C^{s}_{im}(C^{s}_{jm})^*+C^{s-\frac{1}{2}}_{mi}(C^{s-\frac{1}{2}}_{mj})^*\nonumber\\
&A(S_{s,i},\widetilde{\psi}^+,\widetilde{\psi}^+,S_{s,j}):&C^{s-\frac{1}{2}}_{mi}C^{s-\frac{1}{2}}_{mj}&=\frac{2s}{2s+1}C^{s}_{im}C^{s}_{jm}\nonumber\\
&A(S_{s-\frac{1}{2},i},\widetilde{\psi}^+,\widetilde{\psi}^-,S_{s+\frac{1}{2},j}):&C^{s-\frac{1}{2}}_{im}(C^{s}_{mj})^*&=(C^{s-\frac{1}{2}}_{im})^*C^{s}_{mj}\nonumber\\
&A(S_{s-\frac{1}{2},i},\widetilde{\psi}^+,\widetilde{\psi}^+,S_{s+\frac{1}{2},j}):&C^{s-\frac{1}{2}}_{im}C^{s}_{mj}&=0.
\end{align}
Single amplitudes can produce multiple relations because of the presence of multiple independent Lorentz structures that must separately cancel. The amplitudes in which the external matter legs have different spin are ultimately responsible for truncating the size of the supermultiplets and further rely upon the necessary vanishing of the mixed spin gravitational dipole in order to rule-out possible $u$-channel contributions, as discussed around (\ref{MixedSpinGravDi}). In the special case $s=0$, the couplings $C^{s-\frac{1}{2}}$ in (\ref{GravitinoSoftCons}) are to be set to zero and only the single relation 
\begin{align}\label{C0-1/2}
\left(\frac{m}{M_{Pl}}\right)^2\delta_{ij}=C^{0}_{im}(C^{0}_{jm})^*+C^{0}_{jm}(C^{0}_{im})^*
\end{align}
remains. 

These constraints can be solved beginning with the couplings involving scalars ($s=0$) and decomposing them into real and imaginary parts. The coupling $\Re C^{0}_{ij}$ can be singular value decomposed using orthogonal rotations on the scalar and fermion indices to bring it to diagonal form
\begin{align}
\Re C^{0}_{ij}=\frac{m}{M_{Pl}}\begin{pmatrix}
\begin{matrix}
d_1 & & \\
& \ddots & \\
& & d_{\hat{n}}
\end{matrix} & \rvline & \bigzero \\
\hline
\bigzero & \rvline & \bigzero 
\end{pmatrix},
\end{align}
where $\hat{n}\leq\min\{n_0,n_{1/2}\}$ and I generally use $n_s$ to denote the number of particles with spin $s$. Here (\ref{C0-1/2}) implies that $d_i\leq 1/\sqrt{2}$ and I define $d_i>0$ (the special case $\Re C^{0}_{ij}=0$ will still be contained in the following argument). Importantly, the last two equations in (\ref{GravitinoSoftCons}) imply that $\Re C^{0}_{im}\Im C^{\frac{1}{2}}_{mj}=0$, which clearly then implies that $\Im C^{\frac{1}{2}}_{ij}=0$ for $i\leq \hat{n}$. The equations  (\ref{GravitinoSoftCons}) also imply that $\Re C^{\frac{1}{2}}_{im}\Im C^{\frac{1}{2}}_{jm}=2\Re C^{0}_{mi}\Im C^{0}_{mj}$. Using the fact that $\Im C^{\frac{1}{2}}_{jm}=0$ for $j\leq\hat{n}$ then means that $\Im C^{0}_{ij}=0$ for both $i,j\leq\hat{n}$. Further singular value decomposition allows for $\Im C^{0}_{ij}$ to be rotated to the form
\begin{align}\label{ImC0-1/2}
\Im C^{0}_{ij}=\frac{m}{M_{Pl}}\begin{pmatrix}
\bigzero & \rvline & \bf{V}\\
\hline
\bf{U} & \rvline & \bf{F}
\end{pmatrix},
\end{align}
where 
\begin{align}
\bf{F}=
\begin{pmatrix}
\begin{matrix}
f_1 & & \\
& \ddots & \\
& & f_{\hat{m}}
\end{matrix} & \rvline & \bigzero \\
\hline
\bigzero & \rvline & \bigzero 
\end{pmatrix}
\end{align}
for some $\hat{m}\leq\min\{n_{0}-\hat{n},n_{\frac{1}{2}}-\hat{n}\}$ and $\bf{U}$ and $\bf{V}$ are some yet to be determined matrices. Now, the equations in (\ref{GravitinoSoftCons}) imply orthogonality conditions on the rows and columns of $\Im C^{0}_{ij}$. The constraint (\ref{C0-1/2}) implies that the rows must all be orthogonal. Combining the first and second equations of (\ref{GravitinoSoftCons}) gives
\begin{align}\label{InterOGRel}
\left(\frac{m}{M_{Pl}}\right)^2\delta_{ij}=3\Im C^{\frac{1}{2}}_{im}\Im C^{\frac{1}{2}}_{jm}+4\Re C^{0}_{mi}\Re C^{0}_{mj}-2\Im C^{0}_{mi}\Im C^{0}_{mj}.
\end{align}
This implies that the columns of $\Im C^{0}_{ij}$ on the left side of the block partition in (\ref{ImC0-1/2}) must be orthogonal to those on the right side, given what has already been established about $\Im C^{\frac{1}{2}}_{jk}$ and $\Re C^{0}_{ij}$. As a result of these orthogonality results, the entries in (\ref{ImC0-1/2}) vertically above and horizontally left of the non-zero $f_i$ entries must be zero. Equation (\ref{C0-1/2}) therefore sets $f_i=1/\sqrt{2}$ for each $i$. Equation (\ref{InterOGRel}) then reduces to $2\left(\frac{m}{M_{Pl}}\right)^2\delta_{ij}=3\Im C^{\frac{1}{2}}_{im}\Im C^{\frac{1}{2}}_{jm}$ for $\hat{n}<i,j\leq\hat{m}+\hat{n}$. However, the last two equations in (\ref{GravitinoSoftCons}) imply that $\Im C^{0}_{im}\Re C^{\frac{1}{2}}_{mj}=0$ and $\Re C^{0}_{im}\Re C^{\frac{1}{2}}_{mj}=\Im C^{0}_{im}\Im C^{\frac{1}{2}}_{mj}$. Again for $\hat{n}<i,j\leq\hat{m}+\hat{n}$, these imply that $\Re C^{\frac{1}{2}}_{ik}=0$ (for any $k$) and therefore that $\Im C^{\frac{1}{2}}_{ik}=0$, which is a contradiction. This establishes that $\bf{F}=\bf{0}$.

Now, the equations (\ref{GravitinoSoftCons}) and (\ref{C0-1/2}) translate into constraints
\begin{align}
\mathbf{V}_{im}\mathbf{V}_{jm}&=\left(\frac{1}{2}-d_i^2\right)\delta_{ij}\nonumber\\
\mathbf{U}_{im}\mathbf{U}_{jm}&=\frac{1}{2}\delta_{ij}\nonumber\\
\mathbf{U}_{mi}\mathbf{U}_{mj}&=\left(2d_i^2-\frac{1}{2}\right)\delta_{ij}
\end{align}
(there is no sum over $i$). If one of the rows (say row $i$) of $\bf{V}$ is zero, then $d_i=1/\sqrt{2}$ and the corresponding column of $\bf{U}$ is non-zero with a normalisation of $1/\sqrt{2}$. Likewise, if column $i$ of $\bf{U}$ is zero, then $d_i=1/2$ and the corresponding row of $\bf{V}$ is non-zero and has a normalisation of $1/\sqrt{2}$. Since the rows of $\bf{V}$ and the columns of $\bf{U}$ are orthogonal or zero, each block can be respectively singular value decomposed using the residual freedom to apply orthogonal rotations to the remaining fermions and scalars beyond the first $\hat{n}$ ones. Performing this, the normalisation condition on the rows of $\bf{U}$ ensures that any non-zero entry is $1/\sqrt{2}$. The scalar-fermion couplings therefore have the general form:
\begin{align}\label{C0ij}
C^{0}&=\frac{1}{\sqrt{2}}\frac{m}{M_{Pl}}\begin{pmatrix}
\bigI & \rvline & \bigzero & \rvline & \bigzero  & \rvline & \bigzero\\
\hline
\bigzero & \rvline & \frac{1}{\sqrt{2}}\bigI & \rvline &  \frac{i}{\sqrt{2}}\bigI & \rvline & \bigzero\\
\hline
i\bigI & \rvline & \bigzero & \rvline & \bigzero & \rvline & \bigzero
\end{pmatrix}.
\end{align}

Under the gravitino coupling, the states are therefore clearly partitioned into disjoint sets consisting of either a pair of scalars with a single fermion, a pair of fermions with a single scalar, or a miscellaneous fermion. Each type of grouping possesses an identical coupling structure. I will call $\mathbf{n}_0$ the number of sets with a pair of scalars and a single fermion and $\mathbf{n}_{\frac{1}{2}}$ the number with a pair of fermions and a single scalar. Using the aforementioned condition $\Im C^{0}_{im}\Re C^{\frac{1}{2}}_{mj}=0$ implies that $\Re C^{\frac{1}{2}}_{ij}=0$ for $i\leq \mathbf{n}_0$ (and also for $\mathbf{n}_0+\mathbf{n}_{\frac{1}{2}}<i\leq \mathbf{n}_0+2\mathbf{n}_{\frac{1}{2}}$). Thus $C^{\frac{1}{2}}_{ij}=0$ for $i\leq \mathbf{n}_0$ (using the analogous result for the imaginary part established above). The fermions that have minimal couplings to a pair of scalars therefore do not share minimal couplings with vector bosons (or anything else). This establishes the chiral supermultiplet. A complex basis for the scalars in a chiral multiplet can be chosen (using (\ref{ConjBas}) below) in which each helicity of the gravitino couples to only a single state of the particle/anti-particle pair. 

The constraints derived on $C^{\frac{1}{2}}_{ij}$ thus far are that $C^{\frac{1}{2}}_{ij}=0$ for $i\leq \mathbf{n}_0$, $\Im C^{\frac{1}{2}}_{ij}=0$ for $\mathbf{n}_0<i\leq \mathbf{n}_0+\mathbf{n}_{\frac{1}{2}}$ and $\Re C^{\frac{1}{2}}_{ij}=0$ for $\mathbf{n}_0+\mathbf{n}_{\frac{1}{2}}<i\leq \mathbf{n}_0+2\mathbf{n}_{\frac{1}{2}}$. The other consistency conditions can be easily shown to give constraints on the orthogonality of the rows: $\Re C^{\frac{1}{2}}_{ij}=2\Im C^{0}_{im}\Im C^{\frac{1}{2}}_{mj}$ for $\mathbf{n}_0<i\leq \mathbf{n}_0+\mathbf{n}_{\frac{1}{2}}$ and $2\Im C^{\frac{1}{2}}_{im}\Im C^{\frac{1}{2}}_{jm}=\delta_{ij}$ for $\mathbf{n}_0+\mathbf{n}_{\frac{1}{2}}<i\leq \mathbf{n}_0+2\mathbf{n}_{\frac{1}{2}}$. The real and imaginary component matrices can therefore be simultaneously singular value decomposed in these index ranges to give couplings of the general form
\begin{align}\label{C1/2}
C^{\frac{1}{2}}&=\frac{m}{M_{Pl}}\begin{pmatrix}
\bigzero & \rvline & \bigzero \\
\hline
\frac{1}{\sqrt{2}}\bigI & \rvline & \bigzero \\
\hline
\frac{i}{\sqrt{2}}\bigI & \rvline & \bigzero \\
\hline
\bigzero & \rvline & C^{\frac{1}{2}}_{\text{rem}}
\end{pmatrix}.
\end{align}
The bottom right block, denoted as $C^{\frac{1}{2}}_{\text{rem}}$, describes the couplings of the previously miscellaneous fermions that do not have a minimal gravitino coupling with scalars. The fact that the bottom left block in (\ref{C1/2}) is zero has not yet been established. Demonstrating this becomes the next goal, along with showing that the vector bosons indexed by the columns to the left of the block partition do not have further minimal couplings to higher spin particles. This latter requirement follows from the relations $\Re C^{\frac{1}{2}}_{im}\Im C^{1}_{mj}=0$ and $\Im C^{\frac{1}{2}}_{im}\Re C^{1}_{mj}=0$ (which follow from the last two equations in (\ref{GravitinoSoftCons})), which imply that $C^{1}_{ij}=0$ for $i\leq\mathbf{n}_{\frac{1}{2}}$. The former requirement follows from 
\begin{align}
\left(\frac{m}{M_{Pl}}\right)^2\delta_{ij}=\frac{4}{3}\left(\Re C^{1}_{im}\Re C^{1}_{jm}+\Im C^{1}_{im}\Im C^{1}_{jm}\right)+\Re C^{\frac{1}{2}}_{mi}\Re C^{\frac{1}{2}}_{mj}+\Im C^{\frac{1}{2}}_{mi}\Im C^{\frac{1}{2}}_{mj},
\end{align}
which is contained in the first pair of relations in (\ref{GravitinoSoftCons}). For $i,j\leq\mathbf{n}_{\frac{1}{2}}$, the $C^{1}$ terms vanish, while the equation is fully satisfied by the second and third row blocks of $C^{\frac{1}{2}}$ in (\ref{C1/2}), therefore necessitating that the bottom left block vanishes. The gravitino therefore further subdivides each of the remaining scalars into independent sets grouped with two unique fermions and a single unique vector, thereby establishing the massive vector multiplets. 

The remaining fermions and vectors that have minimal gravitino couplings described by $C^{\frac{1}{2}}_{\text{rem}}$ exist in an entirely independent subspace to those with non-zero scalar couplings. The supermultiplet structure of these particles, and likewise that of even higher spin multiplets, can be deduced by repeating the argument just presented, beginning with a set of spin $s$ particles with no minimal gravitino couplings to lower spin particles. This is easier than the case beginning with scalars, as the first two equations in (\ref{GravitinoSoftCons}) with $C^{s-\frac{1}{2}}_{ij}=0$ and $s>0$ are more demanding than (\ref{C0-1/2}). In particular, the constraints degenerate to 
\begin{align}
\Re C^{s}_{im}\Re C^{s}_{jm}=\Im C^{s}_{im}\Im C^{s}_{jm}=\frac{2s+1}{2(2s+2)}\left(\frac{m}{M_{Pl}}\right)^2\delta_{ij},\qquad\qquad
\Re C^{s}_{im}\Im C^{s}_{jm}=0.
\end{align}
It directly follows that, in an appropriate basis, 
\begin{align}\label{Csij}
C^{s}&=\sqrt{\frac{2s+1}{2(2s+2)}}\frac{m}{M_{Pl}}\begin{pmatrix}
\bigI & \rvline & i\bigI & \rvline & \bigzero 
\end{pmatrix}.
\end{align}
The form of the couplings $C^{s+\frac{1}{2}}_{ij}$ is then constrained by the set of equations (\ref{GravitinoSoftCons}) analogously to (\ref{C1/2}) above to be
\begin{align}\label{Cs1/2ij}
C^{s+\frac{1}{2}}&=\frac{m}{M_{Pl}}\begin{pmatrix}
\frac{1}{\sqrt{2}}\bigI & \rvline & \bigzero \\
\hline
\frac{i}{\sqrt{2}}\bigI & \rvline & \bigzero \\
\hline
\bigzero & \rvline & C^{s+1/2}_{\text{rem}}
\end{pmatrix}.
\end{align}
The argument then repeats to determine $C^{s+\frac{1}{2}}_{\text{rem}}$ and establish the supermultiplet structure and couplings with half a spin unit higher. 

A general $\mathcal{N}=1$ massive supermultiplet $\mathcal{S}_s$ therefore consists of two spin $s$ particles (which may be chosen to be mutually conjugate pairs, in which case each couples to only a specific helicity of gravitino), a single spin $s-\frac{1}{2}$ particle and a single spin $s+\frac{1}{2}$ particle. Notably, the gravitino coupling is not quite universal, because the coupling to the lowest spin state in the multiplet is reduced by a factor of $\sqrt{\frac{2s}{2s+1}}$ relative to $\frac{m}{M_{Pl}}$. In Appendix \ref{OSsuperfields}, this same factor arises as a normalisation that accompanies this state in the standard representation of the multiplet as a fermionic coherent state (\ref{N=1GenSpinSM}). The general $\mathcal{N}=1$ SUGRA matter-gravity superamplitudes (generalising those presented in Appendix \ref{N=1SUGRA} like (\ref{N=1VecMatter}) for low spin matter) are given by 
\begin{align}\label{N=1MassiveSpinning}
\mathcal{A}(\mathcal{S}_{s,i},\mathcal{S}_{s,j},H^+)&=\frac{1}{M_{Pl}m^{2(s-1)}}\delta^{(2)}\left(Q^\dagger\right)\left(\delta_{ij}\frac{1}{x^2}\da{\bf{12}}^{2s}+\text{anomalous multipoles}\right)\nonumber\\
\mathcal{A}(\mathcal{S}_{s,i},\mathcal{S}_{s,j},H^-)&=\frac{1}{M_{Pl}m^{2(s-1)}}\delta^{(2)}\left(Q^\dagger\right)\left(\delta_{ij}x^2\ds{\bf{12}}^{2s}F_3+\text{anomalous multipoles}\right).
\end{align}
See \cite{Herderschee:2019ofc} and the subsequent Sections and Appendices below for further explanation of massive $\mathcal{N}=1$ superamplitudes (and \cite{elvang2015scattering} for massless ones). Appendices \ref{OSsuperfields} and \ref{N=1SUGRA} in particular contain relevant background for the specific discussion here. Here $F_3$ is a Grassmann structure introduced further below in Section \ref{sec:SimpleN=1}:
\begin{align}\label{FInvProto}
F_3=\eta_3+\frac{1}{2m_1}\ds{31^I}\eta_{1,I}+\frac{1}{2m_2}\ds{32^J}\eta_{2,J}.
\end{align}
I will defer further explanation to Section \ref{sec:SimpleN=1}. While the overall coupling of these superamplitudes is fixed by SUSY to be the universal gravitational constant (which, of course, is inherited from the presence of the graviton), the non-trivial normalisation factor of the lowest spin massive state must be removed by hand when extracting component amplitudes in which this particle appears as an external leg. 

As is well known (egs. \cite{Ema:2025qgd,Falkowski:2020aso}), the gravitational Compton amplitude for minimally coupled matter can be calculated with BCFW recursion, albeit with a spurious pole that must be eliminated by supplementary contact terms if the spin of the massive matter is $s\geq 2$. The analogous calculation can be performed with $\mathcal{N}=1$ superamplitudes using super-BCFW recursion to obtain the gravitational Compton superamplitude
\begin{align}
\mathcal{A}(\mathcal{S}_{s},H^+,H^-,\mathcal{S}_{s})&=\frac{1}{M_{Pl}^2}\frac{\la{3}p_1\rs{2}^{3-2s}\left(\da{3\bf{4}}\ds{2\bf{1}}+\da{3\bf{1}}\ds{2\bf{4}}\right)^{2s}}{(s-m^2)(t-m^2)u}\nonumber\\
&\qquad\qquad\qquad\qquad\times\delta^{(2)}\left(Q^\dagger\right)\left(\ds{21^I}\eta_{1I}-\ds{24^L}\eta_{4L}\right),
\end{align}
from which the component amplitudes can be extracted. The Compton superamplitude can be used to cross-check the direct construction of the gravitino Compton amplitudes used to derive (\ref{GravitinoSoftCons}) from individual $3$-particle component amplitudes. 

With the supermultiplet structure established with the (almost) universal gravitational coupling, the consistency of the soft gravitino limits directly implies the SWIs for the remaining amplitudes (either $3$-leg or higher). The soft gravitino factors analogous to (\ref{Softino}) above are, in this massive case, 
\begin{align}
A_3(\hat{S}_{r\{M\}}\rightarrow \hat{S}^{\{N\}}_{r'},\hat{\widetilde{\psi}}^+)=\frac{1}{M_{Pl}}\frac{1}{\epsilon z\da{Xs}}\ds{s\mathfrak{k}^{\{N\}}_{Sr'r\{M\}}},
\end{align}
where
\begin{align}
\rs{\mathfrak{k}^{\{N\}}_{Sr'r\{M\}}}=\begin{cases}
\rs{k_M}(\delta_M^N)^{2s},\qquad &r=s+\frac{1}{2},\,r'=s\\
\sqrt{\frac{2s}{2s+1}}\rs{k^N}(\delta_M^N)^{2s-1},\qquad &r=s-\frac{1}{2},\,r'=s\\
\rs{k^N}(\delta_M^N)^{2s},\qquad &r=\bar{s},\,r'=s+\frac{1}{2}\\
\sqrt{\frac{2s}{2s+1}}\rs{k_M}(\delta_M^N)^{2s-1},\qquad &r=\bar{s},\,r'=s-\frac{1}{2}.
\end{cases}
\end{align}
All massive $SU(2)$ spin indices (abbreviated here as $\{M\}$ and $\{N\}$ for the two hard particles) are implicitly symmetrised over, as usual. The massive multiplet $\mathcal{S}_s$ consists of particles $S_{s\pm\frac{1}{2}}$, $S_s$ and its conjugate partner $S_{\bar{s}}$. These states are indexed by the subscripts $r$ and $r'$ which can take on the values $\{s-\frac{1}{2},s,\bar{s},s+\frac{1}{2}\}$. The spinor is zero for choices of $r$ and $r'$ not included in the cases explicitly listed. The SWIs are then 
\begin{align}
\sum_k\rs{\mathfrak{k}^{\{N\}}_{Sr'r{\{M\}}}}\cdot A_n^{(k)}(1,2\ldots,k^{\{M\}}_{Sr},\ldots)=0.
\end{align}
Here, the hard legs in the amplitude are indexed by $k$ and $S_{r'}$ denotes the identity of leg $k$ as a state within a particular massive spinning supermultiplet. In each term in the sum, the identity of the indexed leg $k$ is replaced by a superpartner $S_{r}$ in the hard amplitude. All possible superpartners are summed over as part of the contraction, which also includes the appropriate spin sum. 

These SWIs apply to the full $3$-particle graviton and gravitino-matter amplitudes and force the higher order multipole couplings to be consistent with SUSY as well. Notably, the inconsistency of the anomalous gravitational dipole excludes the partner anomalous gravitino dipoles.

\subsubsection{Extended supergravity}\label{sec:ExtSUGRAder}

The structure of the $\mathcal{N}=2$ supermultiplets can be derived analogously. The substantially new feature of extended SUSY is the possibility of central charges. Having established $\mathcal{N}=1$ SUSY if a massless helicity-$3/2$ particle is present, $\mathcal{N}=1$ on-shell superfields can be used to describe the particle spectra of a theory if a second RS flavour is introduced. The $\mathcal{N}=2$ graviton multiplet, having already been established in Appendix \ref{sec:MasslessSUGRA}, decomposes into $\mathcal{N}=1$ graviton and gravitino multiplets. The standard representation of supermultiplets in $\mathcal{N}=1$ SUGRA as on-shell superfields (fermionic coherent states) and their minimal $3$-particle superamplitudes is summarised in Appendices \ref{OSsuperfields} and \ref{N=1SUGRA} using the conventions established in \cite{Herderschee:2019ofc}. The new gravitino multiplets are represented as on-shell superfields as
\begin{align}
\widetilde{\Psi}^+&=\widetilde{\psi}^++\eta\gamma^+\\
\widetilde{\Psi}^-&=\gamma^-+\eta\widetilde{\psi}^-.
\end{align}
The $\mathcal{N}=1$ gravitino $3$-particle superamplitudes with the graviton are 
\begin{equation}
\begin{split}
\mathcal{A}(\widetilde{\Psi}^+,\widetilde{\Psi}^-,H^-)=\frac{1}{M_{Pl}}\delta^{(2)}(Q^\dagger)\frac{\da{23}^4}{\da{13}\da{12}^2}
\end{split}
\qquad\begin{split}
\mathcal{A}(\widetilde{\Psi}^+,\widetilde{\Psi}^-,H^+)&=\frac{1}{M_{Pl}}\widetilde{\delta}^{(1)}(Q)\frac{\ds{13}^4}{\ds{12}^2\ds{23}}.
\end{split}
\end{equation}
This was effectively established by the analysis in Appendix \ref{sec:MasslessSUGRA}. 

The main point of difference between the gravitino-matter couplings studied above and their $\mathcal{N}=1$ gravitino-matter supermultiplet counterparts is the existence of two distinct SUSY structures incorporating minimal gravitino coupling to massive spinning matter:
\begin{align}\label{gravitinoAmps}
\mathcal{A}(\mathcal{S}_{s,i},\mathcal{S}_{s+\frac{1}{2},j},\widetilde{\Psi}^+)&=C^s_{ij}\,\delta^{(2)}\left(Q^\dagger\right)\frac{1}{m^{2s+1}x}\da{\bf{12}}^{2s}\ds{3\bf{2}}\nonumber\\
\mathcal{A}(\mathcal{S}_{s,i},\mathcal{S}_{s+\frac{1}{2},j},\widetilde{\Psi}^-)&=\left(C^s_{ij}\right)^*\delta^{(2)}\left(Q^\dagger\right)\frac{x}{m^{2s+1}}\ds{\bf{12}}^{2s}\da{3\bf{2}}F_3\\
\mathcal{A}(\mathcal{S}_{s,i},\mathcal{S}_{s,j},\widetilde{\Psi}^+)&=iD^s_{ij}\,\delta^{(2)}\left(Q^\dagger\right)\frac{1}{m^{2s}x}\da{\bf{12}}^{2s}F_3\nonumber\\
\mathcal{A}(\mathcal{S}_{s,i},\mathcal{S}_{s,j},\widetilde{\Psi}^-)&=-\left(iD^s_{ij}\right)^*\delta^{(2)}\left(Q^\dagger\right)\frac{x}{m^{2s}}\ds{\bf{12}}^{2s}
\end{align}
(I omit possible anomalous multipoles, which are not relevant for establishing the supermultiplet structure). The first pair of superamplitudes obviously parallel (\ref{SUSYmin}) and contains only graviphoton couplings beginning at dipole order. The second pair however contain charge monopoles. The couplings $D^s_{ij}=-D^s_{ji}$ by identical particle exchange symmetry and these are expected to interpretable as Abelian Lie algebra generators, as will be shown below. 

The holomorphic soft limit taken in (\ref{MasslessSoftShift}), (\ref{MassSpinShift}) and the resulting master factorisation formula (\ref{MasterSoft}) can be adapted to $\mathcal{N}=1$ superamplitudes. The soft supermultiplet has Grassmann variable $\eta_s$ and corresponding supercharge $\ra{\hat{s}}\eta_s$. This supercharge vanishes in the soft limit because $\ra{\hat{s}}\rightarrow 0$, leaving supercharge conservation among the hard legs in the SUSY version of (\ref{MasterSoft}). The variable $\eta_s$ may nevertheless appear in (\ref{MasterSoft}) as an independent Grassmann structure appearing in the soft superfactors. The recoil of the soft supercharge may be absorbed into shifts of the supercharges of the two hard legs $i$ and $j$, precisely analogous to the soft momenta. This is accounted for by shifting the Grassmann variables in an identical way to the left-handed spinor shifts implicit in (\ref{MasslessSoftShift}) and (\ref{MassSpinShift}), with the uncontracted factor of $\ls{s}$ replaced with $\eta_s$. With this soft supershift established, (\ref{MasterSoft}) generalises to superamplitudes in the obvious way, with a Grassmann integral over the exchanged particles' Grassmann variables to be included in each term.
 
\begin{figure}[h]
\begin{fmffile}{Gravitino4}

\begin{center}
\begin{tabular}{ c c c }
& & \\
 \begin{fmfgraph*}(100,67)
   \fmfleft{i1,i2}
   \fmfright{o1,o2}
   \fmf{dbl_plain}{i1,v1}
   \fmf{dbl_plain}{i1,v1}
   \fmf{dbl_plain}{v1,v2}
   \fmf{dbl_plain}{v1,v2}
   \fmf{dbl_plain}{v2,o1}
   \fmf{dbl_plain}{v2,o1}
   \fmf{plain}{i2,v1}
   \fmf{boson}{i2,v1}
   \fmf{plain}{o2,v2}
   \fmf{boson}{o2,v2} \fmfv{decor.shape=circle,decor.filled=gray50,decor.size=0.15w}{v1,v2}
   \fmflabel{$\widetilde{\Psi}^+$}{i2}
   \fmflabel{$\widetilde{\Psi}^-$}{o2}
   \fmflabel{$s_i$}{i1}
   \fmflabel{$s_j$}{o1} \fmfv{label=$s_m$,label.angle=-25,label.dist=0.15w}{v1}
 \end{fmfgraph*} 
 &\,& \begin{fmfgraph*}(100,67)
   \fmfleft{i1,i2}
   \fmfright{o1,o2}
   \fmf{dbl_plain}{i1,v1}
   \fmf{dbl_plain}{i1,v1}
   \fmf{dbl_plain}{v1,v2}
   \fmf{dbl_plain}{v1,v2}
   \fmf{dbl_plain}{v2,o1}
   \fmf{dbl_plain}{v2,o1}
   \fmf{phantom}{v1,i2}
   \fmf{phantom}{v1,i2}
   \fmf{phantom}{v2,o2}
   \fmf{phantom}{v2,o2}
   \fmf{plain,tension=-0.25}{v2,i2}
   \fmf{plain,tension=-0.25}{v1,o2}
   \fmf{boson,tension=-0.25}{v2,i2}
   \fmf{boson,tension=-0.25}{v1,o2}
   \fmfv{decor.shape=circle,decor.filled=gray50,decor.size=0.15w}{v1,v2}
   \fmflabel{$\widetilde{\Psi}^+$}{i2}
   \fmflabel{$\widetilde{\Psi}^-$}{o2}
   \fmflabel{$s_i$}{i1}
   \fmflabel{$s_j$}{o1}
   \fmfv{label=$s_m$,label.angle=-25,label.dist=0.2w}{v1}
 \end{fmfgraph*}\nonumber\\
 $(s)$ & \, &  $(t)$ 
\end{tabular}
\end{center}
\end{fmffile}
\caption{On-shell diagrams for $\mathcal{A}(\mathcal{S}_{s,i},\widetilde{\Psi}^+,\widetilde{\Psi}^-,\mathcal{S}_{s,j})$ involving central charges.}
\end{figure}
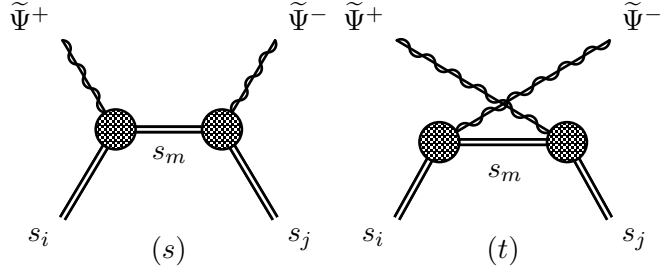

\begin{figure}[h]
\begin{fmffile}{Gravitino5}

\begin{center}
\begin{tabular}{ c c c }
& & \\
 \begin{fmfgraph*}(100,67)
   \fmfleft{i1,i2}
   \fmfright{o1,o2}
   \fmf{dbl_plain}{i1,v1}
   \fmf{dbl_plain}{i1,v1}
   \fmf{dbl_plain}{v1,v2}
   \fmf{dbl_plain}{v1,v2}
   \fmf{dbl_plain}{v2,o1}
   \fmf{dbl_plain}{v2,o1}
   \fmf{plain}{i2,v1}
   \fmf{boson}{i2,v1}
   \fmf{plain}{o2,v2}
   \fmf{boson}{o2,v2} \fmfv{decor.shape=circle,decor.filled=gray50,decor.size=0.15w}{v1,v2}
   \fmflabel{$\widetilde{\Psi}^+$}{i2}
   \fmflabel{$\widetilde{\Psi}^\pm$}{o2}
   \fmflabel{$s_i$}{i1}
   \fmflabel{$(s+\frac{1}{2})_j$}{o1} \fmfv{label=$s(+\frac{1}{2})_m$,label.angle=-35,label.dist=0.07w}{v1}
 \end{fmfgraph*} 
 &\,\,\,\,\,\qquad\qquad\qquad & \begin{fmfgraph*}(100,67)
   \fmfleft{i1,i2}
   \fmfright{o1,o2}
   \fmf{dbl_plain}{i1,v1}
   \fmf{dbl_plain}{i1,v1}
   \fmf{dbl_plain}{v1,v2}
   \fmf{dbl_plain}{v1,v2}
   \fmf{dbl_plain}{v2,o1}
   \fmf{dbl_plain}{v2,o1}
   \fmf{phantom}{v1,i2}
   \fmf{phantom}{v1,i2}
   \fmf{phantom}{v2,o2}
   \fmf{phantom}{v2,o2}
   \fmf{plain,tension=-0.25}{v2,i2}
   \fmf{plain,tension=-0.25}{v1,o2}
   \fmf{boson,tension=-0.25}{v2,i2}
   \fmf{boson,tension=-0.25}{v1,o2}
   \fmfv{decor.shape=circle,decor.filled=gray50,decor.size=0.15w}{v1,v2}
   \fmflabel{$\widetilde{\Psi}^+$}{i2}
   \fmflabel{$\widetilde{\Psi}^\pm$}{o2}
   \fmflabel{$s_i$}{i1}
   \fmflabel{$(s+\frac{1}{2})_j$}{o1}
   \fmfv{label=$s(+\frac{1}{2})_m$,label.angle=-25,label.dist=0.1w}{v1}
 \end{fmfgraph*}\nonumber\\
 $(s)$ & \,\,\,\,\,\qquad\qquad\qquad &  $(t)$ 
\end{tabular}
\end{center}
\end{fmffile}
\caption{On-shell diagrams for $\mathcal{A}(\mathcal{S}_{s,i},\widetilde{\Psi}^+,\widetilde{\Psi}^\pm,\mathcal{S}_{s+\frac{1}{2},j})$ involving central charges.}
\end{figure}
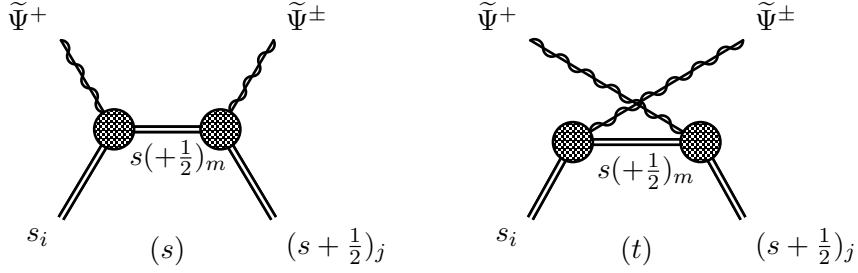
By taking soft $\mathcal{N}=1$ gravitino limits and demanding cancellation of inconsistent terms, the conditions (\ref{GravitinoSoftCons}) are reproduced with the following modification to the elastic, opposite-sign Compton results: 
\begin{align}\label{N=2ModCons}
\left(\frac{m}{M_{Pl}}\right)^2\delta_{ij}&=C^{s}_{im}(C^{s}_{jm})^*+\frac{1}{2s+1}C^{s}_{jm}(C^{s}_{im})^*+C^{s-\frac{1}{2}}_{mj}(C^{s-\frac{1}{2}}_{mi})^*\nonumber\\
&\qquad\qquad\qquad\qquad\qquad\qquad\qquad+\frac{1}{2}\left((D^s_{im})^*D^s_{jm}+D^s_{im}(D^s_{jm})^*\right)\nonumber\\
&=C^{s}_{jm}(C^{s}_{im})^*+\frac{1}{2s+1}C^{s}_{im}(C^{s}_{jm})^*+C^{s-\frac{1}{2}}_{mi}(C^{s-\frac{1}{2}}_{mj})^*\nonumber\\
&\qquad\qquad\qquad\qquad\qquad\qquad\qquad+\frac{1}{2}\left((D^s_{im})^*D^s_{jm}+D^s_{im}(D^s_{jm})^*\right).
\end{align}
The new term arises because $\mathcal{A}(\mathcal{S}_{s,i},\widetilde{\Psi}^+,\widetilde{\Psi}^-,\mathcal{S}_{s,j})$ admits $s$ and $t$-channel exchanges of $\mathcal{S}_{s}$ multiplets in addition to $\mathcal{S}_{s\pm\frac{1}{2}}$. This process additionally requires that
\begin{align}
D_{im}^s(D_{jm}^s)^*=(D_{im}^s)^*D_{jm}^s,
\end{align}
which is consistent with the expectations of photon couplings described in \cite{Trott:2026cjj} (in particular, that they commute with their adjoint). Consistency of mixed Compton superamplitudes involving the gravitino mutliplet and a regular vector multiplet likewise establishes that the couplings $D^s$ and their complex conjugates commute with the vector-matter couplings $(t_A)$ and their conjugates (of course, consistent Compton scattering of gluons off massive matter demands that the $(t_A)$ couplings be Lie algebra generators \cite{Benincasa:2007xk,Trott:2026cjj}, regardless of supersymmetry). 

Compton superamplitudes in which the spin of the massive legs change by $1/2$ are also now possible. Consistency of the superamplitudes $\mathcal{A}(\mathcal{S}_{s,i},\widetilde{\Psi}^+,\widetilde{\Psi}^\pm,\mathcal{S}_{s+\frac{1}{2},j})$ leads to the conditions
\begin{align}\label{ExtraN=2Constraint}
D^s_{im}C^{s}_{mj}&=C^{s}_{im}D^{s+\frac{1}{2}}_{mj}\nonumber\\
(D^s_{im})^*C^{s}_{mj}&=C^{s}_{im}(D^{s+\frac{1}{2}}_{mj})^*.
\end{align}
In deriving these results, it is convenient to make use of the fact that the $F_3$ factors in (\ref{gravitinoAmps}) can always be represented without the Grassmann variable of the exchanged particle, provided that it is massive, by using on-shell supercharge conservation. This trivialises the Grassmann integrals for the $s$ and $t$-channels. Evaluated on the residues of each channel, the $F$ factors are also not affected by the shifts. Supercharge conservation of the hard legs then helps to reduce the number of different Grassmann structures appearing in the supersymmetrised master equation (\ref{MasterSoft}) to four: one of degree $2$ (which is just $\delta^{(2)}\left(Q^\dagger\right)$), one of degree $4$ and two of degree $3$, the latter consisting of one Grassmann structure with $\eta_2$ (which here corresponds to the soft particle) and one without. 

Rearranging equation (\ref{N=2ModCons}) gives 
\begin{align}
&\left(\frac{m}{M_{Pl}}\right)^2\delta_{ij}-\frac{1}{2}\left((D^s_{im})^*D^s_{jm}+D^s_{im}(D^s_{jm})^*\right)\nonumber\\
&\qquad\qquad=\frac{2s+2}{2s+1}\left(\Re C^{s}_{im}\Re C^{s}_{jm}+\Im C^{s}_{im}\Im C^{s}_{jm}\right)+\Re C^{s-\frac{1}{2}}_{mi}\Re C^{s-\frac{1}{2}}_{mj}+\Im C^{s-\frac{1}{2}}_{mi}\Im C^{s-\frac{1}{2}}_{mj},
\end{align}
for which the right-hand side is clearly positive semi-definite (as before, the last two terms are to be dropped for the special $s=0$ case). The charge matrices $D^s$ can be orthogonally rotated to normal form:
\begin{align}\label{CCcoupling}
D^s=\begin{pmatrix}
0 && -(q_1+ip_1) && && &&\\
q_1+ip_1 && 0 && && &&\\
&& && 0 && -(q_2+ip_2) &&\\
&& && q_2+ip_2 && 0 &&\\
&& && && && \ddots
\end{pmatrix},
\end{align}
which diagonalises $(D^s_{im})^*D^s_{jm}=D^s_{im}(D^s_{jm})^*=\text{diag}(|z_1|^2,|z_1|^2,|z_2|^2,|z_2|^2,\dots)_{ij}$, where $z_i=q_i+ip_i$ (a further basis change into charge eigenstates is, of course, also possible, but I will not bother to do that here). This directly implies the BPS bound:
\begin{align}
|z_i|\leq \frac{m}{M_{Pl}}.
\end{align}
If saturated for some $i$, then $C^{s}_{ij}=C^{s-\frac{1}{2}}_{ki}=0$ for any $j,k$ in this basis. These charge conjugate pairs of $\mathcal{N}=1$ multiplets therefore do not change identity under emission of a gravitino and therefore constitute closed $\mathcal{N}=2$ representations. These are the short or BPS multiplets. The charges $z_i$ are evidently expected to be the central charges normalised to the Planck mass. 

In the discussion of perturbatively consistent theories of photons in \cite{Trott:2026cjj}, consistency of Moller scattering ruled-out magnetic charges and it was generally possible to find a basis in which the Abelian generators were purely real (up to an overall factor of $i$). In the present context of $\mathcal{N}=1$ SUSY, this argument envelopes the $D^s$ couplings in (\ref{gravitinoAmps}) along with regular couplings to massless vector multiplets identified as Abelian generators. Define the charge vector for particle $i$ as 
\begin{align}
\mathfrak{q}_i=({q_0}_i+i{p_0}_i,{q_1}_i+i{p_1}_i,{q_2}_i+i{p_2}_i\ldots,{q_n}_i+i{p_n}_i),
\end{align}
where ${q_A}_i+i{p_A}_i$ are its charges under photon species $A$ when the couplings are expressed as matrices with the normal form in (\ref{CCcoupling}). I furthermore choose to identify the $A=0$ component with the specific coupling to the gravitino multiplet in (\ref{CCcoupling}). So as long as the charge vectors (extended to include those in (\ref{CCcoupling})) obey 
\begin{align}\label{ElectricConsist}
\sum_A({q_A}_i+i{p_A}_i)({q_A}_j-i{p_A}_j)
&=\sum_A({q_A}_i-i{p_A}_i)({q_A}_j+i{p_A}_j)\nonumber\\
\Leftrightarrow \qquad\mathfrak{q}_i\cdot\mathfrak{q}_j&=\mathfrak{q}_i\cdot\mathfrak{q}_j
\end{align}
(using the standard definition of the complex dot product) for any $i$ and $j$, there are no problems with Moller scattering (mediated by exchange of photon or gravitino multiplets) and no magnetic charges. Usually the basis choice of photon states is chosen so that one of the charged particles has a purely real charge with respect to a single photon state. Then (\ref{ElectricConsist}) would imply that the imaginary parts of the charges of every other particle must vanish. This reflects the inconsistency of magnetic charges with perturbative amplitudes (see \cite{Csaki:2020inw} for a recent reformulation of the $S$-matrix to accommodate magnetic charges). The point of note here though is that supersymmetry already identifies a preferred basis of photon states by explicitly linking one of them to the gravitino. As a result, the imaginary parts of (\ref{CCcoupling}) cannot, in general, be rotated away without ruining the manifest $\mathcal{N}=1$ SUSY. Both real and imaginary parts of (\ref{CCcoupling}) may therefore potentially represent electric charges and, if they both exist, then a second photon must also exist in some other vector multiplet in order to restore consistency to the Moller (super)amplitudes. Explicit demonstrations of this will be given below in Sections \ref{sec:LowSpinGrav} and \ref{sec:BPSGrav}. The fact that the basis of photons with manifest electric couplings may not be a basis in which the states occupy separate SUSY multiplets reflects part of the off-shell issue of electric-magnetic duality frames. 

More generally, the equations (\ref{GravitinoSoftCons}), modified by (\ref{N=2ModCons}), can be solved similarly to the single gravitino case before, with (\ref{ExtraN=2Constraint}) providing the additional information that fixes the central charge matrices along the way. Ignoring BPS particles, the couplings $C^s_{ij}$ have almost the same form as (\ref{C0ij}), (\ref{C1/2}), (\ref{Csij}), and (\ref{Cs1/2ij}), differing only in the values of their non-zero entries, which will be stated later in this paragraph. The central charge matrices consist of diagonal blocks of the form (\ref{CCcoupling}) that are consistent with the block structure of the $C^s_{ij}$ couplings. In particular, the spin $s$ states that label the rows of (\ref{Csij}) have a corresponding diagonal block in their $D^s_{ij}$ couplings given by (\ref{CCcoupling}). Then the spin $s+\frac{1}{2}$ states that couple to them in (\ref{Csij}) have charge couplings $D^{s+\frac{1}{2}}=D^s\oplus D^s$, while the spin $s+1$ states that couple to these through (\ref{Cs1/2ij}) have charge couplings $D^{s+1}=D^s$. In other words, all states belonging to the same $\mathcal{N}=2$ multiplet have the same central charge. Then, with non-zero charges, the non-zero entries in (\ref{Csij}) are multiplied by new factors of $\sqrt{1-|z_i|^2}$ (for rows $2i-1$ and $2i$, since the charged states come in pairs), as are the corresponding entries in the columns of (\ref{Cs1/2ij}):
\begin{align}
C^{s}&=\sqrt{\frac{2s+1}{2(2s+2)}}\frac{m}{M_{Pl}}\begin{pmatrix}
\sqrt{1-(D^s)^2} & \rvline & i\,\sqrt{1-(D^s)^2} & \rvline & \bigzero 
\end{pmatrix}\\
C^{s+\frac{1}{2}}&=\frac{m}{M_{Pl}}\begin{pmatrix}
\frac{1}{\sqrt{2}}\sqrt{1-(D^s)^2} & \rvline & \bigzero \\
\hline
\frac{i}{\sqrt{2}}\sqrt{1-(D^s)^2} & \rvline & \bigzero \\
\hline
\bigzero & \rvline & C^{s+1/2}_{\text{rem}}
\end{pmatrix},
\end{align}
where 
\begin{align}
\sqrt{1-(D^s)^2}=\begin{pmatrix}
\sqrt{1-|z_1|^2} &&  && && &&\\
 && \sqrt{1-|z_1|^2} && && &&\\
&& && \sqrt{1-|z_2|^2} &&  &&\\
&& &&  && \sqrt{1-|z_2|^2} &&\\
&& && && && \ddots
\end{pmatrix}
\end{align}
(the appropriate adaptation of these results also applies to the spin $0$ case). In summary, it has been shown that non-BPS (or long) $\mathcal{N}=2$ multiplets consist of quartets of $\mathcal{N}=1$ multiplets with analogous structure to the assembly of single particles into $\mathcal{N}=1$ multiplets. The difference here is the possibility of central charges, which introduces an additional mixing between states across the $\mathcal{N}=1$ submultiplets with respect to the second gravitino. 

It is worth noting that not all of the graviphoton's central charge couplings contained in the latter two equations in (\ref{gravitinoAmps}) are minimal and this is a necessary consequence of SUSY. Assuming only a single charge (so $D^s$ is real), then in the supermultiplet $\mathcal{S}_s$, the states $S_s$ and $S_{\bar{s}}$ have minimal coupling, but $S_{s+\frac{1}{2}}$ has the typical ``graviphoton-type'' coupling structure described below in (\ref{graviphoton3pt}), while the lowest spin state $S_{s-\frac{1}{2}}$ couples as
\begin{align}
A\left(S_{s-\frac{1}{2}},\overline{S}_{s-\frac{1}{2}},\gamma^+\right)&=-\frac{1}{m^{2s-2}x}\da{\bf{12}}^{2s-1}-\frac{2s-1}{2s+1}\frac{1}{m^{2s-1}}\da{\bf{12}}^{2s-2}\ds{3\bf{1}}\ds{3\bf{2}}\nonumber\\
A\left(S_{s-\frac{1}{2}},\overline{S}_{s-\frac{1}{2}},\gamma^-\right)&=-\frac{x}{m^{2s-2}}\ds{\bf{12}}^{2s-1}-\frac{2s-1}{2s+1}\frac{1}{m^{2s-1}}\ds{\bf{12}}^{2s-2}\da{3\bf{1}}\da{3\bf{2}}
\end{align}
(dropping the overall factor of the charge).

It remains to be shown that the couplings to the second gravitino are consistent with gravitational universality. This is done simply by finding how the states in the $\mathcal{N}=2$ multiplet arrange into $\mathcal{N}=1$ submultiplets under the second gravitino and showing that they have the expected universal gravitational coupling (up to the spin normalisation factor described around (\ref{N=1MassiveSpinning})). To do this, I will begin by slightly modifying the meaning of the notation employed in this Section. For the remainder of this Section, $S_x$ will be used to denote the spin $x$ state of the $\mathcal{N}=1$ multiplet $\mathcal{S}$, where now $S$ and $\mathcal{S}$ specifically denote the spin number of the multiplet (which I define as the intermediate spin level of the four states in the multiplet) instead of being generic symbols for ``massive spinning particle''. Then the multiplet $\mathcal{S}$ contains states $\{S_s,S_{s+\frac{1}{2}},S_{s-\frac{1}{2}},S_{\bar{s}}\}$, where $S_s$ and $S_{\bar{s}}$ couple respectively only to $\widetilde{\psi}^+$ and $\widetilde{\psi}^-$ in the $\mathcal{N}=1$ graviton multiplet. Working in a complex basis of central charge eigenstates, each long $\mathcal{N}=2$ multiplet with central charges $z$ (which I will normalise to the BPS limit $m/M_{Pl}$ in the following list) consists of four $\mathcal{N}=1$ multiplets: $\{\mathcal{S},\left(\mathcal{S}+\frac{1}{2}\right),\left(\mathcal{S}-\frac{1}{2}\right),\widetilde{\mathcal{S}}\}$, where $\mathcal{S}$ and $\widetilde{\mathcal{S}}$ couple respectively to only $\widetilde{\Psi}^+$ and $\widetilde{\Psi}^-$. Then with respect to the second gravitino, the states rearrange into $\mathcal{N}=1$ multiplets with the composition listed here:
\begin{align}
\left(\mathcal{S}+\frac{1}{2}\right)':\qquad &\left(S+\frac{1}{2}\right)_{s+1}\nonumber\\
&\sqrt{1-|z|^2}\widetilde{S}_{s+\frac{1}{2}}+z\left(S+\frac{1}{2}\right)_{s+\frac{1}{2}}\nonumber\\
&\sqrt{1-|z|^2}S_{s+\frac{1}{2}}+z\left(S+\frac{1}{2}\right)_{\overline{s+\frac{1}{2}}}\nonumber\\
&\frac{(2s+2)|z|^2-1}{2s+1}\left(S+\frac{1}{2}\right)_{s}-(1-|z|^2)\frac{\sqrt{2s(2s+2)}}{2s+1}\left(S-\frac{1}{2}\right)_{s}\nonumber\\
&\qquad\qquad\qquad\qquad\qquad\qquad\qquad+z^*\sqrt{1-|z|^2}\sqrt{\frac{2s+2}{2s+1}}\left(\widetilde{S}_{\bar{s}}-S_s\right)
\nonumber\\
\mathcal{S}':\qquad & -\sqrt{1-|z|^2}\left(S+\frac{1}{2}\right)_{\overline{s+\frac{1}{2}}}+z^*S_{s+\frac{1}{2}}\nonumber\\
&z\sqrt{1-|z|^2}\sqrt{\frac{1}{2s+1}}\left(\sqrt{2s+2}\left(S+\frac{1}{2}\right)_{s}+\sqrt{2s}\left(S-\frac{1}{2}\right)_{s}\right)\nonumber\\
&\qquad\qquad\qquad\qquad\qquad\qquad\qquad+|z|^2S_s+(1-|z|^2)\widetilde{S}_{\bar{s}}\nonumber\\
&S_{\bar{s}}\nonumber\\
&-\sqrt{1-|z|^2}\left(S-\frac{1}{2}\right)_{\overline{s-\frac{1}{2}}}+z^*S_{s-\frac{1}{2}}\nonumber\\
\widetilde{\mathcal{S}}':\qquad &-\sqrt{1-|z|^2}\widetilde{\left(S+\frac{1}{2}\right)}_{s+\frac{1}{2}}+z^*\widetilde{S}_{s+\frac{1}{2}}\nonumber\\
&-z\sqrt{1-|z|^2}\sqrt{\frac{1}{2s+1}}\left(\sqrt{2s+2}\left(S+\frac{1}{2}\right)_{s}+\sqrt{2s}\left(S-\frac{1}{2}\right)_{s}\right)\nonumber\\
&\qquad\qquad\qquad\qquad\qquad\qquad\qquad+(1-|z|^2)S_s+|z|^2\widetilde{S}_{\bar{s}}\nonumber\\
&\widetilde{S}_{s}\nonumber\\
&-\sqrt{1-|z|^2}\widetilde{\left(S-\frac{1}{2}\right)}_{s-\frac{1}{2}}+z^*\widetilde{S}_{s-\frac{1}{2}}\nonumber\\
\left(\mathcal{S}-\frac{1}{2}\right)':\qquad &-(1-|z|^2)\frac{\sqrt{2s(2s+2)}}{2s+1}\left(S+\frac{1}{2}\right)_{s}+
\frac{2s|z|^2+1}{2s+1}\left(S-\frac{1}{2}\right)_{s}\nonumber\\
&\qquad\qquad\qquad\qquad\qquad\qquad\qquad+z^*\sqrt{1-|z|^2}\sqrt{\frac{2s}{2s+1}}\left(\widetilde{S}_{\bar{s}}-S_s\right)
\nonumber\\
&\sqrt{1-|z|^2}\widetilde{S}_{s-\frac{1}{2}}+z\left(S-\frac{1}{2}\right)_{s-\frac{1}{2}}\nonumber\\
&\sqrt{1-|z|^2}S_{s-\frac{1}{2}}+z\left(S-\frac{1}{2}\right)_{\overline{s-\frac{1}{2}}}\nonumber\\
&\left(S-\frac{1}{2}\right)_{s-1}.
\end{align}
These states have the expected coupling structure to the second gravitino (i.e. agree with the couplings derived in Section \ref{sec:ExtSUGRAder}). In the BPS limit ($|z|\rightarrow 1^-$), these converge to the original $\mathcal{N}=1$ multiplets. When $z=0$, each multiplet instead consists of a single component state from each of the original $\mathcal{N}=1$ multiplets, although some mixing between $\left(S-\frac{1}{2}\right)_{s}$ and $\left(S+\frac{1}{2}\right)_{s}$ remains. 

With the supermultiplet structure and their gravitational couplings established, the soft limit consistency tests enforce the $\mathcal{N}=2$ SWIs on the theory. Higher $\mathcal{N}$ SUGRA can be derived likewise by introducing more flavours of gravitinos, but I will move on from this.

\subsection{Goldstini}

In flat spacetime, massless fermions described by the Volkov-Akulov (VA) effective action \cite{Volkov:1972jx} are characterised by analogous soft properties as Goldstone bosons. Amplitudes of both Goldstone bosons $\varphi$ and VA fermions $\chi$ obey universal soft limits of the form $A(\varphi/\chi\ldots)\sim \epsilon $ when the momentum of the soft leg scales to zero with $\epsilon\rightarrow 0$ \cite{Elvang:2018dco}. In particular, the $4$-particle amplitude for the VA theory is 
\begin{align}
A(\chi^+,\chi^-,\chi^+,\chi^-)=\frac{t}{\Lambda^4}\ds{13}\da{24},
\end{align}
where $\Lambda$ represents a dimensionful coupling constant. Through the soft bootstrap, this subsequently generates the higher leg amplitudes. However, when the fermion(s) coupling to gravity is introduced, the graviton exchange contribution to the amplitude is 
\begin{align}
A(\chi^+,\chi^-,\chi^+,\chi^-)=\frac{t^2}{M_{Pl}^2su}\ds{13}\da{24}
\end{align}
This violates the soft scaling property for VA fermions. Gravity spoils the VA theory's shift symmetry. This is analogous to the way in which gauging an unbroken part of the symmetry manifold of a non-linear sigma model (NL$\Sigma$M) hard breaks the full, spontaneously broken global symmetry and introduces violations to the Goldstone bosons' shift invariance through their couplings to the associated vector bosons. The argument just presented demonstrates how exact VA fermions cannot exist in a theory with gravity. They appear in field theory through the Goldstino equivalence of the longitudinal polarisations of massive gravitinos, which is not gravitational at leading order. The Goldstini equivalent to massive gravitinos correspond to the spontaneous breaking of ``gauged'' supersymmetries, but Goldstini cannot otherwise exist as artifacts of spontaneous breaking of some other exact symmetry. This is the on-shell version of the qualitative part of the argument presented \cite{Cheung:2010mc}.

\section{Special 3-Particle Massive Kinematics}\label{3PSK}

A formulation of on-shell massive $3$-particle amplitudes that included cases kinematically inaccessible in Minkowski space was presented in \cite{Arkani-Hamed:2017jhn}. It was shown that the prevalent case of amplitudes between two equal mass and one massless particle exhibited special locality features that constrain the consistency of the theories in which they arise, some of the consequences of which were elucidated in the previous Section. However, the general case of three massive particles was viewed as unremarkable. 

There is nevertheless a massive analogue of the celebrated $3$-leg special complex kinematics of massless amplitudes that applies whenever the particles have a conserved complex mass, of which the two equal mass and one massless amplitude is a special degenerate case. These kinematics can be viewed as a dimensional reduction of massless scattering in $6d$ \cite{Cheung:2009dc} (see also \cite{Ni:2024yrr} for some other recent comments), where the masses are reinterpreted as momenta along the extra dimensions. Such a scenario applies to BPS particles in theories of extended supersymmetry, where the particles' masses are proportional to their central charges which are conserved in scattering processes. When there is more than one central charge, the masses are instead lengths of vectors on a charge lattice. The sum of these vectors must add to zero. In a single $3$-particle amplitude, conservation of such charges implies that there are, at most, two such independent dimensions and hence they can be described by complex numbers. More generally, it is possible for a theory to be interpretable as having a conserved complex mass without having extended SUSY. QED, in which the electron and positron have ``opposite-sign masses'', is such an example (and others may exist in which the masses of the interacting particles merely obey the triangle inequality and decays are kinematically prohibited). Note that, henceforth, I will refer to the ``mass'' of a particle in its standard sense, both verbally and symbolically, even in the context of a theory that admits a conserved complexified mass interpretation. 

In theories with extended supersymmetry, this special kinematics is an essential feature of BPS states that provides a loop-hole crucial  for the consistency of $3$-leg interactions, as will be shown below in Section \ref{SUSYAlgebra}. This underpins the special properties of superamplitudes of BPS particles that leads to self-consistency restrictions on their possible interactions. In the following, I work through the derivation of the special $3$-particle kinematics in detail excessive to the applications needed for much of this study. A briefer summary of the key conclusions and useful results of this Section is given in \cite{Trott:2026cjj}, which the reader may prefer to consult if they want to quickly get to the point. 

So assume that three particles comprising an on-shell scattering amplitude have conserved, generally complex masses. 
The case of real conserved masses represents the special instance in which it is possible for the amplitude can be non-zero for real momentum. This occurs precisely for the purely collinear limit - that is, the heaviest state splits into the two lighter ones entirely at rest. However, if the masses are only conserved when complexified in some sense, then they obey the triangle indequality and the amplitude can only be non-zero on-shell if the momenta are also complex. Conservation of complexified mass implies that
\begin{align}\label{MassCons}
\sum_i m_i e^{i\varphi_i}=0,
\end{align}
for some phases $\varphi_i$, of which one choice is arbitrary (as stated above, I will always use $m_i$ to refer to the regular definition of mass - the ``complexified mass'' is $m_ie^{i\varphi_i}$). It follows from the mass shell condition and momentum conservation that, for any of the three particles $i$ and $j$,
\begin{align}\label{Phase3PSK}
\cos(\varphi_i-\varphi_j)&=-\frac{p_i\cdot p_j}{m_im_j}.
\end{align}

Now, just as chiral helicity spinors may be defined with a manifest massive little group representation \cite{Arkani-Hamed:2017jhn}, the standard Dirac spinors for free particles may likewise be constructed. Incorporating the phases from (\ref{MassCons}), these are
\begin{equation}\label{DiracSpinors}
\begin{split}
u_i^I=
\begin{pmatrix}
\rs{i^I}\\
e^{-i\varphi_i}\ra{i^I}
\end{pmatrix}
\end{split}\qquad
\begin{split}
v_i^I=
\begin{pmatrix}
e^{-i\varphi_i}\rs{i^I}\\
-\ra{i^I}
\end{pmatrix}
\end{split}
\end{equation}
The Dirac conjugates are 
\begin{align}
\bar{u}_I=
\begin{pmatrix}
e^{i\varphi_i}\ls{i_I} && -\la{i_I}
\end{pmatrix}\qquad \bar{v}_I=
\begin{pmatrix}
\ls{i_I} && e^{i\varphi_i}\la{i_I}
\end{pmatrix}.
\end{align}
Generally, setting $\varphi_i=0$ gives the standard $SU(2)$ little group covariant form of the Dirac spinors as defined in textbooks e.g. \cite{Srednicki:2007qs} (in this reference, the translation to little group covariance is $u_s\mapsto u_I$ and $v_s\mapsto v^I$ - note that, more precisely, $\bar{u}^I$ is the conjugate of $u_I$ and the spinors in the expressions above are obtained by then raising/lowering the $SU(2)$ spin index). The phases allow the spinors (\ref{DiracSpinors}) to continuously interpolate between the particle and anti-particle solutions to the free Dirac equation. The usual relations between the Dirac spinors are modified in some cases. For example, $\gamma_5 u^\varphi_{i,I}=-e^{-i\varphi_i}v_{i,I}^{-\varphi}$, where the spinors have been explicitly labeled by the mass phases that they depend upon. This holds for both real and complex momentum. Notably, this expression relates the particle and anti-particle spinors to each other: $u^{\varphi=\pi}=v^{\varphi=0}$ and $v^{\varphi=\pi}=-u^{\varphi=0}$.

The spin sums become 
\begin{align}\label{DiracSpinSums}
    u_I(p)\bar{u}^I(p)=-p^\mu\gamma_\mu + \mathbf{m}\qquad v^I(p)\bar{v}_I(p)=-p^\mu\gamma_\mu - \mathbf{m}^*,
\end{align}
where
\begin{align}
    \mathbf{m}=m\begin{pmatrix}
e^{i\varphi}I && 0\\
0 && e^{-i\varphi}I
\end{pmatrix}
\end{align}
are its $2\times 2$ block components. Henceforth, Dirac spinor index structure will be denoted as $X_A=(X_\alpha,X^{\dot{\beta}})$ and $\bar{Y}^A=(Y^\alpha,Y_{\dot{\beta}})$ where applicable, for any $(\mathbf{\frac{1}{2}},\mathbf{0})\oplus(\mathbf{0},\mathbf{\frac{1}{2}})$ spinors $X$ and conjugates $\bar{Y}$. 

Select two external particles $i$ and $j$ of a general scattering amplitude. The scalar bilinears of their associated Dirac spinors are 
\begin{align}\label{SpinFact}
\overline{v}_i^Iu_j^J&=-e^{i(\varphi_i-\varphi_j)}\overline{u}_j^Jv_i^I=\da{i^Ij^J}e^{i(\varphi_i-\varphi_j)}+\ds{i^Ij^J}.
\end{align}
These bilinears are themselves matrices of the particles' little groups. For general real momenta, they have non-zero, degenerate singular values. However, a complex deformation of the momenta breaks the degeneracy. If the deformation is chosen so that one of the singular values is sent to zero, then the spinor bilinears degenerate into rank-$1$ matrices in their little group indices and can therefore be expressed as a factorised pair of pure little group spinors: 
\begin{align}\label{SpecialSVD}
\overline{v}_i^Iu_j^J&=\mathbf{v}_{j|i}^I\mathbf{u}_{i|j}^J.
\end{align}
Here, $\mathbf{v}_{j|i}^I$ and $\mathbf{u}_{i|j}^J$ are spinors of the little groups under which they are explicitly indexed, but have no other suppressed tensor indices. They are defined through (\ref{SpecialSVD}) analogously to the way that $\rs{i}$ and $\ra{i}$ can be defined through $p_i=-\rs{i}\la{i}$. Likewise, they also carry a $GL(1,\mathbb{C})$ rescaling ambiguity under which $\mathbf{v}_{j|i}^I\rightarrow \lambda\mathbf{v}_{j|i}^I$ and $\mathbf{u}_{i|j}^J\rightarrow \lambda^{-1} \mathbf{u}_{i|j}^J$ for any $\lambda\in\mathbb{C}\backslash\{0\}$. The condition required for the factorisation (\ref{SpecialSVD}) to happen, equivalent to $\left(\overline{v}_i^Iu_j^J\right)\left(\overline{v}_{iI}u_{jJ}\right)=0$, is that the momenta satisfy (\ref{Phase3PSK}), which is the statement that the sum of momenta $p_i+p_j$ satisfies a mass shell condition $\left(p_i+p_j\right)^2=-|m_ie^{i\varphi_i}+m_je^{i\varphi_j}|^2$. This therefore applies to the kinematical case of an on-shell $3$-leg amplitude in which there is a conserved complex mass. 

So for a complex mass-conserving $3$-particle amplitude, there exist six pairs of little group spinors defined through (\ref{SpecialSVD}) (corresponding to each choice of two of the three legs). Each pair of little group spinors $\mathbf{u}_{i|j}^J$ and $\mathbf{v}_{j|i}^I$ have a $GL(1,\mathbb{C})_{i,j}$ projective rescaling redundancy. This is the complexification of the $U(1)_{i,j}$ ``tiny group'' - the Lorentz stabiliser of a pair of momenta \cite{Boels:2012ie} (the subscript $\{i,j\}$ labeling the group denotes the leg pairing). 

The identities (\ref{SpecialSVD}) imply that 
\begin{align*}
    \mathbf{v}_{j|i}^I&=\alpha_{jk}\mathbf{v}_{k|i}^I\qquad&
    \mathbf{u}_{j|i}^I&=\beta_{jk}\mathbf{u}_{k|i}^I\\
    e^{i\varphi_j}\mathbf{u}_{i|j}^J\rs{j_J}&=\frac{-1}{\alpha_{jk}}e^{i\varphi_k}\mathbf{u}_{i|k}^K\rs{k_K}\qquad& e^{-i\varphi_j}\mathbf{v}_{i|j}^J\rs{j_J}&=\frac{-1}{\beta_{jk}}e^{-i\varphi_k}\mathbf{v}_{i|k}^K\rs{k_K}\\
    \mathbf{u}_{i|j}^J\ra{j_J}&=\frac{-1}{\alpha_{jk}}\mathbf{u}_{i|k}^K\ra{k_K}\qquad& \mathbf{v}_{i|j}^J\ra{j_J}&=\frac{-1}{\beta_{jk}}\mathbf{v}_{i|k}^K\ra{k_K},\addtocounter{equation}{1}\tag{\theequation}\label{3PSKMassive}
\end{align*}
where $(i,j,k)$ are chosen to be cyclic orderings of $(1,2,3)$. The variables $\alpha_{jk}$ and $\beta_{jk}$ each have tiny group charges necessary for consistency of the above relations. They must also satisfy the further constraint 
\begin{align}
    \alpha_{12}\alpha_{23}\alpha_{31}=-\beta_{12}\beta_{23}\beta_{31}
\end{align}
for self-consistency.


The tiny group scaling ambiguity in the little group spinors represents a redundancy that the scattering amplitude must be unaffected by. If the little group spinors are used as tensorial building blocks out of which the amplitude is constructed, then they can only appear in tiny group neutral combinations. This constrains the possible functional dependence of the amplitude. The appearance of the tiny group is suggestive of the recently constructed dyon $S$-matrix proposed in \cite{Csaki:2020inw}. I will make some comments about this in Appendix \ref{Dyons}, but most of this study will be focused on the traditional analytic $S$-matrix describing the scattering of purely electric particles.

Since each of the six copies of the tiny group is isomorphic, their indices are annoying and largely unnecessary. To reduce clutter, I will fix the scalings so that $\alpha_{ij}=\alpha_{i+1j+1}=\alpha$, $\beta_{ij}=\beta_{i+1j+1}=-\alpha$ (for each pair of legs $i$ and $j=i+1$ mod $3$) and define 
$\mathbf{u}_i^I=\mathbf{u}_{i+1|i}^I$ and $\mathbf{v}_i^I=\mathbf{v}_{i-1|i}^I$ (again for each $i$ mod $3$). The spinors $\mathbf{u}_i^I$ and $\mathbf{v}_i^I$ now have opposite charges under a single tiny group. The remaining spinors $\mathbf{u}_{i-1|i}^I$ and $\mathbf{v}_{i+1|i}^I$ are redundant. Since a consistent amplitude must be tiny group invariant, they can only appear in a way that always allows them to be converted into the spinors $\mathbf{u}_i^I$ and $\mathbf{v}_i^I$ through the equations in the first line of (\ref{3PSKMassive}). To reiterate, none of the results presented in this study require these fixings, but they lead to much clearer presentation. Only in Appendix \ref{Dyons} will these choices be reconsidered.

Thus, the spinors $\mathbf{u}_i^I$ and $\mathbf{v}_i^I$ are an adequate set of amplitude building blocks and this partial fixing of redundancies is sufficient to preserve restrictions on the structure of (super)amplitudes from tiny group invariance. The defining relation (\ref{SpecialSVD}) can be rewritten as
\begin{align}\label{SpecialSVD2}
\overline{v}_i^Iu_j^J=\mathbf{v}_{i}^I\mathbf{u}_{j}^J
\end{align}
when $i=j+1$ (mod $3$) (and it can be easily shown as well that $\overline{v}_i^Iu_j^J=-\mathbf{v}_{i}^I\mathbf{u}_{j}^J$ when $i=j-1$ (mod $3$)). The relations (\ref{3PSKMassive}) can likewise be rewritten as
\begin{align}\label{Trimmed3PSK}
\begin{split}
\rs{u}&=e^{i\varphi_i}\mathbf{u}_i^I\rs{i_I}=e^{i\varphi_j}\mathbf{u}_j^J\rs{j_J}\\
\ra{u}&=\mathbf{u}_i^I\ra{i_I}=\mathbf{u}_j^J\ra{j_J}
\end{split}
\begin{split}
\rs{v}&=e^{-i\varphi_i}\mathbf{v}_i^I\rs{i_I}=e^{-i\varphi_j}\mathbf{v}_j^J\rs{j_J}\\
\ra{v}&=\mathbf{v}_i^I\ra{i_I}=\mathbf{v}_j^J\ra{j_J}
\end{split}
\end{align}
for any legs $i,j$. These relations define special spacetime spinors $\rs{u}$, $\ra{u}$, $\rs{v}$ and $\ra{v}$. 
The Dirac spinors 
\begin{align}
\mathbf{u}_{i}^Iu_{iI}&=e^{-i\varphi_i}\begin{pmatrix}
\rs{u}\\
\ra{u}
\end{pmatrix}=e^{-i(\varphi_i-\varphi_j)}\mathbf{u}_{j}^Ju_{jJ}
\\
\mathbf{v}_{i}^Iv_{iI}&=
\begin{pmatrix}
\rs{v}\\
-\ra{v}
\end{pmatrix}=\mathbf{v}_{j}^Jv_{jJ}
\end{align}
also align. 

The equations above in (\ref{Trimmed3PSK}) demonstrate that the spacetime spinors projected in the direction of the special little group spinors all align.
This is the massive adaptation of the  well-known special $3$-particle kinematics for massless particles. As is frequently done for $6d$ amplitudes, supplementary (or reference) little group basis spinors $\mathbf{w}_{iI}$ and $\widetilde{\mathbf{w}}_{iI}$ may be introduced, each satisfying $\mathbf{w}_{iI}\mathbf{u}_{i}^I=1$ and $\widetilde{\mathbf{w}}_{iI}\mathbf{v}_{i}^I=1$ (note that $\mathbf{u}_{i}^I$ and $\mathbf{v}_{i}^I$ have mass dimension $1/2$, so the reference spinors must have dimension $-1/2$). Then using
\begin{align}
    \delta^I_J=\mathbf{u}_{i}^I\mathbf{w}_{iJ}-\mathbf{u}_{iJ}\mathbf{w}_{i}^I
\end{align}
(and likewise for the $\mathbf{v}_{i}^I$ spinors), each external massive momentum may be decomposed as 
\begin{align}
p_i=\ra{i^w}\ls{u}e^{-i\varphi_i}-\ra{u}\ls{i^w},
\end{align}
where $\ra{i^w}=\mathbf{w}_{iI}\ra{i^I}$, $\ls{i^w}=\mathbf{w}_{iI}\ls{i^I}$. 
Each of the massive momenta may therefore be decomposed into a sum of two null vectors, the first of which all have aligned left-handed spinor factors and the second all have aligned right-handed spinors. This generalises the massless kinematical configurations in which either the left-handed or right-handed spinors all align. An analogous repetition of these statements also holds with the $\mathbf{u}_{i}^I$ spinors replaced with $\mathbf{v}_{i}^I$. 

The case where the phases $\varphi_i=\varphi_j\, (\text{mod}\,\pi)$ for each $i,j$ will be loosely referred to as the ``real mass case''. This kinematically corresponds to the configuration where mass is conserved and a heavier particle is splitting into two lighter ones. Technically, the bilinears $\overline{v}_i^Iu_j^J=-\overline{v}_j^Ju_i^I$ degenerate and the little group spinors $\mathbf{u}_{i}^I$ and $\mathbf{v}_{i}^I$ become linearly dependent and define the same little group direction: $\mathbf{u}_i^I=y\mathbf{v}_i^I$ for some constant of proportionality $y$. The relations (\ref{Trimmed3PSK}) imply that the same constant of proportionality $y$ relates $\mathbf{u}_i^I$ to $\mathbf{v}_i^I$ for all of the legs $i$. This constant bears some resemblance to the spacetime spinorial analogue $x$ (\ref{Prelimx}), where here $y=\frac{\da{uq}}{\da{vq}}$ for some reference spinor $\ra{q}$ satisfying $\da{vq}\neq 0$. It is little group neutral but carries tiny group charge. However, since scattering (super)amplitudes must always be tiny group invariant, it is practically redundant as a building block. Parenthetically, in general, 
\begin{align}
\mathbf{u}_i^I\mathbf{v}_{iI}=-2im_{i+1}\sin\left(\varphi_i-\varphi_{i+1}\right),
\end{align}
where, as before, the particle labels are mod $3$. 

While giving the three particles a common complex mass phase is geometrically redundant in the description of the $3$-particle kinematics in isolation, it can be necessary for the purposes of computing $4$-particle amplitudes below. For this reason, even in this case I will keep the $\mathbf{u}_{i}^I$ and $\mathbf{v}_{i}^I$ spinors distinct and they will still obey distinct relations with conjugate contributions from the mass phase like (\ref{Trimmed3PSK}). Note that the partial fixing of the tiny groups and definitions made in the paragraphs preceding (\ref{Trimmed3PSK}) still continue to carry through unaffected. 

The real mass case includes the important special configuration in which one of the particles is massless (leaving the other two with equal mass $m$), which pervades the fundamental interactions that generate the $S$-matrices of both gauge theories (with matter) and gravity. Labeling the massless leg particle $3$, its $SU(2)$ little group indices freeze into invariant $U(1)$ helicity indices. Its little group spinors likewise freeze into Lorentz scalars with opposite massless $U(1)$ little group charges: $\mathbf{u}_{3}^\pm$ and $\mathbf{v}_{3}^\pm$. These can combine to form the tiny group neutral, Lorentz scalar variable
\begin{align}\label{xExp}
x=-e^{-i\varphi}\mathbf{u}_{3}^-/\mathbf{u}_{3}^+=-e^{i\varphi}\mathbf{v}_{3}^-/\mathbf{v}_{3}^+=\frac{\ls{q}p_1\ra{3}}{m\rs{q3}}
\end{align}
introduced in \cite{Arkani-Hamed:2017jhn} as a vital ingredient for the construction of amplitudes involving two equal mass particles and one massless particle. The traditional definition of this object was given above in (\ref{Prelimx}). It has been assumed that $\varphi_1=\varphi$, $\varphi_2=\varphi+\pi$ (and the ambiguous $\varphi_3$ is chosen to be zero). Again, as stated above, this kinematics is ``real mass'' in the sense that $\varphi$ can be set to zero from the purely geometric standpoint of (\ref{MassCons}), but there are applications below involving a fourth particle against which such a phase can be compared, so it is appropriate not to commit to this. Additionally, using (\ref{Trimmed3PSK}) and (\ref{SpecialSVD2}), the products of massless little group spinors are
\begin{align}\label{xmass}
\mathbf{v}_{3}^+\mathbf{u}_{3}^-&=me^{i\varphi}\nonumber\\
\mathbf{v}_{3}^-\mathbf{u}_{3}^+&=me^{-i\varphi}.
\end{align}




\section{Extended Supersymmetry and BPS states}\label{SUSYAlgebra}

The central charges of the SUSY algebra are conserved in scattering processes (since they source the massless graviphoton superpartners of the graviton). For a BPS particle, the central charges are effectively a vector with norm given by the particle's mass. The scattering of BPS particles in extended SUSY theories with two active central charges therefore involves the special complex massive kinematics described in Section \ref{3PSK}. This Section is dedicated to examining this interrelation, establishing the formalism for describing scattering superamplitudes of BPS particles and beginning to elucidate the consequences of self-consistency from $S$-matrix (super)factorisation.

The structure of the central charge lattice depends upon the theory and the amount of supersymmetry. I will predominantly focus here upon $\mathcal{N}=4$ SYM as it provides an ideal reference point for demonstrating the mechanics of the superalgebra and the special kinematical properties of the superamplitudes. This theory's $S$-matrix is essentially constructed from the minimal SUSY invariants and superspace ingredients that may be subsequently adapted to theories with more complicated kinematic structures, either with more spin (supergravity) or fewer constraints (less supersymmetry). Once the $\mathcal{N}=4$ superspace and SUSY invariants have been established, I will comment on how this descends to $\mathcal{N}=2$, providing the foundation for the calculations presented in the subsequent Sections. Further details relevant to $\mathcal{N}=2$ theories will be introduced in the contexts in which they are required. Superamplitudes with central charges have been discussed previously in \cite{Caron-Huot:2018ape,Chen:2021hjl,Chen:2021huj}. Scattering in $\mathcal{N}=4$ SYM with a single active central charge was specifically studied in \cite{Herderschee:2019dmc}, but I will be more general here.

\subsection{$\mathcal{N}=4$ on-shell superspace}\label{sec:N=4SS}

The SUSY algebra represented on a single particle state labeled by $i$ is given by \cite{Herderschee:2019ofc}
\begin{align}\label{SUSYAlg}
\{q_{i,a}^I,q_i^{\dagger bJ}\}=-\epsilon^{IJ}\delta^b_a\qquad\{q_{i,a}^I,q^{J}_{i,b}\}=-\epsilon^{IJ}\frac{Z_{i,ab}}{2m_i}\qquad\{q^{\dagger aI}_i,q_i^{\dagger bJ}\}=\epsilon^{IJ}\frac{Z_i^{ab}}{2m_i},
\end{align}
where the spinor-stripped supercharges $q_{i,a}^I$ and $q^{\dagger aI}_i$ are related to the regular supercharges \cite{Srednicki:2007qs} through $Q_{i,a}=-\rs{i_I}q^{I}_{i,a}$ and $Q^{\dagger a}_i=-q^{\dagger aI}_i\la{i_I}$. The labels $a$ and $b$ denote $R$-index. The central charges $Z_{ab}$ are defined conventionally. These definitions of the supercharges differ from the conventional presentation of the SUSY algebra \cite{Srednicki:2007qs} by a factor of $1/\sqrt{2}$.

The BPS bound for a particle can be derived from the supersymmetry algebra using the fact that
\begin{align}
\left(q_{i,Ia}+\frac{1}{2m_i}Z_{i,am}q^{\dagger m}_{i,I}\right)\left(q_{i,Ia}+\frac{1}{2m_i}Z_{i,an}q^{\dagger n}_{i,I}\right)^\dagger
\end{align}
is positive semi-definite. This translates into the constraint
\begin{align}
\delta_a^b-\left(\frac{1}{2m_i}\right)^2Z_{i,am}(Z_{i,bm})^*\succeq 0.
\end{align}
The eigenvalues of $Z_{i,am}(Z_{i,bm})^*$ consist of two independent pairs. The state is $\frac{1}{2}$BPS when both values are $(2m_i)^2$, while it is $\frac{1}{4}$BPS when only one of the eigenvalues saturate the bound. 

A particle's $\mathcal{N}=4$ central charge matrix can be decomposed into $SU(4)_R$ self-dual and anti-self-dual components. Generally, these can be rotated by $SU(4)_R$ transformations to the form:
\begin{align}\label{CentralCharges1/4BPSN=4}
Z_{i,ab}=q_i\begin{bmatrix}
 0 &  0 & -e^{i\varphi_i} & 0 \\ 
 0 &  0 & 0 & -e^{-i\varphi_i} \\ 
 e^{i\varphi_i} &  0 & 0 & 0 \\ 
 0 & e^{-i\varphi_i} & 0 & 0 \\ 
\end{bmatrix}
+p_i\begin{bmatrix}
 0 &  0 & -e^{i\phi_i} & 0 \\ 
 0 &  0 & 0 & e^{-i\phi_i} \\ 
 e^{i\phi_i} &  0 & 0 & 0 \\ 
 0 & -e^{-i\phi_i} & 0 & 0 \\ 
\end{bmatrix}.
\end{align}
Here, $q_i$, $p_i$, $\varphi_i$ and $\phi_i$ are free parameters. The eigenvalues of $Z_{i,am}(Z_{i,bm})^*$ are $q_i^2+p_i^2\pm2q_ip_i\cos(\varphi_i-\phi_i)$. Without loss of generality, I will take $\cos(\varphi_i-\phi_i)\geq 0$. Clearly, $\frac{1}{2}$BPS states require either $q_i=0$, $p_i=0$ or $\varphi_i-\phi_i=\frac{(2n+1)\pi}{2}$ for $n\in\mathbb{Z}$ (that is, the anti-self-dual charge vectors are parallel to the self-dual charge vectors) \cite{Osborn:1979tq}. For the state to be $\frac{1}{4}$BPS, then $q_i^2+p_i^2+2q_ip_i\cos(\varphi_i-\phi_i)=(2m_i)^2$ and the conditions just stated for the multiplet to be $\frac{1}{2}$BPS cannot also hold. This implies that at least two of the anti-self-dual (under the definitions made here) or two of the self-dual central charges are active. I will leave further analysis of $\frac{1}{4}$BPS representations to Section \ref{sec:1/4BPS} and concentrate here on the $\frac{1}{2}$BPS case. 

With two active charges, the central charge matrix for $\frac{1}{2}$BPS states has the general form
\begin{align}\label{1/2BPSCC}
Z_{i,ab}=2m_ie^{i\theta_i}\begin{bmatrix}
 0 &  0 & -e^{i\varphi_i} & 0 \\ 
 0 &  0 & 0 & -e^{-i\varphi_i} \\ 
 e^{i\varphi_i} &  0 & 0 & 0 \\ 
 0 & e^{-i\varphi_i} & 0 & 0 \\ 
\end{bmatrix}.
\end{align}
For a single particle, the central charge matrix can always be $U(4)_R$-rotated into the purely symplectic form $Z_{i,ab}=2m_i\Omega_{ab}$ (in this Section, I will not define the central charges to be normalised to $M_{Pl}$, as was done earlier). $3$- and $4$-leg amplitudes of $\frac{1}{2}$BPS (including massless) particles will be of primary interest here. Within a $3$-particle amplitude, the combination of central charge conservation and the $\frac{1}{2}$BPS condition together for each leg implies that the charge vectors can always be rotated either to the form 
\begin{align}\label{CentralChargesEleN=4}
Z_{i,ab}=2m_i\begin{bmatrix}
 0 &  0 & -e^{i\varphi_i} & 0 \\ 
 0 &  0 & 0 & -e^{-i\varphi_i} \\ 
 e^{i\varphi_i} &  0 & 0 & 0 \\ 
 0 & e^{-i\varphi_i} & 0 & 0 \\ 
\end{bmatrix}
\end{align}
or 
\begin{align}\label{CentralChargesMagN=4}
Z_{i,ab}=2m_ie^{i\theta_i}\Omega_{ab},
\end{align}
where $\Omega_{ab}$ is a $4\times 4$ symplectic form (so either $\theta_i=0$ or $\varphi_i=0$ in (\ref{1/2BPSCC})). It will be argued below that the former case represents purely electric particles with up to two active charges, while the latter case typically describes dyonic particles with only one type of electric and magnetic charge. In this form, the $R$-symmetry is clearly explicitly broken to either $USp(4)$ (if $\theta_i,\varphi_i=0$) or $SU(2)\times SU(2)$ (otherwise). Conservation of central charge translates into conservation of complexified mass described in Section \ref{3PSK} above. The angles $\varphi_i$ or $\theta_i$ are identified with the mass phases described above e.g. 
\begin{align}
\sum_im_ie^{i\varphi_i}=\sum_im_ie^{-i\varphi_i}=0.
\end{align}
The existence of only a single set of active angles in a configuration of three particles is consistent with there being no further available degrees of freedom in the massive special $3$-particle kinematics beyond those associated with the mass triangle. 

I will represent external particles as fermionic coherent states, or, in other words, use the conventional on-shell superspace. This follows from the fact that the SUSY algebra (\ref{SUSYAlg}) can be converted into the form of a set of decoupled fermionic harmonic oscillators. This is already clear in the absence of central charges. In the more general case, the ladder operators are given by
\begin{align}\label{GenLadder}
\overline{q}^I_{i,a}=\frac{1}{\sqrt{D}}\left(q^I_{i,a}+\left(\frac{2m_i}{|z_i|}\right)^2\left(1-\sqrt{1-\left(\frac{|z_i|}{2m_i}\right)^2}\right)\frac{Z_{i,ab}}{2m_i}q^{\dagger b I}_{i}\right),
\end{align}
where
\begin{align}
D=2\left(\left(\frac{2m_i}{|z_i|}\right)^2-1\right)\left(1-\sqrt{1-\left(\frac{|z_i|}{2m_i}\right)^2}\right),
\end{align}
and conjugates. Here, $z_i$ are the singular values of the central charge matrix and the $R$-indices of the supercharges in this expression are restricted to the corresponding subspaces. The general relation between ladder operators and SUSY charges (\ref{GenLadder}) will be needed for describing hypothetical $\frac{1}{4}$BPS particles in Section \ref{sec:1/4BPS}, but will otherwise be made little use of in this study. The $\frac{1}{2}$BPS limit, of most relevance here, will be described next. 

The conventions and definitions for the massive on-shell superspaces (that is, representations of the SUSY algebra on single particle subspaces as fermionic coherent states) that will be utilised here were established in \cite{Herderschee:2019ofc} and \cite{Herderschee:2019dmc} and will not be restated. I will primarily focus on $\frac{1}{2}$BPS particles with two purely anti-self-dual central charges of the form (\ref{CentralChargesEleN=4}) and make some comments later about the other configuration (\ref{CentralChargesMagN=4}). Generally, the BPS condition can be stated as 
\begin{align}\label{BPS}
p^{\dot{\alpha}\alpha}_iQ_{i,\alpha,a}=\frac{1}{2}Z_{i,ab}Q^{\dagger,b\dot{\alpha}}_i,
\end{align}
or, equivalently, 
\begin{align}\label{BPSladder}
q_{i,a}^I=-\frac{1}{2m_i}Z_{i,ab}q_{i}^{\dagger bI}
\end{align}
under which a pair of supercharges clearly degenerates (this incidentally accounts for the otherwise pathological BPS limit of (\ref{GenLadder})). In the $\frac{1}{2}$BPS case, pursued here, representations of the supersymmetry algebra can then be chosen so that half of the supercharges are purely multiplicative in the Grassmann variables and the others are purely derivative. The total supercharges of all the legs in this representation are
\begin{align}\label{supercharges}
\begin{split}
Q^{\dagger a}&=-\sum_{i}\ra{i^I}\eta_{i,I}^a\\
Q_{a+2}&=\sum_{i}\rs{i^I}e^{\pm i\varphi_i}\eta_{i,I}^a
\end{split}
\qquad\begin{split}
Q_a&=\sum_{i}\rs{i_I}\frac{\partial}{\partial\eta_{i,I}^a}\\
Q^{\dagger a+2}&=\sum_{i}\ra{i_I}e^{\mp i\varphi_i}\frac{\partial}{\partial\eta_{i,I}^a}.
\end{split}
\end{align}
The BPS conditions relate the supercharges in the top line to those in the bottom line. These expressions manifests a $U(1)\times U(1)$ subgroup of the $SU(2)\times SU(2)$ unbroken $R$-symmetry. The $R$-index $a$ on the Grassmann variable $\eta_{i,I}^a$ determines the $U(1)$ factor that it is charged under (where now $a\in\{1,2\}$ is no longer the full $U(4)_R$ index). Note that the upper symbol in the $\pm$ and $\mp$ signs corresponds to $a=1$ and the lower symbol to $a=2$. The ``real mass limit'', discussed in \cite{Herderschee:2019dmc}, in which only a single central charge is active and the $R$-symmetry is restored to $USp(4)$, corresponds to the phases $\varphi_i\rightarrow 0,\pi$ (depending on the sign of the central charge). In this case, the supercharges combine into doublets of a manifest $U(2)$ subgroup. In the massless limit, this representation corresponds to the non-chiral superspace \cite{Huang:2011um}, where the Clifford vacuum for the fermionic coherent states is chosen to be an intermediate helicity state, rather than the highest helicity state in the multiplet, which has been the more common choice in studies of massless $\mathcal{N}=4$ SYM. For ease of reference, the supercharges for massless particles are represented as
\begin{align}\label{superchargesMassless}
\begin{split}
Q^{\dagger a}&=\sum_{i}\ra{i}\eta_{i}^a\\
\frac{1}{\sqrt{2}}Q_{a+2}&=\sum_{i}\rs{i}\overline{\eta}_{i}^a
\end{split}
\qquad\begin{split}
Q_a&=\sum_{i}\rs{i}\frac{\partial}{\partial\eta_{i}^a}\\
Q^{\dagger a+2}&=\sum_{i}\ra{i}\frac{\partial}{\partial\overline{\eta}_{i}^a}.
\end{split}
\end{align}
Here, $\eta_i^a$ and $\overline{\eta}_i^a$ are two independent sets of Grassmann variables parameterising the massless coherent states. Obviously, both massless and massive particles can be present in general and the two sets of equations (\ref{supercharges}) and (\ref{superchargesMassless}) should be added together.

There is tension between parity (if it is present) and the inherently chiral nature of the massive on-shell superspace. Dirac spinors are the more natural building blocks of the amplitudes for non-chiral theories, but a choice of chirality needs to be made in selecting which of the supercharges to represent as raising and lowering operators across the supermultiplet. In the present case of $\mathcal{N}=4$ SUSY, this can be compensated for by the breaking of the $R$-symmetry - supercharges with different $R$-indices can be paired into Dirac spinors. These are arguably natural building blocks for the superamplitude. Since manifest $U(4)_R$ invariance is already broken, nothing is lost by grouping different supercharges into Dirac spinors:
\begin{align}\label{N=4Charges}
\begin{split}
\mathcal{Q}^1&=\begin{pmatrix}
Q_3\\
Q^{\dagger 1}
\end{pmatrix}=\sum_iv^{-\varphi,I}_{i}\eta^{1}_{i,I}\\
\mathcal{Q}^2&=\begin{pmatrix}
Q_4\\
Q^{\dagger 2}
\end{pmatrix}=\sum_i v^{\varphi,I}_{i}\eta^{2}_{i,I}
\end{split}\qquad
\begin{split}
\mathcal{C}\overline{\mathcal{Q}}_{1}^T&=\begin{pmatrix}
Q_1\\
Q^{\dagger 3}
\end{pmatrix}=\sum_i u_{i,I}^\varphi\frac{\partial}{\partial\eta^1_{i,I}}\\
\mathcal{C}\overline{\mathcal{Q}}_{2}^T&=\begin{pmatrix}
Q_2\\
Q^{\dagger 4}
\end{pmatrix}=\sum_i u_{i,I}^{-\varphi}\frac{\partial}{\partial\eta^2_{i,I}}.
\end{split}
\end{align}
Here $\mathcal{C}$ is the charge conjugation matrix \cite{Srednicki:2007qs}. This grouping also helps manifest the interpretation of the BPS amplitudes as amplitudes of massless particles in $6d$ SYM and SUGRA \cite{Dennen:2009vk,Dennen:2010dh,Bern:2010qa}, where the Dirac spinors of $4d$ correspond to chiral spinors in $6d$. 


\subsection{3-particle superamplitude and SUSY delta functions}\label{Sec:N=4SYM}

I will begin by assuming that central charge matrices are of the purely anti-self-dual form (\ref{CentralChargesEleN=4}). For three particles, the expressions (\ref{supercharges}) for the supercharges are then fully general (in that a basis in which they have this form can always be found). In this case, the BPS condition (\ref{BPS}) combines with the $3$-particle special kinematics (\ref{Trimmed3PSK}) to ensure that the supercharges projected onto the special spinors degenerate: 
\begin{align}\label{SuperDegen}
\da{v Q^{\dagger 1}}&=\ds{v Q_3}\nonumber\\ \da{u Q^{\dagger 2}}&=\ds{u Q_4},
\end{align}
leaving only a single independent supercharge per pair. This is crucial for the existence of the $3$-particle superamplitudes given below in theories of BPS particles. Ordinarily, with $\mathcal{N}=4$ SUSY, the supersymmetric delta function has Grassmann degree $8$($=2\mathcal{N}$). A Grassmann degree $4n-8$ SUSY invariant can also be obtained by Grassmann Fourier transforming the SUSY delta function in the opposite chirality superspace, where $n$ is the number of external legs. For $n=3$, this is smaller than $8$, which is a contradiction. 

The special relation (\ref{SuperDegen}) for $3$-legs is the loophole. This allows for the existence of a Grassmann degree $6$ SUSY invariant from which the superamplitude can be constructed. The complete invariant delta function is given by a product of two SUSY invariants that I define as:
\begin{align}\label{SuperBPSDelta}
\Delta_v&=\frac{1}{\ds{q_vv}\da{q_vv}}\delta\left(\ds{q_vQ_3}\right)\delta\left(\da{q_vQ^{\dagger 1}}\right)\delta\left(\ds{vQ_3}\right)\nonumber\\
\Delta_u&=\frac{1}{\ds{q_uu}\da{q_uu}}\delta\left(\ds{q_uQ_4}\right)\delta\left(\da{q_uQ^{\dagger 2}}\right)\delta\left(\ds{uQ_4}\right)
\end{align}
(I will also refer to each of these as SUSY delta functions as well). The prefactors are necessary for the expressions to be independent of the reference spinors $\rs{q_v}$, $\ra{q_v}$, $\rs{q_u}$ and $\ra{q_u}$, which are defined to satisfy $\ds{q_vv},\da{q_vv},\ds{q_uu},\da{q_uu}\neq 0$ but are otherwise free (see \cite{Herderschee:2019dmc} for further discussion). Note that the individual delta functions are just defined as $\delta\left(\ds{sQ}\right)=\ds{sQ}$, for any spinor $\ls{s}$ and supercharge $\rs{Q}$ (and likewise for the right-handed counterparts). Notably, the two SUSY invariant factors $\Delta_v$ and $\Delta_u$ have opposite tiny group scaling. 
An alternative expression using the Dirac form of the supercharges is 
\begin{align}
\Delta_v&=\frac{1}{\ds{q_vv}\da{q_vv}}\epsilon\left(q_v,\mathcal{Q}^1,\mathcal{Q}^1,\mathcal{Q}^1\right)\nonumber\\
\Delta_u&=\frac{1}{\ds{q_uu}\da{q_uu}}\epsilon\left(q_u,\mathcal{Q}^2,\mathcal{Q}^2,\mathcal{Q}^2\right),
\end{align}
where $q_v$ and $q_u$ denote Dirac spinors assembled out of the corresponding chiral reference spinors and the Levi-Civita contraction of Dirac spinors is defined as 
\begin{align}
\epsilon(X,Y,Z,W)=\epsilon^{ABCD}X_AY_BZ_CW_D.
\end{align}
This latter form of the supercharges parallels the expressions proposed in $6d$ SYM in \cite{Dennen:2009vk}.


For three $\frac{1}{2}$BPS vector multiplets, each of which are represented as scalars in this superspace (see (\ref{LongMultNtwo})), the superamplitude is simply equal to the product of delta functions and is given by
\begin{align}\label{N=43vec}
\mathcal{A}(\mathcal{W}_A,\mathcal{W}_B,\mathcal{W}_C)=-if_{ABC}\Delta_u\Delta_v,
\end{align}
where an internal flavour index has been assigned to each vector and $f_{ABC}$ are the coupling constants. Bosonic exchange symmetry implies that these constants are fully antisymmetric under index exchanges (since $\Delta_u\Delta_v$ is antisymmetric under exchanges - something that careful attention to the definitions of Section \ref{3PSK} is required to show), while unitarity implies that 
\begin{align}\label{SYMUnitary}
f_{\bar{A}\bar{B}\bar{C}}=(f_{ABC})^*,
\end{align}
where $\bar{A}$ indicates the antiparticle of the vector of flavour $A$ (which must have opposite central charges). If the vector is massless, this may be self-conjugate ($\bar{A}=A$). This superamplitude is both little group and tiny group neutral, as required. The component amplitudes comprising (\ref{N=43vec}) all describe kinematic structures compatible with standard ``spontaneously broken'' Yang-Mills interactions. These are entirely fixed by supersymmetry.

\subsection{Superfactorisation and 4-particle superamplitudes}\label{N=4SuperFact}

Over a factorisation channel, two special $3$-leg supersymmetric delta functions can be combined into a standard supersymmetric delta function expected in a $4$-leg superamplitude. For $\mathcal{N}=4$ SYM, this can be performed systematically as part of a super-BCFW recursion of a $4$-leg superamplitude into a product of $3$-leg superamplitudes, where the special kinematical variables $\mathbf{u}_i^I$ and $\mathbf{v}_i^I$ are unambiguously defined on a residue with complex kinematics. However the details of the shift and the location of the pole are actually unimportant for the calculation. More generally, all that really matters is that the superamplitude be evaluated on an on-shell factorisation channel. The resulting expression can be analytically continued off-shell as was done in \cite{Arkani-Hamed:2017jhn} for regular amplitudes.

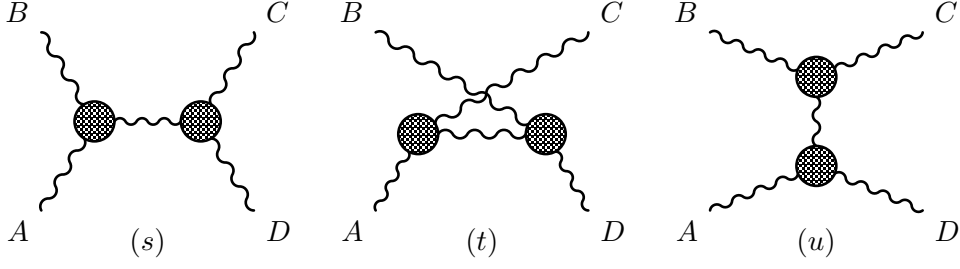
\begin{figure}[h]
\begin{fmffile}{N=4SYMTest}
\begin{center}
\begin{tabular}{ c c c c c }
& & & & \\
\begin{fmfgraph*}(100,67)
   \fmfleft{i1,i2}
   \fmfright{o1,o2}
   \fmf{boson}{i2,v1}
   \fmf{boson}{v2,o2}
   \fmf{boson}{i1,v1}
   \fmf{boson}{v2,o1}
   \fmf{boson}{v1,v2}
   \fmfv{decor.shape=circle,decor.filled=gray50,decor.size=0.15w}{v1,v2}
   \fmflabel{$A$}{i1}
   \fmflabel{$B$}{i2}
   \fmflabel{$C$}{o2}
   \fmflabel{$D$}{o1}
 \end{fmfgraph*} 
 &\,& \begin{fmfgraph*}(100,67)
   \fmfleft{i1,i2}
   \fmfright{o1,o2}
   \fmf{boson}{i1,v1}
   \fmf{boson}{v2,o1}
   \fmf{boson}{v1,v2}
   \fmf{phantom}{v1,i2}
   \fmf{phantom}{v2,o2}
   \fmf{boson,tension=-0.25}{v2,i2}
   \fmf{boson,tension=-0.25}{v1,o2}
   \fmfv{decor.shape=circle,decor.filled=gray50,decor.size=0.15w}{v1,v2}
   \fmflabel{$A$}{i1}
   \fmflabel{$B$}{i2}
   \fmflabel{$C$}{o2}
   \fmflabel{$D$}{o1}
 \end{fmfgraph*} &\,&
 \begin{fmfgraph*}(100,67)
   \fmfleft{i1,i2}
   \fmfright{o1,o2}
   \fmf{boson}{i1,v1,o1}
   \fmf{boson}{i2,v2,o2}
   \fmf{boson}{v1,v2}
   \fmfv{decor.shape=circle,decor.filled=gray50,decor.size=0.15w}{v1,v2}
   \fmflabel{$A$}{i1}
   \fmflabel{$B$}{i2}
   \fmflabel{$C$}{o2}
   \fmflabel{$D$}{o1}
 \end{fmfgraph*}\nonumber\\
 $(s)$ & \,&  $(t)$ & \,&  $(u)$
\end{tabular}
\end{center}
\end{fmffile}
\caption{Superfactorisation channels for BPS vector multiplet scattering.}\label{fig:N=4SYMTest}
\end{figure}

Ignoring the coupling constants (which will be treated further below), then on the $s$-channel residue, the $4$-particle superamplitude can be kinematically determined from factorisation into $3$-particle superamplitudes (\ref{N=43vec}) as:
\begin{align}
\mathcal{A}_s[\mathcal{W},\mathcal{W},\mathcal{W},\mathcal{W}]\sim\frac{-1}{s}\int d^4\eta_{\hat{P}}\Delta_{u}^{(12)}\Delta_{v}^{(12)}\Delta_{u}^{(34)}\Delta_{v}^{(34)}.
\end{align}
Here, the $(12)$ and $(34)$ superscripts (or subscripts below) denote the side of the channel that the corresponding object originates, while $\hat{P}$ labels the exchanged intermediate supermultiplet (see Figure \ref{fig:N=4SYMTest} below for the corresponding on-shell diagram). 

The $3$-particle delta functions on opposite sides of the channel can be represented as 
\begin{align}\label{deltaComb}
    \int \Delta_{u}^{(12)}\Delta_{u}^{(34)}d^2\,\eta_{\hat{P}}=&\frac{1}{\da{qu_{12}}\da{qu_{34}}\ds{qu_{12}}\ds{qu_{34}}}\int\delta\left(\ds{q\hat{Q}_{12}}\right)\left(\da{q\hat{Q}_{12}^\dagger}\right)\delta\left(\da{u_{12}\hat{Q}_{12}^\dagger}\right)\nonumber\\ &\qquad\qquad\qquad\qquad\qquad\times\delta\left(\ds{q\hat{Q}_{34}}\right)\left(\da{q\hat{Q}_{34}^\dagger}\right)\delta\left(\da{u_{34}\hat{Q}_{34}^\dagger}\right)d^2\eta_{\hat{P}},
\end{align}
where the reference spinors $\rs{q}$ and $\ra{q}$ are selected to satisfy both $\da{qu_{12,34}},\ds{qu_{12,34}}\neq 0$ (this is always possible to find). 
The $R$-indices on the supercharges (here $Q_4$ and $Q^{\dagger2}$) are also omitted to reduce clutter. The hats denote that the corresponding objects are evaluated on the factorisation residue (and are ``shifted'' if regarded as a term in a BCFW expansion). The delta functions can be combined most economically by applying the constraints provided by the other deltas in precisely the right order. To begin with, because 
\begin{align}
    \hat{Q}_{12}=\hat{Q}_1+\hat{Q}_2+\hat{Q}_P,\qquad \hat{Q}_{34}=\hat{Q}_3+\hat{Q}_4-\hat{Q}_P
\end{align}
and similarly for the right-handed supercharges, then 
\begin{align}
    \ds{q\hat{Q}_{12}}\sim \ds{qQ}\qquad \da{q\hat{Q}^\dagger_{12}}\sim \da{qQ^\dagger}
\end{align}
on the support of the $(34)$-side delta functions. Next,
\begin{align}\label{QRIntProj}
    \ds{q\hat{Q}_{34}}&=\frac{1}{\ds{u_{12}u_{34}}}\left(\ds{u_{34}\hat{Q}_{34}}\ds{qu_{12}}-\ds{u_{12}\hat{Q}_{34}}\ds{qu_{34}}\right)\nonumber\\
    &\sim \frac{\ds{qu_{34}}}{\ds{u_{34}u_{12}}}\ds{u_{12}\hat{Q}_{34}},
\end{align}
using the constraint from the last delta function in (\ref{deltaComb}) and the special $3$-particle kinematics $\delta\left(\da{u_{34}\hat{Q}_{34}^\dagger}\right)=\delta\left(\ds{u_{34}\hat{Q}_{34}}\right)$. Then using the analogous constraints from the $(12)$-side delta functions,
\begin{align}
    \ds{u_{12}\hat{Q}_{34}}\sim\ds{u_{12}Q}.
\end{align}
Next, 
\begin{align}
    \da{u_{12}\hat{Q}^\dagger_{12}}&=\frac{1}{\da{u_{34} q}}\left(\da{u_{34} \hat{Q}^\dagger_{12}}\da{u_{12}q}-\da{q \hat{Q}^\dagger_{12}}\da{u_{12}u_{34}}\right)\nonumber\\
    &\sim \frac{1}{\da{u_{34} q}}\left(\da{u_{34}Q^\dagger}\da{u_{12}q}-\da{qQ^\dagger}\da{u_{12}u_{34}}\right)\nonumber\\
    &=\da{u_{12}Q^\dagger},
\end{align}
also using the constraints from the remaining $(34)$-side delta functions. Finally, analogous to the relation (\ref{QRIntProj}),
\begin{align}
    \da{q\hat{Q}_{34}^\dagger}\sim\frac{\da{qu_{34}}}{\da{u_{34}u_{12}}}\da{u_{12}\hat{Q}_{34}^\dagger}.
\end{align}
The expression (\ref{deltaComb}) is then 
\begin{align}\label{CombDelta}
    \frac{1}{\ds{u_{34}u_{12}}\da{u_{34}u_{12}}\ds{qu_{12}}\da{qu_{12}}}\int&\delta\left(\ds{qQ}\right)\delta\left(\da{qQ^\dagger}\right)\delta\left(\da{u_{12}Q^\dagger}\right)\delta\left(\ds{u_{12}Q}\right)\delta\left(\da{u_{12}\hat{Q}_{34}^\dagger}\right)\nonumber\\
    &\qquad\qquad\qquad\qquad\qquad\qquad\qquad\qquad\times\delta\left(\ds{u_{34}\hat{Q}_{34}}\right)d^2\eta_{\hat{P}}\nonumber\\
    &= \frac{1}{\ds{u_{34}u_{12}}\da{u_{34}u_{12}}}\delta^{(2)}\left(Q\right)\delta^{(2)}\left(Q^\dagger\right)\la{u_{12}}\hat{P}\rs{u_{34}}e^{-i\varphi_P}\nonumber\\
    &=\frac{1}{e^{2i\varphi_P}\mathbf{u}^{(12)M}_{P}\mathbf{u}^{(34)}_{PM}}\delta^{(2)}\left(Q\right)\delta^{(2)}\left(Q^\dagger\right).
\end{align}
Here, $\mathbf{u}^{(12)M}_{P}$ and $\mathbf{u}^{(34)M}_{P}$ are the little group spinors of the internal exchanged particle on each side of the factorisation channel
and $\varphi_P$ is its mass phase. The combination of the $\Delta_v$ factors across the channel is mostly analogous, but with the phase appearing conjugated (and $\mathbf{u}$ replaced with $\mathbf{v}$). 
So altogether,
\begin{align}\label{CombDelta2}
\int \Delta_{u}^{(12)}\Delta_{u}^{(34)}d^2\,\eta_{\hat{P}}&=\frac{1}{e^{2i\varphi_P}\mathbf{u}^{(12)M}_{P}\mathbf{u}^{(34)}_{PM}}\delta^{(2)}\left(Q_4\right)\delta^{(2)}\left(Q^{\dagger 2}\right)\nonumber\\
\int \Delta_{v}^{(12)}\Delta_{v}^{(34)}d^2\,\eta_{\hat{P}}&=\frac{1}{e^{-2i\varphi_P}\mathbf{v}^{(12)M}_{P}\mathbf{v}^{(34)}_{PM}}\delta^{(2)}\left(Q_3\right)\delta^{(2)}\left(Q^{\dagger 1}\right).
\end{align}
It can be verified that these results still hold when the intermediate particle is massless despite some of the steps above appearing to be invalid in this special case. 

The factor of ${\mathbf{u}_{12}}^M\mathbf{u}_{34M}$ introduces kinematical dependence in the denominator of the amplitude that can generate poles describing other factorisation channels. Precisely how this affects the amplitude depends on the context. The product of cross-channel contractions of little group spinors of both types may be evaluated as 
\begin{align}\label{CrossContr1}
\mathbf{u}^{(12)M}_{P}\mathbf{u}^{(34)}_{PM}\mathbf{v}^{(12)N}_{P}\mathbf{v}^{(34)}_{PN}&=\frac{1}{m_P^2}\mathbf{u}^{(12)M}_{P}\da{P_MP_S}\mathbf{u}^{(34)S}_{P}\mathbf{v}^{(12)N}_{P}\da{P_NP_Q}\mathbf{v}^{(34)Q}_{P}\nonumber\\
&=\frac{1}{m_P^2}\mathbf{u}^{(12)M}_{1}\da{1_M3_S}\mathbf{u}^{(34)S}_{3}\mathbf{v}^{(12)N}_{P}\da{P_NP_Q}\mathbf{v}^{(34)Q}_{P}.
\end{align}
The lingering presence of the little group spinors can be removed with (\ref{SpecialSVD2}) and subsequently BPS complex mass or central charge conservation to give
\begin{align}\label{CrossContr2}
\frac{1}{m_P^2}\mathbf{u}^{(12)M}_{1}\da{1_M3_S}\mathbf{u}^{(34)S}_{3}\mathbf{v}^{(12)N}_{P}\da{P_NP_Q}\mathbf{v}^{(34)Q}_{P}&=\frac{1}{m_P^2}\da{1_M3_S}\da{P_NP_Q}\overline{v}_P^Nu^M_1\overline{v}_P^Qu_3^S\nonumber\\
&=t-m_t^2,
\end{align}
the cross-channel Mandelstam inverse propagator. Here $m_t=|m_2e^{i\varphi_2}+m_4e^{i\varphi_4}|$ (and similarly $m_s=|m_3e^{i\varphi_3}+m_4e^{i\varphi_4}|$ and $m_u=|m_1e^{i\varphi_1}+m_4e^{i\varphi_4}|$). If the factorisation occurs on a residue as part of a BCFW decomposition, then this can be directly identified with a specific unshifted Mandelstam invariant. Otherwise, whether this factor is to be identified with the $t$- or the $u$-channel is ambiguous and remains to be inferred (noting that $s+t+u=m_s^2+m_t^2+m_u^2$ and, on the $s$-channel, $s\rightarrow m_s^2$). So altogether,
\begin{align}\label{CrossContr}
\left(\mathbf{u}^{(12)}_{P}\cdot\mathbf{u}^{(34)}_{P}\right)\left(\mathbf{v}^{(12)}_{P}\cdot\mathbf{v}^{(34)}_{P}\right)=t-m_t^2\sim -(u-m_u^2),
\end{align}
where I use the dot products simply to denote implicit contractions of $SU(2)$ little group indices of the internal particle (and I will also use this notation for the analogous expression when this particle is massless). Once again, this calculation parallels the analogous massless amplitudes in $6d$ \cite{Dennen:2009vk}.

Again, while the steps in the above calculation do not hold in the special case in which the intermediate particle is massless, it is easy to verify that the conclusion remains true. Application of (\ref{xExp}) and (\ref{xmass}) demonstrates that 
\begin{align}\label{CrossContr}
&\left(\mathbf{u}^{(12)+}_P\mathbf{u}^{(34)-}_P-\mathbf{u}^{(12)-}_P\mathbf{u}^{(34)+}_P\right)\left(\mathbf{v}^{(12)+}_P\mathbf{v}^{(34)-}_P-\mathbf{v}^{(12)-}_P\mathbf{v}^{(34)+}_P\right)\nonumber\\
&\qquad\qquad\qquad\qquad\qquad\qquad=m_{12}m_{34}\left(\frac{x_{12}}{x_{34}}+\frac{x_{34}}{x_{12}}-2\cos\left(\varphi_{12}-\varphi_{34}\right)\right),
\end{align}
where $m_{12}=m_1=m_2$, $m_{34}=m_3=m_4$, $\varphi_{12}=\varphi_1=\varphi_2+\pi$ and $\varphi_{34}=\varphi_3=\varphi_4+\pi$. The terms consisting of ratios of $x$ factors constitute the photon exchange residue of Moller scattering in scalar QED, summarised below in (\ref{sQED}). Substituting this in and using central charge conservation affirms the expected result. The appearance of the self-consistent QED residue reflects the absence of dyonic phases in the couplings implicit within the SUSY delta functions defining (\ref{N=43vec}), at least in the case of a pair of $W$-bosons coupled to a photon. 

The arbitrary reference spinors in (\ref{SuperBPSDelta}) provide evidence of the special locality properties of the superamplitudes, similar to that of the massless amplitudes, and must cancel in any consistent computation in which they are glued together into higher leg superamplitudes. This cancellation was clear in the calculation above. For this reason it is useful not to fix them. 

The standard supersymmetric delta function for $4$-legs in non-chiral superspace can be expressed as 
\begin{align}
    \delta^{(2)}\left(Q_{a+2}\right)\delta^{(2)}\left(Q^{\dagger a}\right)=-\epsilon\left(\mathcal{Q}^a,\mathcal{Q}^a,\mathcal{Q}^a,\mathcal{Q}^a\right),
\end{align}
where $\mathcal{Q}^a$ are the Dirac supercharges defined in (\ref{N=4Charges}). I will define the abbreviation
\begin{align}\label{N=4SYMPartial}
\epsilon\left(\mathcal{Q},\mathcal{Q},\mathcal{Q},\mathcal{Q}\right)^2=\prod_a\epsilon\left(\mathcal{Q}^a,\mathcal{Q}^a,\mathcal{Q}^a,\mathcal{Q}^a\right)
\end{align}
specific for $\mathcal{N}=4$. Using (\ref{CrossContr}), the (colour-stripped) $\mathcal{N}=4$ SYM $4$-leg superamplitude is therefore determined from unitarity to be
\begin{align}\label{N=4SYM}
\mathcal{A}[\mathcal{W},\mathcal{W},\mathcal{W},\mathcal{W}]=\frac{-1}{(s-m_s^2)(u-m_u^2)}\epsilon\left(\mathcal{Q},\mathcal{Q},\mathcal{Q},\mathcal{Q}\right)^2.
\end{align}
These expressions have been known in equivalent $6d$ theories \cite{Cheung:2009dc,Dennen:2009vk}. In the present context, they provide and contain simple expressions for amplitudes in theories of massive spinning particles that have manifestly unitary high-energy scaling, a feature that is entirely obscured in the traditional Feynman diagrammatic approach \cite{Trott:2026cjj}. 

The appearance of the cross-channel pole in the residue of the merged delta functions is atypical of cuts of massive amplitudes, but familiar for massless theories. In massless theories, self-consistency conditions arose from requiring the emergent poles in residues be consistent with their own factorisation channels. The most well-known example is the constraint that gluon self-coupling constants obey the Jacobi identity. An analogous $S$-matrix derivation applies here for massive $\frac{1}{2}$BPS vector multiplets with $\mathcal{N}=4$ SUSY. Consistent (super)-factorisation of the $4$-leg superamplitude implies that it must have the form 
\begin{align}\label{SYMgen}
\mathcal{A}(\mathcal{W}_A,\mathcal{W}_B,\mathcal{W}_C,\mathcal{W}_D)&=\epsilon\left(\mathcal{Q},\mathcal{Q},\mathcal{Q},\mathcal{Q}\right)^2\nonumber\\
&\,\,\,\times\left(\frac{X_{ABCD}}{(s-m_s^2)(t-m_t^2)}+\frac{Y_{ABCD}}{(s-m_s^2)(u-m_u^2)}+\frac{Z_{ABCD}}{(t-m_t^2)(u-m_u^2)}\right).
\end{align}
Restoring the coupling constants to the superresidue calculation above, then matching this $4$-particle superamplitude onto the factorisation residues produced across each channel leads to the requirements:
\begin{align}\label{N=4SYMRes}
X_{ABCD}-Y_{ABCD}&=-f_{ABE}f_{\bar{E}CD}\nonumber\\
Y_{ABCD}-Z_{ABCD}&=-f_{BCE}f_{\bar{E}AD}\nonumber\\
Z_{ABCD}-X_{ABCD}&=-f_{CAE}f_{\bar{E}BD}.
\end{align}
The coupling constants must therefore satisfy 
\begin{align}\label{ComplexJac}
f_{ABE}f_{\bar{E}CD}+f_{BCE}f_{\bar{E}AD}+f_{CAE}f_{\bar{E}BD}=0
\end{align}
for consistency. This is the Jacobi identity, albeit in a complex basis of states. The colour indices can be converted to a self-conjugate basis by applying the unitary transformation (\ref{ConjBas}) below (this Takagi diagonalises the way that indices are contracted in (\ref{ComplexJac})). The Lie algebra structure constants are identified as $f_{AB}^{\,\,\,\,\,\,\,\,\,C}=f_{ABC}$. The full colour index antisymmetry automatically implies that the Lie algebra is semi-simple and, if non-Abelian, compact with homogeneously normalised generators. This argument demonstrates that theories of interacting massive $\frac{1}{2}$BPS vectors with $\mathcal{N}=4$ SUSY must be ``spontaneously broken'' super-Yang-Mills in exactly the same way that \cite{Benincasa:2007xk} showed from consistency that self-interacting massless vector bosons must be gluons from Yang-Mills theory.

In contrast, without maximal SUSY, the Lie algebra structure of the couplings merely emerges as a requirement for suppressed unitarity-violating growth of the amplitude at high energies. It is generally not a requirement of causality.
I elaborate extensively upon this in \cite{Trott:2026cjj}. The Higgs mechanism in this $\mathcal{N}=4$ SYM theory is also automatic and must be both entirely determined from and arranged by the SUSY structure of the multiplets and the Lie algebra structure of the couplings (unsurprisingly, this just corresponds to the adjoint Higgs potential). More examples of constraints from consistent super-factorisation will be shown in the subsequent Sections.

To conclude this analysis, I will momentarily return to the alternative central charge configuration (\ref{CentralChargesMagN=4}) consistent with configurations of three $\frac{1}{2}$BPS particles. In this case, the supercharges in the analogous superspace are represented as
\begin{align}\label{Scharge1/2dyon}
\begin{split}
Q^{\dagger a}&=-\sum_{i}\ra{i^I}\eta_{i,I}^a\\
Q_{a+2}&=\sum_{i}\rs{i^I}e^{ i\theta_i}\eta_{i,I}^a.
\end{split}
\qquad\begin{split}
Q_a&=\sum_{i}\rs{i_I}\frac{\partial}{\partial\eta_{i,I}^a}\\
Q^{\dagger a+2}&=\sum_{i}\ra{i_I}e^{- i\theta_i}\frac{\partial}{\partial\eta_{i,I}^a}.
\end{split}
\end{align}
Because the phases appearing in each pair of supercharges are now the same, the relation (\ref{SuperDegen}) instead becomes 
\begin{align}\label{SuperDegenDyon}
\da{v Q^{\dagger 1}}&=\ds{v Q_3}\nonumber\\ \da{v Q^{\dagger 2}}&=\ds{v Q_4}
\end{align}
and the $3$-particle SUSY invariants are now both defined from the $\mathbf{v}$ spinors (I will denote them both as $\Delta_v$). However, $\Delta_v\Delta_v$ is not tiny group neutral. A consistent superamplitude between particles with this central charge configuration therefore requires further factors of little group spinors in order to cancel the tiny group charge of the delta functions. However, this requires spin. The central charge configuration (\ref{CentralChargesMagN=4}) is therefore prohibited for self-consistent self-interacting elementary vector multiplets. In Appendix \ref{Dyons}, I suggest a possible reinterpretation as a superamplitude between dyons. 

A special exception exists when one of the particles is massless. In this case, $\Delta_v\Delta_v \mathbf{v}_3^+\mathbf{v}_3^-$ is supersymmetric, little group and tiny group invariant and can therefore provide the kinematic and SUSY structure of the superamplitude. This will be elaborated upon further below in Section \ref{sec:N=4SUGRAsub}. I note here however that $\int d^2\eta_{P}\Delta_u^{(12)}\Delta_v^{(34)}\mathbf{v}_{P}^{(34)+}\mathbf{v}_{P}^{(34)-}$ cannot be combined into a consistent residue that can be analytically continued off shell into a sensible Lorentz structure, where here $P$ is an exchanged massless particle between two massive particles each with central charge configurations given by (\ref{CentralChargesEleN=4}) and (\ref{CentralChargesMagN=4}). When the particles are vector multiplets, this reflects the fact that the couplings to the exchanged massless vector are relatively dyonic. The electric/magnetic nature of the central charges can be easily verified by extracting the three vector component amplitudes from $\Delta_v\Delta_u$ for (\ref{CentralChargesEleN=4}) and $\Delta_v\Delta_v\frac{\mathbf{v}_3^+\mathbf{v}_3^-}{me^{-i\theta}}$ for (\ref{CentralChargesMagN=4}), when both are interpreted as superamplitudes of the form $\mathcal{A}\left(\mathcal{W},\overline{\mathcal{W}},V\right)$, where $\mathcal{W}$ and its conjugate are charged massive vector multiplets and $V$ is a massless photon multiplet. The latter candidate superamplitude contains dyonic phases while the former does not. 

The self-conjugacy of the massless vector multiplets ensures that there is no further freedom available in the overall coupling constants with which to cancel such a dyonic phase originating from the central charges. Again, this will be explicitly illustrated in Section \ref{sec:N=4SUGRAsub} below and contrasts with the $\mathcal{N}=2$ case, which is described in Section \ref{Sec:LowSpinAmp}. The central charge configurations in (\ref{CentralChargesEleN=4}) and (\ref{CentralChargesMagN=4}) therefore respectively correspond to the cases of, firstly, two active electric charges and, secondly, one electric charge with its magnetic partner. More generally, the anti-self-dual and self-dual components of the central charge matrices are respectively identified as electric and magnetic \cite{Osborn:1979tq}. This argument would also apply to $\frac{1}{4}$BPS states and the structure of their central charges imply that they are necessarily dyonic. In Section \ref{sec:BPSGrav} further below, it is shown that a loophole to these conclusions can arise in the presence of gravity.

\subsection{$\mathcal{N}=2$ extended SUSY}\label{sec:N<4}

For $\mathcal{N}=2$, there are only two possible central charges, so the central charge matrix in general is of the simple form
\begin{align}\label{N=2CC}
Z_{i,ab} = z_i\epsilon_{ab}
\end{align}
(where $z_i$ is complex valued). This clearly preserves an $SU(2)$ subgroup of the $U(2)$ $R$-symmetry, as expected for states in Coulomb branch theories where the central charges are generated by VEVs of scalars belonging to vector multiplets \cite{Fayet:1978ig}. A central charge matrix of this form includes the general dyonic case, where the entries are interpreted as combinations of the possible two electric and two magnetic charges. Because massless vector multiplets are not self-conjugate, a relative dyonic phase can also be introduced directly into their couplings to conjugate pairs of massive ``matter'' particles, which affects the interpretation of the nature of the central charge matrix. This will be elaborated upon in the examples below. I will simply note here that this contrasts with the $\mathcal{N}=4$ SYM case above, where the identification of central charge configurations as electric or magnetic is unambiguous (modulo the $U(1)$ electric-magnetic duality that allows all central charges to be simultaneously multiplied by a common phase - this is the $U(1)$ factor from the $U(4)_R$ $R$-symmetry of the $\mathcal{N}=4$ SUSY algebra). A particle with an anti-self-dual central charge matrix is guaranteed to have purely electric couplings to massless photons (see comments in Appendix \ref{Dyons}) - the possible phases always cancel between the $\Delta_u$ and $\Delta_v$ SUSY delta functions.  

The form of the massive on-shell superfields for $\mathcal{N}=2$ theories are stated in Appendix \ref{N=2Superspace}. For long multiplets in the absence of central charges, the superspace can be simply chosen to be two independent copies of the usual superspace for $\mathcal{N}=1$ theories. For short multiplets, a non-chiral superspace is more natural, in which the supercharges are represented in the same way as a single BPS-related pair of supercharges in the $\mathcal{N}=4$ superspace explained above. Here, I will choose the $\mathcal{N}=2$ supercharges to embed into $\mathcal{N}=4$ as $Q^{\dagger 1}$ and $Q_3$ (and their conjugates). The expressions stated in (\ref{supercharges}) and (\ref{N=4Charges}) for these supercharges are completely general in the $\mathcal{N}=2$ context. The $3$-particle superamplitudes and the way in which they construct the $4$-particle superamplitudes through unitarity will be progressively introduced in the following Sections. A full catalogue of the possible $3$-particle superamplitudes can be found across Sections \ref{sec:LowSpinGrav}, \ref{sec:N=23super}, \ref{sec:N=2SUGRA}, \ref{GenLongVec} and Appendix \ref{Sec:3legAmp} (for spin $s\leq 1$ massive particles and possibly gravitons). 

\section{Low Spin 4-Particle Matter Amplitudes}\label{Sec:LowSpinAmp}

\subsection{Electrodynamics and unification in $\mathcal{N}=2$ SQED}\label{sec:QED}

The general procedure for calculating ``tree-level'' amplitudes from their singularity structure consists of two steps. Firstly, a guess should be made for the terms proportional to the Mandelstam poles in each channel, which by unitarity, have the general form
\begin{align}
A(1,2,3,4)\sim\frac{-1}{s-m^2}A(\rightarrow 12P)A(P\rightarrow 34)
\end{align}
as $s\rightarrow m^2$. I assume here and throughout that all external particles are outgoing (unless stated otherwise). The second step is to supplement these terms with an infinite series of contact terms. The terms from the first step are ambiguous up to terms that vanish on the pole, so there a precise distinction between the two classes is basis dependent. A power counting scheme must be invoked in order to justify the truncation of the amplitude to some particular form.

Of interest here is the 4-particle scattering of massive, spinning particles by the exchange of a massless intermediary (photon, graviphoton or graviton), which I will generally refer to as ``Moller'' scattering. Expressions for the residues on the factorisation channels can be easily determined from the on-shell $3$-particle amplitudes - the problem is then to lift these into a candidate off-shell expression. This can be accomplished by recycling lower spin Moller amplitudes with known factorisation channels, the calculation of which is the goal of this Section. These will then be used to calculate the gravitational amplitudes of higher spin vector bosons in Section \ref{Sec:VecGrav} below. The lowest spin ``seed'' Moller amplitudes can be calculated from on-shell recursion (for scalars), unitarity directly or simply the Feynman rules. This was done in \cite{Trott:2026cjj} and I summarise the results here. Some of these results have appeared before in \cite{Christensen:2022nja,Ema:2024vww}.

In the following, I will omit the overall coupling constants. The minimal coupling $3$-particle QED amplitudes are 
\begin{align}\label{QED3}
\begin{split}
A(\varphi,\overline{\varphi},\gamma^+)&=-(-1)^{n_1+n_3}\frac{m}{x}\\
A(\psi,\overline{\psi},\gamma^+)&=-(-1)^{n_1+n_2+n_3}\frac{1}{x}\da{\bf{12}}
\end{split}
\begin{split}
A(\varphi,\overline{\varphi},\gamma^-)&=-(-1)^{n_1+n_3}mx\\
A(\psi,\overline{\psi},\gamma^-)&=-(-1)^{n_3}x\ds{\bf{12}}.
\end{split}
\end{align}
Scalars are denoted by $\varphi$ and fermions by $\psi$. The factors of $(-1)^{n_i}$ represent sign changes that must be introduced into the expression in order to reinterpret particle $i$ as crossed, otherwise preserving the particle ordering. By ``crossed'', I mean that the particle is incoming, charge conjugated and must have opposite helicity if massless, or have $SU(2)$ spin index heights reversed if massive - again, see \cite{Trott:2026cjj} for the prescription of the way that crossing is handled here. The presence of some of these crossing sign factors assumes that $x$ is defined through the form in (\ref{Prelimx}), independently of whether the particles are incoming or outgoing. The crossed amplitudes are required in order to combine them across factorisation channels to compute higher leg amplitudes. The charge eigenstates can be converted to a self-conjugate basis through
\begin{align}\label{ConjBas}
\mathbb{U}=\frac{1}{\sqrt{2}}\begin{bmatrix}
 1 &  i  \\ 
 1 &  -i  \\ 
\end{bmatrix}:\qquad\qquad\qquad\begin{bmatrix}
 S \\ 
 \overline{S} \\ 
\end{bmatrix}=\mathbb{U} \begin{bmatrix}
 S_1 \\ 
 S_{2} \\ 
\end{bmatrix},
\end{align}
where $S$ and $\overline{S}$ are the charge eigenstates and $S_1$ and $S_2$ are the self-conjugate states. Finally, it is useful to define common spinor structures
\begin{align}\label{SpinStrucDef}
\mathbf{\Pi}_s&=\ds{\bf{12}}\da{\bf{34}}+\ds{\bf{34}}\da{\bf{12}}\nonumber\\
\mathbf{\Pi}_t&=\ds{\bf{13}}\da{\bf{24}}+\ds{\bf{24}}\da{\bf{13}}\nonumber\\
\mathbf{\Pi}_u&=\ds{\bf{14}}\da{\bf{23}}+\ds{\bf{23}}\da{\bf{14}}
\end{align}
for the $4$-particle amplitudes. 

The Moller QED amplitudes and residues are, for the scattering of scalars,
\begin{align}\label{sQED}
A(\varphi,\overline{\varphi},\varphi',\overline{\varphi'})=-\left(t-m^2-M^2\right)\frac{-1}{s}
\end{align}
and 
\begin{align}\label{sQEDRes}
-\left(\frac{x_{34}}{x_{12}}+\frac{x_{12}}{x_{34}}\right)\sim \frac{1}{mM}(m^2+M^2-t)\sim \frac{1}{mM}\frac{(u-t)}{2}.
\end{align}
Here $m$ and $M$ are the masses of $\varphi$ and $\varphi'$ respectively (and, in the following, will be identified as the masses of the first and the second species of particles listed in the amplitude). The second form of the residue is equivalent on the pole and has manifest exchange symmetry. Off-shell, these residues lead to amplitudes that differ by the addition of a constant contact term that represents a possible scalar quartic coupling. For fermions, the Moller amplitude and residue are
\begin{align}\label{FermionQED}
A(\psi,\overline{\psi},\psi',\overline{\psi'})=-\frac{\mathbf{\Pi}_t+\mathbf{\Pi}_u}{s},
\end{align}
and
\begin{align}\label{QEDRes}
-\left(\frac{x_{34}}{x_{12}}\da{\bf{12}}\ds{\bf{34}}+\frac{x_{12}}{x_{34}}\da{\bf{34}}\ds{\bf{12}}\right)\sim\left(\mathbf{\Pi}_t+\mathbf{\Pi}_u\right),
\end{align}
while for mixed scalar and fermion scattering, they are
\begin{align}\label{sfQED}
A(\psi,\overline{\psi},\varphi,\overline{\varphi})=\frac{1}{2}\left(\la{\bf{1}}p_4-p_3\rs{\bf{2}}+\la{\bf{2}}p_4-p_3\rs{\bf{1}}\right)\frac{-1}{s},
\end{align}
and 
\begin{align}\label{sfSQED}
-\left(\frac{x_{34}}{x_{12}}\da{\bf{12}}+\frac{x_{12}}{x_{34}}\ds{\bf{12}}\right)\sim\frac{1}{2M}\left(\la{\bf{1}}p_4-p_3\rs{\bf{2}}+\la{\bf{2}}p_4-p_3\rs{\bf{1}}\right).
\end{align}

These results can be unified with Yukawa amplitudes generated by the couplings
\begin{align}\label{3Yukawa}
A(\psi,\overline{\psi},\phi)&=(-1)^{n_1}\ds{\bf{12}}\nonumber\\
A(\psi,\overline{\psi},\overline{\phi})&=(-1)^{n_2}\da{\bf{12}},
\end{align}
into the $\mathcal{N}=2$ super-QED superamplitude for Moller scattering of BPS hypermultiplets, the structure of which follows immediately by supersymmetry. Alternatively, the superamplitude can also be easily constructed out of the $3$-leg superamplitudes 
\begin{align}\label{N=2SQED3}
\mathcal{A}(K,\overline{K},V^+)&=e^{-i\varphi_1}\Delta_v \mathbf{v}^+_3\nonumber\\
\mathcal{A}(K,\overline{K},V^-)&=(-1)^{n_3}e^{-i\varphi_1}\Delta_v \mathbf{v}^-_3\nonumber\\
\mathcal{A}(K,\overline{K},V^K)&=-e^{-i(\varphi_2+\varphi_3)}\Delta_v \mathbf{v}^K_3.
\end{align}
These are fixed by SUSY, little group covariance and tiny group invariance. The tiny group scaling of the delta function must be cancelled by the presence of the little group spinor. I have included the case in which the vector multiplet is also massive in the last line. In this case, the particles are all BPS and the hypermultiplets are not charge conjugates, but this otherwise has the same structure as the massless photon case. These $\mathcal{N}=2$ superamplitudes will be studied more extensively below in Section \ref{sec:N=2}.
Incidentally, there are no possible $3$-leg superamplitudes involving only massive hypermultiplets (so there is no cubic ``hyperpotential''). 

In (\ref{N=2SQED3}), I have allowed for both central charges to potentially be activated, represented by non-zero mass phases. While these can be eliminated for a single species of hypermultiplet by absorbing them into the definition of the state of the vector multiplet, there can generally be a misalignment of charge vectors between two different hypermultiplets electromagnetically scattering off each other. As alluded to in Section \ref{sec:N<4}, the phases in a particle's central charge matrix translates into a phase in its couplings to photons or gravitons through the SUSY delta function that must necessarily be part of the superamplitude. However, the separation of the opposite helicity photons and gravitons into distinct supermultiplets (in contrast to $\mathcal{N}=4$ SYM) provides the parametric freedom to introduce separate phases into the overall coupling constants to ensure that the interactions remain consistently electric (which is necessary for a perturbative theory). I have done this in the expressions above (\ref{N=2SQED3}) - the explicit phase factors cancel those inside the delta functions that would otherwise make the monopole charges complex, although the Yukawa couplings involving the photon's scalar superpartner now acquire a (harmless) phase relative to (\ref{3Yukawa}).


The $4$-particle superamplitude can now be constructed from unitarity. On the $s$-channel vector exchange, the superamplitude superfactorises as 
\begin{align}\label{SQEDFact}
\mathcal{A}(K,\overline{K},K',\overline{K'})&\sim\frac{-1}{s}\int d^2\eta_P\Delta_v^{(12)}\mathbf{v}_P^{(12)M}e^{-i(\varphi_2+\varphi_P)}(-1)\Delta_v^{(34)}\mathbf{v}_{PM}^{(34)}e^{-i(\varphi_4+\varphi_P)}
\nonumber\\
&=\epsilon\left(\mathcal{Q},\mathcal{Q},\mathcal{Q},\mathcal{Q}\right)e^{-i(\varphi_2+\varphi_4)}\frac{-1}{s}.
\end{align}
The cross-channel contraction of the little group spinors produced by the polarisation sum of the internal vector cancels the factor squeezed-out from combining the SUSY delta functions in (\ref{CombDelta2}), leaving (\ref{SQEDFact}) with no further poles. Note that when the exchanged vector multiplet is massless, there is a relative negative produced between terms in the polarisation sum that arises when the outgoing $V^-$ is crossed to an incoming $V^+$ (so $\mathbf{v}_P^-$ is effectively accompanied by a sign flip). This is easily seen from comparison with the massive case and is consistent with the crossing rules for the $x$-factors implicit in (\ref{QED3}). Note that the overall phase factor in (\ref{SQEDFact}) cancels phases implicit in the SUSY delta function to ensure agreement with the parts of the component amplitudes generated by the photon exchange in (\ref{sQED}), (\ref{FermionQED}) and (\ref{sfQED}). 

\subsection{$\mathcal{N}=1$ Wess-Zumino and SQED}\label{sec:SimpleN=1}

The $\mathcal{N}=1$ superamplitudes can be assembled from these results as well. Chiral supermultiplets have the same structure as the $\mathcal{N}=2$ BPS hypermultiplets. However, with less supersymmetry, the superamplitudes have more freedom. A large fraction of the analysis of this subsection has been substantially performed previously in \cite{Johansson:2023ymb} (which I was unaware of until recently and with which I agree). 

Yukawa interactions between massless chiral multiplets (which I denote by $\Phi$) are 
\begin{align}
\mathcal{A}\left(\Phi^+,\Phi^+,\Phi^+\right)&=\widetilde{\delta}^{(1)}\left(Q\right)\nonumber\\
\mathcal{A}\left(\Phi^-,\Phi^-,\Phi^-\right)&=-\delta^{(2)}\left(Q^\dagger\right).
\end{align}
I will not bother to include coupling constants or internal indices here as I am mostly interested in the kinematic and SUSY structure. The holomorphy in helicity reflects the holomorphy of the superpotential. These are easily combined across a factorisation channel to produce the $4$-leg superamplitude
\begin{align}\label{N=1MasslessYukawa}
\mathcal{A}\left(\Phi^+,\Phi^+,\Phi^-,\Phi^-\right)=\frac{1}{s}\delta^{(2)}\left(Q^\dagger\right)\ds{12},
\end{align}
which is entirely fixed by symmetry anyway. Off-shell, the scalar quartic term in the action is determined from the $3$-particle Yukawa couplings to ensure that there are no additional contact terms in (\ref{N=1MasslessYukawa}). This is reflected in how the cubic terms of the superpotential determine both the scalar quartic and Yukawa vertices. If the external chiral supermultiplets are massive, then a superunitarity calculation can likewise also be performed to infer the superamplitude. 

Alternatively, both the superamplitude for the massless Yukawa exchange and Moeller scattering in $\mathcal{N}=1$ SQED can be easily extracted from (\ref{SQEDFact}) by separating the scalar and photon exchange parts and identifying $\mathcal{N}=1$ SUSY invariants. I will define notation 
\begin{align}\label{GenLegSupercharges}
\ra{\bf{q}^{\dagger \mathnormal{(a)}}_i}=-\ra{i^I}\eta^{(a)}_{i,I}\qquad\qquad
\rs{\bf{q}^{\mathnormal{(a)}}_i}=\pm\rs{i^I}\eta^{(a)}_{i,I}.
\end{align}
Obviously these have the interpretations of being the supercharges of each leg in the relevant superspaces for $\mathcal{N}=2$ and $4$ as defined in Section \ref{SUSYAlgebra}, but I will continue to use these symbols where these objects arise regardless of whether they continue to bear this interpretation, such as in the $\mathcal{N}=1$ context here. The sign in the last equation is chosen to be $(+)$ for particles and $(-)$ for antiparticles, consistent with the requirements of central charges or mass phases in the extended SUSY case. The superscript $(a)$ denotes a theory-dependent $R$-index that is not relevant to the present focus on $\mathcal{N}=1$, but will be required later. In the following, I will adopt the on-shell superspace conventions used in \cite{Herderschee:2019ofc} for both $\mathcal{N}=1$ and extended SUSY without central charges, in which the supercharges $Q^\dagger$ are represented as Grassmann multiplicative raising operators and the conjugate supercharges $Q$ are Grassmann derivative lowering operators on the external fermionic coherent states. It is also convenient to introduce abbreviations for some common Grassmann structures in the superamplitudes (somewhat analogous to (\ref{SpinStrucDef})):
\begin{align}\label{SuperLorentz}
\mathcal{P}_s^{(a)}&=\ds{\bf{q}^\mathnormal{(a)}_1\bf{q}^\mathnormal{(a)}_2}+\ds{\bf{q}^\mathnormal{(a)}_3\bf{q}^\mathnormal{(a)}_4}\nonumber\\
\mathcal{P}_t^{(a)}&=\ds{\bf{q}^\mathnormal{(a)}_1\bf{q}^\mathnormal{(a)}_3}+\ds{\bf{q}^\mathnormal{(a)}_2\bf{q}^\mathnormal{(a)}_4}\nonumber\\
\mathcal{P}_u^{(a)}&=\ds{\bf{q}^\mathnormal{(a)}_1\bf{q}^\mathnormal{(a)}_4}+\ds{\bf{q}^\mathnormal{(a)}_2\bf{q}^\mathnormal{(a)}_3}\nonumber\\
\mathcal{P}_{s12}^{(a)}&=\ds{\bf{q}^\mathnormal{(a)}_1\bf{q}^\mathnormal{(a)}_2}+\frac{1}{2}\left(\ds{\bf{q}^\mathnormal{(a)}_1\bf{q}^\mathnormal{(a)}_1}+\ds{\bf{q}^\mathnormal{(a)}_2\bf{q}^\mathnormal{(a)}_2}\right)\nonumber\\
\mathcal{P}_{s34}^{(a)}&=\ds{\bf{q}^\mathnormal{(a)}_3\bf{q}^\mathnormal{(a)}_4}+\frac{1}{2}\left(\ds{\bf{q}^\mathnormal{(a)}_3\bf{q}^\mathnormal{(a)}_3}+\ds{\bf{q}^\mathnormal{(a)}_4\bf{q}^\mathnormal{(a)}_4}\right).
\end{align}

Extracted in this way, the superamplitude for an exchange of a massless chiral supermultiplet between massive chiral multiplets is then 
\begin{align}\label{N=1YukMassive}
\mathcal{A}\left(\Phi,\overline{\Phi},\Phi',\overline{\Phi'}\right)=\frac{1}{s}\delta^{(2)}\left(Q^\dagger\right)\left(\mathcal{P}_{s12}+\mathcal{P}_{s34}\right).
\end{align}
This is easily verified by unitarity using the relevant $3$-leg superamplitudes
\begin{align}
\mathcal{A}\left(\Phi,\overline{\Phi},\Phi^+\right)&=\delta^{(2)}\left(Q^\dagger\right)F_3\nonumber\\
\mathcal{A}\left(\Phi,\overline{\Phi},\Phi^-\right)&=\delta^{(2)}\left(Q^\dagger\right),
\end{align}
where 
\begin{align}\label{FInv}
F_3=\eta_3+\frac{1}{2m_1}\ds{3\bf{q}_1}-\frac{1}{2m_2}\ds{3\bf{q}_2}\sim\eta_3+\frac{1}{m_1}\ds{3\bf{q}_1}
\end{align}
(the last relation holding on the support of the SUSY delta function), to give
\begin{align}\label{N=1Yuks}
\mathcal{A}\left(\Phi,\overline{\Phi},\Phi',\overline{\Phi'}\right)&=\frac{-1}{s}\int d\eta_P\delta^{(2)}\left(Q^\dagger_{12}\right)\delta^{(2)}\left(Q^\dagger_{34}\right)\left({F}_{P}^{(12)}+{F}_{P}^{(34)}\right)\nonumber\\
&=\frac{1}{s}\delta^{(2)}\left(Q^\dagger\right)\left(\mathcal{P}_{s12}+\mathcal{P}_{s34}\right),
\end{align}
where the two terms in the last equation correspond respectively to evaluations of the two terms in the line above.

For $\mathcal{N}=1$ SQED, the superamplitude is inferred to be
\begin{align}\label{N=1SQED}
\mathcal{A}\left(\Phi,\overline{\Phi},\Phi',\overline{\Phi'}\right)=\frac{1}{s}\delta^{(2)}\left(Q^\dagger\right)\left(\mathcal{P}_{t}+\mathcal{P}_{u}\right).
\end{align}
The decomposition of (\ref{SQEDFact}) into exchanged $\mathcal{N}=1$ supermultiplets is clear from separating the terms corresponding to spin $0$ and $1$ exchange across the channel. The asymmetry between left and right spinors is a consequence of the inherent breaking of parity by the choice of superspace for the $\mathcal{N}=1$ theory. The $3$-leg $\mathcal{N}=1$ SQED superamplitudes are \cite{Herderschee:2019ofc}:
\begin{align}
\mathcal{A}\left(\Phi,\overline{\Phi},V^+\right)&=\frac{1}{x}\delta^{(2)}\left(Q^\dagger\right)\nonumber\\
\mathcal{A}\left(\Phi,\overline{\Phi},V^-\right)&=x\delta^{(2)}\left(Q^\dagger\right) F_3,
\end{align}
where 
\begin{align}\label{SuperBlockMassless}
xF_3=x\left(\eta_3+\frac{1}{2m}\left(\ds{3\bf{q}_1}-\ds{3\bf{q}_2}\right)\right)\sim\frac{-1}{\ds{q3}}\left(\ds{q\bf{q}_1}+\ds{q\bf{q}_2}\right).
\end{align}
The form given in the last equation is closest to the form directly extracted from the $\mathcal{N}=2$ superamplitude and is independent of the reference spinor $\rs{q}$. The $\mathcal{N}=1$ SQED superamplitude superfactorises on the $s$-channel as 
\begin{align}\label{N=1SQEDs}
\mathcal{A}\left(\Phi,\overline{\Phi},\Phi',\overline{\Phi'}\right)&=\frac{1}{s}\int d\eta_P\delta^{(2)}\left(Q^\dagger_{12}\right)\delta^{(2)}\left(Q^\dagger_{34}\right)\left(\frac{x_{34}}{x_{12}}{F}_{P}^{(34)}+\frac{x_{12}}{x_{34}}{F}_{P}^{(12)}\right)\nonumber\\
&=\frac{1}{s}\delta^{(2)}\left(Q^\dagger\right)\left(\frac{x_{34}}{x_{12}}\mathcal{P}_{s34}+\frac{x_{12}}{x_{34}}\mathcal{P}_{s12}\right).
\end{align}

These results can be generalised to the exchange of a massive multiplet. A general superamplitude of massive particles can be constructed from the usual SUSY delta function multiplied by a Grassmann polynomial in each leg's supercharges. The Grassmann polynomial must be invariant under the action of the derivatively represented supercharges and is defined modulo $Q^\dagger\sim 0$ (the support enforced by the SUSY delta function). For any three legs $i,j,k$, all such independent linear polynomials $F_{k;ij}$ satisfying the constraint $QF_{k;ij}=0$ are of the form 
\begin{align}\label{SuperBlock}
F_{k;ij}^{K}&=\frac{1}{m_i}\ds{k^Ki^M}\eta_{iM}-\frac{1}{m_j}\ds{k^Kj^M}\eta_{jM}.
\end{align}
An alternative, but equivalent basis choice of linear Grassmann structures that is more closely analogous to the case with one massless leg (\ref{FInv}) is
\begin{align}\label{SuperBlock2}
F_{k;ij}^{'K}&=\eta_{k}^K+\frac{1}{2m_i}\ds{k^Ki^M}\eta_{iM}+\frac{1}{2m_j}\ds{k^Kj^M}\eta_{jM},
\end{align}
although I will not use this much in what follows. Higher order invariant polynomials can be constructed from these linear blocks. 

For a $3$-leg superamplitude, there is only a single independent combination of the form (\ref{SuperBlock}) and I will abbreviate the notation to simply $F_k^K=F_{k;k+1k-1}^K$ (where subscript particle labels are mod $3$). This can be used to construct any higher Grassmann degree invariant with the appropriate little group tensorial properties that can appear in the superamplitude. Manifest exchange symmetry can be restored to a candidate superamplitude by appropriately symmetrising a complete expression constructed out of this. However, when the $3$-leg superamplitude appears within a factorised $4$-particle superamplitude, the internal leg is distinguished and it is computationally useful to eliminate the appearance of the internal supercharge as in (\ref{SuperBlock}). This is possible by inverting supercharge conservation $Q^\dagger=0$. Eliminating the internal leg's supercharge then trivialises the Grassmann integral over the superfactorisation channel (or, equivalently stated, the state sum) so that it merely merges the SUSY delta functions on each side.

In the special case of $3$-particle superamplitudes in which one of the legs is massless and the others have equal mass, the massless particle's Grassmann variable cannot be determined from $Q^\dagger=0$ without introducing redundancies. Taking the limit $m_3\rightarrow0$ and $m_1,m_2\rightarrow m>0$, the SUSY invariants (\ref{SuperBlock}) and (\ref{SuperBlock2}) become inequivalent, with (\ref{SuperBlock}) converging to $F^{+}_3\rightarrow\frac{1}{m^2}\ls{3}p_1\ra{Q^\dagger}$ and thus degenerating with a right-handed supercharge, while 
\begin{align}
F^{'+}_3\rightarrow \eta_3+\frac{1}{2m}\left(\ds{3\bf{q_1}}-\ds{3\bf{q_2}}\right)\sim\eta_3+\frac{1}{m}\ds{3\bf{q_1}}
\end{align}
becomes the non-trivial SUSY invariant (\ref{FInv}), which has already been used in the examples above. The two forms given in (\ref{SuperBlockMassless}) demonstrate how redundancy must be introduced in order to eliminate the massless Grassmann variable. 

As for $\mathcal{N}=0$ amplitudes, superamplitudes are generally also constructed by adding together superfactorising and contact terms. A classification of a basis of contact terms for superamplitudes is similar to the analogous and better studied problem for merely Lorentz invariant, little group covariant, terms (see, for example, \cite{DeAngelis:2022qco,Dong:2022mcv,Durieux:2020gip}), except that they must also now be SUSY-invariant Grassmann polynomials. However, the choice of on-shell superspace (or, in other words, the choice of Clifford vacua chosen to represent the supermultiplets) necessarily obscures some of the symmetries, and while the choice made in \cite{Herderschee:2019ofc} (and here) preserves both the little group and Lorentz invariance, parity is the victim. The intrinsic chirality of the superspace obscures the parity symmetry of the space of possible SUSY invariants when constructed out of the SUSY delta function and the linear building blocks (\ref{SuperBlock}). The invariants (\ref{SuperBlock}) contain energy dependence that, at high energies $E$, indicates that they should scale as $\sim E$ (the same power as their Grassmann degree), but the energy dependence of possible contact terms is not simply given by the number of such factors that they contain. For example, the SUSY delta function $\delta^{(2)}(Q^\dagger)$ is the unique Grassmann degree $2$ invariant and it scales with energy as $\sim E$. The highest degree Grassmann invariant is also unique, as it can be obtained by Grassmann Fourier transforming the SUSY delta function from the on-shell superspace of opposite chirality. This invariant must also scale with energy as $\sim E$, in spite of the fact that it is a polynomial that may be constructed by multiplying $\delta^{(2)}(Q^\dagger)$ with factors of (\ref{SuperBlock}), all of which contain a single power of energy dependence. Hence the construction of these SUSY invariants out of factors of (\ref{SuperBlock}) must necessarily involve cancellations of energy dependence. The breaking of manifest parity therefore obscures the high energy power counting.

The superamplitude for three massive chiral multiplets is \cite{Herderschee:2019ofc}
\begin{align}\label{WZ3leg}
\mathcal{A}\left(\Phi_i,\Phi_j,\Phi_k\right)=\delta^{(2)}(Q^\dagger)\left(a_{ijk}+\frac{1}{2}\frac{m_1m_2}{m_3^2}b_{ijk}F_{3}^KF_{3K}\right).
\end{align}
The coupling constants are $a_{ijk}$ and $b_{ijk}$. As just mentioned, the higher Grassmann degree term is unique and, despite appearances, is exchange symmetric and does not contain higher energy dependence: 
\begin{align}
-\delta^{(2)}(Q^\dagger)\frac{m_1m_2}{2m_3^2}F_{3}^KF_{3K}&=-\frac{1}{6}\delta^{(2)}(Q^\dagger)\left(\frac{m_2m_3}{m_1^2}F_{1}^KF_{1K}+\frac{m_3m_1}{m_2^2}F_{2}^KF_{2K}+\frac{m_1m_2}{m_3^2}F_{3}^KF_{3K}\right)\nonumber\\
&=\eta_1^2\eta_{2J}\ds{2^J3^K}\eta_{3K}+\eta_2^2\eta_{3K}\ds{3^K1^I}\eta_{1I}+\eta_3^2\eta_{1I}\ds{1^I2^J}\eta_{2J}\nonumber\\&\qquad+m_1\eta_2^2\eta_3^2+m_2\eta_3^2\eta_1^2+m_3\eta_1^2\eta_2^2\nonumber\\
&=\widetilde{\delta}^{(4)}(Q),
\end{align}
where $\eta_i^2=\frac{1}{2}\eta_{iI}\eta_{i}^I$. In the last line, it has been identified that this expression is the Grassmann Fourier transform of the SUSY delta function in the on-shell superspace of conjugate chirality. The exchange symmetry implies that the coupling constants are fully symmetric in their flavour indices. When combined across a factorisation channel, with the internal line's supercharges eliminated in (\ref{WZ3leg}) outside the delta function, the resulting $4$-leg superamplitude is easily computed from the residue 
\begin{align}\label{WZ4fact}
\mathcal{A}\left(\Phi_i,\Phi_j,\Phi_k,\Phi_l\right)&=\frac{1}{s-m_P^2}\delta^{(2)}(Q^\dagger)m_P\left(a_{ijm}+\frac{1}{2}\frac{m_1m_2}{m_P^2}b_{ijm}F_{P}^{(12)M}F_{PM}^{(12)}\right)\nonumber\\
&\qquad\qquad\qquad\qquad\qquad\qquad\times\left(a_{mkl}+\frac{1}{2}\frac{m_3m_4}{m_P^2}b_{mkl}F_{P}^{(34)M}F_{PM}^{(34)}\right)\nonumber\\
&=\frac{1}{s-m_P^2}\delta^{(2)}(Q^\dagger)m_P\left(a_{ijm}-\frac{1}{m_P}b_{ijm}\left(m_2\eta_1^2+m_1\eta_2^2+\eta_{1I}\ds{1^I2^J}\eta_{2J}\right)\right)\nonumber\\
&\qquad\qquad\times\left(a_{mkl}-\frac{1}{m_P} b_{mkl}\left(m_4\eta_3^2+m_3\eta_4^2+\eta_{3K}\ds{3^K4^L}\eta_{4L}\right)\right),
\end{align}
where $m_P$ is the mass of the exchanged chiral multiplet and $m$ its flavour index. 

This superamplitude is ambiguous up to contact terms that must generally be included. These are necessary in the present case where the correctly superfactorising hypothesis (\ref{WZ4fact}) appears to have divergent high energy scaling $\sim E$ in the Grassmann degree $6$ terms proportional to $b_{ijm}b_{mkl}$. Since the $3$-particle amplitudes underpinning this expression alone should constitute a fully unitary theory, it should be possible to tune the contact terms to cancel the divergence and produce a unitarised superamplitude. 

The Grassmann degree $6$ SUSY invariant is unique and is therefore already contained within the relevant terms in (\ref{WZ4fact}). This can be identified by direct expansion as 
\begin{align}
\widetilde{\delta}^{(6)}(Q)&=\eta_1^2\eta_2^2\eta_{3K}\ds{3^K4^L}\eta_{4L}+\eta_3^2\eta_4^2\eta_{1I}\ds{1^I2^J}\eta_{2J}+\eta_1^2\eta_3^2\eta_{2J}\ds{2^J4^L}\eta_{4L}\nonumber\\
&\qquad+\eta_2^2\eta_4^2\eta_{1I}\ds{1^I3^K}\eta_{3K}+\eta_1^2\eta_4^2\eta_{2J}\ds{2^J3^K}\eta_{3K}+\eta_2^2\eta_3^2\eta_{1I}\ds{1^I4^L}\eta_{4L}\nonumber\\
&\qquad+m_1\eta_2^2\eta_3^2\eta_4^2+m_2\eta_3^2\eta_4^2\eta_1^2+m_3\eta_4^2\eta_1^2\eta_2^2+m_4\eta_1^2\eta_2^2\eta_3^2.
\end{align}
Relabeling the invariant combinations 
\begin{align}
F_{ij}^2&=\frac{m_im_j}{2m_P}F_{P,ijM}F_{P,ij}^{\,\,\,\,\,M}\nonumber\\
&=m_i\eta_j^2+m_j\eta_i^2+\eta_{iI}\ds{i^Ij^J}\eta_{jJ},
\end{align}
(which is entirely independent of the particle ``$P$'') then it is simultaneously verifiable that 
\begin{align}
\delta^{(2)}(Q^\dagger)F_{12}^2F_{34}^2=s\,\widetilde{\delta}^{(6)}(Q)
\end{align}
and therefore 
\begin{align}
\widetilde{\delta}^{(6)}(Q)&=\left(\frac{1}{\sum_i m_i^2}\right)\delta^{(2)}(Q^\dagger)\left(F_{12}^2F_{34}^2+F_{13}^2F_{24}^2+F_{14}^2F_{23}^2\right).
\end{align}
This invariant scales as $\sim E$, contrary to appearances on the right-hand side. On the residue of the pole in (\ref{WZ4fact}), $s\sim m_P^2$, so the unitarised superamplitude can be directly inferred to be  
\begin{align}\label{WZ4Uni}
\mathcal{A}\left(\Phi_i,\Phi_j,\Phi_k,\Phi_k\right)&=\frac{1}{s-m_P^2}\delta^{(2)}(Q^\dagger)
\big(m_Pa_{ijm}a_{mkl}-a_{ijm}b_{mkl}F_{34}^2-b_{ijm}a_{mkl}F_{12}^2\nonumber\\
&\qquad\qquad\qquad+\left(\frac{m_P}{\sum_i m_i^2}\right)b_{ijm}b_{mkl}\left(F_{12}^2F_{34}^2+F_{13}^2F_{24}^2+F_{14}^2F_{23}^2\right)\big).
\end{align}
This is also consistent with (\ref{N=1Yuks}) in the limit that $m_P\rightarrow 0$ wherein only the Grassmann degree $4$ term survives. The notation may be made consistent with the identification $\mathcal{P}_{s12}=-F_{12}^2$ and $\mathcal{P}_{s34}=-F_{34}^2$.

With a massive vector, the superamplitude is parametrically fixed as
\begin{align}\label{N=1QEDM3}
\mathcal{A}\left(\Phi_i,\Phi_j,V_A^K\right)=-\frac{1}{2m_3}\left(m_2(t_A)_{ij}-m_1(t_A)_{ji}\right)\delta^{(2)}(Q^\dagger)F_3^K.
\end{align}
I have matched the overall coupling onto the general non-supersymmetric fermion coupling to a massive vector, reproduced below in (\ref{FermionMat3leg}). The coupling must satisfy $-m_2(t_A)_{ij}=m_1(t_A)_{ji}$, which also reflects its parity structure. When combined across a factorisation channel, this produces a superamplitude:
\begin{align}\label{N=1MassiveQED}
\mathcal{A}\left(\Phi_i,\Phi_j,\Phi_k,\Phi_l\right)&=\frac{1}{s-m_P^2}\delta^{(2)}(Q^\dagger)\frac{1}{4m_P}\left(m_2(t_M)_{ij}-m_1(t_M)_{ji}\right)\left(m_4(t_M)_{kl}-m_3(t_M)_{lk}\right)\nonumber\\
&\qquad\qquad\qquad\qquad\qquad\qquad\times F_P^{(12)M}F_{PM}^{(34)}\nonumber\\
&=\frac{1}{s-m_P^2}\delta^{(2)}(Q^\dagger)\frac{1}{4}\left(m_2(t_M)_{ij}-m_1(t_M)_{ji}\right)\left(m_4(t_M)_{kl}-m_3(t_M)_{lk}\right)\nonumber\\
&\qquad\qquad\qquad\qquad\times\left(\frac{\ds{\bf{q_1q_3}}}{m_1m_3}+\frac{\ds{\bf{q_1q_4}}}{m_1m_4}+\frac{\ds{\bf{q_2q_3}}}{m_2m_3}+\frac{\ds{\bf{q_2q_4}}}{m_2m_4}\right).
\end{align}
This is clearly consistent with the massless QED superamplitude (\ref{N=1SQED}) provided that the couplings of the $3$-leg interactions have the mass dependence as stated in (\ref{N=1QEDM3}).

\subsection{Gravity and supergravity}\label{sec:LowSpinGrav}

The calculations in Section \ref{sec:QED} above can be repeated for graviton and graviphoton exchange, which are again unified by scattering of hypermultiplets in $\mathcal{N}=2$ SUGRA. The superamplitude in this case is given by 
\begin{align}\label{HyperSUGRA}
\mathcal{A}(K,\overline{K},K',\overline{K'})=-\frac{(t-u+2mM)}{2M_{Pl}^2}\epsilon\left(\mathcal{Q},\mathcal{Q},\mathcal{Q},\mathcal{Q}\right)\frac{-1}{s},
\end{align}
as may be easily verified with super-BCFW recursion (up to possible contact terms) or directly from unitarity as follows. The hypermultiplet coupling to gravitons is similar to the SQED case, but with extra factors of $x$:
\begin{align}\label{HyperSUGRA3}
\mathcal{A}(K,\overline{K},H^+)&=\frac{m}{M_{Pl}}e^{-i\varphi_1}\Delta_v \frac{1}{x}\mathbf{v}^+_3\nonumber\\
\mathcal{A}(K,\overline{K},H^-)&=\frac{m}{M_{Pl}}e^{-i\varphi_1}\Delta_v x\mathbf{v}^-_3.
\end{align}
Across the $s$-channel, the superamplitude factorises as
\begin{align}
\mathcal{A}(K,\overline{K},K',\overline{K'})&\sim\frac{-1}{s}\frac{mM}{M_{Pl}^2}\int d^2\eta_P\Delta_v^{(12)}\left(\frac{(\mathbf{v}^{(12)+}_P)^2}{\mathbf{v}^{(12)-}_P}\frac{(\mathbf{v}^{(34)-}_P)^2}{\mathbf{v}^{(34)+}_P}-\frac{(\mathbf{v}^{(12)-}_P)^2}{\mathbf{v}^{(12)+}_P}\frac{(\mathbf{v}^{(34)+}_P)^2}{\mathbf{v}^{(34)-}_P}\right)\Delta_v^{(34)}\nonumber\\
&=\frac{-1}{s}\frac{mM}{M_{Pl}^2}\Delta_v\frac{1}{\mathbf{v}^{(12)}\cdot\mathbf{v}^{(34)}}\mathbf{v}^{(12)}\cdot\mathbf{v}^{(34)}\left(\frac{\mathbf{v}^{(12)+}_P\mathbf{v}^{(34)-}_P}{\mathbf{v}^{(12)-}_P\mathbf{v}^{(34)+}_P}+1+\frac{\mathbf{v}^{(12)-}_P\mathbf{v}^{(34)+}_P}{\mathbf{v}^{(12)+}_P\mathbf{v}^{(34)-}_P}\right)\nonumber\\
&=\frac{-1}{s}\epsilon\left(\mathcal{Q},\mathcal{Q},\mathcal{Q},\mathcal{Q}\right)\frac{t-u+2mM}{2M_{Pl}^2}.
\end{align}
I have assumed everywhere here that the central charges of the hypermultiplets are aligned and have set the mass phases to zero. 
This is ideal for illustrating the relationship between the non-SUSY component amplitudes to follow and the elegant generating function (\ref{HyperSUGRA}) from SUGRA. The case of misaligned central charges will be described at the end of this Section. Misaligned central charges in SQED simply required the explicit inclusion of annoying complex phases into the couplings, but in SUGRA they demand a non-minimal extension of the theory. To go from the second to third line, I have used (\ref{xExp}), (\ref{xmass}) and then the scalar QED Moller residue (\ref{sQED}) (so this SUGRA residue is a double copy of $\mathcal{N}=2$ and $\mathcal{N}=0$ scalar QED residues). Finally, just as for the scalar QED Moller amplitude (\ref{sQED}), this superamplitude is ambiguous up to an overall contact term, which corresponds off-shell to the presence of dimension $6$ operators. The form of the residue arrived at by super-BCFW actually involves $t-m^2-M^2$ instead of the more symmetric looking $\frac{1}{2}(t-u)$ (the central charges break the symmetry, so the $t-m^2-M^2$ form is expected).

Non-supersymmetric gravitational scalar scattering may be computed from unitarity as 
\begin{align}\label{ScalarGrav}
A(\varphi,\overline{\varphi},\varphi',\overline{\varphi'})&=\frac{-1}{s}\frac{m^2M^2}{M_{Pl}^2}\left(\left(\frac{x_{34}}{x_{12}}\right)^2+\left(\frac{x_{12}}{x_{34}}\right)^2\right)\nonumber\\
&=\frac{-1}{s}\frac{m^2M^2}{M_{Pl}^2}\left(\left(\frac{x_{34}}{x_{12}}+\frac{x_{12}}{x_{34}}\right)^2-2\right)\nonumber\\
&\sim -\frac{\left(t-(m^2+M^2)\right)^2-2m^2M^2}{M_{Pl}^2s}\sim -\frac{\frac{1}{4}\left(t-u\right)^2-2m^2M^2}{M_{Pl}^2s}.
\end{align}
The second equality manifests a double copy structure, where the first term is the square of the same process with a photon exchange (\ref{sQEDRes}) and the second term is the removal of the contribution from a scalar exchange. This presents the simplest instance of how the gravitational residues can be reduced to simpler electromagnetic ones. See \cite{Bjerrum-Bohr:2013bxa} for simple examples of this for gravitational Compton scattering. The minimal gravitational coupling induces Moller amplitudes with high energy dependence of the form $\sim E^2/M_{pl}^2$, which are of the same order as dimension $6$ contact terms that should be included in general and represent an ambiguity in the correctly factorising proposal above. This is responsible for the last equivalence above (which holds for $s\rightarrow 0$). For future convenience, I repeat the $s$-channel graviton exchange residue:
\begin{align}
\frac{m^2M^2}{M_{Pl}^2}\left(\left(\frac{x_{34}}{x_{12}}\right)^2+\left(\frac{x_{12}}{x_{34}}\right)^2\right)\sim\frac{\frac{1}{4}\left(t-u\right)^2-2m^2M^2}{M_{Pl}^2}.
\end{align}
This is positive in the complexified forward limit $u\rightarrow(m+M)^2$ and $t\rightarrow (m-M)^2$, as required by unitarity \cite{Trott:2026cjj}.

In extended SUGRA, the graviphotons typically do not couple minimally to massive particles with central charges.  
For the fermion amplitudes in (\ref{HyperSUGRA3}), the spinors have opposite chirality to minimal coupling:
\begin{align}
A(\psi,\overline{\psi},\gamma^+)&=-\frac{m}{M_{Pl}}\frac{1}{x}\ds{\bf{12}}\nonumber\\
A(\psi,\overline{\psi},\gamma^-)&=-\frac{m}{M_{Pl}}x\da{\bf{12}}.
\end{align}
These fermions do not have a dipole moment (so the gyromagnetic ratio $g=0$). 
The typical graviphoton couplings to massive vector bosons will be elaborated upon further below in Sections \ref{sec:Review3P}, \ref{sec:BPSGrav} and \ref{sec:N=4SSSB}.

Using either the Feynman rules or constructing from factorisation, the Moller amplitude for fermion scattering by graviphoton exchange is
\begin{align}\label{MollerGraviPhot}
A(\psi,\overline{\psi},\psi',\overline{\psi'})=\frac{\frac{1}{2}(t-u)\mathbf{\Pi}_s+mM\left(\mathbf{\Pi}_t+\mathbf{\Pi}_u\right)}{M_{Pl}^2s}.
\end{align}
Possible contact terms have been dropped - dimension $6$ four fermion contact interactions contribute at the same order in high energy scaling and should be included in general. The $s$-channel residue is 
\begin{align}
-\frac{mM}{M_{Pl}^2}\left(\frac{x_{34}}{x_{12}}\ds{\bf{12}}\da{\bf{34}}+\frac{x_{12}}{x_{34}}\ds{\bf{34}}\da{\bf{12}}\right)\sim-\frac{\frac{1}{2}(t-u)\mathbf{\Pi}_s+mM\left(\mathbf{\Pi}_t+\mathbf{\Pi}_u\right)}{M_{Pl}^2}.
\end{align}
Under crossing fermions $1$ and $2$ to incoming, the term proportional to $\mathbf{\Pi}_s$ flips sign, but the $\mathbf{\Pi}_t+\mathbf{\Pi}_u$ term does not, just as for the SQED component amplitudes. Notably, this latter term has the opposite sign in (\ref{MollerGraviPhot}) than the corresponding Lorentz structure appears with in the minimally coupled QED amplitude (\ref{FermionQED}). It is therefore not correct to identify the $\mathbf{\Pi}_t+\mathbf{\Pi}_u$ and $\mathbf{\Pi}_s$ terms as respective contributions from minimal coupling and anomalous dipole moments. This is instead a feature of the general relation between the residue for graviphoton mediated scattering and minimally coupled electromagnetic (``E/M'') scattering (with the same charges). If the graviphoton has couplings to a spin-$s$ particles $S_s$ of the form
\begin{align}\label{graviphoton3pt}
A(S_s,\overline{S}_s,\gamma^+)&=-\frac{m^{2(1-s)}}{M_{Pl}}\frac{1}{x}\da{\bf{12}}^{2s-1}\ds{\bf{12}}\nonumber\\
A(S_s,\overline{S}_s,\gamma^-)&=-\frac{m^{2(1-s)}}{M_{Pl}}x\ds{\bf{12}}^{2s-1}\da{\bf{12}},
\end{align}
then on the exchange residue,
\begin{align}\label{GraviphotonRec}
A_{\text{graviphoton}}\left(S_s,\overline{S}_s,S_{s'},\overline{S}_{s'}\right)&\sim -A_{\text{E/M}}\left(S_s,\overline{S}_s,S_{s'},\overline{S}_{s'}\right)\nonumber\\
&\qquad\qquad+A_{\text{E/M}}\left(S_{s-\frac{1}{2}},\overline{S}_{s-\frac{1}{2}},S_{{s'}-\frac{1}{2}},\overline{S}_{{s'}-\frac{1}{2}}\right)\mathbf{\Pi}_s.
\end{align}
The couplings (\ref{graviphoton3pt}) are not general (although they frequently appear), I am just using it to illustrate the pattern, which will also appear in Section \ref{sec:BPSGrav}. Parenthetically, for these amplitudes, the anomalous gyromagnetic ratio is $-1/s$ and the highest possible multipole moment is zero. 

From the hypermultiplet superamplitude, the purely gravitational amplitude for fermion scattering may be extracted by subtracting off the graviphoton exchange terms, leaving:
\begin{align}
A(\psi,\overline{\psi},\psi',\overline{\psi'})=\frac{\frac{1}{2}(t-u)(\mathbf{\Pi}_t+\mathbf{\Pi}_u)+mM\mathbf{\Pi}_s}{M_{Pl}^2s}.
\end{align}
Again, this is ambiguous up to the inclusion of dimension-$6$ contact terms. As for the scalar case, it is also possible to easily construct this directly from factorisation and recurse to QED processes. The $s$-channel residue is 
\begin{align}\label{FermionGravRes}
\frac{mM}{M_{Pl}^2}&\left(\left(\frac{x_{34}}{x_{12}}\right)^2\da{\bf{12}}\ds{\bf{34}}+\left(\frac{x_{12}}{x_{34}}\right)^2\da{\bf{34}}\ds{\bf{12}}\right)\nonumber\\
&=\frac{mM}{M_{Pl}^2}\left(\left(\frac{x_{34}}{x_{12}}\da{\bf{12}}\ds{\bf{34}}+\frac{x_{12}}{x_{34}}\da{\bf{34}}\ds{\bf{12}}\right)\left(\frac{x_{34}}{x_{12}}+\frac{x_{12}}{x_{34}}\right)-\left(\ds{\bf{12}}\da{\bf{34}}+\da{\bf{12}}\ds{\bf{34}}\right)\right)\nonumber\\
&\sim -\frac{\frac{1}{2}(t-u)(\mathbf{\Pi}_t+\mathbf{\Pi}_u)+mM\mathbf{\Pi}_s}{M_{Pl}^2}.
\end{align}
This makes the double copy origin of the kinematic structure of the residue clear - the leading spin term is the product of photon exchange residues between fermions and scalars, while the second term is the removal of the scalar exchanges expected to be produced by the double copy.

Gravitational fermion-scalar scattering can be extracted from (\ref{HyperSUGRA}). The contribution from the delta function to this component amplitude is given by the corresponding residue in the SQED case (which contains contributions from both photon and scalar exchange):
\begin{align}\label{sfHyperSUGRA}
A(\psi,\overline{\psi},\varphi,\overline{\varphi})=(\left(t-u\right)+2mM)\left(M\left(\ds{\bf{12}}+\da{\bf{12}}\right)+\la{\bf{1}}p_4\rs{\bf{2}}+\la{\bf{2}}p_4\rs{\bf{1}}\right)\frac{1}{2M_{Pl}^2s}.
\end{align}
The graviphoton exchange is responsible for the spin-$1$ terms 
\begin{align}
A(\psi,\overline{\psi},\varphi,\overline{\varphi})=\left(\left(\frac{1}{2}\left(t-u\right)\right)\left(\ds{\bf{12}}+\da{\bf{12}}\right)+m\left(\la{\bf{1}}p_4\rs{\bf{2}}+\la{\bf{2}}p_4\rs{\bf{1}}\right)\right)\frac{M}{M_{Pl}^2s}
\end{align}
and the graviton exchange terms are the remainder
\begin{align}
A(\psi,\overline{\psi},\varphi,\overline{\varphi})=\left(\frac{1}{2}(t-u)(\la{\bf{1}}p_4\rs{\bf{2}}+\la{\bf{2}}p_4\rs{\bf{1}})+mM^2(\ds{\bf{12}}+\da{\bf{12}})\right)\frac{1}{M_{Pl}^2s}.
\end{align}
The residue of the latter is given by
\begin{align}\label{MixedGravRes}
\frac{mM^2}{M_{Pl}^2}\left(\left(\frac{x_{34}}{x_{12}}\right)^2\da{\bf{12}}+\left(\frac{x_{12}}{x_{34}}\right)^2\ds{\bf{12}}\right)&\sim-\frac{t-u}{2M_{Pl}^2}(\la{\bf{1}}p_4\rs{\bf{2}}+\la{\bf{2}}p_4\rs{\bf{1}})\nonumber\\
&\qquad\qquad\qquad\qquad\qquad-\frac{mM^2}{M_{Pl}^2}\left(\ds{\bf{12}}+\da{\bf{12}}\right).
\end{align}

The $\mathcal{N}=1$ SUGRA superamplitude for chiral multiplet scattering can be extracted from the $\mathcal{N}=2$ superamplitude in an analogous way. This is given by
\begin{align}
\mathcal{A}\left(\Phi,\overline{\Phi},\Phi',\overline{\Phi'}\right)&=\frac{-1}{M_{Pl}^2s}\delta^{(2)}\left(Q^\dagger\right)\left(\frac{1}{2}\left(t-u\right)\left(\mathcal{P}_t+\mathcal{P}_u\right)+mM\left(\mathcal{P}_{s12}+\mathcal{P}_{s34}\right)\right).
\end{align}
The factorisation channel on-shell has the same form as (\ref{N=1SQEDs}), differing by the squaring of the $x$ factors (and an overall factor of $1/M_{Pl}^2$). The residue has the same double-copy structure as the $\mathcal{N}=0$ cases explained above. 

I will end this Section by explaining how $\mathcal{N}=2$ SUGRA accommodates having two central charges despite having only a single graviphoton in the graviton multiplet. When the central charge has a non-zero phase, the phase factors present in the SQED superamplitudes (\ref{N=2SQED3}) must also be introduced into (\ref{HyperSUGRA3}) (where they have already been stated) in order to ensure that the matter retains its universal and consistent coupling to the graviton. However, while in SQED, the phase imprints itself harmlessly in the Yukawa couplings, in SUGRA it analogously appears in the graviphoton's couplings, indicating that they are dyonic. This is to be expected: in $\mathcal{N}=2$ SUGRA, there is only a single graviphoton in the graviton multiplet, so only one real parameter in the central charge can be electric (the other must be magnetic). Perturbative consistency can be restored, however, by introducing a new graviphoton that can provide the necessary degrees of freedom to ``gauge'' a second central charge. This must be part of a new vector multiplet and have a coupling to matter with a strength and phase chosen to cancel the imaginary part of the original graviphoton's coupling. As mentioned in Section \ref{sec:ExtSUGRAder}, this is an example in which the SUSY eigenbasis of photon states is not one where the electric charges are represented as real. An ``electric-magnetic duality'' transformation is required to convert the states to a basis in which the couplings are manifestly electric. 

\begin{figure}[h]
\begin{fmffile}{HyperMoller}
\begin{center}
\begin{tabular}{ c c c }
& & \\
\begin{fmfgraph*}(120,80)
   \fmfleft{i1,i2}
   \fmfright{o1,o2}
   \fmf{plain}{i2,v1}
   \fmf{plain}{v2,o2}
   \fmf{plain}{i1,v1}
   \fmf{plain}{v2,o1}
   \fmf{dbl_wiggly,label=$\mp\quad\pm$}{v1,v2}
   \fmfv{decor.shape=circle,decor.filled=gray50,decor.size=0.15w}{v1,v2}
   \fmflabel{$i$}{i1}
   \fmflabel{$j$}{i2}
   \fmflabel{$k$}{o2}
   \fmflabel{$l$}{o1}
   \fmfv{label=$\delta_{i\bar{j}}$,label.dist=0.1w}{v1}
   \fmfv{label=$\delta_{k\bar{l}}$,label.dist=0.1w}{v2}
   \fmfv{decor.shape=circle,decor.filled=gray50,decor.size=0.15w}{v1,v2}
 \end{fmfgraph*} &\,&
\begin{fmfgraph*}(120,80)
   \fmfleft{i1,i2}
   \fmfright{o1,o2}
   \fmf{plain}{i2,v1}
   \fmf{plain}{v2,o2}
   \fmf{plain}{i1,v1}
   \fmf{plain}{v2,o1}
   \fmf{zigzag,label=$\mp\quad\pm$}{v1,v2}
   \fmfv{decor.shape=circle,decor.filled=gray50,decor.size=0.15w}{v1,v2}
   \fmflabel{$i$}{i1}
   \fmflabel{$j$}{i2}
   \fmflabel{$k$}{o2}
   \fmflabel{$l$}{o1}
   \fmfv{label=$\delta_{i\bar{j}}$,label.dist=0.1w}{v1}
   \fmfv{label=$\delta_{k\bar{l}}$,label.dist=0.1w}{v2}
   \fmfv{decor.shape=circle,decor.filled=gray50,decor.size=0.15w}{v1,v2}
 \end{fmfgraph*}\\
 \,\\
\end{tabular}
\end{center}
\end{fmffile}
\caption{Superfactorisation channels for Moller scattering.}
\end{figure}
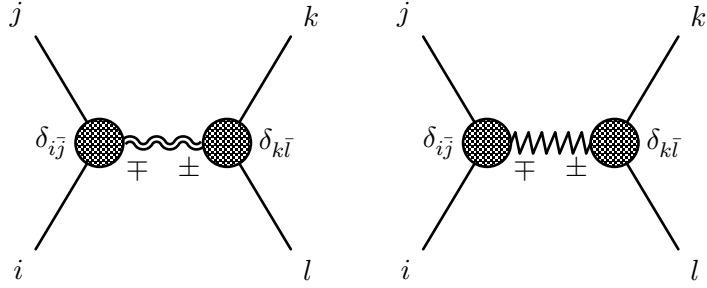

The Moller scattering calculation for $\mathcal{N}=2$ SUGRA above in (\ref{HyperSUGRA}) can now be repeated for the case of hypermultiplets with misaligned central charges. This requires keeping the explicit factors of $e^{-i\varphi_1}$ in (\ref{HyperSUGRA3}). Note that the factor $x$ is related to the little group spinors through (\ref{xExp}), which also contains a factor of the mass phase. The graviton exchange residue can be written as
\begin{align}\label{GenGravExHyper}
&\mathcal{A}(K,\overline{K},K',\overline{K'})\nonumber\\
&\sim\frac{-1}{s}\frac{mM}{M_{Pl}^2}\epsilon\left(\mathcal{Q},\mathcal{Q},\mathcal{Q},\mathcal{Q}\right)\frac{e^{-i\left(\varphi_1+\varphi_3\right)}}{\mathbf{v}^{(12)}\cdot\mathbf{v}^{(34)}}\nonumber\\
&\qquad\qquad\qquad\times\bigg(\mathbf{v}^{(12)}\cdot\mathbf{v}^{(34)}\left(\frac{x_{34}}{x_{12}}+\frac{x_{12}}{x_{34}}+2\cos\left(\varphi_1-\varphi_3\right)\right)\nonumber\\
&\qquad\qquad\qquad\qquad\qquad\qquad\qquad+e^{i\left(\varphi_1-\varphi_3\right)}\mathbf{v}^{(12)+}\mathbf{v}^{(34)-}-e^{-i\left(\varphi_1-\varphi_3\right)}\mathbf{v}^{(12)-}\mathbf{v}^{(34)+}\bigg).
\end{align}
The phases in the last line obstruct the accompanying little group spinors from canceling, unlike in the line above. The residue is therefore, by itself, inconsistent. However, other photon multiplet exchanges can be introduced in order to cancel these phases. 

The couplings of these massless vectors to the hypermultiplets $K$ and $\overline{K}$ are still as they are in (\ref{N=2SQED3}), but with further factors of coupling constants $q_ie^{\pm i\phi_i}$ for each photon multiplet $i$ with helicity $\pm$, where $q_i$ are real (a basis of photon states always exists in which the couplings have this form, see comments in Section \ref{sec:ExtSUGRAder} and \cite{Trott:2026cjj}). I will likewise denote the analogous charges of the $K'$ pair of hypermultiplets as $q'_ie^{\pm i\phi'_i}$. Including these vector exchanges produces residue
\begin{align}
&\mathcal{A}(K,\overline{K},K',\overline{K'})\nonumber\\
&\sim\frac{-1}{s}\epsilon\left(\mathcal{Q},\mathcal{Q},\mathcal{Q},\mathcal{Q}\right)\frac{e^{-i\left(\varphi_1+\varphi_3\right)}}{\mathbf{v}^{(12)}\cdot\mathbf{v}^{(34)}}\nonumber\\
&\qquad\times\left(\left(\sum_iq_iq'_ie^{i\left(\phi_i-\phi'_i\right)}\right)\mathbf{v}^{(12)+}\mathbf{v}^{(34)-}-\left(\sum_iq_iq'_ie^{-i\left(\phi_i-\phi'_i\right)}\right)\mathbf{v}^{(12)-}\mathbf{v}^{(34)+}\right)
\end{align}
which should be added to (\ref{GenGravExHyper}). Consistency is restored on the condition that the explicit presence of the little group spinors can be removed, or equivalently, that the combined coefficients of $\mathbf{v}^{(12)+}\mathbf{v}^{(34)-}$ and $-\mathbf{v}^{(12)-}\mathbf{v}^{(34)+}$ in the numerator be equal. Assembling the couplings into complex charge vectors of the form
\begin{align}\label{ChargeVec}
\mathfrak{q}=(\frac{m}{M_{Pl}}e^{-i\varphi_1},q_1e^{-i\phi_1},q_2e^{-i\phi_2}\ldots)\qquad\text{and}\qquad \mathfrak{q}'=(\frac{M}{M_{Pl}}e^{-i\varphi_3},q'_1e^{-i\phi'_1},q'_2e^{-i\phi'_2}\ldots),
\end{align}
then consistency demands
\begin{align}\label{ElectricCharge}
\mathfrak{q}\cdot \mathfrak{q}'=\mathfrak{q}'\cdot \mathfrak{q},
\end{align}
This is simply the condition that the charges are purely electric, as required by the consistency of the non-supersymmetric Moller scattering residue \cite{Trott:2026cjj}. This can be expressed as the requirement that there exists some unitary matrix $U\in U(1+n_\gamma)$ (where $n_\gamma\geq 1$ is the number of massless vector multiplets) with respect to which the couplings for all of the hypermultiplets have the form
\begin{align}
\mathfrak{q}=U(q_0,q_1,q_2,\ldots)
\end{align}
where $q_i\in\mathbb{R}$ (I am ignoring possible requirements of charge quantisation here). Accepting (\ref{ElectricCharge}), the complete superamplitude then reduces to 
\begin{align}\label{FullHyperMoller}
\mathcal{A}(K,\overline{K},K',\overline{K'})=
\frac{1}{s}\epsilon\left(\mathcal{Q},\mathcal{Q},\mathcal{Q},\mathcal{Q}\right)e^{-i(\varphi_1+\varphi_3)}\left(\frac{t-u+2mM\cos\left(\varphi_1-\varphi_3\right)}{2M_{Pl}^2}+\mathfrak{q}\cdot\mathfrak{q}'\right).
\end{align}

\section{Vector Bosons and $\mathcal{N}=2$ Supersymmetry}\label{sec:N=2}

In \cite{Trott:2026cjj}, special classes of theories with massive vector bosons were identified from the conditions under which their $4$-particle scattering amplitudes had tempered high-energy dependence. These theories constitute a privileged subset of a broader EFT landscape in within which they are distinguished by enhanced descriptive power at short distances. Some of the properties (such as the Yang-Mills Lie algebra and assembling of particles into unitary representations thereof) of these special theories agreed with those required for massless particles, for which they arose as a direct requirement of consistency with Lorentz invariance, unitarity and locality in the form of consistent factorisation of tree-level amplitudes. Others, however, were special to massive vector bosons (GCS couplings, non-semisimple or non-compact Lie algebra, emergent covariant Yukawa couplings, the Higgs mechanism). With the highly constraining $\mathcal{N}=4$ SUSY, it was shown in Section \ref{N=4SuperFact} that the Lie algebraic Yang-Mills structure was instead a necessary consequence of consistent superfactorisation of superamplitudes of BPS vector supermultiplets. It is then natural to inquire about the compromising case of $\mathcal{N}=2$ SUSY. With less SUSY, the vector boson couplings are less restricted, but the superamplitudes of BPS multiplets still possess the intriguing factorisation properties described in Section \ref{N=4SuperFact}, meaning that they should still be constrained by consistent superfactorisation. In this Section, I will calculate the $4$-particle superamplitude residues for massive BPS vector and hypermultiplets from unitarity and derive self-consistency constraints on the $3$-particle couplings, thereby determining the extent to which the results of \cite{Trott:2026cjj} become obligatory.

First however, I will remark on what is not studied in this Section. The permissible $3$-particle $\mathcal{N}=2$ superamplitudes of long vector multiplets are examined in Appendix \ref{N=2Rigid}. SUSY fixes the three vector component amplitude to have the same structure as in standard spontaneously broken Yang-Mills, despite allowing for the inclusion of some additional effective interactions between the other states. I will leave a calculation of the $4$-particle superamplitudes in this theory for further work, as well as a derivation of the $\mathcal{N}=2$ supersymmetrisation of the Higgs mechansim. However, such a calculation would simply assemble and combine a simplification of the general $\mathcal{N}=0$ results derived in \cite{Trott:2026cjj} into a unified (and elegant) ensemble. The Lie algebra, its representations and the Higgs mechanism would still emerge from demands for supressed high-energy scaling and not from infrared self-consistency. Similarly, some further comments on $\mathcal{N}=1$ superamplitudes are made in Appendix \ref{N=1Rigid}, but I will not embark here on a study adapting the results of \cite{Trott:2026cjj} to $\mathcal{N}=1$ SUSY. Unlike the case of long $\mathcal{N}=2$ multiplets, many of the distinct $3$-particle structures in the general $\mathcal{N}=0$ amplitudes are still permitted in this case.

\subsection{Review of non-supersymmetric 3-particle amplitudes}\label{sec:Review3P}

For convenience, I reproduce and summarise here the general non-supersymmetric $3$-particle amplitudes of massive vector bosons \cite{Trott:2026cjj}. I assume in these expressions that the particles are self-conjugate, although a complex basis of central charge eigenstates will be assumed in remainder of this Section into which the relevant expressions can be converted using (\ref{ConjBas}). The general three vector amplitude is given by
\begin{align}\label{3legVec}
&A(W_A,W_B,W_C)=\frac{ig_{ABC}}{\Lambda^2}\ds{\bf{12}}\ds{\bf{23}}\ds{\bf{31}}+\frac{i(g_{ABC})^*}{\Lambda^2}\da{\bf{12}}\da{\bf{23}}\da{\bf{31}}\nonumber\\
&\qquad\qquad\qquad\quad+(-1)^{n_1+n_2}\frac{i{f}_{AB}^{\,\,\,\,\,\,\,\,\,C}}{2m_1m_2}\ds{\bf{23}}\ds{\bf{13}}\da{\bf{12}}+(-1)^{n_2+n_3}\frac{i{f}_{BC}^{\,\,\,\,\,\,\,\,\,A}}{2m_2m_3}\ds{\bf{31}}\ds{\bf{21}}\da{\bf{23}}\nonumber\\&\qquad\qquad\qquad\quad+(-1)^{n_1+n_3}\frac{i{f}_{CA}^{\,\,\,\,\,\,\,\,\,B}}{2m_3m_1}\ds{\bf{12}}\ds{\bf{32}}\da{\bf{31}}+(-1)^{n_1+n_2}\frac{i({f}_{AB}^{\,\,\,\,\,\,\,\,\,C})^*}{2m_1m_2}\da{\bf{23}}\da{\bf{13}}\ds{\bf{12}}\nonumber\\
&\qquad\qquad\qquad\quad+(-1)^{n_2+n_3}\frac{i({f}_{BC}^{\,\,\,\,\,\,\,\,\,A})^*}{2m_2m_3}\da{\bf{31}}\da{\bf{21}}\ds{\bf{23}}+(-1)^{n_1+n_3}\frac{i({f}_{CA}^{\,\,\,\,\,\,\,\,\,B})^*}{2m_3m_1}\da{\bf{12}}\da{\bf{32}}\ds{\bf{31}},
\end{align}
where $g_{ABC}$ is fully antisymmetric in its indices, 
\begin{align}\label{YMantisym}
{f}_{AB}^{\,\,\,\,\,\,\,\,\,C}=-{f}_{BA}^{\,\,\,\,\,\,\,\,\,C}
\end{align}
and the amplitude is invariant under
\begin{align}\label{GCSShiftInvar}
\Im f_{AB}^{\,\,\,\,\,\,\,\,\,C}\mapsto \Im f_{AB}^{\,\,\,\,\,\,\,\,\,C}+a_{ABC}\qquad\Im f_{BC}^{\,\,\,\,\,\,\,\,\,A}\mapsto \Im f_{BC}^{\,\,\,\,\,\,\,\,\,A}+a_{ABC}\qquad\Im f_{CA}^{\,\,\,\,\,\,\,\,\,B}\mapsto \Im f_{CA}^{\,\,\,\,\,\,\,\,\,B}+a_{ABC}
\end{align}
for any fully antisymmetric collection of constants $a_{ABC}\in\mathbb{R}$. This representation of the amplitude has been chosen to make manifest unitarity and particle exchange symmetries. Other representations are possible by making use of the identity \cite{Durieux:2020gip}
\begin{align}\label{3PMassRed}
&(-1)^{n_2+n_3}m_1\ds{\bf{12}}\da{\bf{23}}\ds{\bf{31}}+(-1)^{n_1+n_3}m_2\ds{\bf{23}}\da{\bf{31}}\ds{\bf{12}}+(-1)^{n_1+n_2}m_3\ds{\bf{31}}\da{\bf{12}}\ds{\bf{23}}\nonumber\\
-&(-1)^{n_2+n_3}m_1\da{\bf{12}}\ds{\bf{23}}\da{\bf{31}}-(-1)^{n_1+n_3}m_2\da{\bf{23}}\ds{\bf{31}}\da{\bf{12}}-(-1)^{n_1+n_2}m_3\da{\bf{31}}\ds{\bf{12}}\da{\bf{23}}\nonumber\\
&\qquad\qquad\qquad\qquad\qquad\qquad\qquad\qquad\qquad\qquad\qquad\qquad\qquad\qquad\qquad\qquad\qquad=0.
\end{align}
For example, this is required to obtain the representation of the amplitude embedded within the $\mathcal{N}=1$ three vector superamplitude presented in Appendix \ref{N=1Rigid}. The vector boson component amplitudes contained within some of the gravitino multiplet interactions described in Section \ref{sec:N=4SSSB} provide another example. The standard spontaneously broken Yang-Mills amplitude is obtained through the condition
\begin{align}\label{StandardLA}
f_{AB}^{\,\,\,\,\,\,\,\,\,M}=f_{BM}^{\,\,\,\,\,\,\,\,\,A}=f_{MA}^{\,\,\,\,\,\,\,\,\,B}=f_{ABM},
\end{align}
which is fully antisymmetric (and $g_{ABC}=0$). This implies that the imaginary parts of the $f$-type couplings are zero. Of course, YM theory still requires that these couplings obey the Jacobi identity, but this is not apparent from the $3$-particle amplitude alone.

When one of the vectors is massless, the three vector amplitude becomes 
\begin{align}\label{2Mass1MasslessVec}
A\left(W_A,W_B,g_C^+\right)&=i(-1)^{n_1+n_3}\,\Re f_{BC}^{\,\,\,\,\,\,\,\,\,A}\frac{1}{mx}\da{\bf{12}}^2\nonumber\\
&\qquad\qquad\qquad-i(-1)^{n_1+n_2}\left(f_{BC}^{\,\,\,\,\,\,\,\,\,A}-f_{AB}^{\,\,\,\,\,\,\,\,\,C}\right)\frac{1}{2m^2}\ds{3\bf{2}}\ds{3\bf{1}}\da{\bf{12}}\nonumber\\
A\left(W_A,W_B,g_C^-\right)&=i(-1)^{n_1+n_3}\,\Re f_{BC}^{\,\,\,\,\,\,\,\,\,A}\frac{x}{m}\ds{\bf{12}}^2\nonumber\\
&\qquad\qquad\qquad-i(-1)^{n_1+n_2}\left((f_{BC}^{\,\,\,\,\,\,\,\,\,A})^*-(f_{AB}^{\,\,\,\,\,\,\,\,\,C})^*\right)\frac{1}{2m^2}\da{3\bf{2}}\da{3\bf{1}}\ds{\bf{12}},
\end{align}
where I omit the less interesting $g_{ABC}$ terms. This requires that the couplings further obey
\begin{align}\label{PartAntiSym}
f_{CB}^{\,\,\,\,\,\,\,\,\,A}=-f_{CA}^{\,\,\,\,\,\,\,\,\,B}.
\end{align}

For a general theory of massive vector bosons, it was shown in \cite{Trott:2026cjj} that, demanding suppressed, but not necessarily unitarised high energy growth of the $4$-particle amplitude, the parity-conserving parts of the couplings $\Re f_{AB}^{\,\,\,\,\,\,\,\,\,C}$ had to obey the Jacobi identity (but not necessarily (\ref{StandardLA})), while the parity-violating parts $\Im f_{AB}^{\,\,\,\,\,\,\,\,\,C}$ had to obey the non-Abelian Generalised Chern-Simons constraint (this will be stated further below). When a lowered external colour index in these relations corresponds to a massless vector boson, these properties are instead demanded by consistent factorisation of $4$-particle amplitudes. This can be established using the soft limit procedure described earlier in Section \ref{sec:MassiveMatter}. 

Several archetypical Lie algebra structures that commonly arise in the simple SUGRA examples below can be exemplified in (\ref{2Mass1MasslessVec}). In case of standard spontaneously broken YM with $\mathfrak{su}_2\rightarrow\mathfrak{u}_1$ and (\ref{StandardLA}), only the ``minimal coupling'' terms are non-zero in (\ref{2Mass1MasslessVec}). For the analogous non-compact case $\mathfrak{su}_{1,1}\rightarrow\mathfrak{u}_1$ with otherwise homogeneous generator normalisations, then ignoring the possible GCS terms, $\Re f_{AB}^{\,\,\,\,\,\,\,\,\,C}=-\Re f_{BC}^{\,\,\,\,\,\,\,\,\,A}$ and so the Lorentz structures in (\ref{2Mass1MasslessVec}) reduce to $A\left(W_A,W_B,g_C^\pm\right)\propto x^{\mp 1}\da{\bf{12}}\ds{\bf{12}}$. The special case of a three-dimensional Lie algebra with $\Re f_{AB}^{\,\,\,\,\,\,\,\,\,C}=0$ (and $\Re f_{BC}^{\,\,\,\,\,\,\,\,\,A}\neq 0$) corresponds to the non-semisimple $\mathfrak{e}_{2}\rightarrow\mathfrak{u}_1$, where $\mathfrak{e}_2$ is the isometry algebra of the Euclidean plane (the massive vectors correspond to the translation generators and the massless vector is the rotation). Vectors corresponding to central extensions couple exclusively through multipole moments. However, the converse need not be true. Generally, the vector bosons can be identified with mixtures of generators for which no unitary transformation exists in which they decompose into separate ideals or subalgebras. 

The general amplitude between a scalar and a pair of vector bosons is
\begin{align}\label{2vec1scalar}
A\left(W_A,W_B,\varphi_i\right)=\frac{c_{AB}^i}{\Lambda}\ds{\bf{12}}^2+\frac{(c_{AB}^i)^*}{\Lambda}\da{\bf{12}}^2+(-1)^{n_1+n_2}\frac{1}{\sqrt{2}}\lambda^i_{AB}\ds{\bf{12}}\da{\bf{12}}.
\end{align}
The last term describes Higgs-like couplings, while the first two describe effective interactions of the schematic form $\phi F^2$. Finally, the coupling of a massive vector to fermions is
\begin{align}\label{FermionMat3leg}
A(\psi_i,\psi_j,W_A)&=-(-1)^{n_3}(t_A)_{ij}\frac{1}{m_3}\da{\bf{13}}\ds{\bf{23}}+(-1)^{n_1+n_2+n_3}((t_{A})_{ij})^*\frac{1}{m_3}\da{\bf{23}}\ds{\bf{13}}\nonumber\\
&\qquad-(-1)^{n_1}\frac{2(m_A)_{ij}}{\Lambda}\ds{\bf{13}}\ds{\bf{23}}-(-1)^{n_2}\frac{2((m_A)_{ji})^*}{\Lambda}\da{\bf{13}}\da{\bf{23}}.
\end{align}
The first line of terms describe vector and axial vector currents, which are respectively determined by the couplings $\Im(t_A)_{ij}=-\Im(t_A)_{ji}$ and $\Re(t_A)_{ij}=\Re(t_A)_{ji}$, while the second line determines the dipole moments.

\subsection{3-particle superamplitudes}\label{sec:N=23super}

The possible massless $3$-particle superamplitudes with $\mathcal{N}=2$ SUSY are listed in Appendix \ref{N=2Rigid}, along with the superamplitudes for long massive multiplets. In this Section I will present the $3$-particle superamplitudes of massive BPS multiplets. Amplitudes and the double copy of theories with non-compact Lie algebras have recently been studied in $\mathcal{N}=2$ SUGRA in $5d$ in \cite{Chiodaroli:2023tvo}, but I am unaware of any previous work discussing the amplitudes of the general massive $\mathcal{N}=2$ theories here in $4d$.

Just as for $\mathcal{N}=4$ ``broken'' SYM, the $3$-particle $\mathcal{N}=2$ superamplitudes of BPS particles involve the special $3$-particle massive kinematics described in Section \ref{3PSK} and require the special SUSY delta functions described in Section \ref{Sec:N=4SYM}. As explained in Section \ref{sec:N<4}, I will choose to construct the $\mathcal{N}=2$ on-shell superspace so that it embeds in $\mathcal{N}=4$ with the identification of $Q_1$ and $Q_3$ in (\ref{supercharges}) as the two supercharges. 
This identifies the $3$-leg SUSY invariant required for these superamplitudes as $\Delta_v$. The remaining kinematic factors must neutralise the overall tiny group scaling while accounting for the little groups of the external particles. 

The superamplitudes involving hypermultiplets are:
\begin{align}\label{HyperQCD}
\begin{array}{l}
\mathcal{A}\left(K_i,K_j,\mathcal{W}_A\right)\\
\mathcal{A}\left(K_i,k_j,\mathcal{W}_A\right)
\end{array}\Bigg\}&=-(T_A)_{ij}e^{-i(\varphi_3+\varphi_2)}\Delta_v\mathbf{v}_3\nonumber\\
\mathcal{A}\left(K_i,K_j,V_A^+\right)&=-(T_A)_{ij}e^{-i\varphi_2}\Delta_v\mathbf{v}_3^+\nonumber\\
\mathcal{A}\left(K_i,K_j,V_A^-\right)&=-(T_A)_{ij}e^{-i\varphi_2}\Delta_v\mathbf{v}_3^-.
\end{align}
These are uniquely fixed by symmetries (little $k$ indicates a massless hypermultiplet). The SQED superamplitudes presented earlier  in Section \ref{sec:QED} are special cases of these. The massive hypermultiplets are central charge eigenstates, so are not self-conjugate, but identical bosonic exchange symmetry implies that $(T_A)_{ij}=-e^{-i(\varphi_1-\varphi_2)}(T_A)_{ji}$. The phases of the central charges evidently determine the parity structure of these couplings. The kinematic parts of the vector-fermion component amplitudes in the all massive case can be extracted using the map
\begin{align}
\Delta_v \mathbf{v}_3\rightarrow \frac{e^{i\varphi_3}}{m_3}\left(e^{i\varphi_2}\da{\bf{13}}\ds{\bf{23}}-e^{i\varphi_1}\da{\bf{23}}\ds{\bf{13}}\right).
\end{align}
This allows for the coupling constants of the superamplitudes to be matched onto those of their components in (\ref{FermionMat3leg}) (once converted into a basis of central charge eigenstates). In the cases where the vector multiplet is massless, performing this matching allows the couplings $(T_A)_{ij}$ to be identified with self-adjoint Lie algebra generators (which is assumed in the corresponding expressions above), although the fact that they are generators still remains to be shown in the massive case (and will be demonstrated in the next subsection). It is nevertheless possible to conclude, from unitarity of the component amplitudes, that $(T_{\bar{A}})_{\bar{i}\bar{j}}=((T_A)_{ji})^*$. 

The three vector superamplitudes admit multiple possible Lorentz structures, so require a particular basis choice. I will choose to represent these structures as combinations of a single little group spinor (to cancel the tiny group charge of the SUSY delta function) and a regular spacetime spinor bilinear. The possible Lorentz structures are then linear combinations of terms of the form $\mathbf{v}_i\da{\bf{jk}}$ or $\mathbf{v}_i\ds{\bf{jk}}$ for any distinct legs $i,j,k$. These terms contain two redundancies represented by the identities
\begin{align}\label{S3PKRed}
&m_1\mathbf{v}_1\da{\bf{23}}+m_2\mathbf{v}_2\da{\bf{31}}+m_3\mathbf{v}_3\da{\bf{12}}=0\nonumber\\
&m_1e^{-i\varphi_1}\mathbf{v}_1\ds{\bf{23}}+m_2e^{-i\varphi_2}\mathbf{v}_2\ds{\bf{31}}+m_3e^{-i\varphi_3}\mathbf{v}_3\ds{\bf{12}}=0,
\end{align}
leaving four remaining independent Lorentz structures altogether. Extracting the three vector component amplitude and matching these terms onto (\ref{3legVec}) (adapted to a complex basis) fully identifies the coupling constants in the superamplitude, which can be represented most symmetrically as
\begin{align}\label{N=23vec}
&\mathcal{A}\left(\mathcal{W}_A,\mathcal{W}_B,\mathcal{W}_C\right)\nonumber\\
&\,=\frac{-i}{6}e^{-i(\varphi_1+\varphi_2+\varphi_3)}\Delta_v\nonumber\\
&\quad\times\bigg(\mathbf{v}_1\da{\bf{23}}\left(\frac{e^{i\varphi_3}}{m_2}({f}_{\bar{A}\bar{B}}^{\,\,\,\,\,\,\,\,\,\bar{C}})^*+\frac{e^{i\varphi_2}}{m_3}({f}_{\bar{A}\bar{C}}^{\,\,\,\,\,\,\,\,\,\bar{B}})^*\right)-\mathbf{v}_1\ds{\bf{23}}\left(\frac{e^{i\varphi_2}}{m_2}{f}_{AB}^{\,\,\,\,\,\,\,\,\,C}+\frac{e^{i\varphi_3}}{m_3}{f}_{AC}^{\,\,\,\,\,\,\,\,\,B}\right)\nonumber\\
&\quad\quad\quad+\mathbf{v}_2\da{\bf{31}}\left(\frac{e^{i\varphi_1}}{m_3}({f}_{\bar{B}\bar{C}}^{\,\,\,\,\,\,\,\,\,\bar{A}})^*+\frac{e^{i\varphi_3}}{m_1}({f}_{\bar{B}\bar{A}}^{\,\,\,\,\,\,\,\,\,\bar{C}})^*\right)-\mathbf{v}_2\ds{\bf{31}}\left(\frac{e^{i\varphi_3}}{m_3}{f}_{BC}^{\,\,\,\,\,\,\,\,\,A}+\frac{e^{i\varphi_1}}{m_1}{f}_{BA}^{\,\,\,\,\,\,\,\,\,C}\right)\nonumber\\
&\quad\quad\quad+\mathbf{v}_3\da{\bf{12}}\left(\frac{e^{i\varphi_2}}{m_1}({f}_{\bar{C}\bar{A}}^{\,\,\,\,\,\,\,\,\,\bar{B}})^*+\frac{e^{i\varphi_1}}{m_2}({f}_{\bar{C}\bar{B}}^{\,\,\,\,\,\,\,\,\,\bar{A}})^*\right)-\mathbf{v}_3\ds{\bf{12}}\left(\frac{e^{i\varphi_1}}{m_1}{f}_{CA}^{\,\,\,\,\,\,\,\,\,B}+\frac{e^{i\varphi_2}}{m_2}{f}_{CB}^{\,\,\,\,\,\,\,\,\,A}\right)\bigg),
\end{align}
and also reveals a new constraint:
\begin{align}\label{N=2VacRels}
m_1e^{-i\varphi_1}{f}_{BC}^{\,\,\,\,\,\,\,\,\,A}+m_2e^{-i\varphi_2}{f}_{CA}^{\,\,\,\,\,\,\,\,\,B}+m_3e^{-i\varphi_3}{f}_{AB}^{\,\,\,\,\,\,\,\,\,C}=0
\end{align}
that originates from the supersymmetry. In standard YM, (\ref{StandardLA}) implies that (\ref{N=2VacRels}) trivially reduces to central charge conservation, but away from this special case it extends (\ref{PartAntiSym}) to general BPS mass configurations. The identities (\ref{S3PKRed}) can be used to eliminate some terms in (\ref{N=23vec}) to make them easier to combine across factorisation channels when constructing higher leg superamplitudes. Note that, in this complex basis of central charge eigenstates, the parity symmetric and violating parts of the couplings are $\frac{1}{2}\left({f}_{AB}^{\,\,\,\,\,\,\,\,\,C}+({f}_{\bar{A}\bar{B}}^{\,\,\,\,\,\,\,\,\,\bar{C}})^*\right)$ and $\frac{1}{2i}\left({f}_{AB}^{\,\,\,\,\,\,\,\,\,C}-({f}_{\bar{A}\bar{B}}^{\,\,\,\,\,\,\,\,\,\bar{C}})^*\right)$.

On the Coulomb branches of spontaneously broken SYM, the masses of the pairs of $W$ bosons are free parameters (represented in field theory by the VEVs of certain vector multiplets), provided that their couplings obey central charge conservation. This theory is nevertheless identified as a special point (characterised by unitary high energy scaling) in a broader space of EFTs in which the strength and phase of the interactions among the massive vectors (identifiable as the normalisation and signs of generators appearing in the Lie algebra's commutation relations) are also adjustable parameters. Away from (\ref{StandardLA}), (\ref{N=2VacRels}) implies that changing the mass or central charge of a $W$ pair must be further accommodated by an adjustment to the generator normalisations or GCS terms, as would be expected by shifting VEVs in a gauged NL$\Sigma$M. The moduli spaces of vacua are thus partially lifted and shaped by the effective interactions.

Away from the special case of spontaneously broken super-Yang-Mills characterised by (\ref{StandardLA}), the superamplitude (\ref{N=23vec}) contains dimension-$5$ effective amplitudes involving scalars and fermions of the form of those in (\ref{2vec1scalar}) and (\ref{FermionMat3leg}) described in field theory by $\phi F^2$ or $\chi F\chi$ operators. A comparison between this superamplitude and its $\mathcal{N}=1$ counterpart (\ref{N=13vec}) is somewhat analogous to that between the $3$-particle superamplitudes for $\mathcal{N}=4$ SYM and Higgs branch $\mathcal{N}=2$ SYM, reviewed in Appendix \ref{N=2Rigid}. The vector multiplets have the same particle content in each pair of theories. However, the partner amplitudes to the three vector component amplitude are different, in particular the scalar trilinear and possible Higgs couplings. Finally, I mention that in the special case in which one of the vectors is massless, the superamplitude (\ref{N=23vec}) becomes
\begin{align}\label{MasslessVec2MassiveVec}
\mathcal{A}\left(\mathcal{W}_A,\mathcal{W}_B,V_C^+\right)&=\frac{i}{2m}e^{-i\varphi_2}\Delta_v\mathbf{v}_3^+\nonumber\\
&\quad\times\left(\left({f}_{BC}^{\,\,\,\,\,\,\,\,\,A}+({f}_{\bar{B}\bar{C}}^{\,\,\,\,\,\,\,\,\,\bar{A}})^*\right)\da{\bf{12}}-\frac{x}{m}\left({f}_{BC}^{\,\,\,\,\,\,\,\,\,A}-{f}_{AB}^{\,\,\,\,\,\,\,\,\,C}\right)\ds{3\bf{2}}\ds{3\bf{1}}\right)\nonumber\\
\mathcal{A}\left(\mathcal{W}_A,\mathcal{W}_B,V_C^-\right)&=\frac{i}{2m}e^{-i\varphi_2}\Delta_v\mathbf{v}_3^-\nonumber\\
&\quad\times\left(\left({f}_{BC}^{\,\,\,\,\,\,\,\,\,A}+({f}_{\bar{B}\bar{C}}^{\,\,\,\,\,\,\,\,\,\bar{A}})^*\right)\ds{\bf{12}}-\frac{1}{mx}\left(({f}_{\bar{B}\bar{C}}^{\,\,\,\,\,\,\,\,\,\bar{A}})^*-({f}_{\bar{A}\bar{B}}^{\,\,\,\,\,\,\,\,\,\bar{C}})^*\right)\da{3\bf{2}}\da{3\bf{1}}\right).
\end{align}


\subsection{Consistent superfactorisation of 4-particle superamplitudes}\label{sec:N=24Vec}

The simple ``$4$-particle test'' for massive vector multiplets with $\mathcal{N}=4$ SUSY was presented in Section (\ref{N=4SuperFact}), while the well-known examples with massless gluons and other particles have been explained in \cite{Benincasa:2007xk,McGady:2013sga,Arkani-Hamed:2017jhn,Trott:2026cjj}. Formulated generally, a (tree-level) (super)amplitude can be expressed as
\begin{align}\label{GenAmp}
\mathcal{A}=&\frac{N_{stu}}{(s-m_s^2)(t-m_t^2)(u-m_u^2)}\nonumber\\
&\,+\frac{N_{st}}{(s-m_s^2)(t-m_t^2)}+\frac{N_{tu}}{(t-m_t^2)(u-m_u^2)}+\frac{N_{us}}{(u-m_u^2)(s-m_s^2)}\nonumber\\
&\,+\frac{N_{s}}{(s-m_s^2)}+\frac{N_{t}}{(t-m_t^2)}+\frac{N_{u}}{(u-m_u^2)}+\text{contact terms}.
\end{align}
Here $N_{stu}$ etc are the numerators with some dependence on kinematics and coupling constants. Generally, the separation of terms in (\ref{GenAmp}) into those with three, two, one or zero poles is dependent upon the basis choice for the Lorentz structures appearing in the numerators and contact terms. In the examples presented in this Section however, manifest particle exchange symmetries provide guidance toward a privileged basis with respect to which this separation is unambiguous. 

If a factorisation residue contains two cross-channel poles, then consistent factorisation of the three pole term in (\ref{GenAmp}) immediately demands that both cross-channel residues must agree, producing the constraint
\begin{align}\label{GenConsFactTriple}
(t-m_t^2)(u-m_u^2)R_s'=(u-m_u^2)(s-m_s^2)R_t'=(s-m_s^2)(t-m_t^2)R_u'.
\end{align}
Here, $R_i'$ denotes the part of the residue in channel $i=s,t,u$ containing two other Mandelstam poles. These are the terms that $N_{stu}$ in (\ref{GenAmp}) must specifically match onto in each limit. 
Residues containing two poles are absent from most of the examples considered in this study (including those in this Section). They appear only in some facets of the reconstruction of the $\mathcal{N}=4$ super-Higgs mechanism presented in Section \ref{sec:N=4SSSB} below.

Forgoing the three pole $N_{stu}$ structures, consistent factorisation of the two pole $N_{st}$ etc structures necessitates the constraint
\begin{align}\label{GenConsFact}
(u-m_u^2)R_s+(s-m_s^2)R_t+(t-m_t^2)R_u=0.
\end{align}
Here $R_i$ denotes the channel $i$ residue specifically contributing to the two pole terms in (\ref{GenAmp}) (that is, the terms in the residue that contain a single cross-channel pole). In a basis manifesting full particle exchange symmetry, this latter condition amounts to dropping all terms proportional to inverse Mandelstam propagators that would appear if the full residues were substituted into (\ref{GenConsFact}) instead (these would contribute to the single pole terms in (\ref{GenAmp}) that have no influence on cross-channel factorisation). This will be illustrated in the examples below. 
The Mandelstam prefactors in (\ref{GenConsFact}) therefore just cancel the cross-channel poles appearing in the residues.
Amplitudes of $\mathcal{N}=2$ BPS massive vector multiplets are more complicated than the examples analysed previously because the numerators and residues contain multiple independent Lorentz structures. This can consequently yield multiple constraints on the theory, but also makes their extraction nontrivial. 

I will begin by testing Compton scattering of a massive hypermultiplet with massive vectors $\mathcal{A}(\mathcal{W}_A,\mathcal{W}_B,K_i,K_j)$. I do not have a complete derivation of all of the sign flips produced by crossing particles in the superamplitudes relevant here (and a complete derivation is always necessary to correctly derive a binary number, which is why negative signs are a nuisance). However, in this particular example, there is only a single relative sign in the calculation where this could potentially be a problem (and I find that it is not), while for four vector scattering further below, the exchange symmetries are sufficiently stringent to limit the potentially ambiguous signs to a small number that can be feasibly sanity checked against the expected results. The definition of the little group spinors (\ref{SpecialSVD2}), the Dirac bilinears (\ref{SpinFact}) and the subsequent identities in Section \ref{3PSK} are largely unchanged under reinterpreting legs as incoming, although factors of $(-1)^{n_i}$ should be introduced in (\ref{Trimmed3PSK}) to accompany each respective massive spinor $\rs{i_I}$ or $\ra{i_I}$.

\begin{figure}[h]
\begin{fmffile}{N=2ComptonTest}
\begin{center}
\begin{tabular}{ c c c c c }
& & & & \\
 \begin{fmfgraph*}(100,67)
   \fmfleft{i1,i2}
   \fmfright{o1,o2}
   \fmf{boson}{i2,v1}
   \fmf{plain}{v2,o2}
   \fmf{boson}{i1,v1}
   \fmf{plain}{v2,o1}
   \fmf{boson}{v1,v2}
   \fmfv{decor.shape=circle,decor.filled=gray50,decor.size=0.15w}{v1,v2}
   \fmflabel{$A$}{i1}
   \fmflabel{$B$}{i2}
   \fmflabel{$i$}{o2}
   \fmflabel{$j$}{o1}
 \end{fmfgraph*} 
&\,& \begin{fmfgraph*}(100,67)
   \fmfleft{i1,i2}
   \fmfright{o1,o2}
   \fmf{plain}{o1,v1,v2,o2}
   \fmf{phantom}{v1,i1}
   \fmf{phantom}{v2,i2}
   \fmf{boson,tension=-0.25}{v2,i1}
   \fmf{boson,tension=-0.25}{v1,i2}
   \fmfv{decor.shape=circle,decor.filled=gray50,decor.size=0.15w}{v1,v2}
   \fmflabel{$A$}{i1}
   \fmflabel{$B$}{i2}
   \fmflabel{$i$}{o2}
   \fmflabel{$j$}{o1}
 \end{fmfgraph*} &\,&
 \begin{fmfgraph*}(100,67)
   \fmfleft{i1,i2}
   \fmfright{o1,o2}
   \fmf{boson}{i1,v1}
   \fmf{boson}{i2,v2}
   \fmf{plain}{o1,v1,v2,o2}
   \fmfv{decor.shape=circle,decor.filled=gray50,decor.size=0.15w}{v1,v2}
   \fmflabel{$A$}{i1}
   \fmflabel{$B$}{i2}
   \fmflabel{$i$}{o2}
   \fmflabel{$j$}{o1}
 \end{fmfgraph*}\nonumber\\
 $(s)$ & \,&  $(t)$ & \,&  $(u)$
 \end{tabular}
\end{center} 
\end{fmffile}
\caption{Superfactorisation channels for Compton scattering of BPS multiplets.}
\end{figure}
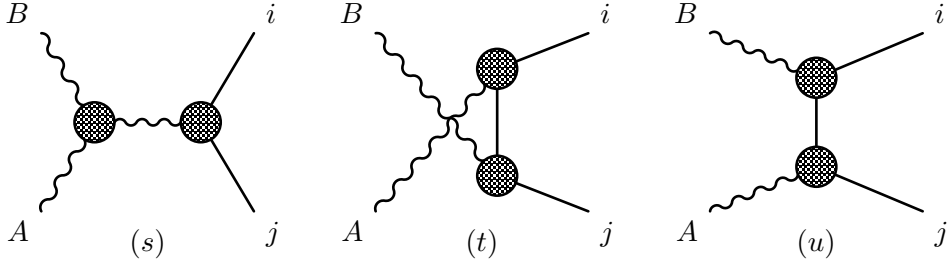
Dropping the overall SUSY delta function for brevity, the $t$-channel residue for the Compton superamplitude is given by
\begin{align}\label{N=2CompRes}
(s-m_s^2)R_t&\sim -e^{i(\varphi_t-\varphi_1-\varphi_2-\varphi_4)}(T_A)_{im}(T_B)_{\bar{m}j}\nonumber\\
&\qquad\qquad\times\bigg(\frac{1}{2}e^{i(\varphi_2-\varphi_t)}\ls{\bf{1}}p_4-p_3\ra{\bf{2}}-\frac{1}{2}e^{i(\varphi_1-\varphi_t)}\ls{\bf{2}}p_4-p_3\ra{\bf{1}}\nonumber\\
&\quad\qquad\qquad\qquad+e^{-i\varphi_t}\left(e^{i\varphi_t}m_t+\frac{1}{2}e^{i\varphi_1}m_1-\frac{1}{2}e^{i\varphi_2}m_2\right)\ds{\bf{12}}\nonumber\\
&\quad\qquad\qquad\qquad-e^{i(\varphi_1+\varphi_2-\varphi_t)}\left(e^{-i\varphi_t}m_t+\frac{1}{2}e^{-i\varphi_1}m_1-\frac{1}{2}e^{-i\varphi_2}m_2\right)\da{\bf{12}}\bigg).
\end{align}
The $u$-channel residue is given by exchanging $1\leftrightarrow 2$ in the expression above. The $s$-channel residue is
\begin{align}\label{N=2Comptons}
(u-m_u^2)R_s&\sim \frac{-i}{4}e^{-i(\varphi_1+\varphi_2+\varphi_4)}(T_{\bar{M}})_{ij}\left({f}_{AB}^{\,\,\,\,\,\,\,\,\,M}+({f}_{\bar{A}\bar{B}}^{\,\,\,\,\,\,\,\,\,\bar{M}})^*\right)\nonumber\\
&\qquad\times\bigg(-e^{i\varphi_2}\ls{\bf{1}}p_4-p_3\ra{\bf{2}}+e^{i\varphi_1}\ls{\bf{2}}p_4-p_3\ra{\bf{1}}\nonumber\\
&\qquad\qquad+\left(e^{i\varphi_1}m_1+e^{i\varphi_2}m_2+2e^{i\varphi_3}m_3\right)\ds{\bf{12}}\nonumber\\
&\qquad\qquad-e^{i(\varphi_1+\varphi_2)}\left(e^{-i\varphi_1}m_1+e^{-i\varphi_2}m_2+2e^{-i\varphi_3}m_3\right)\da{\bf{12}}\bigg).
\end{align}
The phases $\varphi_{s,t,u}$ are those of the exchanged internal multiplet in each respective channel. I define the internal central charges analogously to the internal momenta: $m_se^{i\varphi_s}=m_3e^{i\varphi_3}+m_4e^{i\varphi_4}$, $m_te^{i\varphi_t}=m_2e^{i\varphi_2}+m_4e^{i\varphi_4}$ and $m_ue^{i\varphi_u}=m_1e^{i\varphi_1}+m_4e^{i\varphi_4}$. Note that e.g. $m_se^{i\varphi_s}$ flips sign under an exchange of particles $1$ and $2$ with particles $3$ and $4$. In these residues, I have used (\ref{SpinFact}) and (\ref{SpecialSVD2}) to eliminate the little group spinors. The natural basis of Lorentz structures, chosen to represent the expressions above, is one that manifests the exchange symmetries of the external particles (note that the vector multiplets are fermionic). The terms proportional to inverse Mandelstam propagators can be dropped for the purposes here. Finally, I have applied central charge conservation and (\ref{N=2VacRels}) to simplify the $s$-channel residue. 

Now, reading off the coefficients of $\la{\bf{2}}p_4-p_3\rs{\bf{1}}$ in (\ref{GenConsFact}) gives
\begin{align}\label{N=2LAGen}
(T_A)_{im}(T_B)_{\bar{m}j}-(T_B)_{im}(T_A)_{\bar{m}j}-\frac{i}{2}\left({f}_{AB}^{\,\,\,\,\,\,\,\,\,M}+({f}_{\bar{A}\bar{B}}^{\,\,\,\,\,\,\,\,\,\bar{M}})^*\right)(T_{\bar{M}})_{ij}=0,
\end{align}
which is the expected Lie algebra commutator (assuming that the vector self-couplings are Lie algebra structure constants, which will be established shortly). The three remaining independent Lorentz structures in the residues give constraints that reduce to the same result. The hypermultiplets must therefore arrange into unitary representations of the Lie algebra underpinning self-interactions of the vector multiplets. 

I will now proceed to the superamplitude of four vector multiplets $\mathcal{A}(\mathcal{W}_A,\mathcal{W}_B,\mathcal{W}_C,\mathcal{W}_D)$. Manifest particle exchange symmetries identify privileged bases of Lorentz structures. These consist of terms that, at dimension two, are simply pairs of spinor bilinears, while at dimension three, are the $\mathbf{L}_i$ structures introduced in \cite{Trott:2026cjj}:
\begin{align}\label{L}
\mathbf{L}_1/m_1&=\la{\bf{1}}p_2-p_3\rs{\bf{4}}\ds{\bf{23}}+\la{\bf{1}}p_2-p_4\rs{\bf{3}}\ds{\bf{24}}+\la{\bf{1}}p_4-p_3\rs{\bf{2}}\ds{\bf{43}}\nonumber\\
&\qquad+\la{\bf{1}}p_4-p_2\rs{\bf{3}}\ds{\bf{42}}+\la{\bf{1}}p_3-p_2\rs{\bf{4}}\ds{\bf{32}}+\la{\bf{1}}p_3-p_4\rs{\bf{2}}\ds{\bf{34}},
\end{align}
with $\mathbf{L}_2=\mathbf{L}_1|_{1\leftrightarrow 2}$, $\mathbf{L}_3=\mathbf{L}_1|_{1\leftrightarrow 3}$ and $\mathbf{L}_4=\mathbf{L}_1|_{1\leftrightarrow 4}$. The structure $\mathbf{L}_i$ is explicitly symmetric under exchanges of any pair of particles besides $i$. Parity conjugate structures $\hat{\mathbf{L}}_i$ are defined analogously with the bracket shapes switched.

Higher mass dimension terms are given by dressing these structures with Mandlestam invariants. However, terms with Mandlestam zeros are to be excluded from (\ref{GenConsFact}), so will be dropped from the residues when computed from unitarity. The complication though is that the syzygy 
\begin{align}\label{OldSyz2}
&(t-\frac{1}{3}\sum_im_i^2)\ds{\bf{14}}\ds{\bf{32}}+(u-\frac{1}{3}\sum_im_i^2)\ds{\bf{13}}\ds{\bf{42}}=\frac{1}{12}\sum_i\mathbf{L}_i
\end{align}
implies that the combination $\sum_i \mathbf{L}_i$ is also redundant, in that it is a linear combination of terms with Mandelstam zeros (which should be dropped for the purposes here) and lower dimensional structures. When presenting the residue below in (\ref{RsN=2}), I will retain this combination as it arises in the computation from unitarity, but note that it must be ultimately replaced in the residues in (\ref{GenConsFact}) with the dimension $2$ ``holomorphic'' terms  appearing in (\ref{OldSyz2}).

The $s$-channel residue is 
\begin{align}\label{RsN=2}
&(u-m_u^2)R_s\nonumber\\
&\sim\frac{1}{4}e^{-i\sum_j\varphi_j}\nonumber\\
&\quad\times\Bigg(\bigg(\frac{1}{12}e^{i\sum_j\varphi_j}\frac{e^{-i\varphi_1}}{m_1}\frac{e^{-i\varphi_s}}{m_s}\left({f}_{AB}^{\,\,\,\,\,\,\,\,\,M}+({f}_{\bar{M}\bar{A}}^{\,\,\,\,\,\,\,\,\,\bar{B}})^*\right)\left(({f}_{M\bar{C}}^{\,\,\,\,\,\,\,\,\,\bar{D}})^*+({f}_{M\bar{D}}^{\,\,\,\,\,\,\,\,\,\bar{C}})^*\right)\mathbf{L}_1+(1\leftrightarrow 2)
\nonumber\\
&\qquad\qquad\qquad\qquad\quad+\frac{1}{12}\frac{e^{i\varphi_1}}{m_1}\frac{e^{i\varphi_s}}{m_s}\left(({f}_{\bar{A}\bar{B}}^{\,\,\,\,\,\,\,\,\,\bar{M}})^*+{f}_{MA}^{\,\,\,\,\,\,\,\,\,B}\right)\left({f}_{\bar{M}C}^{\,\,\,\,\,\,\,\,\,D}+{f}_{\bar{M}D}^{\,\,\,\,\,\,\,\,\,C}\right)\hat{\mathbf{L}}_1+(1\leftrightarrow 2)\bigg)\nonumber\\
&\qquad\qquad\qquad\qquad\qquad\qquad\qquad\qquad\qquad\qquad\qquad\qquad\qquad\qquad+(1\leftrightarrow 3)\land(2\leftrightarrow 4)\nonumber\\
&\qquad\qquad+\frac{1}{2}e^{i(\varphi_1+\varphi_3)}\Big(\left({f}_{AB}^{\,\,\,\,\,\,\,\,\,M}+({f}_{\bar{A}\bar{B}}^{\,\,\,\,\,\,\,\,\,\bar{M}})^*\right)\left(({f}_{M\bar{D}}^{\,\,\,\,\,\,\,\,\,\bar{C}})^*-{f}_{\bar{M}C}^{\,\,\,\,\,\,\,\,\,D}\right)\nonumber\\
&\qquad\qquad\qquad\qquad\qquad\qquad+\left({f}_{CD}^{\,\,\,\,\,\,\,\,\,M}+({f}_{\bar{C}\bar{D}}^{\,\,\,\,\,\,\,\,\,\bar{M}})^*\right)\left(({f}_{M\bar{B}}^{\,\,\,\,\,\,\,\,\,\bar{A}})^*-{f}_{\bar{M}A}^{\,\,\,\,\,\,\,\,\,B}\right)\Big)\da{\bf{13}}\ds{\bf{24}}\nonumber\\
&\qquad\qquad\qquad\qquad\qquad\qquad\qquad\qquad+(1\leftrightarrow 2)+(3\leftrightarrow 4)+(1\leftrightarrow 2)\land(3\leftrightarrow 4)\nonumber\\
&\qquad\qquad+\frac{1}{2}e^{i(\varphi_1+\varphi_2)}\Big(\left({f}_{AB}^{\,\,\,\,\,\,\,\,\,M}+({f}_{\bar{A}\bar{B}}^{\,\,\,\,\,\,\,\,\,\bar{M}})^*\right)\left({f}_{\bar{M}C}^{\,\,\,\,\,\,\,\,\,D}-{f}_{\bar{M}D}^{\,\,\,\,\,\,\,\,\,C}\right)\nonumber\\
&\qquad\qquad\qquad\qquad\qquad\qquad+\left({f}_{CD}^{\,\,\,\,\,\,\,\,\,\bar{M}}+({f}_{\bar{C}\bar{D}}^{\,\,\,\,\,\,\,\,\,M})^*\right)\left(({f}_{\bar{M}\bar{A}}^{\,\,\,\,\,\,\,\,\,\bar{B}})^*-({f}_{\bar{M}\bar{B}}^{\,\,\,\,\,\,\,\,\,\bar{A}})^*\right)\nonumber\\
&\qquad\qquad\qquad\qquad\qquad\qquad\qquad+2{f}_{AB}^{\,\,\,\,\,\,\,\,\,M}({f}_{\bar{C}\bar{D}}^{\,\,\,\,\,\,\,\,\,M})^*-2{f}_{CD}^{\,\,\,\,\,\,\,\,\,\bar{M}}({f}_{\bar{A}\bar{B}}^{\,\,\,\,\,\,\,\,\,M})^*\Big)\da{\bf{12}}\ds{\bf{34}}\nonumber\\
&\qquad\qquad\qquad\qquad\qquad\qquad\qquad\qquad\qquad\qquad\qquad+(1\leftrightarrow 3)\land(2\leftrightarrow 4)\nonumber\\
&\qquad\qquad+\frac{e^{i\sum_j\varphi_j}}{m_s^2}\bigg(\left(\da{\bf{13}}\da{\bf{24}}+\da{\bf{14}}\da{\bf{23}}\right)\nonumber\\
&\qquad\qquad\qquad\times\bigg(\frac{1}{2}m_s^2e^{-2i\varphi_s}\left(({f}_{\bar{M}\bar{A}}^{\,\,\,\,\,\,\,\,\,\bar{B}})^*+({f}_{\bar{M}\bar{B}}^{\,\,\,\,\,\,\,\,\,\bar{A}})^*\right)\left(({f}_{\bar{M}\bar{C}}^{\,\,\,\,\,\,\,\,\,\bar{D}})^*+({f}_{\bar{M}\bar{D}}^{\,\,\,\,\,\,\,\,\,\bar{C}})^*\right)\nonumber\\
&\qquad\qquad\qquad\qquad-\frac{1}{3}\Big(m_3m_4e^{-i(\varphi_3+\varphi_4)}\left(({f}_{\bar{M}\bar{A}}^{\,\,\,\,\,\,\,\,\,\bar{B}})^*+({f}_{\bar{M}\bar{B}}^{\,\,\,\,\,\,\,\,\,\bar{A}})^*\right)\left({f}_{\bar{M}C}^{\,\,\,\,\,\,\,\,\,D}+{f}_{\bar{M}D}^{\,\,\,\,\,\,\,\,\,C}\right)\nonumber\\
&\qquad\qquad\qquad\qquad\qquad\qquad+m_1m_2e^{-i(\varphi_1+\varphi_2)}\left({f}_{MA}^{\,\,\,\,\,\,\,\,\,B}+{f}_{MB}^{\,\,\,\,\,\,\,\,\,A}\right)\left(({f}_{M\bar{C}}^{\,\,\,\,\,\,\,\,\,\bar{D}})^*+({f}_{M\bar{D}}^{\,\,\,\,\,\,\,\,\,\bar{C}})^*\right)\Big)\nonumber\\
&\qquad\qquad\qquad\qquad+\frac{1}{6}m_se^{-i\varphi_s}\bigg(\left(({f}_{M\bar{C}}^{\,\,\,\,\,\,\,\,\,\bar{D}})^*+({f}_{M\bar{D}}^{\,\,\,\,\,\,\,\,\,\bar{C}})^*\right)\nonumber\\
&\qquad\qquad\qquad\qquad\times\left(m_1e^{-i\varphi_1}\left(({f}_{\bar{M}\bar{A}}^{\,\,\,\,\,\,\,\,\,\bar{B}})^*-{f}_{MB}^{\,\,\,\,\,\,\,\,\,A}\right)+m_2e^{-i\varphi_2}\left(({f}_{\bar{M}\bar{B}}^{\,\,\,\,\,\,\,\,\,\bar{A}})^*-{f}_{MA}^{\,\,\,\,\,\,\,\,\,B}\right)\right)\nonumber\\
&\qquad\qquad\qquad\qquad\qquad\qquad\qquad-\left(({f}_{\bar{M}\bar{A}}^{\,\,\,\,\,\,\,\,\,\bar{B}})^*+({f}_{\bar{M}\bar{B}}^{\,\,\,\,\,\,\,\,\,\bar{A}})^*\right)\nonumber\\
&\qquad\qquad\qquad\qquad\times\left(m_3e^{-i\varphi_3}\left(({f}_{M\bar{C}}^{\,\,\,\,\,\,\,\,\,\bar{D}})^*-{f}_{\bar{M}D}^{\,\,\,\,\,\,\,\,\,C}\right)+m_4e^{-i\varphi_4}\left(({f}_{M\bar{D}}^{\,\,\,\,\,\,\,\,\,\bar{C}})^*-{f}_{\bar{M}C}^{\,\,\,\,\,\,\,\,\,D}\right)\right)\bigg)\nonumber\\
&\qquad\qquad\qquad\qquad+\da{\bf{12}}\da{\bf{34}}\nonumber\\
&\qquad\qquad\qquad\times\bigg(\frac{1}{2}m_s^2e^{-2i\varphi_s}\left(({f}_{\bar{M}\bar{A}}^{\,\,\,\,\,\,\,\,\,\bar{B}})^*-({f}_{\bar{M}\bar{B}}^{\,\,\,\,\,\,\,\,\,\bar{A}})^*\right)\left(({f}_{\bar{M}\bar{C}}^{\,\,\,\,\,\,\,\,\,\bar{D}})^*-({f}_{\bar{M}\bar{D}}^{\,\,\,\,\,\,\,\,\,\bar{C}})^*\right)\nonumber\\
&\qquad\qquad\qquad\qquad+\left(m_1e^{-i\varphi_1}({f}_{\bar{M}\bar{A}}^{\,\,\,\,\,\,\,\,\,\bar{B}})^*-m_2e^{-i\varphi_2}({f}_{\bar{M}\bar{B}}^{\,\,\,\,\,\,\,\,\,\bar{A}})^*\right)\nonumber\\
&\qquad\qquad\qquad\qquad\qquad\times\left(m_4e^{-i\varphi_4}{f}_{\bar{M}C}^{\,\,\,\,\,\,\,\,\,D}-m_3e^{-i\varphi_3}{f}_{\bar{M}D}^{\,\,\,\,\,\,\,\,\,C}\right)\nonumber\\
&\qquad\qquad\qquad\qquad+\left(m_3e^{-i\varphi_3}({f}_{M\bar{C}}^{\,\,\,\,\,\,\,\,\,\bar{D}})^*-m_4e^{-i\varphi_4}({f}_{M\bar{D}}^{\,\,\,\,\,\,\,\,\,\bar{C}})^*\right)\nonumber\\
&\qquad\qquad\qquad\qquad\qquad\times\left(m_2e^{-i\varphi_2}{f}_{MA}^{\,\,\,\,\,\,\,\,\,B}-m_1e^{-i\varphi_1}{f}_{MB}^{\,\,\,\,\,\,\,\,\,A}\right)\nonumber\\
&\qquad\qquad\qquad\qquad+\frac{1}{2}m_se^{-i\varphi_s}\bigg(\left(({f}_{M\bar{C}}^{\,\,\,\,\,\,\,\,\,\bar{D}})^*-({f}_{M\bar{D}}^{\,\,\,\,\,\,\,\,\,\bar{C}})^*\right)\nonumber\\
&\qquad\qquad\qquad\qquad\times\left(m_1e^{-i\varphi_1}\left(({f}_{\bar{M}\bar{A}}^{\,\,\,\,\,\,\,\,\,\bar{B}})^*+{f}_{MB}^{\,\,\,\,\,\,\,\,\,A}\right)-m_2e^{-i\varphi_2}\left(({f}_{\bar{M}\bar{B}}^{\,\,\,\,\,\,\,\,\,\bar{A}})^*+{f}_{MA}^{\,\,\,\,\,\,\,\,\,B}\right)\right)\nonumber\\
&\qquad\qquad\qquad\qquad\qquad\qquad\qquad-\left(({f}_{\bar{M}\bar{A}}^{\,\,\,\,\,\,\,\,\,\bar{B}})^*-({f}_{\bar{M}\bar{B}}^{\,\,\,\,\,\,\,\,\,\bar{A}})^*\right)\nonumber\\
&\qquad\qquad\qquad\qquad\times\left(m_3e^{-i\varphi_3}\left(({f}_{M\bar{C}}^{\,\,\,\,\,\,\,\,\,\bar{D}})^*+{f}_{\bar{M}D}^{\,\,\,\,\,\,\,\,\,C}\right)+m_4e^{-i\varphi_4}\left(({f}_{M\bar{D}}^{\,\,\,\,\,\,\,\,\,\bar{C}})^*+{f}_{\bar{M}C}^{\,\,\,\,\,\,\,\,\,D}\right)\right)\bigg)\bigg)\nonumber\\
&\qquad\qquad\qquad\qquad+\text{opposite shape brackets with complex conjugate coefficients}\Bigg).
\end{align}
The residues of the other channels are given by permuting particles. In both deriving (\ref{RsN=2}) and the results that will follow, extensive use has been made of central charge conservation and (\ref{N=2VacRels}). 

Because $\sum_i\mathbf{L}_i$ is redundant and can be reduced to lower dimension Lorentz structures, the set $\{\mathbf{L}_i\}$ is not linearly independent. This means that the coefficients of each of the $\mathbf{L}_i$ structures in the sum of the residues (\ref{GenConsFact}) must be equated, rather than set to zero, in order to satisfy (\ref{GenConsFact}). Equating them (e.g. $\mathbf{L}_3$ and $\mathbf{L}_4$) leads to the condition 
\begin{align}\label{ChiralGCS}
&\left({f}_{BC}^{\,\,\,\,\,\,\,\,\,M}+({f}_{\bar{B}\bar{C}}^{\,\,\,\,\,\,\,\,\,\bar{M}})^*\right)({f}_{\bar{A}M}^{\,\,\,\,\,\,\,\,\,\bar{D}})^*+\left({f}_{DA}^{\,\,\,\,\,\,\,\,\,M}+({f}_{\bar{D}\bar{A}}^{\,\,\,\,\,\,\,\,\,\bar{M}})^*\right)({f}_{\bar{B}M}^{\,\,\,\,\,\,\,\,\,\bar{C}})^*\nonumber\\
&\qquad\qquad+{f}_{BC}^{\,\,\,\,\,\,\,\,\,M}({f}_{\bar{A}\bar{D}}^{\,\,\,\,\,\,\,\,\,M})^*-{f}_{AD}^{\,\,\,\,\,\,\,\,\,M}({f}_{\bar{B}\bar{C}}^{\,\,\,\,\,\,\,\,\,M})^*\nonumber\\
&+\left({f}_{BD}^{\,\,\,\,\,\,\,\,\,M}+({f}_{\bar{B}\bar{D}}^{\,\,\,\,\,\,\,\,\,\bar{M}})^*\right)({f}_{\bar{A}M}^{\,\,\,\,\,\,\,\,\,\bar{C}})^*+\left({f}_{CA}^{\,\,\,\,\,\,\,\,\,M}+({f}_{\bar{C}\bar{A}}^{\,\,\,\,\,\,\,\,\,\bar{M}})^*\right)({f}_{\bar{B}M}^{\,\,\,\,\,\,\,\,\,\bar{D}})^*\nonumber\\
&\qquad\qquad+{f}_{BD}^{\,\,\,\,\,\,\,\,\,M}({f}_{\bar{A}\bar{C}}^{\,\,\,\,\,\,\,\,\,M})^*-{f}_{AC}^{\,\,\,\,\,\,\,\,\,M}({f}_{\bar{B}\bar{D}}^{\,\,\,\,\,\,\,\,\,M})^*\nonumber\\
&+\left({f}_{AB}^{\,\,\,\,\,\,\,\,\,M}+({f}_{\bar{A}\bar{B}}^{\,\,\,\,\,\,\,\,\,\bar{M}})^*\right)\left(({f}_{\bar{C}M}^{\,\,\,\,\,\,\,\,\,\bar{D}})^*+({f}_{\bar{D}M}^{\,\,\,\,\,\,\,\,\,\bar{C}})^*\right)=0,
\end{align}
in which the complex conjugates of the couplings appear in every term. The parity conjugate $\hat{\mathbf{L}}_i$ terms produce the complex conjugates of these constraints. Adding (\ref{ChiralGCS}) to its complex conjugate gives the equation $J_D=-J_C$, where
\begin{align}
J_D&=\left({f}_{BC}^{\,\,\,\,\,\,\,\,\,M}+({f}_{\bar{B}\bar{C}}^{\,\,\,\,\,\,\,\,\,\bar{M}})^*\right)\left({f}_{\bar{M}A}^{\,\,\,\,\,\,\,\,\,D}+({f}_{M\bar{A}}^{\,\,\,\,\,\,\,\,\,\bar{D}})^*\right)+\left({f}_{AB}^{\,\,\,\,\,\,\,\,\,M}+({f}_{\bar{A}\bar{B}}^{\,\,\,\,\,\,\,\,\,\bar{M}})^*\right)\left({f}_{\bar{M}C}^{\,\,\,\,\,\,\,\,\,D}+({f}_{M\bar{C}}^{\,\,\,\,\,\,\,\,\,\bar{D}})^*\right)\nonumber\\
&\qquad\qquad+\left({f}_{CA}^{\,\,\,\,\,\,\,\,\,M}+({f}_{\bar{C}\bar{A}}^{\,\,\,\,\,\,\,\,\,\bar{M}})^*\right)\left({f}_{\bar{M}B}^{\,\,\,\,\,\,\,\,\,D}+({f}_{M\bar{B}}^{\,\,\,\,\,\,\,\,\,\bar{D}})^*\right)
\end{align}
and $J_C$ is defined as $J_C=J_D|_{D\leftrightarrow C}$. These are the combinations of couplings expected to appear in the Jacobi identity. 

The non-holomorphic dimension two terms combine in (\ref{GenConsFact}) to give a new set of independent constraints. Choosing e.g. $\da{\bf{12}}\ds{\bf{34}}$ gives the identity $J_D=J_C$, which combines with the result from above to establish the Jacobi relation $J_D=0$. Having identified the Jacobi identity, adding (\ref{ChiralGCS}) to its complex conjugate produces the non-Abelian Generalised Chern-Simons constraint \cite{deWit:1984rvr,Andrianopoli:2004sv,Anastasopoulos:2006cz} in the central charge eigenbasis. The remaining holomorphic dimension two terms (of which there are two independent combinations that must be chosen for each chirality) do not provide any new information. I note that significant algebraic manipulation of the coefficients generated by the terms in (\ref{RsN=2}) is required to derive these results, in particular making extensive use of both central charge conservation and (\ref{N=2VacRels}).

It has therefore been established that consistent superfactorisation necessitates that trivalent interactions of BPS vector multiplets must be structure constants of a symmetry algebra (of which standard spontaneously broken SYM is a special case) or GCS terms. These were the conditions identified in \cite{Trott:2026cjj} from the partial unitarisation of the general four vector boson scattering amplitude and are characteristic of ``gauged'' non-linear sigma models (NL$\Sigma$Ms). The high-energy dependence of the amplitude under these conditions is tempered to that of gravity, which is consistent with their prevalent appearances in extended SUGRA (see e.g. \cite{VanProeyen:2003zj,DAuria:1998emc}). Graviton exchanges can also be introduced in the analysis and will be the focus of the next Section. 

In the special case of spontaneously broken Yang-Mills, the superamplitude (\ref{N=23vec}) reduces to 
\begin{align}\label{N=23vecYM}
&\mathcal{A}\left(\mathcal{W}_A,\mathcal{W}_B,\mathcal{W}_C\right)\nonumber\\
&\,=\frac{-i}{6}e^{-i(\varphi_1+\varphi_2+\varphi_3)}\frac{f_{ABC}}{m_1m_2m_3}\epsilon\left(\mathcal{Q},\mathcal{Q},\mathcal{Q},\mathcal{Q}\right)\nonumber\\
&\quad\times\bigg(\left(m_3e^{i\varphi_3}-m_2e^{i\varphi_2}\right)\left(m_1\mathbf{v}_1\da{\bf{23}}-e^{i(\varphi_1+\varphi_2+\varphi_3)}m_1e^{-i\varphi_1}\mathbf{v}_1\ds{\bf{23}}\right)\nonumber\\
&\quad\quad\quad+\left(m_1e^{i\varphi_1}-m_3e^{i\varphi_3}\right)\left(m_2\mathbf{v}_2\da{\bf{31}}-e^{i(\varphi_1+\varphi_2+\varphi_3)}m_2e^{-i\varphi_2}\mathbf{v}_2\ds{\bf{31}}\right)\nonumber\\
&\quad\quad\quad+\left(m_2e^{i\varphi_2}-m_1e^{i\varphi_1}\right)\left(m_3\mathbf{v}_3\da{\bf{12}}-e^{i(\varphi_1+\varphi_2+\varphi_3)}m_3e^{-i\varphi_3}\mathbf{v}_3\ds{\bf{12}}\right)\bigg).
\end{align}
In this case, the theory can be embedded directly within $\mathcal{N}=4$ SYM (at tree level). When decomposed into $\mathcal{N}=2$ submultiplets, the $\mathcal{N}=4$ vector superfield has the form 
\begin{align}
\mathcal{W}_{\mathcal{N}=4}=K_{\mathcal{N}=2}+\eta_{I}^{(2)}\mathcal{W}^I_{\mathcal{N}=2}+\frac{1}{2}\eta_{I}^{(2)}\eta^{I(2)}\widetilde{K}_{\mathcal{N}=2}.
\end{align}
It is clear that (\ref{N=43vec}) can only contain $\mathcal{N}=2$ subsuperamplitudes involving an even number of hypermultiplets. This agrees with the fact that there is no distinct Higgs multiplet for ``Coulomb branch'' $\mathcal{N}=2$ SYM, which is clear from the fact that there is no $3$-particle superamplitude between two BPS vector multiplets and a hypermultiplet. The $\mathcal{N}=2$ four vector superamplitude can therefore be directly extracted from (\ref{N=4SYM}) as 
\begin{align}
\mathcal{A}[\mathcal{W},\mathcal{W},\mathcal{W},\mathcal{W}]=\frac{1}{(s-m_s^2)(u-m_u^2)}\epsilon\left(\mathcal{Q},\mathcal{Q},\mathcal{Q},\mathcal{Q}\right)\epsilon_\varphi(1,2,3,4),
\end{align}
where
\begin{align}
\epsilon_\varphi(1,2,3,4)=&\da{\bf{12}}\ds{\bf{34}}e^{i(\varphi_3+\varphi_4)}+\da{\bf{34}}\ds{\bf{12}}e^{i(\varphi_1+\varphi_2)}+\da{\bf{14}}\ds{\bf{23}}e^{i(\varphi_2+\varphi_3)}\nonumber\\
&+\da{\bf{23}}\ds{\bf{14}}e^{i(\varphi_1+\varphi_4)}+\da{\bf{31}}\ds{\bf{24}}e^{i(\varphi_2+\varphi_4)}+\da{\bf{24}}\ds{\bf{31}}e^{i(\varphi_1+\varphi_3)},
\end{align}
which is manifestly unitarised. The supersymmetric Higgs mechanism for these $\mathcal{N}=2$ theories is trivial once the standard YM properties of the vector couplings are assumed. This assumption is likewise sufficient to unitarise the hypermultiplet Compton scattering superamplitude, which is consistent with the fact that Yukawa couplings between three hypermultiplets are not permitted by SUSY.

\section{Vector Boson Scattering in Gravity and Supergravity}\label{Sec:VecGrav}

\subsection{Gravitational scattering of massive vector bosons}\label{sec:pureGrav}

Gravitational scattering of massive vector bosons is simpler than scattering with vector mediators. In this Section, I will study the tree-level scattering of spin $\leq 1$ particles minimally coupled to gravity. Contributions to the amplitude from possible quadrupole moments are deferred to Appendix \ref{Quadrupoles}. These are prohibited in supersymmetric theories. 

In the cases described here, as will be shown, single channel residues themselves have high energy scaling that can be reduced to $\sim E^2/M_{Pl}^2$ without interference from other channels. These optimally tame residues are naturally identified by making use of a double copy structure that reduces them to products of residues in tree-unitary theories. Of course, tree-level gravitational scattering of spin $\leq 1$ particles is not new and carries the status of a textbook exercise by contemporary standards (e.g. \cite{Giddings:2011xs}), although I expect that the exposition presented here will still be illuminating and fits cohesively into the exposition of this study. It will be useful to define Lorentz structures 
\begin{align}\label{SpinStrucDef2}
\mathbf{P}_s&=\da{\bf{12}}\ds{\bf{12}}\da{\bf{34}}\ds{\bf{34}}\nonumber\\
\mathbf{P}_t&=\da{\bf{13}}\ds{\bf{13}}\da{\bf{24}}\ds{\bf{24}}\nonumber\\
\mathbf{P}_u&=\da{\bf{14}}\ds{\bf{14}}\da{\bf{23}}\ds{\bf{23}}.
\end{align}

Beginning with vector scattering off a scalar, the amplitude is determined by the $s$-channel:
\begin{align}\label{vsgrav}
&A(W,\overline{W},\varphi,\overline{\varphi})\nonumber\\
&\qquad=\frac{-1}{s}\left(\frac{M}{M_{Pl}}\right)^2\left(\left(\frac{x_{34}}{x_{12}}\right)^2\da{\bf{12}}^2+\left(\frac{x_{12}}{x_{34}}\right)^2\ds{\bf{12}}^2\right)\nonumber\\
&\qquad=\frac{-1}{s}\left(\frac{M}{M_{Pl}}\right)^2\Bigg(\left(\frac{x_{34}}{x_{12}}\da{\bf{12}}+\frac{x_{12}}{x_{34}}\ds{\bf{12}}\right)\left(\frac{x_{34}}{x_{12}}\da{\bf{12}}+\frac{x_{12}}{x_{34}}\ds{\bf{12}}\right)-2\ds{\bf{12}}\da{\bf{12}}\Bigg)\nonumber\\
&\qquad=\frac{-1}{s}\left(\frac{1}{M_{Pl}}\right)^2\left(\left(\la{\bf{1}}p_4\rs{\bf{2}}+\la{\bf{2}}p_4\rs{\bf{1}}\right)^2-2M^2\ds{\bf{12}}\da{\bf{12}}\right).
\end{align}
In the second line, the residue is recursed to the square of fermion-scalar scattering in QED (\ref{sfSQED}) minus a Higgs exchange. It is always possible, in this and subsequent examples, to recurse the spinning gravitational residues to expressions determined by lower spin seed residues by factorising it in different ways. For example,
\begin{align}
&\left(\frac{x_{34}}{x_{12}}\right)^2\da{\bf{12}}^2+\left(\frac{x_{12}}{x_{34}}\right)^2\ds{\bf{12}}^2\nonumber\\
&=\left(\left(\frac{x_{12}}{x_{34}}\right)^2\da{\bf{12}}+\left(\frac{x_{34}}{x_{12}}\right)^2\ds{\bf{12}}\right)\left(\ds{\bf{12}}+\da{\bf{12}}\right)-\da{\bf{12}}\ds{\bf{12}}\left(\left(\frac{x_{12}}{x_{34}}\right)^2+\left(\frac{x_{34}}{x_{12}}\right)^2\right),
\end{align}
which reduces the residue to lower spin gravitational residues already determined above. As described in the overview of the program at the beginning of Section \ref{Sec:LowSpinAmp}, the resulting proposals for the amplitude generally require supplementation with contact terms that potentially could affect its high energy scaling. Because the QED residues have weaker energy dependence than gravitational residues, the particular form proposed in (\ref{vsgrav}) is automatically guaranteed to have the optimal high energy scaling and is therefore the distinguished choice. While dimension 6 contact terms should be generally added to (\ref{vsgrav}) for consistency in the power counting of the gravitational EFT, none are available that can reduce the high energy scaling of the amplitude below $\sim E^2$.

Vector scattering off a fermion is likewise determined to be
\begin{align}\label{vfgrav}
A(W,\overline{W},\psi,\overline{\psi})&=\frac{-1}{s}\frac{M}{M_{Pl}^2}\left(\left(\frac{x_{34}}{x_{12}}\right)^2\da{\bf{12}}^2\ds{\bf{34}}+\left(\frac{x_{12}}{x_{34}}\right)^2\ds{\bf{12}}^2\ds{\bf{34}}\right)\nonumber\\
&=\frac{-1}{s}\frac{M}{M_{Pl}^2}\Bigg(\left(\frac{x_{34}}{x_{12}}\da{\bf{12}}\ds{\bf{34}}+\frac{x_{12}}{x_{34}}\ds{\bf{12}}\da{\bf{34}}\right)\left(\frac{x_{34}}{x_{12}}\da{\bf{12}}+\frac{x_{12}}{x_{34}}\ds{\bf{12}}\right)\nonumber\\
&\qquad\qquad\qquad\qquad\qquad\qquad\qquad\qquad\qquad\qquad-\da{\bf{12}}\ds{\bf{12}}\left(\ds{\bf{34}}+\da{\bf{34}}\right)\Bigg)\nonumber\\
&=\frac{-1}{s}\frac{1}{M_{Pl}^2}\left(\left(\mathbf{\Pi}_t+\mathbf{\Pi}_u\right)\left(\la{\bf{1}}p_4\rs{\bf{2}}+\la{\bf{2}}p_4\rs{\bf{1}}\right)-M\da{\bf{12}}\ds{\bf{12}}\left(\ds{\bf{34}}+\da{\bf{34}}\right)\right),
\end{align}
while vector scattering off another vector is
\begin{align}\label{vvgrav}
A(W,\overline{W},W',\overline{W'})&=\frac{-1}{s}\frac{1}{M_{Pl}^2}\left(\left(\frac{x_{34}}{x_{12}}\right)^2\da{\bf{12}}^2\ds{\bf{34}}^2+\left(\frac{x_{12}}{x_{34}}\right)^2\ds{\bf{12}}^2\da{\bf{34}}^2\right)\nonumber\\
&=\frac{-1}{s}\frac{1}{M_{Pl}^2}\left(\left(\frac{x_{34}}{x_{12}}\da{\bf{12}}\ds{\bf{34}}+\frac{x_{12}}{x_{34}}\ds{\bf{12}}\da{\bf{34}}\right)^2-2\mathbf{P}_s\right)\nonumber\\
&=\frac{-1}{s}\frac{1}{M_{Pl}^2}\left(\left(\mathbf{\Pi}_t+\mathbf{\Pi}_u\right)^2-2\mathbf{P}_s\right).
\end{align}
The QED Moller residue (\ref{QEDRes}) has been used here. In contrast to electromagnetic scattering of massive vector bosons, where the correct high energy scaling can only be made manifest by combining distinct factorisation channels, gravitational scattering with $s\leq 1$ has only a single channel. The double copy is sufficient to identify the residue that yields the weakest high energy scaling.

These results are all unified by the scattering of long vector multiplets in $\mathcal{N}=2$ SUGRA with no central charges. Here, the vector supermultiplets are represented by scalars and the superamplitude has an analogous factorisation structure to pure gravitational scalar scattering. While SUSY is not sufficient to uniquely fix the gravitational interaction to minimal coupling, the possible anomalous gravitational dipole that it does allow for is already ruled-out by inconsistency (see Section \ref{sec:Multi}). The gravitational vector coupling is therefore determined to be 
\begin{align}
\mathcal{A}(\mathcal{W},\overline{\mathcal{W}},H^+)&=\frac{1}{M_{Pl}}\delta^{(4)}(Q^\dagger)\frac{1}{x^2}\nonumber\\
\mathcal{A}(\mathcal{W},\overline{\mathcal{W}},H^-)&=\frac{1}{M_{Pl}}\delta^{(4)}(Q^\dagger)x^2\frac{1}{2}\epsilon_{ab}{F}_3^{a}{F}_3^{b}.
\end{align}
I am here using the usual chiral superspace in which the Clifford vacua of the massless coherent states are chosen to be the states with highest helicity, rather than the non-chiral superspace that is natural for describing BPS multiplets. The generalisation to non-zero central charges is examined below in Section \ref{GenLongVec}. The $s$-channel residue is therefore
\begin{align}
\mathcal{A}(\mathcal{W},\overline{\mathcal{W}},\mathcal{W}',\overline{\mathcal{W}'})&=\frac{-1}{s}\frac{1}{M_{Pl}^2}\int d^2\eta_P\delta^{(4)}(Q^\dagger_{12})\delta^{(4)}(Q^\dagger_{34})\frac{1}{2}\epsilon_{ab}\nonumber\\
&\qquad\qquad\qquad\times\left(\left(\frac{x_{34}}{x_{12}}\right)^2{F}_{34}^a{F}_{34}^b+\left(\frac{x_{12}}{x_{34}}\right)^2{F}_{12}^a{F}_{12}^b\right)\nonumber\\
&=\frac{-1}{s}\frac{1}{M_{Pl}^2}\int d^2\eta_P\delta^{(4)}(Q^\dagger_{12})\delta^{(4)}(Q^\dagger_{34})\frac{1}{2}\epsilon_{ab}\nonumber\\
&\qquad\qquad\times\left(\left(\frac{x_{34}}{x_{12}}{F}_{34}^a+\frac{x_{12}}{x_{34}}{F}_{12}^a\right)\left(\frac{x_{34}}{x_{12}}{F}_{34}^b+\frac{x_{12}}{x_{34}}{F}_{12}^b\right)-2{F}_{12}^a{F}_{34}^b\right),
\end{align}
where ${F}_{ij}^a$ is just a relabeling of (\ref{SuperBlockMassless}) by the two massive legs (and the appropriate $R$-index of the supercharges). This has the same double copy structure as (\ref{ScalarGrav}), where each term factorises into a product of $\mathcal{N}=1$ SQED (the first term) or Yukawa (the second term) residues. Using (\ref{N=1YukMassive}), (\ref{N=1Yuks}), (\ref{N=1SQED}) and (\ref{N=1SQEDs}), this can be evaluated as 
\begin{align}\label{N=2vecGravLong}
\mathcal{A}(\mathcal{W},\overline{\mathcal{W}},\mathcal{W}',\overline{\mathcal{W}'})&=\frac{-1}{M_{Pl}^2s}\delta^{(4)}(Q^\dagger)\frac{1}{2}\epsilon_{ab}\left(
\left(\mathcal{P}_t^a+\mathcal{P}_u^a\right)\left(\mathcal{P}_t^b+\mathcal{P}_u^b\right)-2\mathcal{P}_{s12}^{a}\mathcal{P}_{s34}^{b}\right).
\end{align}
Because the graviphoton does not couple particles with their conjugates, this $\mathcal{N}=2$ superamplitude contains the pure gravitational amplitudes of each component of the vector multiplet. These can be easily extracted and have demonstrable agreement with the results derived thus far.

The $\mathcal{N}=1$ gravitational scattering superamplitudes can be likewise determined. Again using the SQED and Yukawa residues (\ref{N=1YukMassive}), (\ref{N=1Yuks}), (\ref{N=1SQED}) and (\ref{N=1SQEDs}), the mixed vector-chiral multiplet superamplitude is 
\begin{align}
&\mathcal{A}(\mathcal{W},\overline{\mathcal{W}},\Phi,\overline{\Phi})\nonumber\\
&=\frac{-1}{s}\frac{M}{M_{Pl}^2}\int d\eta_P\delta^{(2)}(Q^\dagger_{12})\delta^{(2)}(Q^\dagger_{34})\left(\left(\frac{x_{34}}{x_{12}}\right)^2\da{\bf{12}}{F}_{34}+\left(\frac{x_{12}}{x_{34}}\right)^2\ds{\bf{12}}{F}_{12}\right)\nonumber\\
&=\frac{-1}{s}\frac{M}{M_{Pl}^2}\int d\eta_P\delta^{(2)}(Q^\dagger_{12})\delta^{(2)}(Q^\dagger_{34})\nonumber\\&\qquad\qquad\qquad\times\Bigg(\left(\frac{x_{34}}{x_{12}}{F}_{34}+\frac{x_{12}}{x_{34}}{F}_{12}\right)\left(\frac{x_{34}}{x_{12}}\da{\bf{12}}+\frac{x_{12}}{x_{34}}\ds{\bf{12}}\right)-(\da{\bf{12}}{F}_{12}+\ds{\bf{12}}{F}_{34})\Bigg)\nonumber\\
&=\frac{1}{M_{Pl}^2s}\delta^{(2)}(Q^\dagger)(\left(\la{\bf{1}}p_4\rs{\bf{2}}+\la{\bf{2}}p_4\rs{\bf{1}}\right)\left(\mathcal{P}_{t}+\mathcal{P}_{u}\right)-M\left(\ds{\bf{12}}\mathcal{P}_{s34}+\da{\bf{12}}\mathcal{P}_{s12}\right))
\end{align}
and vector-vector scattering is
\begin{align}\label{N=1VecGravitonEx}
&\mathcal{A}(\mathcal{W},\overline{\mathcal{W}},\mathcal{W}',\overline{\mathcal{W}'})\nonumber\\
&=\frac{-1}{s}\frac{1}{M_{Pl}^2}\int d\eta_P \delta^{(2)}(Q^\dagger_{12})\delta^{(2)}(Q^\dagger_{34})\left(\left(\frac{x_{34}}{x_{12}}\right)^2\da{\bf{12}}\ds{\bf{34}}{F}_{34}+\left(\frac{x_{12}}{x_{34}}\right)^2\ds{\bf{12}}\da{\bf{34}}{F}_{12}\right)\nonumber\\
&=\frac{-1}{s}\frac{M}{M_{Pl}^2}\int d\eta_P\delta^{(2)}(Q^\dagger_{12})\delta^{(2)}(Q^\dagger_{34})\nonumber\\
&\qquad\qquad\qquad\times\Bigg(\left(\frac{x_{34}}{x_{12}}{F}_{34}+\frac{x_{12}}{x_{34}}{F}_{12}\right)\left(\frac{x_{34}}{x_{12}}\da{\bf{12}}\ds{\bf{34}}+\frac{x_{12}}{x_{34}}\ds{\bf{12}}\da{\bf{34}}\right)\nonumber\\
&\qquad\qquad\qquad\qquad\qquad\qquad\qquad\qquad\qquad-(\da{\bf{12}}\ds{\bf{34}}{F}_{12}+\ds{\bf{12}}\da{\bf{34}}{F}_{34})\Bigg)\nonumber\\
&=\frac{1}{M_{Pl}^2s}\delta^{(2)}(Q^\dagger)(\left(\mathbf{\Pi}_t+\mathbf{\Pi}_u\right)\left(\mathcal{P}_t+\mathcal{P}_u\right)-\left(\ds{\bf{12}}\da{\bf{34}}\mathcal{P}_{s34}+\da{\bf{12}}\ds{\bf{34}}\mathcal{P}_{s12}\right)),
\end{align}
in agreement with the other results presented here.

\subsection{Scattering of BPS vector multiplets}\label{sec:BPSGrav}

Central charges introduce couplings to the graviphoton, so unification of gravitational amplitudes of BPS vectors in extended SUGRA necessarily incorporates trivalent vector boson couplings. These generally lead to worse high-energy scaling than that appearing in the gravitational amplitudes constructed in Section \ref{sec:pureGrav} above. Suppressing this to the level of gravity requires the existence of additional cross-channel particle exchanges, which is again a further complication to the simple, purely single-channel graviton exchanges. The graviphotons therefore provide more of an obstacle to consistency and unitarity than the gravitons. 

In this Section I will construct superamplitudes for scattering of BPS vectors in extended SUGRA. The graviphoton exchange will be associated with the generation of terms with two Mandelstam poles, just as for $\mathcal{N}=4$ SYM in (\ref{SYMgen}). These superamplitudes therefore participate in the tests of consistent superfactorisation that otherwise yielded the Jacobi identity and the Lie algebra structure of the vector boson interactions. This will lead to the emergence of a ``gauged'' structure in supergravity - theories of SUGRA in which massive BPS vectors, the graviphotons and possibly other Yang-Mills subsectors are assembled into adjoint representations of a non-Abelian Lie algebra for which their interactions are identified with its structure constants. The sector of the Lie algebra involving the graviphotons' coupling to massive vectors is typically non-compact. The component amplitudes for the massive vectors will also be decomposed into contributions from each exchanged state (graviton, vectors and scalars) in order to both cross-check the calculations and demonstrate how they combine into the elegant generating function. 

As commented upon earlier, residues describing graviphoton exchange in which the photon couples according to the amplitudes (\ref{graviphoton3pt}) can be decomposed as (\ref{GraviphotonRec}). For vector bosons, this is specifically
\begin{align}\label{vecgraviphot}
&A_{\text{graviphoton}}(W,\overline{W},W',\overline{W'})\nonumber\\
&\qquad=\frac{1}{s}\frac{1}{M_{Pl}^2}\left(\frac{x_{34}}{x_{12}}+\frac{x_{12}}{x_{34}}\right)\mathbf{P}_s\nonumber\\
&\qquad=\frac{-1}{s}\frac{1}{M_{Pl}^2}\bigg(\mathbf{\Pi}_s\left(\mathbf{\Pi}_t+\mathbf{\Pi}_u\right)+\left(\frac{x_{34}}{x_{12}}\da{\bf{12}}^2\ds{\bf{34}}^2+\frac{x_{12}}{x_{34}}\ds{\bf{12}}^2\da{\bf{34}}^2\right)\bigg)\nonumber\\
&\qquad=\frac{-1}{s}\frac{1}{M_{Pl}^2}\mathbf{\Pi}_s\left(\mathbf{\Pi}_t+\mathbf{\Pi}_u\right)-A_{\text{min}}(W,\overline{W},W',\overline{W'}),
\end{align}
where $A_{\text{min}}$ is the amplitude (or, more precisely, the residue) for exchange of a minimally coupled photon between the pair of vectors with electric charges given by $m/M_{Pl}$ and $M/M_{Pl}$. Without the presence of cross-channel processes, the residues induced by the large anomalous multipole moments have no worse high energy scaling than the standard three vector minimal coupling. This expression conveniently isolates the potential $\mathcal{O}(E^3)$ high energy dependence entirely to the residue for purely minimal coupling, which can be substituted for using the results of ununitarised, partially or fully unitarised expressions depending on the contents of the theory. 
Of course, the graviphoton coupling can also be decomposed into minimal electromagnetism plus large anomalous multipoles. Upon doing so, the $4$-leg amplitude can also be decomposed as 
\begin{align}
A_{\text{graviphoton}}(W,\overline{W},W',\overline{W'})&=A_{\text{min}}(W,\overline{W},W',\overline{W'})\nonumber\\
&\qquad+\frac{1}{sM_{Pl}^2}\bigg(mM\left(\mathbf{\Pi}_t+\mathbf{\Pi}_u\right)\left(\ds{\bf{12}}\ds{\bf{34}}+\da{\bf{12}}\da{\bf{34}}+\mathbf{\Pi}_s\right)\nonumber\\
&\qquad\qquad+M\da{\bf{12}}\ds{\bf{12}}\left(\da{\bf{34}}+\ds{\bf{34}}\right)\left(\la{\bf{4}}p_1\rs{\bf{3}}+\la{\bf{3}}p_1\rs{\bf{4}}\right)\nonumber\\
&\qquad\qquad+m\da{\bf{34}}\ds{\bf{34}}\left(\da{\bf{12}}+\ds{\bf{12}}\right)\left(\la{\bf{2}}p_3\rs{\bf{1}}+\la{\bf{1}}p_3\rs{\bf{2}}\right)\bigg).
\end{align}
However, this decomposition is less useful here as the poor high-energy scaling is not contained to the $A_{\text{min}}(W,\overline{W},W',\overline{W'})$ term.

\subsubsection{$\mathcal{N}=4$ SUGRA}\label{sec:N=4SUGRAsub}

Gravitational scattering of vector bosons in the theories listed above is simple and largely self-contained. The structure of the vector bosons' self-interactions was an irrelevance. It is natural to then investigate how these conclusions extend to particles with central charges. The simplest testing ground is provided by incorporating gravity into the 
analysis in Section \ref{SUSYAlgebra} of scattering of $\frac{1}{2}$BPS vector multiplets with $\mathcal{N}=4$ SUSY. As for SYM, the $\frac{1}{2}$BPS vectors are represented as scalars. The gravitational $3$-leg superamplitudes have the same form as their SYM counterparts with additional $x$ factors to account for the helicity of the graviton:
\begin{align}\label{3legN=4SUGRA}
\mathcal{A}\left(\mathcal{W}_A,\mathcal{W}_B,H^\pm\right)=\frac{m_A}{M_{Pl}}\delta_{A\bar{B}}\Delta_v\Delta_u x^{\mp 1}.
\end{align}
I am returning to labeling the vectors with colour indices (which I also use to label the corresponding masses to emphasise their association with a particular particle identity/colour). 

As for the analysis of $\mathcal{N}=4$ SYM in Section \ref{SUSYAlgebra}, I will assume here that the central charge matrices of the vector multiplets are manifestly electric (\ref{CentralChargesEleN=4}), but will subsequently reconsider the naively dyonic alternative (\ref{CentralChargesMagN=4}). In either case, the entries of the $\mathcal{N}=4$ central charge matrix directly correspond to the couplings to the respective graviphotons in the graviton multiplet. There is no freedom through electric-magnetic duality, as there was in $\mathcal{N}=2$ SQED, to reinterpret the anti-self-dual and self-dual components as anything other than the electric and magnetic charges respectively (apart from an overall phase) without the introduction of additional multiplets of graviphotons, just as was argued for in the case of $\mathcal{N}=2$ SUGRA in Section \ref{Sec:LowSpinAmp}.

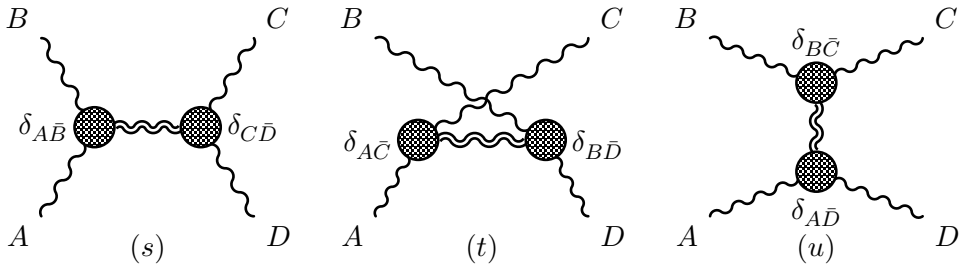
\begin{figure}[h]
\begin{fmffile}{N=4SUGRATest}
\begin{center}
\begin{tabular}{ c c c c c }
& & & & \\
\begin{fmfgraph*}(100,67)
   \fmfleft{i1,i2}
   \fmfright{o1,o2}
   \fmf{boson}{i2,v1}
   \fmf{boson}{v2,o2}
   \fmf{boson}{i1,v1}
   \fmf{boson}{v2,o1}
   \fmf{dbl_wiggly}{v1,v2}   
   \fmfv{label=$\delta_{A\bar{B}}$,label.dist=0.1w}{v1}
   \fmfv{label=$\delta_{C\bar{D}}$,label.dist=0.1w}{v2}
   \fmfv{decor.shape=circle,decor.filled=gray50,decor.size=0.15w}{v1,v2}
   \fmflabel{$A$}{i1}
   \fmflabel{$B$}{i2}
   \fmflabel{$C$}{o2}
   \fmflabel{$D$}{o1}
 \end{fmfgraph*} 
&\,& \begin{fmfgraph*}(100,67)
   \fmfleft{i1,i2}
   \fmfright{o1,o2}
   \fmf{boson}{i1,v1}
   \fmf{boson}{v2,o1}
   \fmf{dbl_wiggly}{v1,v2}
   \fmf{phantom}{v1,i2}
   \fmf{phantom}{v2,o2}
   \fmf{boson,tension=-0.25}{v2,i2}
   \fmf{boson,tension=-0.25}{v1,o2}
   \fmfv{label=$\delta_{A\bar{C}}$,label.dist=0.1w}{v1}
   \fmfv{label=$\delta_{B\bar{D}}$,label.dist=0.1w}{v2}
   \fmfv{decor.shape=circle,decor.filled=gray50,decor.size=0.15w}{v1,v2}
   \fmflabel{$A$}{i1}
   \fmflabel{$B$}{i2}
   \fmflabel{$C$}{o2}
   \fmflabel{$D$}{o1}
 \end{fmfgraph*} &\,&
 \begin{fmfgraph*}(100,67)
   \fmfleft{i1,i2}
   \fmfright{o1,o2}
   \fmf{boson}{i1,v1,o1}
   \fmf{boson}{i2,v2,o2}
   \fmf{dbl_wiggly}{v1,v2}
   \fmfv{label=$\delta_{A\bar{D}}$,label.dist=0.1w}{v1}
   \fmfv{label=$\delta_{B\bar{C}}$,label.dist=0.1w}{v2}
   \fmfv{decor.shape=circle,decor.filled=gray50,decor.size=0.15w}{v1,v2}
   \fmflabel{$A$}{i1}
   \fmflabel{$B$}{i2}
   \fmflabel{$C$}{o2}
   \fmflabel{$D$}{o1}
 \end{fmfgraph*}\nonumber\\
 $(s)$ & \,&  $(t)$ & \,&  $(u)$
 \end{tabular}
\end{center} 
\end{fmffile}
\caption{Superfactorisation channels for BPS vector multiplet scattering by graviton exchange.}
\end{figure}
When amalgamated across the $s$-channel, the delta functions in (\ref{3legN=4SUGRA}) combine in the same way as for $\mathcal{N}=4$ SYM to produce a cross-channel Mandelstam invariant, while the remaining factors describing the graviton's helicity combine in the same way as for electromagnetic scalar scattering (\ref{sQEDRes}). This gives, for scattering of identical vector supermultiplets, an $s$-channel residue
\begin{align}\label{N=4SUGRAVecGravRes}
R_s\sim -\frac{m_A^2}{M_{Pl}^2}\frac{1}{ u}\left(\frac{x_{12}}{x_{34}}+\frac{x_{34}}{x_{12}}\right)\epsilon\left(\mathcal{Q},\mathcal{Q},\mathcal{Q},\mathcal{Q}\right)^2\sim -\frac{1}{M_{Pl}^2}\frac{t-2m_A^2}{u}\epsilon\left(\mathcal{Q},\mathcal{Q},\mathcal{Q},\mathcal{Q}\right)^2
\end{align}
and complete superamplitude 
\begin{align}\label{N=4SUGRAmg}
\mathcal{A}\left(\mathcal{W}_A,\mathcal{W}_{\bar{A}},\mathcal{W}_A,\mathcal{W}_{\bar{A}}\right)&=\frac{1}{M_{Pl}^2}\frac{t-2m_A^2}{s u}\epsilon\left(\mathcal{Q},\mathcal{Q},\mathcal{Q},\mathcal{Q}\right)^2\nonumber\\
&=\frac{-1}{M_{Pl}^2}\left(\left(\frac{1}{s}+\frac{1}{u}\right)-\frac{2m_A^2}{s u}\right)\epsilon\left(\mathcal{Q},\mathcal{Q},\mathcal{Q},\mathcal{Q}\right)^2.
\end{align}
I use the fact that the $u$-channel must be identical to the $s$-channel to shortcut the computation of the full superamplitude from the $s$-channel residue (the exchange symmetry fixes possible contact ambiguities in the evaluation of the scalar QED residue). The second term in the last line above contains both $s$ and $u$ poles that cannot be split apart. This appears to be the SYM superamplitude and could be suspected as representing a contribution from graviphoton exchange, but has the wrong overall sign. In spite of appearances, the graviphoton still mediates a repulsive force and (\ref{N=4SUGRAmg}) mirrors the component relation (\ref{vecgraviphot}). In the massless limit, the two-pole term disappears, as does the graviphoton's monopole coupling to the vectors contained in the $3$-leg superamplitude. This limit represents the ``ungauging'' of the theory. 

The single pole terms in the superamplitude (\ref{N=4SUGRAmg}) are uniquely determined from graviton exchange. However, the graviton multiplet exchange in (\ref{N=4SUGRAmg}) should in general be added to the vector supermultiplet exchange residue in (\ref{SYMgen}). In this case, the terms with two Mandelstam poles are not uniquely fixed by the residues and consistent factorisation in all channels is not automatic. The $s$-channel residue terms on the right-hand side of (\ref{N=4SYMRes}) are modified to $uR_s\sim -f_{A\bar{A}E}(f_{A\bar{A}E})^*+\frac{2m_A^2}{M_{Pl}^2}$ (and the $u$-channel residue similarly), which effectively incorporates the graviphoton couplings into the Jacobi identity (\ref{ComplexJac}) that is demanded by consistent superfactorisation. However, since the gravitational couplings appear with opposite sign to the YM couplings, they cannot be interpreted as fully antisymmetric structure constants. Instead, the structure constants can be identified as ${f}_{\bar{A}H}^{\,\,\,\,\,\,\,\,\,A}=-i\sqrt{2}m_A/M_{Pl}$ and ${f}_{A\bar{A}}^{\,\,\,\,\,\,\,\,\,H}=i\sqrt{2}m_A/M_{Pl}$, where $H$ is the graviton multiplet (the factors of $-i$ are present because the massive vectors are complex charge eigenstates - in a self-conjugate basis the structure constants would be real). I will make further remarks to establish this further below. The identity of the Lie algebra can be determined by extracting the graviphoton component amplitudes from (\ref{3legN=4SUGRA}) and comparing them to the general expressions in (\ref{2Mass1MasslessVec}). In the simplest case of a single conjugate pair of $\frac{1}{2}$BPS vectors, the Lie algebra is $\mathfrak{su}_{1,1}$ broken to $\mathfrak{u}_1$ (by which I just mean a one dimensional Abelian subalgebra). In the flat spacetime limit $M_{Pl}\rightarrow\infty$, this leaves a free $\frac{1}{2}$BPS vector. 

A near minimal extension is to couple spontaneously broken $\mathcal{N}=4$ SYM with Lie algebra $\mathfrak{su}_{2}$ broken to $\mathfrak{u}_1$ to SUGRA. In this case, the gauge coupling $g$, which I will define as the coupling constant for the superamplitude between the three vector multiplets, is a further free parameter. Both the ``matter'' photon (from the massless vector multiplet) and the graviphoton (from the graviton multiplet) couple to the massive $W$ bosons (of mass $m$). In standard broken YM, it would always be possible to find a mixture of photons in which one decouples from the $W$s. Here however, this cannot be accomplished by a unitary rotation of the states. A photon can always be identified with respect to which the $W$-bosons do not carry a monopole electric charge, but it will still couple to them with large multipole moments (see (\ref{2Mass1MasslessVec}) in Section \ref{sec:N=2} above, also Appendix C in \cite{Trott:2026cjj}). By ``large'' I mean comparable to $m/M_{Pl}$ (not even smaller quantum corrections induced from loops of graviphotons and other particles). The Lie algebra of the full theory is identifiable as either $\mathfrak{su}_{2}\oplus\mathfrak{u}_1$ or $\mathfrak{su}_{1,1}\oplus\mathfrak{u}_1$, the signature being determined by the relative size of $g$ to $m/M_{Pl}$ (or the special non-semisimple boundary case of $g=\sqrt{2}m/M_{Pl}$ is $\mathfrak{e}_2\oplus\mathfrak{u}_1$). The non-Abelian ideal is broken to $\mathfrak{u}_1$. The active photons in the massless vector and graviton multiplets are associated with mixtures of the two unbroken $\mathfrak{u}_1$ generators. However, unlike for regular YM theory, a basis consisting of unbroken generators from each separate ideal cannot be arrived at by applying a unitary rotation. As a result, a basis of photons does not exist in which one of the photons fully decouples from the $W$-bosons.
Both photons necessarily have components identified with the ``unbroken'' non-Abelian generator, so the Abelian ideal does not imply the existence of a photon with no tree-level couplings to the $W$-bosons. Note that the marginal case in which $g=\sqrt{2}m/M_{Pl}$ also corresponds to the Lie algebra known as the $4d$ oscillator algebra $\mathfrak{h}_3\oplus\mathfrak{u}_1$, which contains the non-semisimple Heisenberg algebra (see e.g. \cite{BIGGS2014199}). In this case, the additional photon exchange cancels the two-pole term in (\ref{N=4SUGRAmg}). 

If the pairs of massive vectors in the superamplitude are distinct, then to calculate $\mathcal{A}\left(\mathcal{W}_A,\mathcal{W}_{\bar{A}},\mathcal{W}_B,\mathcal{W}_{\bar{B}}\right)$ with $A\neq B$ from the graviton exchange residue, the term with the single $u$-pole in (\ref{N=4SUGRAmg}) can be consistently dropped, but not the term with both poles. The $s$-channel residue is still given by (\ref{N=4SUGRAVecGravRes}) and the 
\begin{align}\label{N=4SUGRAVecGravRes2}
R_s&\sim -\frac{m_Am_B}{M_{Pl}^2}\frac{1}{ u}\left(\frac{x_{12}}{x_{34}}+\frac{x_{34}}{x_{12}}\right)\epsilon\left(\mathcal{Q},\mathcal{Q},\mathcal{Q},\mathcal{Q}\right)^2\nonumber\\
&\sim \frac{-1}{M_{Pl}^2}\frac{t-m_t^2+2m_Am_B\cos(\varphi_A-\varphi_B)}{u}\epsilon\left(\mathcal{Q},\mathcal{Q},\mathcal{Q},\mathcal{Q}\right)^2,
\end{align}
where $m_t=|m_Ae^{i\varphi_A}+m_Be^{i\varphi_B}|$ and $m_u=|m_Ae^{i\varphi_A}-m_Be^{i\varphi_B}|$ (here $\varphi_A-\varphi_B$ is just the relative angle between the two central charge vectors of the vector multiplets). When incorporated into the $4$-particle test, the part of this $s$-channel residue containing a cross-channel Mandelstam pole that must be added to (\ref{N=4SYMRes}) to give $uR_s\sim f_{A\bar{A}E}f_{\bar{E}B\bar{B}}+\frac{2m_Am_B\cos(\varphi_A-\varphi_B)}{M_{Pl}^2}$. As above, this effectively incorporates the graviphoton couplings into the emergent Jacobi identity ($E$ indexes the exchanged vector multiplets). When the central charges are misaligned, the graviton couplings must be interpreted as producing two separate Lie algebra generators appearing in the internal sum of the emergent Jacobi identity. These correspond to the two active graviphotons. The two generators, denoted here as $H_1$ and $H_2$, must have structure constants 
\begin{align}\label{GravStruc}
{f}_{\bar{A}H_1}^{\,\,\,\,\,\,\,\,\,A}&=-{f}_{A\bar{A}}^{\,\,\,\,\,\,\,\,\,H_1}=-i\sqrt{2}m_A/M_{Pl}\cos\varphi_A\nonumber\\
{f}_{\bar{A}H_2}^{\,\,\,\,\,\,\,\,\,A}&=-{f}_{A\bar{A}}^{\,\,\,\,\,\,\,\,\,H_2}=-i\sqrt{2}m_A/M_{Pl}\sin\varphi_A.
\end{align}

In order to justify the interpretation of the graviphoton couplings as Lie algebra structure constants, it still remains to show that the Jacobi identities hold for configurations in which some external indices are identified as graviphotons. However, given the identification (\ref{GravStruc}), it is easy to check that central charge conservation ensures that these remaining identities are automatically satisfied. 

The appearance of a cross-channel Mandelstam pole in the residue and its tension with consistent factorisation contrasts with the pure gravity calculations in the subsection above, where the $s$-channel graviton exchange is manifestly local in itself. In a non-SUSY theory, the graviphoton exchange would require the existence of other vector bosons with the full Lie algebra structure of their couplings for the high-energy dependence of its amplitudes to be suppressed to the level of gravity ($\sim E^2$). Here however, just like the examples in Sections \ref{N=4SuperFact} and \ref{sec:N=2} for $\mathcal{N}=4$ and $\mathcal{N}=2$ SUSY, this structure is again enforced by consistency with locality in the IR, rather than as an additional demand arising from high-energy scaling. 

One of the goals of this Section is to demonstrate the assemblage and unification of graviton, graviphoton and scalar exchange amplitudes in extended SUGRA theories. I will present a candidate superamplitude describing the gravitational scattering of a pair of $\frac{1}{2}$BPS vector multiplets under the assumption of the absence of a $t$-channel pole for the purpose of writing a concrete expression that can be easily dissected. This superamplitude is therefore given by
\begin{align}\label{N=4SUGRAgauged}
&\mathcal{A}\left(\mathcal{W}_A,\mathcal{W}_{\bar{A}},\mathcal{W}_B,\mathcal{W}_{\bar{B}}\right)=\frac{-1}{M_{Pl}^2}\left(\frac{1}{s}-\frac{2m_Am_B\cos(\varphi_A-\varphi_B)}{s\left(u-m_u^2\right)}\right)\epsilon\left(\mathcal{Q},\mathcal{Q},\mathcal{Q},\mathcal{Q}\right)^2.
\end{align}
This should be added to the general vector exchange terms from Section \ref{N=4SuperFact}, which affects the precise interpretation of the singularity structure associated with vector boson component exchanges. 

I note that there are then two possible consequences for the term with two poles in (\ref{N=4SUGRAgauged}) when combined with the contributions from vector multiplet exchanges. Either the cross-channel pole should be accepted and accounted for by a corresponding cross-channel vector exchange, or a $s$-channel vector exchange can occur with couplings that exactly cancel this term altogether, leaving only the single pole term in the superamplitude. Both options correspond to different Lie algebras. The latter case is described by the oscillator algebra mentioned above, where additional pairs of $W$-bosons can be included and correspond to increasing the oscillator algebra's dimension \cite{BIGGS2014199}. This theory describes a collection of massive vector multiplets that exclusively either scatter elastically or pair annihilate/create.

As for (\ref{HyperSUGRA}), the superamplitude (\ref{N=4SUGRAgauged}) can be used to infer or cross-check both component amplitudes and contributions from individual exchanged states derived earlier, for which there is agreement. The final ingredient in the $\mathcal{N}=4$ graviton multiplet exchange not thus far discussed is the modulus. Adding the $s$-channel exchange of the modulus 
\begin{align}\label{vvmod}
A_{\text{mod}}(W,\overline{W},W',\overline{W'})=\frac{-1}{s}\frac{1}{M_{Pl}^2}\left(\ds{\bf{12}}^2\da{\bf{34}}^2+\da{\bf{12}}^2\ds{\bf{34}}^2\right)
\end{align}
to (\ref{vvgrav}) and (\ref{vecgraviphot}) reproduces the four vector component amplitude contained in (\ref{N=4SUGRAgauged}) (which can be easily derived by just replacing $\epsilon\left(\mathcal{Q},\mathcal{Q},\mathcal{Q},\mathcal{Q}\right)$ with $\epsilon\left(1,2,3,4\right)=\mathbf{\Pi}_s+\mathbf{\Pi}_t+\mathbf{\Pi}_u$, setting the central charge phases to zero for simplicity). Note that the second term in (\ref{N=4SUGRAgauged}) corresponds to the second term in (\ref{vecgraviphot}) (which is just the $\mathcal{N}=4$ electromagnetic $W$-boson scattering residue). However, (\ref{vecgraviphot}) includes a contribution from a non-existent Higgs exchange (as is automatically part of the $\mathcal{N}=4$ SYM superamplitude) that must be removed. This effectively reverses the sign of the term proportional to $\mathbf{P}_s$ in (\ref{vvgrav}) and combines with (\ref{vvmod}) to produce the term $\propto\mathbf{\Pi}_s^2$ extracted in the first term in (\ref{N=4SUGRAgauged}).

Section \ref{Sec:LowSpinAmp} explained the conditions under which the single complex central charge of the $\mathcal{N}=2$ SUSY algebra could be interpreted as two purely electric couplings. In $\mathcal{N}=2$ SQED, the phases of the coupling constants involving the photon multiplet can be independently parametrically controlled to ensure that coupling of the BPS particles to the photon is always electric. Because of the self-conjugacy of the vector multiplet, this option was not available in spontaneously broken $\mathcal{N}=4$ SYM. As a result, the electric and magnetic charges could be directly identified with the self-dual and anti-self-dual components of the central charge matrix. In SUGRA, the complex $\mathcal{N}=2$ central charge required the introduction of an additional photon multiplet in order to be interpreted as consistent electrical couplings. Such an option could also be potentially viable in $\mathcal{N}=4$ SUGRA for particles with the central charge configuration given in (\ref{CentralChargesMagN=4}) (that would otherwise be dyonic in the context of SYM). Only ``real mass'' interactions would be permitted without the introduction of higher spin multiplets, since, as mentioned in Section \ref{Sec:N=4SYM}, it is otherwise not possible to construct a tiny group neutral Lorentz scalar involving the SUSY invariant $\Delta_v\Delta_v$ that would be necessary for a superamplitude involving three massive vector multiplets. The only $3$-particle central charge configurations that would be otherwise dyonic in SYM but potentially resuscitatable in SUGRA are those involving a massless leg. 

For a conjugate pair of $W$-multiplets with the general $\frac{1}{2}$BPS central charge configuration (\ref{1/2BPSCC}) (that is, both $\varphi$ and $\theta$ active) coupled to a massless multiplet, the angles $\theta\pm\varphi$ each appear in one of the two BPS constraints that relates pairs of supercharges (\ref{BPS}). There is some freedom in precisely how the SUSY delta functions are represented in this case. Explicitly labeling the little group spinors by the mass phases with respect to which they are defined, then (\ref{SpecialSVD2}) implies that 
\begin{align}\label{LGspinorPhases}
\begin{split}
\mathbf{v}_{3}^{(\alpha+\beta)+}&=\mathbf{v}_{3}^{(\alpha)+}\\
\mathbf{v}_{3}^{(\alpha+\beta)-}&=e^{-i\beta}\mathbf{v}_{3}^{(\alpha)-}
\end{split}\qquad
\begin{split}
\mathbf{u}_{3}^{(\alpha+\beta)+}&=\mathbf{u}_{3}^{(\alpha)+}
\\
\mathbf{u}_{3}^{(\alpha+\beta)-}&=e^{i\beta}\mathbf{u}_{3}^{(\alpha)-},
\end{split}
\end{align}
where, as usual, I label the massless particle as particle $3$. Here, $\alpha$, $\beta$ and $\alpha+\beta$ are possible assignments of a mass phase to the two massive particles. In this case, the phase itself does not affect the tiny group properties. Continuing to use the central charge phase as a label, the appropriate SUSY delta functions underpinning the $3$-particle superamplitudes are then $\Delta_v^{(\varphi+\theta)}\Delta_u^{(\varphi-\theta)}$ (that is, the $\mathbf{v}$ spinors in the first delta function are defined for mass phase $\varphi+\theta$, while the $\mathbf{u}$ spinors in the second delta function are defined for mass phase $\varphi-\theta$). The coupling to the graviton is then given by
\begin{align}\label{GravVec2Angles}
\mathcal{A}\left(\mathcal{W}_A,\mathcal{W}_B,H^\pm\right)=\frac{m_A}{M_{Pl}}\delta_{A\bar{B}}e^{-3 i\theta_A}\Delta_v^{(\varphi+\theta)}\Delta_u^{(\varphi-\theta)}e^{\pm i\theta_A}x^{\mp 1}.
\end{align}
When $\theta_A=0$, this agrees with (\ref{3legN=4SUGRA}). The overall phase is fixed to ensure that the graviton component amplitude is free of spurious dyonic phases. However, it can be easily verified that one of the graviphotons necessarily carries a dyonic coupling to the matter states. 

Gravitationally scattering two species of these massive vector multiplets off each other produces a residue for the superamplitude $\mathcal{A}\left(\mathcal{W}_A,\mathcal{W}_{\bar{A}},\mathcal{W}_B,\mathcal{W}_{\bar{B}}\right)$ given by
\begin{align}\label{DyonicGravEx}
R_s&\sim \frac{1}{M_{Pl}^2}e^{-3i(\theta_A+\theta_B)}\epsilon\left(\mathcal{Q},\mathcal{Q},\mathcal{Q},\mathcal{Q}\right)^2\frac{m_Am_B}{(\mathbf{v}_P^{(12)}\cdot \mathbf{v}_P^{(34)})(\mathbf{u}_P^{(12)}\cdot \mathbf{u}_P^{(34)})}\nonumber\\
&\qquad\qquad\qquad\qquad\qquad\qquad\qquad\qquad\times\left(\frac{x_{12}}{x_{34}}e^{-i(\theta_A-\theta_B)}+\frac{x_{34}}{x_{12}}e^{i(\theta_A-\theta_B)}\right)\nonumber\\
&=\frac{1}{M_{Pl}^2}e^{-2i(\theta_A+\theta_B)}\epsilon\left(\mathcal{Q},\mathcal{Q},\mathcal{Q},\mathcal{Q}\right)^2\frac{m_Am_B}{\left(t-\left(m_t^{(\varphi+\theta)}\right)^2\right)\left(t-\left(m_t^{(\varphi-\theta)}\right)^2\right)}\nonumber\\
&\,\,\times\left(\frac{x_{12}}{x_{34}}e^{i(\theta_A-\theta_B)}+\frac{x_{34}}{x_{12}}e^{-i(\theta_A-\theta_B)}-2\cos(\varphi_A-\varphi_B)\right)\left(\frac{x_{12}}{x_{34}}e^{-i(\theta_A-\theta_B)}+\frac{x_{34}}{x_{12}}e^{i(\theta_A-\theta_B)}\right)\nonumber\\
&=\frac{1}{M_{Pl}^2}e^{-2i(\theta_A+\theta_B)}\epsilon\left(\mathcal{Q},\mathcal{Q},\mathcal{Q},\mathcal{Q}\right)^2\frac{(m_Am_B)^2}{\left(t-\left(m_t^{(\varphi+\theta)}\right)^2\right)\left(t-\left(m_t^{(\varphi-\theta)}\right)^2\right)}\nonumber\\
&\qquad\quad\times\bigg(\frac{1}{(m_Am_B)^2}\left((t-m_A^2-m_B^2)^2-2m_A^2m_B^2\right)+2\cos(2(\theta_A-\theta_B))\nonumber\\
&\qquad\qquad\qquad\qquad\qquad\qquad\qquad-2\cos(\varphi_A-\varphi_B)\left(\frac{x_{12}}{x_{34}}e^{-i(\theta_A-\theta_B)}+\frac{x_{34}}{x_{12}}e^{i(\theta_A-\theta_B)}\right)\bigg).
\end{align}
In the first line, I drop the explicit labels of the phases on the little group spinors in order to avoid clutter. It is left implicit that the $\mathbf{v}_P$ spinors are defined with the $\varphi+\theta$ phase and the $\mathbf{u}_P$ spinors the $\varphi-\theta$ phase. In the second line, I have invoked (\ref{CrossContr}) and applied (\ref{xExp}), (\ref{xmass}) and (\ref{LGspinorPhases}). I do label the internal masses in the emergent Mandelstam poles by the central charge phase configuration that determines them:
$m_t^{\varphi+\theta}=|m_Ae^{i(\varphi_A+\theta_A)}+m_Be^{i(\varphi_B+\theta_B)}|$ and $m_t^{\varphi-\theta}=|m_Ae^{i(\varphi_A-\theta_A)}+m_Be^{i(\varphi_B-\theta_B)}|$. Since this residue is valid only on the $s$-pole, $t-\left(m_t^{(\varphi\pm\theta)}\right)^2\sim -u+\left(m_u^{(\varphi\pm\theta)}\right)^2$, where $m_u^{(\varphi\pm\theta)}$ is defined analogously. Finally, in the last line, I have recycled the graviton exchange residue between scalars (\ref{ScalarGrav}).

There are two obvious problems with this residue. Firstly, it contains two cross-channel poles. These must be consistent with potential cross-channel particle exchanges, but, as mentioned above, exchanges of BPS particles are not possible in either channel because there are no permissible $3$-particle superamplitudes (excluding trivial limits equivalent to cases already considered above). The second problem is that the part of the residue resembling that of QED (\ref{sQEDRes}) is inconsistent because of the presence of dyonic phases. 

Both of these problems can be solved by adding the residue produced by massless vector exchange. The $3$-particle superamplitude describing the couplings is simply
\begin{align}\label{PhotVec2Angles}
\mathcal{A}\left(\mathcal{W}_A,\mathcal{W}_{\bar{A}},V_i\right)=\frac{q_{iA}m_A}{M_{Pl}}e^{-3 i\theta_A}\Delta_v^{(\varphi+\theta)}\Delta_u^{(\varphi-\theta)},
\end{align}
where $q_{iA}$ are some real-valued charges. This time, the overall phase is fixed by the self-conjugacy of the superamplitude (which is clear from extracting the components). I include the normalisation factor of $\frac{m_A}{M_{Pl}}$ for convenience. The exchange residue that they produce is 
\begin{align}
R_s&\sim \frac{-1}{M_{Pl}^2}e^{-3i(\theta_A+\theta_B)}\epsilon\left(\mathcal{Q},\mathcal{Q},\mathcal{Q},\mathcal{Q}\right)^2\frac{m_Am_B \sum_i q_{iA}q_{iB}}{(\mathbf{v}_P^{(12)}\cdot \mathbf{v}_P^{(34)})(\mathbf{u}_P^{(12)}\cdot \mathbf{u}_P^{(34)})}\nonumber\\
&=\frac{1}{M_{Pl}^2}e^{-2i(\theta_A+\theta_B)}\epsilon\left(\mathcal{Q},\mathcal{Q},\mathcal{Q},\mathcal{Q}\right)^2\frac{(m_Am_B)^2\sum_i q_{iA}q_{iB}}{\left(t-\left(m_t^{(\varphi+\theta)}\right)^2\right)\left(t-\left(m_t^{(\varphi-\theta)}\right)^2\right)}\nonumber\\
&\qquad\qquad\qquad\times\left(\frac{x_{12}}{x_{34}}e^{i(\theta_A-\theta_B)}+\frac{x_{34}}{x_{12}}e^{-i(\theta_A-\theta_B)}-2\cos(\varphi_A-\varphi_B)\right).
\end{align}
The vector exchange residue is clearly inconsistent by itself for the same reason as the graviton exchange (which justifies the comments about it made in Section \ref{N=4SuperFact}). However, adding it to the graviton exchange residue (\ref{DyonicGravEx}) and tuning the $3$-particle coupling constants so that $\sum_iq_{iA}q_{iB}=2\cos(\varphi_A-\varphi_B)$ (so there must be two species of massless vectors with respect to which each $W$-boson must have respective charges $q_{1A}=\cos\varphi_A$ and $q_{2A}=\sin\varphi_A$), then the spurious components of the residues cancel to leave a fully consistent residual:
\begin{align}\label{GravScatVec2Ang}
\mathcal{A}\left(\mathcal{W}_A,\mathcal{W}_{\bar{A}},\mathcal{W}_B,\mathcal{W}_{\bar{B}}\right)=\frac{-1}{s}\frac{1}{M_{Pl}^2}e^{-2i(\theta_A+\theta_B)}\epsilon(\mathcal{Q},\mathcal{Q},\mathcal{Q},\mathcal{Q})^2.
\end{align}
This has no cross-channel poles and is perfectly self-consistent, as required. 

Thus similarly to $\mathcal{N}=2$ case, reinterpreting the self-dual components of the central charge matrix as electric requires the introduction of new graviphoton multiplets that are sourced by these charges. However, this has stringent implications for the structure of the Lie algebra. Different sectors of massive vector bosons with ordinarily dyonic phases in their central charge matrices are not permitted to couple directly. A simple example of this is an oscillator gauge algebra, as described in the paragraph below (\ref{N=4SUGRAgauged}). In this case, an arbitrary number of conjugate $W$-boson pairs can be added independently that do not couple directly to each other. In the rigid spacetime limit, these freeze to free BPS multiplets.

\subsubsection{$\mathcal{N}=2$ SUGRA}\label{sec:N=2SUGRA}

Gravitational scattering of $\mathcal{N}=2$ BPS vector multiplets can be calculated similarly, either from superunitarity or extraction from the $\mathcal{N}=4$ result (\ref{N=4SUGRAgauged}). The $3$-particle vector-graviton superamplitude admits both minimal coupling and an anomalous gravitational dipole term, but again I drop the latter because of inconsistency with factorisation, leaving
\begin{align}
\mathcal{A}\left(\mathcal{W}_A,\mathcal{W}_B,H^+\right)&=\frac{1}{M_{Pl}}\delta_{A\bar{B}}e^{-i\varphi_A}\Delta_v\frac{1}{x}\mathbf{v}_3^+\da{\bf{12}}\nonumber\\
\mathcal{A}\left(\mathcal{W}_A,\mathcal{W}_B,H^-\right)&=\frac{1}{M_{Pl}}\delta_{A\bar{B}}e^{-i\varphi_A}\Delta_v x\mathbf{v}_3^-\ds{\bf{12}}.
\end{align}
As for the hypermultiplet superamplitudes in Section \ref{Sec:LowSpinAmp}, a pair of massive vectors with misaligned central charges can only have a purely electric coupling if there is an extra graviphoton multiplet present. Across the $s$-channel, the graviton exchange residue for the superamplitude $\mathcal{A}\left(\mathcal{W}_A,\mathcal{W}_{\bar{A}},\mathcal{W}_B,\mathcal{W}_{\bar{B}}\right)$ factorises as 
\begin{align}\label{N=2VecGravRes}
&R_s=\frac{1}{M_{Pl}^2}\int d^2\eta_P\Delta_v^{(12)}\Delta_v^{(34)}\nonumber\\
&\qquad\qquad\times\left(e^{-2i\varphi_A}\frac{(\mathbf{v}_P^{(12)+}\mathbf{v}_P^{(34)-})^2}{\mathbf{v}_P^{(12)-}\mathbf{v}_P^{(34)+}}\da{\bf{12}}\ds{\bf{34}}-e^{-2i\varphi_B}\frac{(\mathbf{v}_P^{(12)-}\mathbf{v}_P^{(34)+})^2}{\mathbf{v}_P^{(12)+}\mathbf{v}_P^{(34)-}}\ds{\bf{12}}\da{\bf{34}}\right)\nonumber\\
&\quad=\frac{1}{M_{Pl}^2}\epsilon\left(\mathcal{Q},\mathcal{Q},\mathcal{Q},\mathcal{Q}\right)\frac{m_Am_Be^{-i\left(\varphi_A+\varphi_B\right)}}{(\mathbf{v}_P^{(12)}\cdot\mathbf{v}_P^{(34)})(\mathbf{u}_P^{(12)}\cdot\mathbf{u}_P^{(34)})}\nonumber\\
&\qquad\qquad\times\Bigg(-\left(\left(\frac{x_{34}}{x_{12}}\right)^2\da{\bf{12}}\ds{\bf{34}}+\left(\frac{x_{12}}{x_{34}}\right)^2\ds{\bf{12}}\da{\bf{34}}\right)\nonumber\\
&\qquad\qquad\qquad\qquad\qquad+e^{i\left(\varphi_A-\varphi_B\right)}\frac{x_{34}}{x_{12}}\da{\bf{12}}\ds{\bf{34}}+e^{-i\left(\varphi_A-\varphi_B\right)}\frac{x_{12}}{x_{34}}\ds{\bf{12}}\da{\bf{34}}\Bigg).
\end{align}
The terms proportional to squares of $x$ variables have common phase factors and are given by the graviton exchange residue between fermions in (\ref{FermionGravRes}). The terms proportional to single powers of $x$ variables would be given by photon exchange between fermions if it weren't for the dyonic phase when $\varphi_A\neq\varphi_B$ (mod $\pi$). When the central charges are aligned, the residue is self-consistent and the phases can be dropped (for simplicity) to give 
\begin{align}\label{N=2GravVecAlign}
&\mathcal{A}\left(\mathcal{W}_A,\mathcal{W}_{\bar{A}},\mathcal{W}_B,\mathcal{W}_{\bar{B}}\right)\nonumber\\
&\quad=\frac{-1}{s(u-m_u^2)}\frac{1}{M_{Pl}^2}\epsilon\left(\mathcal{Q},\mathcal{Q},\mathcal{Q},\mathcal{Q}\right)\nonumber\\
&\qquad\qquad\left(\left(s+u-|m_A-m_B|^2-2m_Am_B\right)\left(\mathbf{\Pi}_t+\mathbf{\Pi}_u\right)+m_Am_B\left(\mathbf{\Pi}_t+\mathbf{\Pi}_u-\mathbf{\Pi}_s\right)\right).
\end{align}
I have written this as a superamplitude for the purposes of having a concrete comparison to (\ref{N=4SUGRAgauged}), but it is better regarded as a residue that must be combined with the analysis of the four vector superamplitude in Section \ref{sec:N=24Vec}. As for the $\mathcal{N}=4$ SUGRA case above, the two-pole terms contribute to the consistency conditions derived in Section \ref{sec:N=24Vec} and provide new contributions to the Jacobi identity, thereby drawing the gravitational coupling into the Lie algebra. 

Just as for the scattering of hypermultiplets in Section \ref{sec:LowSpinGrav}, the dyonic phases from the central charges in the graviton exchange residue can be made electric by the introduction of a new graviphoton multiplet. The possible couplings of these massless vectors to a conjugate pair of massive BPS vectors is implicit in the general expression given above in (\ref{MasslessVec2MassiveVec}). The minimal coupling parts are specifically given by 
\begin{align}\label{GraviphotonVecAmps}
\mathcal{A}\left(\mathcal{W}_A,\mathcal{W}_B,V^+_i\right)&=\frac{q_{iA}e^{i\phi_{iA}}}{m_A}\delta_{A\bar{B}}e^{-i\varphi_A}\Delta_v\mathbf{v}_3^+\da{\bf{12}}+\ldots\nonumber\\
\mathcal{A}\left(\mathcal{W}_A,\mathcal{W}_B,V^-_i\right)&=\frac{q_{iA}e^{-i\phi_{iA}}}{m_A}\delta_{A\bar{B}}e^{-i\varphi_{A}}\Delta_v \mathbf{v}_3^-\ds{\bf{12}}+\ldots,
\end{align}
where $q_{iA}e^{i\phi_{iA}}$ are the charges of the vector $\mathcal{W}_A$ under photon multiplet $V_i$. I omit writing the possible additional dipole couplings. These are not affected by the problems associated with the dyonic phases, but are still subject to the consistency requirements of cross-channel factorisation, as discussed at length in Section \ref{sec:N=2}. 

A notable specific case is given by the inclusion of a single new photon multiplet with couplings to the massive vectors given by
\begin{align}\label{N=2GravPhotonFromN=4}
\mathcal{A}\left(\mathcal{W}_A,\mathcal{W}_B,V^+\right)&=\frac{1}{M_{Pl}}\delta_{A\bar{B}}e^{-2i\varphi_A}\Delta_v\mathbf{v}_3^+\ds{\bf{12}}\nonumber\\
\mathcal{A}\left(\mathcal{W}_A,\mathcal{W}_B,V^-\right)&=\frac{1}{M_{Pl}}\delta_{A\bar{B}}\Delta_v \mathbf{v}_3^-\da{\bf{12}}.
\end{align}
This arises in the decomposition of the $\mathcal{N}=4$ superamplitudes (\ref{3legN=4SUGRA}) above into $\mathcal{N}=2$ submultiplets. These couplings induce the residue $4$-particle superamplitude $\mathcal{A}\left(\mathcal{W}_A,\mathcal{W}_{\bar{A}},\mathcal{W}_B,\mathcal{W}_{\bar{B}}\right)$
\begin{align}\label{N=2SUGRAVec}
&R_s=\frac{1}{M_{Pl}^2}\int d^2\eta_P\Delta_v^{(12)}\Delta_v^{(34)}\nonumber\\
&\qquad\qquad\qquad\qquad\times\left(-e^{-2i\varphi_A}\mathbf{v}_P^{(12)+}\mathbf{v}_P^{(34)-}\ds{\bf{12}}\da{\bf{34}}+e^{-2i\varphi_B}\mathbf{v}_P^{(12)-}\mathbf{v}_P^{(34)+}\da{\bf{12}}\ds{\bf{34}}\right)\nonumber\\
&\qquad=\frac{1}{M_{Pl}^2}\epsilon\left(\mathcal{Q},\mathcal{Q},\mathcal{Q},\mathcal{Q}\right)\frac{m_A m_B}{(\mathbf{v}_P^{(12)}\cdot\mathbf{v}_P^{(34)})(\mathbf{u}_P^{(12)}\cdot\mathbf{u}_P^{(34)})}\nonumber\\
&\qquad\qquad\qquad\times\left(e^{-2i\varphi_A}\frac{x_{34}}{x_{12}}\ds{\bf{12}}\da{\bf{34}}+e^{-2i\varphi_B}\frac{x_{12}}{x_{34}}\da{\bf{12}}\ds{\bf{34}}-e^{-i(\varphi_A+\varphi_B)}\Pi_s\right).
\end{align}
When the central charges are aligned, then 
\begin{align}
&\mathcal{A}\left(\mathcal{W}_A,\mathcal{W}_{\bar{A}},\mathcal{W}_B,\mathcal{W}_{\bar{B}}\right)\nonumber\\
&\qquad\qquad=\frac{-1}{s(u-m_u^2)}\frac{1}{M_{Pl}^2}\epsilon\left(\mathcal{Q},\mathcal{Q},\mathcal{Q},\mathcal{Q}\right)\nonumber\\
&\qquad\qquad\qquad\times\left(\left(s+u-|m_A-m_B|^2-2m_Am_B\right)\mathbf{\Pi}_s-m_Am_B\left(\mathbf{\Pi}_t+\mathbf{\Pi}_u-\mathbf{\Pi}_s\right)\right).
\end{align}
This adds to (\ref{N=2GravVecAlign}) to reproduce the $\mathcal{N}=4$ SUGRA result (\ref{N=4SUGRAgauged}), as expected. This calculation illustrates how, under an $\mathcal{N}=2$ decomposition of the $\mathcal{N}=4$ process, the graviphoton exchanged between the massive vectors is split between the $\mathcal{N}=2$ graviton multiplet and a separate graviphoton multiplet. 

When the central charges are not aligned, then (\ref{N=2SUGRAVec}) can be combined with (\ref{N=2VecGravRes}) to give a consistent residue that would induce the superamplitude
\begin{align}
&\mathcal{A}\left(\mathcal{W}_A,\mathcal{W}_{\bar{A}},\mathcal{W}_B,\mathcal{W}_{\bar{B}}\right)=\frac{-e^{-i(\varphi_A+\varphi_B)}}{s(u-m_u^2)}\frac{1}{M_{Pl}^2}\epsilon\left(\mathcal{Q},\mathcal{Q},\mathcal{Q},\mathcal{Q}\right)\nonumber\\
&\qquad\qquad\qquad\qquad\qquad\times\left(s+u-|m_Ae^{i\varphi_A}-m_Be^{i\varphi_B}|^2-2m_Am_B\cos(\varphi_A-\varphi_B)\right)\nonumber\\
&\qquad\qquad\qquad\qquad\qquad\times\left(\mathbf{\Pi}_t+\mathbf{\Pi}_u+e^{i(\varphi_A-\varphi_B)}\da{\bf{12}}\ds{\bf{34}}+e^{i(\varphi_B-\varphi_A)}\ds{\bf{12}}\da{\bf{34}}\right),
\end{align}
again recovering the full $\mathcal{N}=4$ expression (\ref{N=4SUGRAgauged}). A distinctly $\mathcal{N}=2$ alternative is to give the vector pairs $\mathcal{W}_A$ and $\mathcal{W}_B$ opposite sign charges, which outright cancels the problematic terms to leave
\begin{align}
&\mathcal{A}\left(\mathcal{W}_A,\mathcal{W}_{\bar{A}},\mathcal{W}_B,\mathcal{W}_{\bar{B}}\right)\nonumber\\
&\qquad=\frac{-e^{-i(\varphi_A+\varphi_B)}}{s(u-m_u^2)}\frac{1}{M_{Pl}^2}\epsilon\left(\mathcal{Q},\mathcal{Q},\mathcal{Q},\mathcal{Q}\right)\nonumber\\
&\qquad\qquad\times\Big(\left(s+u-|m_Ae^{i\varphi_A}-m_Be^{i\varphi_B}|^2-2m_Am_B\cos(\varphi_A-\varphi_B)\right)\left(\mathbf{\Pi}_t+\mathbf{\Pi}_u\right)\nonumber\\
&\qquad\qquad\qquad\qquad\qquad\qquad\qquad\qquad\qquad\qquad\qquad\qquad\qquad\qquad\qquad-2m_Am_B\mathbf{\Pi}_s\Big).
\end{align}
I repeat for emphasis that the residues in these superamplitudes should generally be combined with the results of Section \ref{sec:N=24Vec} where they are incorporated into the self-consistency conditions derived there. The choice of writing the superamplitudes with a cross-channel $u$-pole has been made simply for illustrating the decomposition of a concrete expression into less supersymmetric components. 

More generally, the couplings to a set of additional photon multiplets given in (\ref{GraviphotonVecAmps}) can be added to the general graviton exchange residue (\ref{N=2VecGravRes}) to restore consistency analogously to the hypermultiplet superamplitude (\ref{FullHyperMoller}). The contributions to the $s$-channel residue from vector exchange with purely minimal coupling is given by 
\begin{align}\label{N=2PhotonExVec}
R_s&=\epsilon\left(\mathcal{Q},\mathcal{Q},\mathcal{Q},\mathcal{Q}\right)\frac{-e^{-i\left(\varphi_A+\varphi_B\right)}}{(\mathbf{v}_P^{(12)}\cdot\mathbf{v}_P^{(34)})(\mathbf{u}_P^{(12)}\cdot\mathbf{u}_P^{(34)})}\times\mathbf{u}_P^{(12)}\cdot\mathbf{u}_P^{(34)}\nonumber\\
&\qquad\qquad\times\Bigg(\left(\sum_iq_{iA}q_{iB}e^{i\left(\phi_{iA}-\phi_{iB}\right)}\right)\mathbf{v}_P^{(12)+}\mathbf{v}_P^{(34)-}\da{\bf{12}}\ds{\bf{34}}\nonumber\\
&\qquad\qquad\qquad\qquad\qquad\qquad\qquad\quad+\left(\sum_iq_{iA}q_{iB}e^{-i\left(\phi_{iA}-\phi_{iB}\right)}\right)\mathbf{v}_P^{(12)-}\mathbf{v}_P^{(34)+}\ds{\bf{12}}\da{\bf{34}}\Bigg)\nonumber\\
&=\epsilon\left(\mathcal{Q},\mathcal{Q},\mathcal{Q},\mathcal{Q}\right)\frac{e^{-i\left(\varphi_A+\varphi_B\right)}}{(\mathbf{v}_P^{(12)}\cdot\mathbf{v}_P^{(34)})(\mathbf{u}_P^{(12)}\cdot\mathbf{u}_P^{(34)})}\nonumber\\
&\quad\times\Bigg(\left(\sum_iq_{iA}q_{iB}e^{i\left(\phi_{iA}-\phi_{iB}\right)}\right)\frac{x_{34}}{x_{12}}\da{\bf{12}}\ds{\bf{34}}+\left(\sum_iq_{iA}q_{iB}e^{-i\left(\phi_{iA}-\phi_{iB}\right)}\right)\frac{x_{12}}{x_{34}}\ds{\bf{12}}\da{\bf{34}}\nonumber\\
&\quad\quad-\Bigg(e^{i\left(\varphi_{A}-\varphi_{B}\right)}\left(\sum_iq_{iA}q_{iB}e^{i\left(\phi_{iA}-\phi_{iB}\right)}\right)\da{\bf{12}}\ds{\bf{34}}\nonumber\\
&\qquad\qquad\qquad\qquad\qquad\qquad\quad\quad+e^{-i\left(\varphi_{A}-\varphi_{B}\right)}\left(\sum_iq_{iA}q_{iB}e^{-i\left(\phi_{iA}-\phi_{iB}\right)}\right)\ds{\bf{12}}\da{\bf{34}}\Bigg)\Bigg).
\end{align}
The charge vectors can be defined again as in (\ref{ChargeVec}). Consistency requires that the charges all be electric (\ref{ElectricCharge}). Combining (\ref{N=2PhotonExVec}) with the graviton exchange residue gives the general expression for the residue 
\begin{align}\label{FullN=2Graviton}
&\mathcal{A}\left(\mathcal{W}_A,\mathcal{W}_{\bar{A}},\mathcal{W}_B,\mathcal{W}_{\bar{B}}\right)\nonumber\\
&=\frac{-1}{s}\epsilon\left(\mathcal{Q},\mathcal{Q},\mathcal{Q},\mathcal{Q}\right)\frac{e^{-i\left(\varphi_A+\varphi_B\right)}}{u-m_u^2}\nonumber\\
&\quad\times\Bigg(-\frac{m_Am_B}{M_{Pl}^2}\left(\frac{t-u}{2m_Am_B}\left(\mathbf{\Pi}_t+\mathbf{\Pi}_u\right)+\mathbf{\Pi}_s\right)+\mathfrak{q}_A\cdot\mathfrak{q}_B\left(\mathbf{\Pi}_t+\mathbf{\Pi}_u\right)\nonumber\\
&\qquad\qquad\qquad+e^{i\left(\varphi_A-\varphi_B\right)}\left(\mathfrak{q}_B\cdot\mathfrak{q}_A-\frac{m_Am_B}{M_{Pl}^2}e^{i\left(\varphi_A-\varphi_B\right)}\right)\da{\bf{12}}\ds{\bf{34}}\nonumber\\
&\qquad\qquad\qquad\qquad\qquad+e^{-i\left(\varphi_A-\varphi_B\right)}\left(\mathfrak{q}_A\cdot\mathfrak{q}_B-\frac{m_Am_B}{M_{Pl}^2}e^{-i\left(\varphi_A-\varphi_B\right)}\right)\ds{\bf{12}}\da{\bf{34}}\Bigg).
\end{align}
Again, I have omitted possible contributions from independent dipole couplings of the photon multiplets to the massive vectors. The combined graviton and photon exchange residue (\ref{FullN=2Graviton}), once the dyonic phases are removed, subsequently participates in the $4$-particle test as explained in Section \ref{sec:N=2}. The possible dipole contributions to (\ref{FullN=2Graviton}) should generally be included for these purposes. The gravitational couplings can be reinterpreted as structure constants in the Lie algebra of the same form as described in the $\mathcal{N}=4$ case above around (\ref{GravStruc}). The freedom to include parametrically independent dipole moments to the graviphoton multiplets' couplings allows for a wider range of possible Lie algebras (or ``gaugings'') in addition to GCS interactions. 

While not entirely obvious, each linearly independent term in the residue (\ref{FullN=2Graviton}) affects the emergent Jacobi identity in similar ways. The absence of dimension $3$ Lorentz structures ensures that the GCS relation (\ref{ChiralGCS}) is unaffected. Removing the single pole terms in (\ref{FullN=2Graviton}) amounts to replacing $t-u\mapsto m_t^2-m_u^2=4m_Am_B\cos\left(\varphi_A-\varphi_B\right)$. The terms proportional to $\mathbf{\Pi}_t+\mathbf{\Pi}_u$ then have coefficient 
\begin{align}\label{GravResCharges}
&\mathfrak{q}_A\cdot\mathfrak{q}_B-2\frac{m_Am_B}{M_{Pl}^2}\cos\left(\varphi_A-\varphi_B\right)\nonumber\\
&=\frac{m_Am_B}{M_{Pl}^2}\cos\left(\varphi_A-\varphi_B\right)+q_{A1}q_{B1}\cos\left(\phi_{A1}-\phi_{B1}\right)+\ldots-2\frac{m_Am_B}{M_{Pl}^2}\cos\left(\varphi_A-\varphi_B\right)\nonumber\\
&=-\frac{m_Am_B}{M_{Pl}^2}\cos\left(\varphi_A-\varphi_B\right)+q_{A1}q_{B1}\cos\left(\phi_{A1}-\phi_{B1}\right)+\ldots.
\end{align}
The coefficients of the terms proportional to $\da{\bf{12}}\ds{\bf{34}}$ and $\ds{\bf{12}}\da{\bf{34}}$ arrange in a similar way, differing only in overall phase factors. The reversal of the sign of the gravitational term in (\ref{GravResCharges}) now agrees with the combination of couplings appearing in the residue for the $\mathcal{N}=4$ case stated just below (\ref{N=4SUGRAVecGravRes2}). When introduced into the analysis in Section \ref{sec:N=24Vec}, the modification of the non-holomorphic dimension-$2$ structures leads to the same reinterpretation of (\ref{GravResCharges}) in terms of Lie algebra structure constants as identified in (\ref{GravStruc}). The Jacobi identities in which the graviphoton generators are external uncontracted indices follow automatically from both central charge conservation and the identity (\ref{N=2VacRels}). 

The gravitational contribution to the $\mathcal{N}=2$ Compton superamplitude between vectors and hypermultiplets can likewise be computed. 
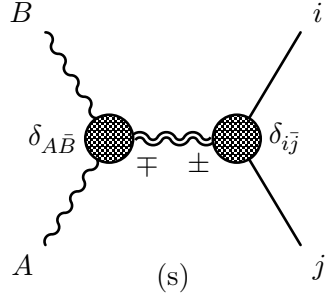
\begin{figure}[h]
\begin{fmffile}{N=2ComptonTestSUGRA}
\begin{center}
\begin{tabular}{ c }
\\
\begin{fmfgraph*}(120,80)
   \fmfleft{i1,i2}
   \fmfright{o1,o2}
   \fmf{boson}{i2,v1}
   \fmf{plain}{v2,o2}
   \fmf{boson}{i1,v1}
   \fmf{plain}{v2,o1}
   \fmf{dbl_wiggly,label=$\mp\quad\pm$}{v1,v2}
   \fmfv{decor.shape=circle,decor.filled=gray50,decor.size=0.15w}{v1,v2}
   \fmflabel{$A$}{i1}
   \fmflabel{$B$}{i2}
   \fmflabel{$i$}{o2}
   \fmflabel{$j$}{o1}
   \fmfv{label=$\delta_{A\bar{B}}$,label.dist=0.1w}{v1}
   \fmfv{label=$\delta_{i\bar{j}}$,label.dist=0.1w}{v2}
   \fmfv{decor.shape=circle,decor.filled=gray50,decor.size=0.15w}{v1,v2}
 \end{fmfgraph*}\nonumber\\
 (s) \nonumber
 \end{tabular}
 \end{center} 
\end{fmffile}
\caption{Superfactorisation channel for Compton scattering of BPS multiplets by graviton exchange.}
\end{figure}
The $s$-channel graviton exchange contribution to $\mathcal{A}(\mathcal{W}_A,\mathcal{W}_{\bar{A}},K_i,K_{\bar{i}})$ is identical to (\ref{N=2VecGravRes}) but with the factors of $\ds{\bf{34}}$ and $\da{\bf{34}}$ replaced with $m_i$ (the mass of the hypermultiplet). With aligned central charges, using (\ref{sfSQED}) and (\ref{MixedGravRes}), this gives
\begin{align}
&\mathcal{A}\left(\mathcal{W}_A,\mathcal{W}_{\bar{A}},K_i,K_{\bar{i}}\right)\nonumber\\
&\quad=\frac{-1}{s(u-m_u^2)}\frac{1}{2M_{Pl}^2}\epsilon\left(\mathcal{Q},\mathcal{Q},\mathcal{Q},\mathcal{Q}\right)\nonumber\\
&\qquad\qquad\times\bigg(\left(s+u-|m_A-m_i|^2-2m_Am_i\right)\left(\la{\bf{1}}p_4-p_3\rs{\bf{2}}+\la{\bf{2}}p_4-p_3\rs{\bf{1}}\right)\nonumber\\
&\qquad\qquad\qquad\qquad\qquad+m_Am_i\left(\la{\bf{1}}p_4-p_3\rs{\bf{2}}+\la{\bf{2}}p_4-p_3\rs{\bf{1}}+2m_i(\ds{\bf{12}}+\da{\bf{12}})\right)\bigg).
\end{align}
With misaligned central charges, then the graviphoton multiplet exchange needs to be included. In the special case of the $\mathcal{N}=4$ descended couplings (\ref{N=2GravPhotonFromN=4}), the exchange residue is again analogous to (\ref{N=2SUGRAVec}) with the substitution described above. This leads to 
\begin{align}
&\mathcal{A}\left(\mathcal{W}_A,\mathcal{W}_{\bar{A}},K_i,K_{\bar{i}}\right)\nonumber\\
&=\frac{-e^{-i(\varphi_A+\varphi_i)}}{s(u-m_u^2)}\frac{1}{M_{Pl}^2}\epsilon\left(\mathcal{Q},\mathcal{Q},\mathcal{Q},\mathcal{Q}\right)\left(s+u-|m_Ae^{i\varphi_A}-m_ie^{i\varphi_i}|^2-2m_Am_i\cos(\varphi_A-\varphi_i)\right)\nonumber\\
&\qquad\qquad\times\left(\la{\bf{1}}p_4-p_3\rs{\bf{2}}+\la{\bf{2}}p_4-p_3\rs{\bf{1}}+2m_i\left(e^{i(\varphi_A-\varphi_i)}\da{\bf{12}}+e^{i(\varphi_i-\varphi_A)}\ds{\bf{12}}\right)\right).
\end{align}
In the more general case of (\ref{GraviphotonVecAmps}), the residue is analogous to (\ref{FullN=2Graviton}):
\begin{align}
&\mathcal{A}\left(\mathcal{W}_A,\mathcal{W}_{\bar{A}},K_i,K_{\bar{i}}\right)\nonumber\\
&=\frac{-1}{s}\epsilon\left(\mathcal{Q},\mathcal{Q},\mathcal{Q},\mathcal{Q}\right)\frac{m_Am_ie^{-i\left(\varphi_A+\varphi_i\right)}}{u-m_u^2}\nonumber\\
&\quad\times\Bigg(-\frac{1}{M_{Pl}^2}\left(\frac{t-u}{4m_Am_i^2}\left(\la{\bf{1}}p_4-p_3\rs{\bf{2}}+\la{\bf{2}}p_4-p_3\rs{\bf{1}}\right)+\ds{\bf{12}}+\da{\bf{12}}\right)\nonumber\\
&\qquad\qquad\qquad+\frac{\mathfrak{q}_A\cdot\mathfrak{q}_i}{2m_i}\left(\la{\bf{1}}p_4-p_3\rs{\bf{2}}+\la{\bf{2}}p_4-p_3\rs{\bf{1}}\right)\nonumber\\
&\qquad\qquad\qquad\qquad\qquad+m_ie^{i\left(\varphi_A-\varphi_i\right)}\left(\mathfrak{q}_i\cdot\mathfrak{q}_A-\frac{m_Am_i}{M_{Pl}^2}e^{i\left(\varphi_A-\varphi_i\right)}\right)\da{\bf{12}}\nonumber\\
&\qquad\qquad\qquad\qquad\qquad\qquad\qquad+m_ie^{-i\left(\varphi_A-\varphi_i\right)}\left(\mathfrak{q}_A\cdot\mathfrak{q}_i-\frac{m_Am_i}{M_{Pl}^2}e^{-i\left(\varphi_A-\varphi_i\right)}\right)\ds{\bf{12}}\Bigg).
\end{align}
In either case, both of these residues contribute to (\ref{N=2Comptons}) and consequently to the Lie algebra commutator (\ref{N=2LAGen}), which is now extended to include the generators identified with the central charges.

The inclusion of the generator corresponding to the graviphoton into the Lie algebra commutation relations is completed by the mixed Compton scattering superamplitude in which a graviton and a vector multiplet scatter off a hypermultiplet. Combined with the extension of the structure constants to include graviphoton indices in the case above, this demonstrates that the representation of the full Lie algebra into which the hypermultiplets assemble must also include the graviphoton. This can be non-compact and, since the representation must be unitary, would necessitate an infinite tower of states. This isn't a problem for massless hypermultiplets, since the vanishing of their central charges places them in trivial representations of the non-compact generators.

\section{Soft Breaking of Extended Supergravity}\label{sec:ExtendedSSSB}

\subsection{Kaluza-Klein $\mathcal{N}=8$ SUGRA}

The lowest spin massive multiplets in $\mathcal{N}=8$ SUGRA are $\frac{1}{2}$BPS and contain particles of spin from $0$ to $2$. I will denote them by $\mathcal{H}$. These admit $3$-particle superamplitudes 
\begin{align}\label{MassiveN=8}
\mathcal{A}(\mathcal{H}_{\mathfrak{q}_1A},\mathcal{H}_{\mathfrak{q}_2B},\mathcal{H}_{\mathfrak{q}_3C})=h_{\{\mathfrak{q}_1A\}\{\mathfrak{q}_2B\}\{\mathfrak{q}_3C\}}\Delta_u^2\Delta_v^2.
\end{align}
I denote by $\mathfrak{q}_i$ a vector of central charges for particle $i$, while the $A,B,C$ indices represent coordinates for other possible internal quantum numbers. The central charges must be conserved, as usual, so $\mathfrak{q}_1+\mathfrak{q}_2+\mathfrak{q}_3=0$, while the coupling $h_{\{\mathfrak{q}_1A\}\{\mathfrak{q}_2B\}\{\mathfrak{q}_3C\}}$ must be fully symmetric under particle exchanges. Unitarity requires that $(h_{\{\mathfrak{q}_1A\}\{\mathfrak{q}_2B\}\{\mathfrak{q}_3C\}})^*=h_{\{-\mathfrak{q}_1\bar{A}\}\{-\mathfrak{q}_2\bar{B}\}\{-\mathfrak{q}_3\bar{C}\}}$. This general superamplitude includes the special case in which one of the multiplets is a massless graviton. Calling particle $3$ massless with $\mathfrak{q}_3=0$, then $\mathfrak{q}_2=-\mathfrak{q}_1$ and $B=\bar{A}$, while $h_{\{\mathfrak{q}_1A\}\{-\mathfrak{q}_1B\}\{0\}}=\frac{1}{M_{Pl}}\delta_{A\bar{B}}$, as usual. The massless graviton is, as always, unique. For ease of reference, the $3$-graviton superamplitude is 
\begin{align}\label{N=83graviton}
\mathcal{A}(H,H,H)=\frac{1}{M_{Pl}}\frac{1}{(\da{12}\da{23}\da{31})^2}\delta^{(8)}(Q^\dagger)\tilde{\delta}^{(4)}(Q^\dagger)+\frac{1}{M_{Pl}}\frac{1}{(\ds{12}\ds{23}\ds{31})^2}\delta^{(8)}(Q)\tilde{\delta}^{(4)}(Q)
\end{align}
(only one term can be non-zero for any analytic continuation).

When the four SUSY delta functions in (\ref{MassiveN=8}) are amalgamated across a factorisation channel, they must produce both distinct cross-channel Mandelstam poles. Gravitationally scattering a pair of massive particles with the same central charges must produce the $4$-leg superamplitude
\begin{align}\label{4legMassiveN=8Single}
\mathcal{A}(\mathcal{H}_{\mathfrak{q}A},\mathcal{H}_{-\mathfrak{q}\bar{A}},\mathcal{H}_{\mathfrak{q}B},\mathcal{H}_{-\mathfrak{q}\bar{B}})=\frac{1}{M_{Pl}^2}\frac{1}{(s-m_s^2)(t-m_t^2)(u-m_u^2)}\epsilon\left(\mathcal{Q},\mathcal{Q},\mathcal{Q},\mathcal{Q}\right)^4,
\end{align}
where, as before, in general $m_s=|m_3e^{i\varphi_3}+m_4e^{i\varphi_4}|$, $m_t=|m_2e^{i\varphi_2}+m_4e^{i\varphi_4}|$ and $m_u=|m_1e^{i\varphi_1}+m_4e^{i\varphi_4}|$, while here specifically $m_s=m_u=0$ and $m_t=2m$, where $m$ is the mass of the external multiplets. The $s$-channel represents the graviton exchange. The necessary existence of the massless $u$-channel implies that this must also be generated by a graviton exchange. This is only possible if $B=A$. The central charges must therefore uniquely identify each $\frac{1}{2}$BPS spin-$2$ multiplets. The possible internal indices can therefore by dropped (I will do this henceforth). Finally, the necessary presence of the $t$-channel implies that a multiplet with central charge $2\mathfrak{q}$ must exist, given the presence of a charge $\mathfrak{q}$ multiplet, and have a gravitational strength coupling $h_{\{\mathfrak{q}\}\{\mathfrak{q}\}\{-2\mathfrak{q}\}}=\frac{1}{M_{Pl}}e^{i\phi}$ for some phase $\phi$. The phase can be freely removed by absorption into the definition of the $\mathcal{H}_{2\mathfrak{q}}$ state.

The charge $\mathfrak{q}$ multiplet can then be gravitationally off the new $\mathcal{H}_{2\mathfrak{q}}$ multiplet to produce a similar residue:
\begin{align}
\mathcal{A}(\mathcal{H}_{\mathfrak{q}},\mathcal{H}_{-\mathfrak{q}},\mathcal{H}_{2\mathfrak{q}},\mathcal{H}_{-2\mathfrak{q}})=\frac{1}{M_{Pl}^2}\frac{1}{(s-m_s^2)(t-m_t^2)(u-m_u^2)}\epsilon\left(\mathcal{Q},\mathcal{Q},\mathcal{Q},\mathcal{Q}\right)^4,
\end{align}
where here $m_s=0$ but $m_t=3m$ and $m_u=m$. The $u$-pole is already accounted for by the interaction derived in the previous paragraph. However, the $t$-poles demands the existence of new spin-$2$ multiplets with central charge $3\mathfrak{q}$ and coupling $h_{\{\mathfrak{q}\}\{2\mathfrak{q}\}\{-3\mathfrak{q}\}}=\frac{1}{M_{Pl}}$ (fixing the phase again by fixing that of the $\mathcal{H}_{3\mathfrak{q}}$ state). 

The argument continues by scattering different multiplets off each other, ultimately reconstructing an infinite lattice of spin-$2$ multiplets with central charges given by integer multiples of $\mathfrak{q}$. Each multiplet is uniquely identified by its central charge and has universal gravitational coupling to each of the others wherever permitted by central charge conservation. At least in the absence of higher-spin particles, this establishes the uniqueness and universality of Kaluza-Klein gravity from consistent factorisation of $\frac{1}{2}$BPS spin-$2$ multiplets in the presence of $\mathcal{N}=8$ SUGRA. 

Introducing a second particle with a misaligned central charge forces the charge lattice of states to extend into a new dimension. As long as a state representing a root of the lattice is present, an entire subspace will necessarily be generated by consistency with gravitational scattering. The general $4$-leg superamplitude is
\begin{align}\label{4legMassiveN=8}
\mathcal{A}(\mathcal{H}_{\mathfrak{q}_1},\mathcal{H}_{\mathfrak{q}_2},\mathcal{H}_{\mathfrak{q}_3},\mathcal{H}_{\mathfrak{q}_4})=\frac{1}{M_{Pl}^2}\frac{1}{(s-m_s^2)(t-m_t^2)(u-m_u^2)}\epsilon\left(\mathcal{Q},\mathcal{Q},\mathcal{Q},\mathcal{Q}\right)^4.
\end{align}
The $\mathcal{N}=8$ KK gravity theory has structural resemblances to spontaneously broken $\mathcal{N}=4$ SYM described above in Section \ref{SUSYAlgebra}. The KK gravitons and the massive $W$-bosons in each respective theory are represented as scalar fields. In both cases, the (perturbative) theories could be entirely reconstructed from consistent superfactorisation. Furthermore, the $4$-particle amplitudes determined from factorisation and are automatically ``unitarised'' (that is, for SYM, does not grow with high particle energy $E$, while for SUGRA, diverge as $\sim E^2/M_{Pl}^2$ and no stronger) without a Higgs mechanism. This is immediately manifest from SUSY and factorisation, in contrast to the form produced from the Feynman rules or related on-shell technique \cite{Ema:2024vww,Gherghetta:2024tob} that constructs the non-SUSY component amplitudes from single pole factorisation terms and generally produces expressions with faster, possibly spurious high energy growth. These observations are consistent with the reinterpretation of each theory as higher dimensional SYM and SUGRA. Both $\mathcal{N}=8$ KK SUGRA and $\mathcal{N}=4$ spontaneously broken SYM are therefore distinguished as especially simple theories in these respects. See \cite{Caron-Huot:2018ape} for some further discussion of special properties of massive $\mathcal{N}=8$ superamplitudes. The focus of this Section will be predominantly superamplitudes of spin-$3/2$ multiplets with similar simple characteristics. See \cite{Hang:2022rjp,Hang:2024uny,Bonifacio:2019ioc,Chivukula:2023qrt} for some recent work on unitarisation of scattering amplitudes in theories of KK gravitons.

\subsection{$\mathcal{N}=6$ supergravity and the simplest gravitino $S$-matrix}

With $\mathcal{N}=6$ SUGRA, the only way to introduce a gravitino without other massive higher spin particles is within a $\frac{1}{2}$BPS multiplet. In this case, the gravitino must carry central charge and have a charge conjugate antiparticle. In the standard on-shell superspace representation, the gravitino multiplet is represented as a scalar. Its central charge may be rotated to a purely symplectic form $Z_{ab}\propto\Omega_{ab}$. No angles need to be introduced. The superspace itself has an identical structure to the $\mathcal{N}=4$ case, but now with an extra pair of supercharges related by the BPS condition. See \ref{sec:GravitinoSS} for details of the multiplets and superspace for $\mathcal{N}=6$ SUGRA. The superamplitude coupling the gravitinos to gravity is 
\begin{align}\label{N=6Gravitino}
    \mathcal{A}\left(\Psi,\overline{\Psi},H^\pm\right)=\frac{1}{M_{Pl}}\Delta_u\Delta^2_v\mathbf{v}_3^\pm
\end{align}
(clearly a double copy of the $\mathcal{N}=4$ SYM kinematic factor and $\mathcal{N}=2$ QED). The superamplitude for gravitational scattering of two gravitinos off each other is easily calculated by amalgamating the $3$-particle superamplitudes: 
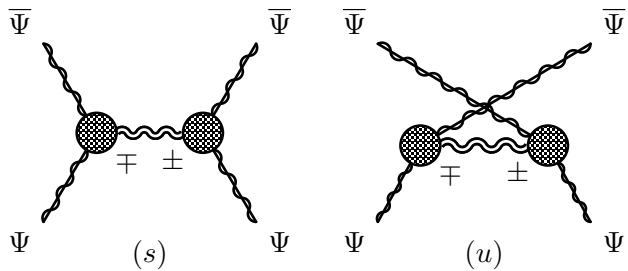
\begin{figure}[h]
\begin{fmffile}{N=6Gravitino2}
\begin{center}
\begin{tabular}{ c c c }
& & \\
 \begin{fmfgraph*}(100,67)
   \fmfleft{i1,i2}
   \fmfright{o1,o2}
   \fmf{plain}{i1,v1}
   \fmf{boson}{i1,v1}
   \fmf{dbl_wiggly}{v1,v2}
   \fmf{dbl_wiggly}{v1,v2}
   \fmf{plain}{v2,o1}
   \fmf{boson}{v2,o1}
   \fmf{plain}{i2,v1}
   \fmf{boson}{i2,v1}
   \fmf{plain}{o2,v2}
   \fmf{boson}{o2,v2} \fmfv{decor.shape=circle,decor.filled=gray50,decor.size=0.15w}{v1,v2}
   \fmflabel{$\overline{\Psi}$}{i2}
   \fmflabel{$\overline{\Psi}$}{o2}
   \fmflabel{$\Psi$}{i1}
   \fmflabel{$\Psi$}{o1}
   \fmfv{label=$\mp$,label.angle=-45,label.dist=0.1w}{v1}
   \fmfv{label=$\pm$,label.angle=-135,label.dist=0.1w}{v2}
 \end{fmfgraph*} 
 &\,& \begin{fmfgraph*}(100,67)
   \fmfleft{i1,i2}
   \fmfright{o1,o2}
   \fmf{plain}{i1,v1}
   \fmf{boson}{i1,v1}
   \fmf{dbl_wiggly}{v1,v2}
   \fmf{dbl_wiggly}{v1,v2}
   \fmf{plain}{v2,o1}
   \fmf{boson}{v2,o1}
   \fmf{phantom}{v1,i2}
   \fmf{phantom}{v1,i2}
   \fmf{phantom}{v2,o2}
   \fmf{phantom}{v2,o2}
   \fmf{plain,tension=-0.25}{v2,i2}
   \fmf{plain,tension=-0.25}{v1,o2}
   \fmf{boson,tension=-0.25}{v2,i2}
   \fmf{boson,tension=-0.25}{v1,o2}
   \fmfv{decor.shape=circle,decor.filled=gray50,decor.size=0.15w}{v1,v2}
   \fmflabel{$\overline{\Psi}$}{i2}
   \fmflabel{$\overline{\Psi}$}{o2}
   \fmflabel{$\Psi$}{i1}
   \fmflabel{$\Psi$}{o1}
   \fmfv{label=$\pm$,label.angle=-135,label.dist=0.1w}{v2}
   \fmfv{label=$\mp$,label.angle=-45,label.dist=0.1w}{v1}
 \end{fmfgraph*}\nonumber\\
 $(s)$ & \, &  $(u)$ 
\end{tabular}
\end{center}
\end{fmffile}
\caption{Superfactorisation channels for BPS gravitino scattering.}
\label{Fig:BPSgravitinoGravEx}
\end{figure}
\begin{align}\label{4legN=6}
    \mathcal{A}\left(\Psi,\overline{\Psi},\Psi,\overline{\Psi}\right)=\frac{1}{M_{Pl}^2}\frac{1}{su}\epsilon\left(\mathcal{Q},\mathcal{Q},\mathcal{Q},\mathcal{Q}\right)^3.
\end{align}
This superamplitude is manifestly unitarised (by gravitational standards) and has the correct pole structure. The super-Higgs mechanism in this theory is therefore trivial - no new particles are required either for consistency or to suppress the high-energy scaling. The graviton multiplet exchange scales as $\sim E^2$ in the limit of high particle energy instead of the $\sim E^4$ dependence that typically occurs without the super-Higgs mechanism \cite{Volkov:1972jx}. The particles in the $\mathcal{N}=6$ graviton multiplet combine with the gravitino multiplets to complete a $\mathcal{N}=8$ graviton multiplet. In this sense, this is the simplest theory of SUGRA breaking. What has been described here is the Cremmer-Scherk-Schwarz (CSS) spontaneous SUSY breaking pattern of $\mathcal{N}=8\rightarrow 6$ identified originally in the context of extra dimensional compactification \cite{Scherk:1978ta,Cremmer:1979uq,Andrianopoli:2002rm}. 

It is therefore unsurprising that this theory of $\mathcal{N}=6$ SUGRA with a pair of gravitinos can be embedded at tree level in maximal KK SUGRA. The $\mathcal{N}=6$ graviton multiplets are extracted from their $\mathcal{N}=8$ counterparts as
\begin{align}
H^+_{\mathcal{N}=6}=\frac{\partial}{\partial\overline{\eta}_4}H_{\mathcal{N}=8}\Big|_{\eta_4=0}\qquad\qquad H^-_{\mathcal{N}=6}=\frac{\partial}{\partial\eta_4}H_{\mathcal{N}=8}\Big|_{\overline{\eta}_4=0},
\end{align}
while the massive gravitinos are
\begin{align}\label{N=6ProjectMassive}
\Psi_{\mathcal{N}=6}=\mathcal{H}_{\mathcal{N}=8}\Big|_{\eta_4^I=0}\qquad\qquad \overline{\Psi}_{\mathcal{N}=6}=2\frac{\partial^2}{\partial\eta_4^I\partial\eta_{4I}}\overline{\mathcal{H}}_{\mathcal{N}=8}.
\end{align}
Here, $\mathcal{H}$ is a $\frac{1}{2}$BPS KK graviton resonance with mass and central charge that the $\mathcal{N}=6$ gravitino is intended to inherit. When decomposed into $\mathcal{N}=6$ submultiplets, the $\mathcal{N}=8$ KK graviton has the same structure as the $\mathcal{N}=1$ chiral multiplet (\ref{N=1Chiral}). The other $\mathcal{N}=6$ submultiplets in the $\mathcal{N}=8$ supermultiplets are intended to be deleted (including the second pair of massive gravitinos contained within $\mathcal{H}$ and $\overline{\mathcal{H}}$).

The only $\mathcal{N}=6$ $3$-particle superamplitudes contained within the $\mathcal{N}=8$ massless SUGRA one (\ref{N=83graviton}) are
\begin{align}
\mathcal{A}(H^+,H^+,H^-)&=\frac{1}{M_{Pl}}\frac{1}{(\da{13}\da{23})^2}\delta^{(6)}(Q^\dagger)\tilde{\delta}^{(3)}(Q^\dagger)\nonumber\\
\mathcal{A}(H^-,H^-,H^+)&=\frac{1}{M_{Pl}}\frac{1}{(\ds{13}\ds{23})^2}\delta^{(6)}(Q)\tilde{\delta}^{(3)}(Q)
\end{align}
and $\mathcal{A}(\widetilde{G},\overline{\widetilde{G}},H^\pm)$, where $\widetilde{G}$ and $\overline{\widetilde{G}}$ are mutually conjugate massless $\mathcal{N}=6$ gravitino multiplets. I will not bother to present these latter amplitudes here since I want to eliminate them. Analogous statements hold for the decomposition of the $\mathcal{N}=8$ KK graviton superamplitude $\mathcal{A}(\mathcal{H},\overline{\mathcal{H}},H)$ into $\mathcal{N}=6$ component subamplitudes, except that here it is only one of the gravitino component superamplitudes (\ref{N=6Gravitino}) that is to be retained and the $\mathcal{N}=6$ massive graviton component should be dropped. 

The massive graviton superamplitude (\ref{4legMassiveN=8Single}) contains a pole that corresponds to the exchange of a higher mass spin-$2$ particle, demonstrating that it is not possible to consistently truncate the $\mathcal{N}=8$ theory to a single charge conjugate pair of massive multiplets. However, this pole is canceled in the extraction of the $\mathcal{N}=6$ subcomponent superamplitude describing gravitino scattering, validating that this subsector of the theory can be consistently truncated. 
If the only external particles permitted are those identified above as belonging to the $\mathcal{N}=6$ CSS theory, then there are no trivalent interactions present in which massive spin-$2$ particles are produced.

Furthermore, in the full $\mathcal{N}=8$ theory, there are no trivalent component superamplitudes involving a pair of multiplets in the $\mathcal{N}=6$ CSS theory and a third that is not. There are therefore no $4$-particle component superamplitudes consisting of only a single leg that is not in the $\mathcal{N}=6$ CSS theory (as no such amplitude can be constructed from the $3$-particle amplitudes, precluding contact interactions with stronger high energy dependence). It can therefore be inductively inferred that there are no higher leg $\mathcal{N}=6$ component superamplitudes within the $\mathcal{N}=8$ theory that have only a single leg that does not belong to the $\mathcal{N}=6$ CSS theory, as they have no permissible factorisation channels. As a result, any superamplitude with legs entirely within the $\mathcal{N}=6$ CSS theory cannot contain factorisation channels corresponding to exchanges of other particles from the $\mathcal{N}=8$ theory. Thus the $\mathcal{N}=6$ CSS theory can be obtained by consistent truncation of the $\mathcal{N}=8$ KK theory through the extraction rules above. 

The tree-level $\mathcal{N}=6$ CSS $S$-matrix is therefore contained within that of compactified maximal SUGRA and therefore inherits much of its simplicity, like BCFW constructibility. This is unsurprising given the extra dimensional interpretation of Scherk-Schwarz SUSY breaking. Spontaneously broken $\mathcal{N}=4$ SYM is analogous in the sense that it likewise can be interpreted as the low energy limit of a compactified SYM theory with fluxes about the extra dimensions. Its superamplitudes are frequently represented as embeddings within a higher-dimensional theory \cite{Dennen:2009vk,Huang:2011um,Scherk:1979zr}. 

Of course, the gravitinos carry charge under one of the ``gauged'' graviphotons in the graviton multiplet. This is elaborated upon further below. Combined also with an exchange of one of the scalars, all of the ingredients of a super-Higgs mechanism for a charged gravitino pair are present and elegantly combined together in a single superamplitude. The gravitino component amplitude can be extracted from (\ref{4legN=6}) to give
\begin{align}
    A(\widetilde{\psi},\overline{\widetilde{\psi}},\widetilde{\psi},\overline{\widetilde{\psi}})=\frac{1}{M_{Pl}^2}\frac{1}{su}\epsilon\left(1,2,3,4\right)^3.
\end{align}

Attempting to add a second gravitino pair to the theory necessarily leads to towers. Gravitational scattering of the two gravitinos off each other requires a cross-channel pole for consistency, but this can only be accommodated by a new massive spin-$2$ multiplet with central charge given by the sum or difference of the gravitinos' central charges. Its coupling to the spin-$3/2$ multiplets (calling $\Psi'$ the new gravitino) needs to be
\begin{align}
\mathcal{A}(\Psi,\Psi',\mathcal{H})=\frac{1}{M_{Pl}}\Delta_u\Delta_v^2\mathbf{v}_3
\end{align}
(a relative angle between gravitino central charges is possible, but this just appears within the $\mathcal{N}=4$ $\Delta_u\Delta_v$ subfactor and takes care of itself), while its gravitational coupling is 
\begin{align}
\mathcal{A}(\mathcal{H},\overline{\mathcal{H}},H^+)&=\frac{1}{m_{\mathcal{H}}M_{Pl}}\Delta_u\Delta_v^2\da{\bf{12}}\mathbf{v}_3^+\nonumber\\
\mathcal{A}(\mathcal{H},\overline{\mathcal{H}},H^-)&=\frac{1}{m_{\mathcal{H}}M_{Pl}}\Delta_u\Delta_v^2\ds{\bf{12}}\mathbf{v}_3^-,
\end{align}
where $m_{\mathcal{H}}$ is the mass of $\mathcal{H}$. Through a similar argument to the $\mathcal{N}=8$ case, this massive spin-$2$ multiplet generates, through consistent factorisation, an entire lattice of unique massive spin-$2$ and spin-$3/2$ states characterised entirely from their central charges. For example, the $4$-particle superamplitude for massive graviton scattering is
\begin{align}
\mathcal{A}(\mathcal{H},\overline{\mathcal{H}},\mathcal{H},\overline{\mathcal{H}})=\frac{1}{M_{Pl}^2}\frac{1}{su(t-4m_{\mathcal{H}}^2)}\Delta_u\Delta_v^2\,\epsilon(1,2,3,4).
\end{align}
This expression is clearly the $\mathcal{N}=6$ extraction from (\ref{4legMassiveN=8}). The scattering of the massive gravitons off the gravitinos similarly forces the existence higher mass gravitino resonances. Therefore only one massive gravitino pair is permitted in a decoupled IR theory.


\subsection{(Not) $\mathcal{N}=5$ SUGRA}

The $\mathcal{N}=5$ graviton multiplet differs from the $\mathcal{N}=6$ graviton multiplet by the deletion of a $\mathcal{N}=5$ gravitino submultiplet. A massive gravitino multiplet without higher spin particles must be $\frac{2}{5}$BPS and this has identical particle content to the $\frac{1}{2}$BPS gravitinos from $\mathcal{N}=6$ SUGRA above. The BPS condition here relates two pairs of supercharges, just as described for $\mathcal{N}=4$ SUSY in Section \ref{SUSYAlgebra}. This leaves behind a single remaining supercharge that must be represented with a regular $\mathcal{N}=1$ on-shell superspace (see Appendix \ref{OSsuperfields} and Section \ref{sec:SimpleN=1} above). Augmenting the non-chiral $\mathcal{N}=4$ on-shell superspace with the regular choice of $\mathcal{N}=1$ superspace, the Clifford vacuum representing the right-handed graviton multiplet is identified as a right-handed graviphoton, while the left-handed graviton is represented by a left-handed fermion:
\begin{align}
H^+_{\mathcal{N}=5}=\frac{\partial}{\partial\overline{\eta}_3}H^+_{\mathcal{N}=6}\qquad\qquad H^-_{\mathcal{N}=5}=H^-_{\mathcal{N}=6}\Big|_{\overline{\eta}_3=0}.
\end{align}
The coupling of the gravitinos to the graviton is given by superamplitudes
\begin{align}
\mathcal{A}\left(\Psi,\overline{\Psi},H^+\right)&=\frac{1}{M_{Pl}}\Delta_v\Delta_u\delta^{(2)}(Q^\dagger)\frac{1}{x}\nonumber\\
\mathcal{A}\left(\Psi,\overline{\Psi},H^-\right)&=\frac{1}{M_{Pl}}\Delta_v\Delta_u\delta^{(2)}(Q^\dagger)xF_3.
\end{align}
These may be simply extracted from the corresponding $\mathcal{N}=6$ superamplitudes above using
\begin{align}\label{SimpleExtract}
\frac{\partial}{\partial\overline{\eta}_3}\Delta_v\mathbf{v}^+_3&=\delta^{(2)}(Q^\dagger)\frac{1}{x}\qquad\qquad \Delta_v\mathbf{v}^-_3\Big|_{\overline{\eta}_3=0}=\delta^{(2)}(Q^\dagger)xF_3.
\end{align}

When combined across a factorisation channel, the graviton exchange induces an $s$-channel residue for the superamplitude $\mathcal{A}\left(\Psi,\overline{\Psi},\Psi,\overline{\Psi}\right)$ given by
\begin{align}
R_s&\sim\frac{1}{M_{Pl}^2}\epsilon\left(\mathcal{Q},\mathcal{Q},\mathcal{Q},\mathcal{Q}\right)^2\frac{1}{u}\int d\eta_P\delta^{(2)}(Q^\dagger_{12})\delta^{(2)}(Q^\dagger_{34})\left(\frac{x_{34}}{x_{12}}F_{34}+\frac{x_{12}}{x_{34}}F_{12}\right)\nonumber\\
&=\frac{1}{M_{Pl}^2}\epsilon\left(\mathcal{Q},\mathcal{Q},\mathcal{Q},\mathcal{Q}\right)^2\frac{1}{u}\delta^{(2)}\left(Q^\dagger\right)\left(\mathcal{P}_t+\mathcal{P}_u\right),
\end{align}
recycling the $\mathcal{N}=1$ SQED superresidue (see (\ref{N=1SQED}) and (\ref{N=1SQEDs})). The problem with this expression is the presence of the cross-channel pole. By assumption of the absence of higher-spin multiplets, the only other possible factorisation channel is provided by the $u$-channel graviton exchange. However, the $u$-channel residue must be proportional to the SUSY structure $\mathcal{P}_t+\mathcal{P}_s$ instead of $\mathcal{P}_t+\mathcal{P}_u$ (as is clear by the identical particle $1\leftrightarrow 3$ exchange symmetry). This clearly violates of the $4$-particle test (\ref{GenConsFact}), since the residues clearly do not add to zero. This demonstrates that $\mathcal{N}=5$ SUGRA cannot be obtained by the soft breaking of a higher-$\mathcal{N}$ SUGRA theory (at least without the presence of higher-spin particles), in agreement with the expectation from state counting \cite{Andrianopoli:2002rm}.

Consistency can be restored by the inclusion of a $\frac{2}{5}$BPS KK graviton exchange in the $t$-channel. These couple to the gravitinos through the $3$-particle superamplitude
\begin{align}
\mathcal{A}\left(\Psi,\Psi,\mathcal{H}\right)=\frac{1}{\sqrt{2}M_{Pl}}\Delta_u\Delta_v\delta^{(2)}(Q^\dagger)F_{3},
\end{align}
where this time $F_3$ is given by (\ref{SuperBlock}). The $t$-channel exchange residue for $\mathcal{A}\left(\Psi,\overline{\Psi},\Psi,\overline{\Psi}\right)$ is 
\begin{align}
R_t\sim \frac{1}{M_{Pl}^2}\epsilon\left(\mathcal{Q},\mathcal{Q},\mathcal{Q},\mathcal{Q}\right)^2\frac{-1}{s}\delta^{(2)}\left(Q^\dagger\right)\left(\mathcal{P}_s-\mathcal{P}_u\right),
\end{align}
recycling (\ref{N=1MassiveQED}). This restores consistency to the $4$-particle test, since now 
\begin{align}
uR_s+(t-m_{\mathcal{H}}^2)R_u+sR_t\propto\left(\mathcal{P}_t+\mathcal{P}_u\right)-\left(\mathcal{P}_t+\mathcal{P}_s\right)+\left(\mathcal{P}_s-\mathcal{P}_u\right)=0.
\end{align}
So it is still possible for BPS gravitinos to exist, they just cannot be decoupled from a tower in the UV. A gapped or ``gauged'' theory with a massive gravitino is prohibited by causality.

\subsection{$\mathcal{N}=4$ SUGRA and the super-Higgs mechanism}\label{sec:N=4SSSB}

Things are more complicated if only $\mathcal{N}=4$ supersymmetries are unbroken. In this case, massive spin-$3/2$ particles can be introduced without higher spin particles if they exist in $\frac{1}{2}$BPS or $\frac{1}{4}$BPS multiplets. The case of $\frac{1}{2}$BPS gravitinos will be the predominant subject of this Section. It will be shown that consistent superfactorisation requires the super-Higgs mechanism and is sufficient to reconstruct the full perturbative structure of gauged maximal supergravity (on a flat background). In SYM, the $\frac{1}{4}$BPS case would be necessarily dyonic and could not describe elementary particles (see e.g. \cite{Bergman:1997yw}). However, it was argued in Section \ref{sec:N=4SUGRAsub} that naively dyonic central charge configurations could actually be electric under certain circumstances in SUGRA. For this reason, I will first consider the possibility of elementary $\frac{1}{4}$BPS gravitino multiplets and test whether they are compatible with a consistent $S$-matrix.

\subsubsection{Superspace for long $\mathcal{N}=2$ supermultiplets}\label{GenLongVec}

A superspace representation for the $\frac{1}{4}$BPS gravitino must first be developed. The particle content of the multiplet is equivalent to the long $\mathcal{N}=3$ gravitino (\ref{N=3Gravitino}) (also the $\frac{1}{2}$BPS $\mathcal{N}=6$ gravitino above) and it is most naturally represented by a scalar Clifford vacuum. As explained in Section \ref{sec:N=4SS}, only a single pair of supercharges degenerate under the BPS constraint. The supercharges therefore decompose into two $\mathcal{N}=2$ pairs - one pair related by the BPS condition and the other pair described by a long representation. The former pair, choosing these as $Q^{\dagger1}$ and $Q_3$, are represented as in the Sections above for $\mathcal{N}=2$ BPS theories. A $3$-leg superamplitude between particles that are either BPS or massless with respect to these supercharges must contain a factor of $\Delta_v$ as the appropriate SUSY delta function. This then needs to be augmented with the delta functions for the remaining non-BPS pair of supercharges $Q^{\dagger2}$ and $Q_4$. I choose to represent these sets of supercharges with the non-chiral superspace, rather than the same-chirality superspace as was done for the long $\mathcal{N}=2$ multiplets in Section \ref{Sec:VecGrav} and Appendix \ref{sec:N=2App}, for the convenience that they would remain homogeneous in Grassmann variables and derivatives when directly added to the supercharges of $\frac{1}{2}$BPS states if present. 

For $\mathcal{N}=1$ massive multiplets, the usual choice of superspace employed elsewhere in this study is defined by a representation of the SUSY algebra given by the first line of (\ref{supercharges}). In the opposite chirality superspace, the supercharges are instead represented as 
\begin{align}
Q=\sum_i\rs{i^I}\overline{\eta}_{i,I}\qquad\qquad\qquad Q^\dagger=\sum_i\ra{i_I}\frac{\partial}{\partial \overline{\eta}_{i,I}}.
\end{align}
For massless multiplets, the supercharges in the conjugate superspace likewise have the same form as those given in the second line of (\ref{superchargesMassless}). The $\mathcal{N}=1$ superamplitudes of Section \ref{sec:SimpleN=1} can be converted into the conjugate superspace by Grassmann Fourier transform. In practice, for the $3$-particle superamplitudes of SQED and SUGRA required here, this simply amounts to replacing the delta function $\delta^{(2)}(Q^\dagger)$ with $\delta^{(2)}(Q)$, erasing the Grassmann $F$-structures (\ref{FInvProto}) or (\ref{SuperBlock}) and placing their conjugate counterparts in the $3$-particle superamplitudes in which the original $F$-structures did not appear. I will call these conjugate structures
\begin{align}\label{ConjF}
\overline{F}_3=\overline{\eta}_3-\frac{1}{2m_1}\da{31^I}\overline{\eta}_{1I}-\frac{1}{2m_2}\da{32^J}\overline{\eta}_{2J}
\end{align}
(for the case involving a single massless particle, which is all that will be required). In this conjugate superspace, the factorisation superresidues for the $4$-particle superamplitudes given in Section \ref{sec:SimpleN=1} can be easily inferred from these replacement rules. They evaluate to analogous expressions in which the bracket shapes are switched and the $\eta$ variables are replaced with $\overline{\eta}$.

Now returning to $\mathcal{N}=2$ SUSY, the central charge matrix is given generally by (\ref{N=2CC}), with $z=|z|e^{i\phi_z}$. The fermionic coherent state representations for long, non-BPS multiplets with central charges have been constructed before in \cite{Chen:2021huj,Chen:2021hjl}, although I will take a different direction here. Define angle $\alpha$
\begin{align}
\sin\alpha=\frac{|z|}{2m}
\end{align}
parameterising the saturation of the BPS limit. Then the algebra (\ref{SUSYAlg}) can be converted into the form of a set of decoupled fermionic harmonic oscillators through the definition (\ref{GenLadder}):
\begin{align}\label{SUSYAlgRot}
\{\overline{q}_a^I,\overline{q}^{\dagger bJ}\}=-\epsilon^{IJ}\delta^b_a\qquad\{\overline{q}_a^I,\overline{q}^{J}_b\}=0\qquad\{\overline{q}^{\dagger aI},\overline{q}^{\dagger bJ}\}=0
\end{align}
with ladder operators given by 
\begin{align}\label{LongMixedLadder}
\overline{q}^I_a=\sqrt{\frac{1+\cos\alpha}{2\cos^2\alpha}}\left(q_a^I+e^{i\phi_z}\sqrt{\frac{1-\cos\alpha}{1+\cos\alpha}}\epsilon_{ab}q^{\dagger bI}\right)
\end{align}
Generally, a long $\mathcal{N}=2$ multiplet with a central charge can be represented as a fermionic coherent state of the algebra (\ref{SUSYAlgRot}) with some choice of Clifford vacuum. 

The SUSY delta functions can remain homogeneous Grassmann polynomials (as they conventionally are for superamplitudes of long multiplets in the absence of central charges - see Appendix \ref{N=2Rigid}) provided that $q^{\dagger 2I}$ and $q_4^I$ are represented homogeneously as Grassmann variables and not derivatives. In this case, the SUSY delta functions are simply the usual $\delta^{(2)}(Q^{\dagger2})\delta^{(2)}(Q_4)$, but the supercharges contain a mixture of Grassmann variables. In such a representation,
\begin{align}\label{SUSYRepsLongGen}
q^{\dagger 2I}&=-\sqrt{\frac{1+\cos\alpha}{2}}\left(\eta^{2I}+e^{-i\phi_z}\sqrt{\frac{1-\cos\alpha}{1+\cos\alpha}}\overline{\eta}^{4I}\right)\nonumber\\
q^{I}_4&=\sqrt{\frac{1+\cos\alpha}{2}}\left(e^{i\phi_z}\sqrt{\frac{1-\cos\alpha}{1+\cos\alpha}}\eta^{2I}+\overline{\eta}^{4I}\right).
\end{align}
The complete SUSY delta function for a $\mathcal{N}=4$ $\frac{1}{4}$BPS superamplitude is therefore the combination $\Delta_v\delta^{(2)}(Q^{\dagger 2})\delta^{(2)}(Q_{4})$. 

In addition to the delta function, the non-BPS $\mathcal{N}=2$ factor of the superamplitudes may also contain further SUSY invariant Grassmann polynomials, analogous to the $\mathcal{N}=1$ $F$-structures introduced in Section \ref{sec:SimpleN=1}. Again, these should be invariant under the action of the derivatively represented supercharges:
\begin{align}
q_2^I&=-\sqrt{\frac{1+\cos\alpha}{2}}\left(\frac{\partial}{\partial\eta^2_I}+e^{i\phi_z}\sqrt{\frac{1-\cos\alpha}{1+\cos\alpha}}\frac{\partial}{\partial\overline{\eta}^4_I}\right)\nonumber\\
q^{\dagger 4I}&=-\sqrt{\frac{1+\cos\alpha}{2}}\left(e^{-i\phi_z}\sqrt{\frac{1-\cos\alpha}{1+\cos\alpha}}\frac{\partial}{\partial\eta^2_I}+\frac{\partial}{\partial\overline{\eta}^4_I}\right).
\end{align}
Defining new Grassmann variables
\begin{align}\label{XGrass}
\xi^{2I}&=\sqrt{\frac{1+\cos\alpha}{2}}\left(\eta^{2I}+e^{-i\phi_z}\sqrt{\frac{1-\cos\alpha}{1+\cos\alpha}}\overline{\eta}^{4I}\right)\nonumber\\
\xi^{4I}&=\sqrt{\frac{1+\cos\alpha}{2}}\left(e^{i\phi_z}\sqrt{\frac{1-\cos\alpha}{1+\cos\alpha}}\eta^{2I}+\overline{\eta}^{4I}\right),
\end{align}
the multiplicatively represented supercharges become $Q_i^{2\dagger}=-\ra{i^I}\xi_{i,I}^2$ and $Q_{i,4}=\rs{i^I}\xi_{i,I}^4$. However, it is only under the further change of variables
\begin{align}\label{ZGrass}
\zeta^{2I}&=\frac{1}{\cos^2\alpha}\left(\xi^{2I}-e^{-i\phi_z}\sin\alpha\,\xi^{4I}\right)\nonumber\\
\zeta^{4I}&=\frac{1}{\cos^2\alpha}\left(-e^{i\phi_z}\sin\alpha\,\xi^{2I}+\xi^{4I}\right)
\end{align}
that the derivatively represented supercharges take the simple form $Q_{i,2}=\rs{i_I}\frac{\partial}{\partial\zeta^{2}_{i,I}}$ and $Q_{i}^{\dagger 4}=\ra{i_I}\frac{\partial}{\partial\zeta^{4}_{i,I}}$ expected for the regular $\mathcal{N}=1$ superspace. It immediately follows that the analogous SUSY-invariant linear Grassmann polynomials to (\ref{FInvProto}), (\ref{ConjF}) and generally (\ref{SuperBlock}) that are applicable to the superamplitudes here are given simply by replacing $\eta^2$ and $\overline{\eta}^4$ with $\zeta^2$ and $\zeta^4$ respectively for each of the $\frac{1}{4}$BPS legs. I will continue to denote these structures by the same symbols $F$ and $\overline{F}$ (and append them with appropriate $R$-indices). Clearly, the mixing between $R$-sectors disappears when $\alpha\rightarrow 0$.

As a warm-up for the application to $\mathcal{N}=4$ $\frac{1}{4}$BPS particles, I will first apply this superspace to a brief re-examination of the gravitational superamplitudes of long massive vector multiplets with $\mathcal{N}=2$ SUGRA (discussed earlier in Section \ref{sec:pureGrav}), this time allowing for non-BPS saturating central charges. For ease of comparison with the ensuing analysis of $\mathcal{N}=4$ $\frac{1}{4}$BPS particles, I will continue to denote the two supercharges as $Q^{\dagger2}$ and $Q_4$. The superamplitudes in which long vector multiplets carrying central charges couple to the graviton are given by
\begin{align}\label{GenLongVecAmp}
\mathcal{A}\left(\mathcal{W},\overline{\mathcal{W}},H^+\right)&=\frac{1}{M_{Pl}\cos^2\alpha}\delta^{(2)}(Q^{\dagger 2})\delta^{(2)}(Q_4)\frac{1}{x}\left(\frac{1}{x}\overline{F}^{4}_3-e^{i\phi_z}\sin\alpha F_3^2\right)\nonumber\\
\mathcal{A}\left(\mathcal{W},\overline{\mathcal{W}},H^-\right)&=\frac{1}{M_{Pl}\cos^2\alpha}\delta^{(2)}(Q^{\dagger 2})\delta^{(2)}(Q_4)x\left(x F_3^2-e^{-i\phi_z}\sin\alpha \overline{F}^{4}_3\right).
\end{align}
The SUSY structure is fixed by the preclusion of inconsistent gravitational dipole moments. This coincides with the elimination of the terms mixing the $R$-sectors for the massive legs: 
\begin{align}\label{GenLongSUSYBlocks}
\frac{1}{x}\overline{F}^{4}_3-e^{i\phi_z}\sin\alpha F_3^2&=\frac{1}{x}\left(\overline{\eta}_3^4-\frac{1}{2m_\Psi}\left(\da{31^I}\xi^4_{1,I}+\da{32^J}\xi^4_{2,J}\right)\right)-e^{i\phi_z}\sin\alpha \eta_3^2\nonumber\\
x F_3^2-e^{-i\phi_z}\sin\alpha \overline{F}^{4}_3&=x\left(\eta_3^2+\frac{1}{2m_\Psi}\left(\ds{31^I}\xi^2_{1,I}+\ds{32^J}\xi^2_{2,J}\right)\right)-e^{-i\phi_z}\sin\alpha\overline{\eta}^4_3.
\end{align}
These evidently have the same Grassmann structure as the chargeless case with the exception of the last term.

The transformation from $\eta$ to $\xi$ type Grassmann variables (\ref{XGrass}) is non-unitary. As a result, when the fermionic coherent state or superamplitude is written as a polynomial in $\xi$ basis variables, normalisation factors of $\sqrt{\cos\alpha}$ must be manually included for each $\xi$ variable when extracting the component states or amplitudes from the coefficients. These normalisation factors cancel the overall factor of $1/\cos^2\alpha$ in (\ref{GenLongVecAmp}) when extracting the component amplitudes involving the graviton. 

In the BPS limit $\alpha\rightarrow \frac{\pi}{2}$, the Grassmann variables $\xi^2$ and $\xi^4$ degenerate. Defining implicitly another basis of Grassmann variables $\Xi$ and $Z$ through 
\begin{align}
\xi^2&=\frac{1}{2}\left(\left(\sqrt{1+\cos\alpha}+\sqrt{1-\cos\alpha}\right)\Xi+\left(\sqrt{1+\cos\alpha}-\sqrt{1-\cos\alpha}\right)Z\right)\nonumber\\
\xi^4&=\frac{e^{i\phi}}{2}\left(\left(\sqrt{1+\cos\alpha}+\sqrt{1-\cos\alpha}\right)\Xi-\left(\sqrt{1+\cos\alpha}-\sqrt{1-\cos\alpha}\right)Z\right),
\end{align}
then in the BPS limit, the $\xi$ variables both clearly converge to $\Xi$. The $\Xi$ variables become the Grassmann parameters of the BPS coherent states that emerge in the limit, while the residual $Z$ variables grade the decomposition of the long multiplet into BPS multiplets. The long vector multiplets each decompose into a BPS vector multiplet and a (full) hypermultiplet. The superamplitudes (\ref{GenLongVecAmp}) can be verified to match onto the expected expressions in Section \ref{sec:N=2SUGRA} for BPS vector and hypermultiplets coupling to gravity. They are accompanied by fully quadratic factors in the residual $Z$-type Grassmann variables. The overall factor of $1/\cos^2\alpha$ in (\ref{GenLongVecAmp}) cancels the zeros that emerge at leading order from the supercharges degenerating in the delta functions (as in (\ref{SuperDegen})) and from the overall SUSY invariant factors (\ref{GenLongSUSYBlocks}) vanishing. 

I will leave the construction of the general $4$-particle superamplitude describing the gravitational scattering of long vector multiplets to a future study. My interest in this general case here is its application to testing the consistency of $\frac{1}{4}$BPS particles in $\mathcal{N}=4$ SUGRA below. However, I will note here that, just as for the chargeless case described in Section \ref{sec:pureGrav}, the graviton exchange superresidue is still free of cross-channel poles. Consistency of the residue for ``Moller'' scattering of multiplets with a relative central charge phase still requires the presence of an additional graviphoton multiplet for the same reasons as already described in the Sections above. Consistency of Compton scattering of the graviphotons also demands that the central charge be conserved. However, besides these, there are no further constraints emerging from applying the $4$-particle test to long massive vector scattering, with or without gravity. A Lie algebra structure for the vector boson self-couplings is not a consequence of causality but instead a property of a special subset of theories with partially suppressed high energy scaling. This extends to the inclusion of the graviphoton, which generally induces $\mathcal{O}(E^3)$ scaling terms in the superamplitude that are not present in the chargeless case (\ref{N=2vecGravLong}). It is only in the BPS limit that the Lie algebra structure, which includes the graviphoton, becomes obligatory and the high energy divergence is suppressed to $\sim E^2$. 

\subsubsection{(Not) $\frac{1}{4}$BPS gravitinos}\label{sec:1/4BPS}

With the superspace and Grassmann structure established, the $3$-particle superamplitudes involving the candidate $\frac{1}{4}$BPS RS particles (or ``gravitinos'') can be constructed. The couplings to the graviton are given by
\begin{align}
\mathcal{A}\left(\Psi,\overline{\Psi},H^+\right)&=\frac{e^{-i\varphi}}{M_{Pl}m_{\Psi}\cos^2\alpha}\Delta_v\delta^{(2)}(Q^{\dagger 2})\delta^{(2)}(Q_4)\mathbf{v}_3^+\left(\frac{1}{x}\overline{F}^{4}_3-e^{i\phi_z}\sin\alpha F_3^2\right)\nonumber\\
\mathcal{A}\left(\Psi,\overline{\Psi},H^-\right)&=\frac{e^{-i\varphi}}{M_{Pl}m_{\Psi}\cos^2\alpha}\Delta_v\delta^{(2)}(Q^{\dagger 2})\delta^{(2)}(Q_4)\mathbf{v}_3^-\left(x F_3^2-e^{-i\phi_z}\sin\alpha \overline{F}^{4}_3\right).
\end{align}
The phase $\varphi$ is that of the BPS-saturating central charge and is also the associated mass phase with respect to which the little group spinors $\mathbf{v}_3^\pm$ are defined. The mass of the RS particle is $m_\Psi$. As usual, I am precluding possible SUSY structures that would produce deviations to the gravitational dipole moments. 

The next step is to compute the superresidue for the gravitational scattering of the $\frac{1}{4}$BPS RS multiplet off itself and test for self-consistency. The superamplitudes factorise into a $\mathcal{N}=2$ BPS subsector and a long subsector. The $R$-sectors of the latter generally mix for $\alpha\neq 0$, as is clear from (\ref{GenLongSUSYBlocks}). I will begin by performing the calculation for the special case of $\alpha=0$. In this case, the long subsector further factorises into $\mathcal{N}=1$ subsectors and the residues of $\mathcal{N}=1$ superamplitudes given in Section \ref{sec:SimpleN=1} can be recycled identically. 
When $\alpha\neq 0$, the mixing between $R$-sectors mildly complicates the calculation, but I will argue that the conclusion is unchanged.

If two $\frac{1}{4}$BPS gravitinos scatter off each other, then the $s$-channel graviton exchange residue for the superamplitude $\mathcal{A}\left(\Psi,\overline{\Psi},\Psi,\overline{\Psi}\right)$ is
\begin{align}\label{1/4BPSGravEx}
&R_s\sim\frac{e^{-2i\varphi}}{(M_{Pl}m_\Psi)^2}\epsilon\left(\mathcal{Q},\mathcal{Q},\mathcal{Q},\mathcal{Q}\right)\frac{1}{(\mathbf{v}_P^{(12)}\cdot\mathbf{v}_P^{(34)})}\nonumber\\
&\qquad\qquad\times\int d^2\eta_P\delta^{(2)}(Q^{\dagger 2}_{12})\delta^{(2)}(Q^{\dagger 2}_{34})\delta^{(2)}(Q_{4,12})\delta^{(2)}(Q_{4,34})\nonumber\\
&\qquad\qquad\qquad\qquad\times\left(\mathbf{v}^{(12)+}_P\mathbf{v}^{(34)-}_P\frac{x_{34}}{x_{12}}F^{2}_{34}\overline{F}^{4}_{12}+\mathbf{v}^{(12)+}_P\mathbf{v}^{(34)-}_P\frac{x_{12}}{x_{34}}F^{2}_{12}\overline{F}^{4}_{34}\right)\nonumber\\
&=\frac{e^{-2i\varphi}}{M_{Pl}^2}\epsilon\left(\mathcal{Q},\mathcal{Q},\mathcal{Q},\mathcal{Q}\right)\frac{1}{u}\nonumber\\
&\quad\times\int d^2\eta_P\delta^{(2)}(Q^{\dagger 2}_{12})\delta^{(2)}(Q^{\dagger 2}_{34})\delta^{(2)}(Q_{4,12})\delta^{(2)}(Q_{4,34})\nonumber\\
&\qquad\times\bigg(\left(\frac{x_{34}}{x_{12}}F^2_{34}+\frac{x_{12}}{x_{34}}F^2_{12}\right)\left(\frac{x_{34}}{x_{12}}\overline{F}^4_{12}+\frac{x_{12}}{x_{34}}\overline{F}^4_{34}\right)\nonumber\\
&\qquad\qquad-\frac{1}{2}\left(\frac{x_{34}}{x_{12}}F^2_{34}+\frac{x_{12}}{x_{34}}F^2_{12}\right)\left(\overline{F}^4_{34}+\overline{F}^4_{12}\right)-\frac{1}{2}\left(F^2_{34}+F^2_{12}\right)\left(\frac{x_{34}}{x_{12}}\overline{F}^4_{12}+\frac{x_{12}}{x_{34}}\overline{F}^4_{34}\right)\nonumber\\
&\qquad\qquad\qquad\qquad\qquad\qquad\qquad\qquad\qquad+\left(\frac{1}{2}\left(\frac{x_{34}}{x_{12}}+\frac{x_{12}}{x_{34}}\right)-1\right)\left(F^2_{12}\overline{F}^4_{12}+F^2_{34}\overline{F}^4_{34}\right)\bigg)\nonumber\\
&=\frac{e^{-2i\varphi}}{M_{Pl}^2}\epsilon\left(\mathcal{Q},\mathcal{Q},\mathcal{Q},\mathcal{Q}\right)\delta^{(2)}(Q^{\dagger 2})\delta^{(2)}(Q_4)\frac{1}{u}\nonumber\\
&\qquad\times\bigg(-\frac{u}{2m_\Psi^2}\left(\mathcal{P}_{s12}^2\overline{\mathcal{P}}_{s12}^4+\mathcal{P}_{s34}^2\overline{\mathcal{P}}_{s34}^4\right)+\left(\mathcal{P}_t^2+\mathcal{P}_u^2\right)\left(\overline{\mathcal{P}}_t^4+\overline{\mathcal{P}}_u^4\right)\nonumber\\
&\qquad\qquad\qquad+\frac{1}{2}\left(\mathcal{P}_t^2+\mathcal{P}_u^2\right)\left(\overline{\mathcal{P}}_{s12}^4+\overline{\mathcal{P}}_{s34}^4\right)+\frac{1}{2}\left(\mathcal{P}_{s12}^2+\mathcal{P}_{s34}^2\right)\left(\overline{\mathcal{P}}_t^4+\overline{\mathcal{P}}_u^4\right)\bigg).
\end{align}
I have recycled the $\mathcal{N}=1$ QED and WZ superresidues (\ref{N=1SQEDs}) and (\ref{N=1Yuks}). The SUSY structures were defined in Section \ref{sec:SimpleN=1}. 
The parity conjugate structures $\overline{\mathcal{P}}$ are defined analogously to (\ref{SuperLorentz}) but with the bracket shapes swapped. The $u$-channel residue is given by exchanging the identical (bosonic) gravitino multiplets $1\leftrightarrow3$. Clearly, the single pole terms in each channel are linearly independent and cannot cancel in the $4$-particle test. The ``$d$-wave'' term above proportional to $\mathcal{P}_u^2\overline{\mathcal{P}}_u^4$ is a notable example of this. 

The gravitinos can also have possible couplings to massless vector multiplets:
\begin{align}
\mathcal{A}\left(\Psi,\overline{\Psi},V_A\right)&=\frac{e^{-i\varphi}}{M_{Pl}m_{\Psi}\cos^2\alpha}\Delta_v\delta^{(2)}(Q^{\dagger 2})\delta^{(2)}(Q_4)\nonumber\\
&\qquad\times\left(f_A\mathbf{v}_3^+\left(\overline{F}^{4}_3-e^{i\phi_z}\sin\alpha xF^2_3\right)+(f_A)^*\mathbf{v}_3^-\left(F^2_3-e^{-i\phi_z}\sin\alpha\frac{1}{x}\overline{F}^{4}_3\right)\right)
\end{align}
for some constants $f_A\in\mathbb{C}$. Again with $\alpha=0$, the vector exchange residue is given by:
\begin{align}\label{1/4BPSVecEx}
R_s&\sim\frac{e^{-2i\varphi}}{(M_{Pl}m_\Psi)^2}\epsilon\left(\mathcal{Q},\mathcal{Q},\mathcal{Q},\mathcal{Q}\right)\frac{1}{(\mathbf{v}_P^{(12)}\cdot\mathbf{v}_P^{(34)})}\nonumber\\
&\qquad\qquad\times\int d^2\eta_P\delta^{(2)}(Q^{\dagger 2}_{12})\delta^{(2)}(Q^{\dagger 2}_{34})\delta^{(2)}(Q_{4,12})\delta^{(2)}(Q_{4,34})\nonumber\\
&\qquad\qquad\qquad\qquad\times\left(f_A\mathbf{v}_P^{(12)+}\overline{F}^{4}_{12}+(f_A)^*\mathbf{v}_P^{(12)-}F^2_{12}\right)\left(f_A\mathbf{v}_P^{(34)+}\overline{F}^{4}_{34}+(f_A)^*\mathbf{v}_P^{(34)-}F^2_{34}\right)\nonumber\\
&=\frac{e^{-2i\varphi}|f_A|^2}{M_{Pl}^2}\epsilon\left(\mathcal{Q},\mathcal{Q},\mathcal{Q},\mathcal{Q}\right)\frac{1}{u}\nonumber\\
&\quad\times\int d^2\eta_P\delta^{(2)}(Q^{\dagger 2}_{12})\delta^{(2)}(Q^{\dagger 2}_{34})\delta^{(2)}(Q_{4,12})\delta^{(2)}(Q_{4,34})\nonumber\\
&\qquad\times\bigg(\frac{1}{2}\left(\frac{x_{34}}{x_{12}}F^2_{34}+\frac{x_{12}}{x_{34}}F^2_{12}\right)\left(\overline{F}^4_{34}+\overline{F}^4_{12}\right)+\frac{1}{2}\left(F^2_{34}+F^2_{12}\right)\left(\frac{x_{34}}{x_{12}}\overline{F}^4_{12}+\frac{x_{12}}{x_{34}}\overline{F}^4_{34}\right)\nonumber\\
&\qquad\qquad\qquad\qquad\qquad\qquad\qquad\qquad-\left(\frac{1}{2}\left(\frac{x_{34}}{x_{12}}+\frac{x_{12}}{x_{34}}\right)-1\right)\left(F^2_{12}\overline{F}^4_{12}+F^2_{34}\overline{F}^4_{34}\right)\bigg)\nonumber\\
&=\frac{e^{-2i\varphi}|f_A|^2}{M_{Pl}^2}\epsilon\left(\mathcal{Q},\mathcal{Q},\mathcal{Q},\mathcal{Q}\right)\frac{1}{u}\nonumber\\
&\qquad\times\bigg(\frac{u}{2m_\Psi^2}\left(\mathcal{P}_{s12}^2\overline{\mathcal{P}}_{s12}^4+\mathcal{P}_{s34}^2\overline{\mathcal{P}}_{s34}^4\right)\nonumber\\
&\qquad\qquad\qquad-\frac{1}{2}\left(\mathcal{P}_t^2+\mathcal{P}_u^2\right)\left(\overline{\mathcal{P}}_{s12}^4+\overline{\mathcal{P}}_{s34}^4\right)-\frac{1}{2}\left(\mathcal{P}_{s12}^2+\mathcal{P}_{s34}^2\right)\left(\overline{\mathcal{P}}_t^4+\overline{\mathcal{P}}_u^4\right)\bigg).
\end{align}
again recycling $\mathcal{N}=1$ SQED residues. Again, the $u$-channel residue can be easily inferred from particle exchanges and it is clear that these residues by themselves are inconsistent (which justifies the claims made earlier that objects with $\frac{1}{4}$BPS central charges are expected to be dyonic, or at least not particles). However, it is also clear that, combining the vector and graviton exchange residues, it is not possible to tune the couplings $f_A$ to ensure that the $4$-particle test is satisfied. This is clear from the absence of the single pole $d$-wave term in (\ref{1/4BPSGravEx}) from (\ref{1/4BPSVecEx}). This superamplitude therefore still fails the $4$-particle test. 

If $\alpha\neq0$, then the long subsector of the superresidues are modified. In this case, the complete expressions for the SUSY structures (\ref{GenLongSUSYBlocks}) are required. In the $4$-particle graviton exchange superresidue, the non-BPS subfactor is modified to 
\begin{align}\label{GenIntCC}
&\frac{1}{\cos^4\alpha}\int d^2\eta_P\delta^{(2)}(Q^{\dagger2}_{12})\delta^{(2)}(Q^{\dagger2}_{34})\delta^{(2)}(Q_{4,12})\delta^{(2)}(Q_{4,34})\nonumber\\
&\quad\times\bigg(\left(\frac{x_{34}}{x_{12}}\right)^2F_{34}^2\overline{F}_{12}^4+\left(\frac{x_{12}}{x_{34}}\right)^2F_{12}^2\overline{F}_{34}^4+\sin^2\alpha\left(\frac{x_{34}}{x_{12}}+\frac{x_{12}}{x_{34}}\right)\eta^2_P\overline{\eta}^4_P\nonumber\\
&\qquad\qquad\quad-e^{i\phi_z}\sin\alpha\left(\frac{x_{34}^2}{x_{12}}F_{34}^2+\frac{x_{12}^2}{x_{34}}F_{12}^2\right)\eta_P^2-e^{-i\phi_z}\sin\alpha\left(\frac{x_{34}^2}{x_{12}}\overline{F}_{34}^4+\frac{x_{12}^2}{x_{34}}\overline{F}_{12}^4\right)\overline{\eta}_P^4\bigg).
\end{align}
Here, I define the $F$-structures as in (\ref{FInvProto}) and (\ref{ConjF}) but with $\xi$ variables instead of $\eta$ variables. The first pair of terms in the integrand are the same as those appearing in (\ref{1/4BPSGravEx}) (except for the overall factor of $1/\cos^4\alpha$). The new terms in the last line have different Grassmann degrees in each $R$-sector. They therefore cannot interfere with the inconsistency of the superresidue already established in the $\alpha=0$ case. Finally, the remaining new terms integrate and evaluate to 
\begin{align}
\frac{\sin^2\alpha}{\cos^4\alpha}\frac{u-2m_\Psi^2}{4m_\psi^2}\delta^{(2)}(Q^{\dagger 2})\delta^{(2)}(Q_4)\da{Q^{\dagger 2}_{12}Q^{\dagger 2}_{34}}\ds{Q_{4,12}Q_{4,34}}.
\end{align}
I define $Q^{\dagger 2}_{12}$ and $Q_{4,12}$ as the sum of the appropriate supercharges of particles $1$ and $2$ without that of the internal particle (and analogously for $Q^{\dagger 2}_{34}$ and $Q_{4,34}$). This new term is an independent SUSY/Lorentz structure compared to anything appearing in the residues (\ref{1/4BPSGravEx}) and (\ref{1/4BPSVecEx}). It therefore also cannot interfere with the inconsistency of these results. Allowing for $\alpha\neq0$ in the vector exchange residue simply has the effect of replacing $f_A$ with $(f_A-(f_A)^*e^{-i(\phi+\varphi)}\sin\alpha)/\cos^2\alpha$. The SUSY structure is unchanged. The allowance for a second non-zero central charge (provided that it is not BPS saturating) therefore does not change the conclusions of $4$-particle test in the $\alpha=0$ case. The $\frac{1}{4}$BPS gravitino remains inconsistent. 

It remains to examine whether a possible $t$-channel exchange process can restore consistency. A particle exchanged in the $t$-channel must have double the central charges of $\Psi$. This must therefore either be long or $\frac{1}{4}$BPS. Just as for the $s$- or $u$-channels, a long multiplet will only couple to the $\frac{1}{4}$BPS particles through the regular (non-BPS) SUSY delta functions. The exchange residues that they induce do not contain potentially spurious poles and therefore do not interfere with the $4$-particle test. This leaves the possibility of a new $\frac{1}{4}$BPS particle. Precluding higher spin particles, then this would be another $\frac{1}{4}$BPS gravitino (call it $\Psi_2$). The resulting $3$-leg superamplitude is determined by SUSY, tiny group and Lorentz invariance as
\begin{align}\label{1/4BPS3part}
\mathcal{A}\left(\Psi,\Psi,\overline{\Psi}_{2A}\right)&=\Delta_v\delta^{(2)}(Q^{\dagger 2})\delta^{(2)}(Q_4)\nonumber\\
&\qquad\times\left(\mathbf{v}_{3K}F^{2K}_{12}\left(\alpha_A+\beta_A\overline{F}^{4}_{12M}\overline{F}^{4M}_{12}\right)+\mathbf{v}_{3K}\overline{F}^{4K}_{12}\left(\gamma_A+\delta_AF^{2}_{12M}F^{2M}_{12}\right)\right).
\end{align}  
Here, $F^{2K}_{12}$ is defined as in (\ref{SuperBlock}) with the variables $\zeta^{2}$ in (\ref{ZGrass}) replacing the $\eta$ variables in that equation, whereas $\overline{F}^{4K}_{12}$ is analogous with $\zeta^{4}$ variables instead and opposite bracket shapes. These are the SUSY building blocks for fully massive $3$-particle superamplitudes. The subscript $A$ is a possible species index and $\alpha_A$, $\beta_A$, $\gamma_A$ and $\delta_A$ are thus far undetermined coupling constants. Rewritten in the $\xi$ basis (\ref{XGrass}),
\begin{align}
F_{12}^2&=\frac{1}{\cos^2\alpha}\left(F_{\xi12}^2-e^{-i\phi_z}\sin\alpha\left(\overline{F}_{\xi12}^4-\frac{1}{2m_\Psi}\la{\bf{3}}p_1-p_2\rs{Q_{4,12}}\right)\right)\nonumber\\
\overline{F}_{12}^4&=\frac{1}{\cos^2\alpha}\left(\overline{F}_{\xi12}^4-e^{i\phi_z}\sin\alpha\left(F_{\xi12}^2-\frac{1}{2m_\Psi}\ls{\bf{3}}p_1-p_2\ra{Q_{12}^{\dagger2}}\right)\right).
\end{align}
Here $F_{\xi12}^2$ is just (\ref{SuperBlock}) in which the Grassmann variables are just $\xi^2$ instead of $\eta$ (and analogously for $\overline{F}_{\xi12}^4$). The superamplitude (\ref{1/4BPS3part}) can then be rewritten as
\begin{align}\label{1/4BPS3part2}
&\mathcal{A}\left(\Psi,\Psi,\overline{\Psi}_{2A}\right)\nonumber\\
&=\frac{1}{\cos^2\alpha}\Delta_v\delta^{(2)}(Q^{\dagger 2})\delta^{(2)}(Q_4)\nonumber\\
&\quad\times\bigg(\alpha_A\mathbf{v}_{3}\cdot\left(F_{\xi12}^{2}-e^{-i\phi_z}\sin\alpha\overline{F}^{4}_{\xi12}\right)+\gamma_A\mathbf{v}_{3}\cdot\left(\overline{F}^{4}_{\xi12}-e^{i\phi_z}\sin\alpha F_{\xi12}^{2}\right)\nonumber\\
&\quad\quad\quad+\frac{1}{\cos^2\alpha}\beta_A\bigg(\frac{2}{m_\Psi}e^{2i\phi_z}\tan^2\alpha\da{Q^{\dagger2}_{12}Q^{\dagger2}_{12}}\mathbf{v}_{3}\cdot\left(F_{\xi12}^{2}-e^{-i\phi_z}\sin\alpha\overline{F}^{4}_{\xi12}\right)\nonumber\\
&\qquad\qquad\qquad\quad\qquad+\left(\overline{F}^{4}_{\xi12}\cdot \overline{F}^{4}_{\xi12}\right)\left(\mathbf{v}_3\cdot F^2_{\xi 12}\right)+e^{i\phi_z}\sin\alpha\left(F^{2}_{\xi12}\cdot F^{2}_{\xi12}\right)\left(\mathbf{v}_3\cdot \overline{F}^4_{\xi 12}\right)\nonumber\\
&\qquad\qquad\qquad\quad\qquad+\frac{1}{m_\Psi^2}e^{i\phi_z}\sin\alpha\left(F^{2}_{\xi12}\cdot\ls{\bf{3}}p_1-p_2\ra{Q^{\dagger 2}_{12}}\right)\left(\mathbf{v}_3\cdot \overline{F}^4_{\xi 12}\right)\bigg)\nonumber\\
&\quad\quad\quad+\frac{1}{\cos^2\alpha}\delta_A\bigg(-\frac{2}{m_\Psi}e^{2i\phi_z}\tan^2\alpha\ds{Q_{4,12}Q_{4,12}}\mathbf{v}_{3}\cdot\left(\overline{F}^{4}_{\xi12}-e^{i\phi_z}\sin\alpha F_{\xi12}^{2}\right)\nonumber\\
&\qquad\qquad\qquad\qquad\quad+\left(F^2_{\xi 12}\cdot F^2_{\xi 12}\right)\left(\mathbf{v}_3\cdot \overline{F}^{4}_{\xi12}\right)+e^{-i\phi_z}\sin\alpha\left(\overline{F}^4_{\xi 12}\cdot \overline{F}^4_{\xi 12}\right)\left(\mathbf{v}_3\cdot F^2_{\xi 12}\right)\nonumber\\
&\qquad\qquad\qquad\qquad\quad+\frac{1}{m_\Psi^2}e^{-i\phi_z}\sin\alpha\left(\overline{F}^4_{\xi 12}\cdot\la{\bf{3}}p_1-p_2\rs{Q_{4,12}}\right)\left(\mathbf{v}_3\cdot F^{2}_{\xi12}\right)\bigg)\bigg),
\end{align}
where I am abbreviating the internal $SU(2)$ little group contraction by the dot product. 

Now, if the $t$-channel exchange is to restore consistency, then it must interfere with the uncanceled terms from the graviton exchange residues. This always contains terms that are Grassman degree $2$ in both $\xi^2$ and $\xi^4$ variables. From (\ref{1/4BPS3part2}), it is clear that all terms of this form arising in a potential $t$-channel exchange residue must contain a factor of either $\left(\mathbf{v}_P^{(13)}\cdot F^{2}_{\xi13}\right)\left(\mathbf{v}_P^{(24)}\cdot F^{2}_{\xi24}\right)$ or $\left(\mathbf{v}_P^{(13)}\cdot \overline{F}^{4}_{\xi13}\right)\left(\mathbf{v}_P^{(24)}\cdot \overline{F}^{4}_{\xi24}\right)$. Combining the BPS sector delta functions across the factorisation channel produces an overall factor of $\int d^2\eta_P\Delta_v^{(13)}\Delta_v^{(24)}\sim-\frac{1}{s}\epsilon\left(\mathcal{Q},\mathcal{Q},\mathcal{Q},\mathcal{Q}\right)\mathbf{u}_P^{(13)}\cdot\mathbf{u}_P^{(24)}$ (as usual). Using (\ref{SpecialSVD2}) and (\ref{Trimmed3PSK}), then
\begin{align}\label{EvenDegRes}
&\mathbf{u}_P^{(13)}\cdot\mathbf{u}_P^{(24)}\left(\mathbf{v}_P^{(13)}\cdot F^{2}_{\xi13}\right)\left(\mathbf{v}_P^{(24)}\cdot F^{2}_{\xi24}\right)\nonumber\\
&\qquad\qquad=\xi_{1I}^2\left(\frac{1}{2}\left(\la{1^I}p_3-p_4\rs{2^J}+\ls{1^I}p_3-p_4\ra{2^J}\right)-m_\Psi\left(\da{1^I2^J}+\ds{1^I2^J}\right)\right)\xi_{2J}^2\nonumber\\
&\qquad\qquad\qquad\qquad\qquad-(1\leftrightarrow 3)-(2\leftrightarrow 4)+(1\leftrightarrow 3)\land(2\leftrightarrow 4).
\end{align}
In this specific context, the syzygy (\ref{OldSyz2}) can be supersymmetrised to $\mathcal{N}=1$ form
\begin{align}
&\frac{2}{m}\delta^{(2)}(Q^\dagger)\bigg(\left(s-\frac{4}{3}m^2\right)\left(\eta_{1I}\da{1^I4^L}\eta_{4L}+\eta_{2J}\da{2^J3^K}\eta_{3K}\right)\nonumber\\
&\qquad\qquad\qquad\qquad-\left(u-\frac{4}{3}m^2\right)\left(\eta_{1I}\da{1^I2^J}\eta_{2J}+\eta_{3K}\da{3^K4^L}\eta_{4L}\right)\bigg)\nonumber\\
&=\delta^{(2)}(Q^\dagger)\bigg(\eta_{1I}\ls{1^I}p_3-p_4\ra{2^J}\eta_{2J}+\eta_{1I}\la{1^I}p_3-p_4\rs{2^J}\eta_{2J}\nonumber\\
&\qquad\qquad\quad+\eta_{3K}\ls{3^K}p_1-p_2\ra{4^L}\eta_{4L}+\eta_{3K}\la{3^K}p_1-p_2\rs{4^L}\eta_{4L}\nonumber\\
&\qquad\qquad\quad+\eta_{1I}\ls{1^I}p_2-p_3\ra{4^L}\eta_{4L}+\eta_{1I}\la{1^I}p_2-p_3\rs{4^L}\eta_{4L}\nonumber\\
&\qquad\qquad\quad+\eta_{2J}\ls{2^J}p_1-p_4\ra{3^K}\eta_{3K}+\eta_{2J}\la{2^J}p_1-p_4\rs{3^K}\eta_{3K}\nonumber\\
&\qquad\qquad-2m\left(\eta_{1I}\ds{1^I2^J}\eta_{2J}+\eta_{3K}\ds{3^K4^L}\eta_{4L}-\eta_{1I}\ds{1^I4^L}\eta_{4L}-\eta_{2J}\ds{2^J3^K}\eta_{3K}\right)\nonumber\\
&\qquad\qquad+\frac{2}{3}m\left(\eta_{1I}\da{1^I2^J}\eta_{2J}+\eta_{3K}\da{3^K4^L}\eta_{4L}-\eta_{1I}\da{1^I4^L}\eta_{4L}-\eta_{2J}\da{2^J3^K}\eta_{3K}\right)\bigg).
\end{align}
I have chosen a form without full exchange symmetry that is appropriate for the $t$-channel partitioning here. I also assume that the masses $m$ are all equal. Applying this to (\ref{EvenDegRes}) (identifying $\eta$ as $\xi^2$) results in the complete cancellation of the pole in the residue. As a result, these terms do not interfere with the $4$-particle test. A $\frac{1}{4}$BPS gravitino is therefore inconsistent, at least without higher spin particles. 

A SUSY delta function describing a pair of $\frac{1}{2}$BPS states coupling to a $\frac{1}{4}$BPS dyon may be constructed by adding the supercharges from Section \ref{SUSYAlgebra} to that of the $\frac{1}{4}$BPS state. This could potentially be of relevance in describing the decay amplitude of a $\frac{1}{4}$BPS dyon into a pair of $\frac{1}{2}$BPS objects. See Appendix \ref{Dyons} for some further cursory comments.

\subsubsection{Single gravitino pair and $\mathcal{N}=6\rightarrow 4$}\label{sec:SingleGravitino}

Discarding $\frac{1}{4}$BPS gravitinos leaves only the option of $\frac{1}{2}$BPS multiplets and therefore the same central charge structure as the vectors in spontaneously broken SYM. The gravitino multiplets have natural coherent state representations in which they have structure given by (\ref{N=2Gravitino}) (so they are fermions). The superamplitudes coupling the gravitinos to the graviton are fixed to be
\begin{align}
\mathcal{A}(\Psi,\overline{\Psi},H^+)&=\frac{1}{M_{Pl}}\Delta_u\Delta_v\frac{1}{x}\da{\bf{12}}\nonumber\\
\mathcal{A}(\Psi,\overline{\Psi},H^-)&=\frac{1}{M_{Pl}}\Delta_u\Delta_vx\ds{\bf{12}}
\end{align}
given that the anomalous gravitational dipole moment is forbidden. Now, the graviton exchange contribution to the superamplitude $\mathcal{A}(\Psi,\overline{\Psi},\Psi,\overline{\Psi})$ has $s$-channel residue (dropping again the SUSY delta functions for brevity)
\begin{align}\label{N=4GravitinoSres}
(u-m_u^2)R_s&\sim \frac{-1}{M_{Pl}^2}\left(\frac{x_{34}}{x_{12}}\da{\bf{12}}\ds{\bf{34}}+\frac{x_{12}}{x_{34}}\ds{\bf{12}}\da{\bf{34}}\right)\nonumber\\
&\sim \frac{1}{M_{Pl}^2}\left(\mathbf{\Pi}_t+\mathbf{\Pi}_u\right)
\end{align}
(using the Moller scattering residue (\ref{QEDRes})) and the $u$-channel residue is given by exchanging $2\leftrightarrow 4$:
\begin{align}
(t-m_t^2)R_u&\sim -\frac{1}{M_{Pl}^2}\left(\mathbf{\Pi}_t+\mathbf{\Pi}_s\right).
\end{align}
Fig \ref{Fig:BPSgravitinoGravEx} displays applicable the on-shell diagrams. The sum of these terms is clearly not zero, so this theory, by itself, is not consistent. 

The only available option that does not involve introducing higher spin particles is to allow for vector multiplets to couple to the gravitinos. There is only one possible coupling that can contribute to tree-exchanges in $\mathcal{A}(\Psi,\overline{\Psi},\Psi,\overline{\Psi})$. This is
\begin{align}\label{GravitinoVectorCoup}
\mathcal{A}(\Psi,\overline{\Psi},V_A)&=\Delta_u\Delta_v\left(\alpha_A\da{\bf{12}}+(\alpha_A)^*\ds{\bf{12}}\right),
\end{align}
where the vector multiplets $V_A$ are self-conjugate. Combining these potential interactions into residues of the superamplitude $\mathcal{A}(\Psi,\overline{\Psi},\Psi,\overline{\Psi})$ gives
\begin{figure}[h]
\begin{fmffile}{N=4Gravitino2}
\begin{center}
\begin{tabular}{ c c c }
& & \\
 \begin{fmfgraph*}(100,67)
   \fmfleft{i1,i2}
   \fmfright{o1,o2}
   \fmf{plain}{i1,v1}
   \fmf{boson}{i1,v1}
   \fmf{zigzag}{v1,v2}
   \fmf{zigzag}{v1,v2}
   \fmf{plain}{v2,o1}
   \fmf{boson}{v2,o1}
   \fmf{plain}{i2,v1}
   \fmf{boson}{i2,v1}
   \fmf{plain}{o2,v2}
   \fmf{boson}{o2,v2} \fmfv{decor.shape=circle,decor.filled=gray50,decor.size=0.15w}{v1,v2}
   \fmflabel{$\overline{\Psi}$}{i2}
   \fmflabel{$\overline{\Psi}$}{o2}
   \fmflabel{$\Psi$}{i1}
   \fmflabel{$\Psi$}{o1}
 \end{fmfgraph*} 
 &\,& \begin{fmfgraph*}(100,67)
   \fmfleft{i1,i2}
   \fmfright{o1,o2}
   \fmf{plain}{i1,v1}
   \fmf{boson}{i1,v1}
   \fmf{zigzag}{v1,v2}
   \fmf{zigzag}{v1,v2}
   \fmf{plain}{v2,o1}
   \fmf{boson}{v2,o1}
   \fmf{phantom}{v1,i2}
   \fmf{phantom}{v1,i2}
   \fmf{phantom}{v2,o2}
   \fmf{phantom}{v2,o2}
   \fmf{plain,tension=-0.25}{v2,i2}
   \fmf{plain,tension=-0.25}{v1,o2}
   \fmf{boson,tension=-0.25}{v2,i2}
   \fmf{boson,tension=-0.25}{v1,o2}
   \fmfv{decor.shape=circle,decor.filled=gray50,decor.size=0.15w}{v1,v2}
   \fmflabel{$\overline{\Psi}$}{i2}
   \fmflabel{$\overline{\Psi}$}{o2}
   \fmflabel{$\Psi$}{i1}
   \fmflabel{$\Psi$}{o1}
 \end{fmfgraph*}\nonumber\\
 $(s)$ & \, &  $(u)$  
\end{tabular}
\end{center}
\end{fmffile}
\caption{Superfactorisation channel for BPS gravitino scattering by vector exchange.}
\end{figure}
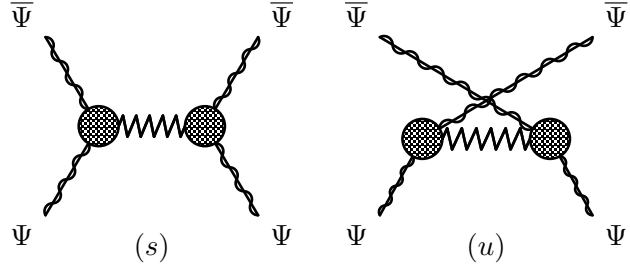
\begin{align}
(u-m_u^2)R_s\sim\left(\alpha_A\da{\bf{12}}+(\alpha_A)^*\ds{\bf{12}}\right)\left(\alpha_A\da{\bf{34}}+(\alpha_A)^*\ds{\bf{34}}\right)
\end{align}
and analogously for the $u$-channel. 

There can be no $t$-channel contributions from the exchange of massive vector multiplets. Repeating a familiar story, such an exchange would require coupling a vector to two identical gravitinos:
\begin{align}
\mathcal{A}(\Psi,\Psi,V_A)&=\Delta_u\Delta_v\left(\beta_A\da{\bf{12}}+\gamma_A\ds{\bf{12}}\right),
\end{align}
where the vector has twice the mass of the gravitino and is a central charge eigenstate (as usual), while $\beta_A$ and $\gamma_A$ are two sets of coupling constants. The gravitino multiplets are represented here as fermions, but this superamplitude is even under their exchange (remembering that $\Delta_v\Delta_u$ is odd under particle exchanges). The superamplitude therefore contains the wrong gravitino exchange statistics and is prohibited. 

The permissible vector exchange residues should be combined with the gravitational ones so that the $4$-particle test $(u-m_u^2)R_s+(t-m_t^2)R_u=0$ is obeyed. Demanding such cancellation implies that $\alpha_A\alpha_A=0$ and $\alpha_A(\alpha_A)^*=\frac{1}{M_{Pl}^2}$. These subsequently imply that a basis of massless vectors can be chosen so that $\alpha_A=\frac{1}{\sqrt{2}M_{Pl}}(1,i)$, or equivalently, that there is a complex basis of vectors with couplings 
\begin{align}\label{N=4SuperHiggs1}
\mathcal{A}(\Psi,\overline{\Psi},V)&=\frac{1}{M_{Pl}}\Delta_u\Delta_v\da{\bf{12}}\nonumber\\
\mathcal{A}(\Psi,\overline{\Psi},\overline{V})&=\frac{1}{M_{Pl}}\Delta_u\Delta_v\ds{\bf{12}}.
\end{align}
This complex pair of vector multiplets are the super-Higgs bosons for this theory and their existence is forced by consistent superfactorisation. 

The complete superamplitude is then 
\begin{align}\label{4legN=4from6}
\mathcal{A}(\Psi,\overline{\Psi},\Psi,\overline{\Psi})=\frac{1}{M_{Pl}^2}\frac{1}{su}\epsilon\left(\mathcal{Q},\mathcal{Q},\mathcal{Q},\mathcal{Q}\right)^2\epsilon(1,2,3,4),
\end{align}
the assemblage of which is similar to the $\mathcal{N}=2$ SQED amplitudes. The amplitude (\ref{4legN=4from6}) is adequately unitarised, as expected. Its connection to (\ref{4legN=6}) is obvious. In the high-energy limit, this matches onto pure $\mathcal{N}=6$ SUGRA. The gravitino mass is the only parameter that breaks the additional two supersymmetries. The super-Higgs construction derived here corresponds to the decoupled SS SUSY breaking (or ``gauging'') from $\mathcal{N}=6\rightarrow 4$ \cite{Andrianopoli:2002rm}. 

Like the $\mathcal{N}=6$ CSS theory above, this $\mathcal{N}=4$ SS theory can be embedded in toroidally compactified $\mathcal{N}=6$ SUGRA. The $\mathcal{N}=6$ graviton decomposes into $\mathcal{N}=4$ submultiplets as 
\begin{align}
H^+_{\mathcal{N}=4}&=\frac{\partial}{\partial\overline{\eta}_3}H^+_{\mathcal{N}=6}\Big|_{\eta_3=0}\qquad\qquad H^-_{\mathcal{N}=4}=\frac{\partial}{\partial\eta_3}H^-_{\mathcal{N}=6}\Big|_{\overline{\eta}_3=0}\nonumber\\
V_{\mathcal{N}=4}&=\frac{\partial}{\partial\eta_3}H^+_{\mathcal{N}=6}\Big|_{\overline{\eta}_3=0}\qquad\qquad \overline{V}_{\mathcal{N}=4}=\frac{\partial}{\partial\overline{\eta}_3}H^-_{\mathcal{N}=6}\Big|_{\eta_3=0}.
\end{align}
The gravitino is contained in the KK graviton multiplet just as above in (\ref{N=6ProjectMassive}):
\begin{align}
\Psi_{\mathcal{N}=4}=\mathcal{H}_{\mathcal{N}=6}\Big|_{\eta_3^I=0}\qquad\qquad\overline{\Psi}_{\mathcal{N}=4}=2\frac{\partial^2}{\partial\eta^I_3\partial\eta_{3I}}\overline{\mathcal{H}}_{\mathcal{N}=6}.
\end{align}
The $\mathcal{N}=6\rightarrow 4$ SS theory is contained in the $\mathcal{N}=6$ KK gravity analogously to the way in which the $\mathcal{N}=6$ SS theory above could be extracted from $\mathcal{N}=8$ KK gravity. The separate graviphoton multiplets in the present case couple to the same combination of submultiplets as the gravitons in both the projected and embedded theories, so do not change the argument. 

A potential alternative to introducing massless vectors into the theory is to allow for a massive spin-$2$ exchange in the $t$-channel. However, from the usual argument, this necessarily leads to towers (and the massless vectors are required as well anyway). The theory constructed above is the only option for a decoupled IR. Additional vector multiplets also cannot be included in the theory. This is because they must be able to gravitationally scatter off the gravitinos, but this residue necessarily contains a cross-channel pole. This can only be accounted for if the vectors couple to the gravitinos. However, this is not possible without messing up the four gravitino superamplitude above unless higher spin particles are introduced or there is a new species of gravitino. I will investigate this latter possibility next, but it is clear in any case that this theory of softly broken SUSY cannot be coupled to a separate SYM sector.

\subsubsection{Two gravitino pairs and the structure of $\mathcal{N}=8\rightarrow4$ gauged SUGRA}

Next introduce a second species of gravitino $\Psi'$. The argument above applies independently to both of the gravitino species. So there must exist a pair of massless vectors that couple to each gravitino pair consistent with (\ref{N=4SuperHiggs1}). How the vectors super-Higgsing each gravitino pair are related is to be determined. Consistency can be further tested by scattering the two gravitino species off each other, described by the superamplitude $\mathcal{A}(\Psi,\overline{\Psi},\Psi',\overline{\Psi'})$. 

At this point, the central charge structure of the second gravitino becomes important. Generally, the central charge of the first gravitino can be $R$-rotated to the purely symplectic form $Z_{\Psi,AB}=2m_{\Psi}\Omega_{AB}$, while the second gravitino's central charge has the form (\ref{1/2BPSCC}). The graviton coupling to the second gravitino is then analogous to gravitational coupling of the $\frac{1}{2}$BPS vector multiplet described in (\ref{GravVec2Angles}), but with additional spin structure:
\begin{align}\label{GravitinoGravCoup2Ang}
\mathcal{A}\left(\Psi',\overline{\Psi'},H^+\right)&=\frac{1}{M_{Pl}}e^{-3 i\theta}\Delta_v^{(\varphi+\theta)}\Delta_u^{(\varphi-\theta)}e^{ i\theta}\frac{1}{x}\da{\bf{12}}\nonumber\\
\mathcal{A}\left(\Psi',\overline{\Psi'},H^-\right)&=\frac{1}{M_{Pl}}e^{-3 i\theta}\Delta_v^{(\varphi+\theta)}\Delta_u^{(\varphi-\theta)}e^{ -i\theta}x\ds{\bf{12}}.
\end{align}
Paralleling the calculation in (\ref{DyonicGravEx}), the $s$-channel graviton exchange residue for $\mathcal{A}(\Psi,\overline{\Psi},\Psi',\overline{\Psi'})$ is then
\begin{align}
R_s&\sim \frac{1}{M_{Pl}^2}e^{-2i\theta}\epsilon\left(\mathcal{Q},\mathcal{Q},\mathcal{Q},\mathcal{Q}\right)^2\frac{m_\Psi m_{\Psi'}}{\left(t-\left(m_t^{(\varphi+\theta)}\right)^2\right)\left(t-\left(m_t^{(\varphi-\theta)}\right)^2\right)}\nonumber\\
&\,\,\times\left(\frac{x_{12}}{x_{34}}e^{-i\theta}+\frac{x_{34}}{x_{12}}e^{i\theta}-2\cos\varphi\right)\left(\frac{x_{12}}{x_{34}}e^{i\theta}\ds{\bf{12}}\da{\bf{34}}+\frac{x_{34}}{x_{12}}e^{-i\theta}\da{\bf{12}}\ds{\bf{34}}\right)\nonumber\\
&=\frac{1}{M_{Pl}^2}e^{-2i\theta}\epsilon\left(\mathcal{Q},\mathcal{Q},\mathcal{Q},\mathcal{Q}\right)^2\frac{m_\Psi m_{\Psi'}}{\left(t-\left(m_t^{(\varphi+\theta)}\right)^2\right)\left(t-\left(m_t^{(\varphi-\theta)}\right)^2\right)}\nonumber\\
&\qquad\quad\times\bigg(\frac{u-t}{2m_\Psi m_{\Psi'}}\left(\mathbf{\Pi}_t+\mathbf{\Pi}_u\right)-\mathbf{\Pi}_s+e^{2i\theta}\ds{\bf{12}}\da{\bf{34}}+e^{-2i\theta}\da{\bf{12}}\ds{\bf{34}}\nonumber\\
&\qquad\qquad\qquad\qquad\qquad\qquad\qquad-2\cos\varphi\left(\frac{x_{12}}{x_{34}}e^{i\theta}\ds{\bf{12}}\da{\bf{34}}+\frac{x_{34}}{x_{12}}e^{-i\theta}\da{\bf{12}}\ds{\bf{34}}\right)\bigg),
\end{align}
where I have recycled the graviton exchange residue between fermions (\ref{FermionGravRes}) this time. The gravitino masses are $m_{\Psi}$ and $m_{\Psi'}$. The possible couplings of the new gravitinos to the old vector multiplets $V$ and $\overline{V}$ introduced in the single gravitino species case above are given by 
\begin{align}\label{GravitinoVecCoup2Ang}
\mathcal{A}\left(\Psi',\overline{\Psi'},V\right)&=e^{-3 i\theta}\Delta_v^{(\varphi+\theta)}\Delta_u^{(\varphi-\theta)}\left(\alpha\ds{\bf{12}}+\beta\da{\bf{12}}\right)\nonumber\\
\mathcal{A}\left(\Psi',\overline{\Psi'},\overline{V}\right)&=e^{-3 i\theta}\Delta_v^{(\varphi+\theta)}\Delta_u^{(\varphi-\theta)}\left(\beta^*\ds{\bf{12}}+\alpha^*\da{\bf{12}}\right),
\end{align}
for some constants $\alpha$ and $\beta$ yet to be determined. The vector exchange residue is therefore given by
\begin{align}
R_s&\sim \frac{1}{M_{Pl}}e^{-2i\theta}\epsilon\left(\mathcal{Q},\mathcal{Q},\mathcal{Q},\mathcal{Q}\right)^2\frac{m_\Psi m_{\Psi'}}{\left(t-\left(m_t^{(\varphi+\theta)}\right)^2\right)\left(t-\left(m_t^{(\varphi-\theta)}\right)^2\right)}\nonumber\\
&\,\,\times\left(\frac{x_{12}}{x_{34}}e^{-i\theta}+\frac{x_{34}}{x_{12}}e^{i\theta}-2\cos\varphi\right)\left(\ds{\bf{12}}\left(\alpha\ds{\bf{34}}+\beta\da{\bf{34}}\right)+\da{\bf{12}}\left(\beta^*\ds{\bf{34}}+\alpha^*\da{\bf{34}}\right)\right).
\end{align}

Unlike the case of vector multiplet scattering in (\ref{GravScatVec2Ang}), it is immediately clear here that the vector exchange residue cannot cancel the graviton exchange residue up to terms free of poles, since they both contain linearly independent Lorentz structures. Consistent superfactorisation therefore requires cross-channel particle exchanges. Unlike the single gravitino superamplitude, here there is no possible $u$-channel graviton exchange, and as long as the central charges of the two gravitino species are not degenerate, then the particles exchanged in the cross-channels must also carry central charge and be massive. Continuing to forbid the existence of higher spin particles, then these must be $\frac{1}{2}$BPS massive vector multiplets. As explained in Section \ref{sec:N=4SS}, the central charges of three $\frac{1}{2}$BPS particles interacting through a $3$-particle superamplitude must be either of the form (\ref{CentralChargesEleN=4}) or (\ref{CentralChargesMagN=4}) (that is, either $\theta=0$ or $\varphi=0$ for the second gravitino). There are therefore two cases to be considered. I will begin with the $\theta=0$ case, which corresponds to manifestly electric graviphoton couplings, and then afterward move on to the $\varphi=0$, $\theta\neq 0$ case.

In the $\theta=0$ case, all appearances of the central charge angle $\varphi$ are contained in the SUSY delta functions. The $s$-channel graviton exchange residue for $\mathcal{A}(\Psi,\overline{\Psi},\Psi',\overline{\Psi'})$ therefore has the same appearance as the single species case above in (\ref{N=4GravitinoSres}). However, this time there is no $u$-channel graviton exchange. Consistency therefore again requires vector exchanges (assuming that higher spin particles are not present). The possible couplings of the vectors $V$ and $\overline{V}$ to the new gravitino simply become
\begin{align}\label{Gravitino2Coup}
\mathcal{A}(\Psi',\overline{\Psi'},V)&=\Delta_u\Delta_v\left(\alpha\ds{\bf{12}}+\beta\da{\bf{12}}\right)\nonumber\\
\mathcal{A}(\Psi',\overline{\Psi'},\overline{V})&=\Delta_u\Delta_v\left(\beta^*\ds{\bf{12}}+\alpha^*\da{\bf{12}}\right).
\end{align}
\begin{figure}[h]
\begin{fmffile}{N=4Gravitino2flavour}
\begin{center}
\begin{tabular}{ c c c c c}
& & & & \\
 \begin{fmfgraph*}(100,67)
   \fmfleft{i1,i2}
   \fmfright{o1,o2}
   \fmf{plain}{i1,v1}
   \fmf{boson}{i1,v1}
   \fmf{dbl_wiggly}{v1,v2}
   \fmf{dbl_wiggly}{v1,v2}
   \fmf{plain}{v2,o1}
   \fmf{boson}{v2,o1}
   \fmf{plain}{i2,v1}
   \fmf{boson}{i2,v1}
   \fmf{plain}{o2,v2}
   \fmf{boson}{o2,v2} \fmfv{decor.shape=circle,decor.filled=gray50,decor.size=0.15w}{v1,v2}
   \fmflabel{$\overline{\Psi}$}{i2}
   \fmflabel{$\overline{\Psi}'$}{o2}
   \fmflabel{$\Psi$}{i1}
   \fmflabel{$\Psi'$}{o1}
   \fmfv{label=$\mp$,label.angle=-45,label.dist=0.1w}{v1}
   \fmfv{label=$\pm$,label.angle=-135,label.dist=0.1w}{v2}
 \end{fmfgraph*} &\,&
 \begin{fmfgraph*}(100,67)
   \fmfleft{i1,i2}
   \fmfright{o1,o2}
   \fmf{plain}{i1,v1}
   \fmf{boson}{i1,v1}
   \fmf{zigzag}{v1,v2}
   \fmf{zigzag}{v1,v2}
   \fmf{plain}{v2,o1}
   \fmf{boson}{v2,o1}
   \fmf{plain}{i2,v1}
   \fmf{boson}{i2,v1}
   \fmf{plain}{o2,v2}
   \fmf{boson}{o2,v2} \fmfv{decor.shape=circle,decor.filled=gray50,decor.size=0.15w}{v1,v2}
   \fmflabel{$\overline{\Psi}$}{i2}
   \fmflabel{$\overline{\Psi}'$}{o2}
   \fmflabel{$\Psi$}{i1}
   \fmflabel{$\Psi'$}{o1}
 \end{fmfgraph*} &\,& \begin{fmfgraph*}(100,67)
   \fmfleft{i1,i2}
   \fmfright{o1,o2}
   \fmf{plain}{i1,v1}
   \fmf{boson}{i1,v1}
   \fmf{boson}{v1,v2}
   \fmf{boson}{v1,v2}
   \fmf{plain}{v2,o1}
   \fmf{boson}{v2,o1}
   \fmf{phantom}{v1,i2}
   \fmf{phantom}{v1,i2}
   \fmf{phantom}{v2,o2}
   \fmf{phantom}{v2,o2}
   \fmf{plain,tension=-0.25}{v2,i2}
   \fmf{plain,tension=-0.25}{v1,o2}
   \fmf{boson,tension=-0.25}{v2,i2}
   \fmf{boson,tension=-0.25}{v1,o2}
   \fmfv{decor.shape=circle,decor.filled=gray50,decor.size=0.15w}{v1,v2}
   \fmflabel{$\overline{\Psi}$}{i2}
   \fmflabel{$\overline{\Psi}'$}{o2}
   \fmflabel{$\Psi$}{i1}
   \fmflabel{$\Psi'$}{o1}
 \end{fmfgraph*}\nonumber\\
 $(s)$ & \, &  $(s)$ & \, &  $(u)$
\end{tabular}
\end{center}
\end{fmffile}
\caption{Superfactorisation channel for BPS gravitino scattering of different flavours (omitting the $t$-channel).}
\end{figure}
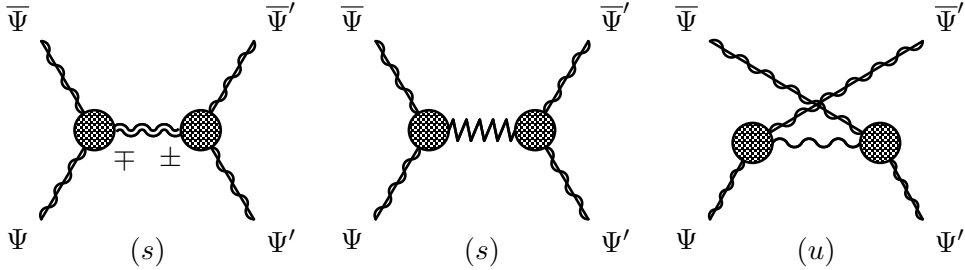
The $s$-channel vector exchange residue contributing to $\mathcal{A}(\Psi,\overline{\Psi},\Psi',\overline{\Psi'})$ is then 
\begin{align}\label{sVecExMixedGravitino}
(u-m_u^2)R_s\sim\frac{1}{M_{Pl}}\left(\ds{\bf{12}}\left(\alpha\ds{\bf{34}}+\beta\da{\bf{34}}\right)+\da{\bf{12}}\left(\beta^*\ds{\bf{34}}+\alpha^*\da{\bf{34}}\right)\right).
\end{align}
(remembering that $V$ and $\overline{V}$ are the only vectors that can couple to $\Psi$ and $\overline{\Psi}$). Cross-channel vector exchanges have a similar $s$-wave structure. There is no way for them to cancel the terms proportional to $\beta$, so $\beta=0$. 

Two other sets of pairs of vectors $\mathcal{W}_A$ and $\mathcal{W}'_A$ can be introduced with the couplings
\begin{align}\label{N=4SSCross}
\mathcal{A}(\Psi,\Psi',\mathcal{W}_A)&=\Delta_u\Delta_v\left(\gamma_A\ds{\bf{12}}+\varepsilon_A\da{\bf{12}}\right)\nonumber\\
\mathcal{A}(\Psi,\overline{\Psi'},\mathcal{W}'_A)&=\Delta_u\Delta_v\left(\zeta_A\ds{\bf{12}}+\eta_A\da{\bf{12}}\right)
\end{align}
and likewise for conjugate superamplitudes of conjugate particles. The coupling constants to be determined are $\gamma_A$, $\varepsilon_A$, $\zeta_A$ and $\eta_A$. The new vector multiplets have masses determined by central charge conservation. I will assume in the following that both $\Psi$ and $\Psi'$ have different central charges so that $\mathcal{W}_A$ and $\mathcal{W}'_A$ are both massive. Despite appearing to introduce some superficial additional details, the special case of degenerate gravitinos proceeds similarly and ultimately leads to the same conclusion, so I will not bother to separately present it. 

Before proceeding with the mixed $4$-particle gravitino amplitude, further constraints upon the vector-gravitino couplings can be inferred from the mixed vector-gravitino superamplitude. As mentioned above, the residue for the gravitational scattering of a hypothetical new massless vector (one that is neither $V$ or $\overline{V}$) off the first gravitino $\Psi$ contains a cross-channel Mandelstam pole. This necessitates that the new vector also couple to $\Psi$ through (\ref{GravitinoVectorCoup}). However, this would ruin the established consistency of the $\mathcal{A}(\Psi,\overline{\Psi},\Psi,\overline{\Psi})$ superamplitude, so is not permitted. Since there are no further massless vectors in the theory, then repeating the $4$-particle test for the new gravitino $\mathcal{A}(\Psi',\overline{\Psi'},\Psi',\overline{\Psi'})$ with coupling (\ref{Gravitino2Coup}) implies that $\alpha=\frac{1}{M_{Pl}}e^{i\phi}$, where $\phi$ is some undetermined phase (in the single gravitino case, this was removed by fixing the phase of $V$, but that freedom is no longer unavailable here). The couplings of the massless vectors $V$ and $\overline{V}$ to the new gravitino multiplets are therefore determined to be 
\begin{align}
\mathcal{A}(\Psi',\overline{\Psi'},V)&=\frac{e^{i\phi}}{M_{Pl}}\Delta_u\Delta_v\ds{\bf{12}}\nonumber\\
\mathcal{A}(\Psi',\overline{\Psi'},\overline{V})&=\frac{e^{-i\phi}}{M_{Pl}}\Delta_u\Delta_v\da{\bf{12}},
\end{align}
which have the opposite chirality to the analogous couplings involving $\Psi$ and $\overline{\Psi}$. 

Returning now to the mixed gravitino $\mathcal{A}(\Psi,\overline{\Psi},\Psi',\overline{\Psi'})$ superamplitude, the massive vector exchanges contribute to the $t$ and $u$-channel residues. Combining the $3$-particle superamplitudes (\ref{N=4SSCross}), the residues are 
\begin{align}
(s-m_s^2)R_t&\sim -\left(\gamma_A\ds{\bf{13}}+\varepsilon_A\da{\bf{13}}\right)\left(\gamma_A^*\da{\bf{24}}+\varepsilon_A^*\ds{\bf{24}}\right)\nonumber\\
(t-m_t^2)R_u&\sim -\left(\zeta_A\ds{\bf{14}}+\eta_A\da{\bf{14}}\right)\left(\zeta_A^*\da{\bf{23}}+\eta_A^*\ds{\bf{23}}\right).
\end{align}
Combining with the $s$-channel graviton exchange residues and demanding cancellation implies that $\gamma_A(\gamma_A)^*=\varepsilon_A(\varepsilon_A)^*=\zeta_A(\zeta_A)^*=\eta_A(\eta_A)^*=\frac{1}{M_{Pl}^2}$ (sum over $A$ is implied), while cancellation with the $s$-channel vector exchange residue (\ref{sVecExMixedGravitino}) implies that $\gamma_A\varepsilon_A^*=-\zeta_A\eta_A^*=\frac{1}{M_{Pl}^2}e^{i\phi}$. A basis for the vectors $\mathcal{W}_A$ and $\mathcal{W}'_A$ can be chosen so that $\gamma_A=\zeta_A=\frac{1}{M_{Pl}}\delta_{A1}$, from which it follows that $\varepsilon_A=-\eta_A=\frac{e^{-i\phi}}{M_{Pl}}\delta_{A1}$ (so there is only a single pair of $\mathcal{W}$ and $\mathcal{W}'$ and the hypothetical flavour index $A$ will henceforth be dropped). The couplings of the massive vector multiplets to the gravitinos are thus fully determined.

Superamplitudes with external vector multiplets contain further information required to specify the theory. I will test the consistency of the superamplitude $\mathcal{A}(V,\mathcal{W},\Psi,\Psi')$ (other mixed gravitino-vector superamplitudes lead to similar conclusions).
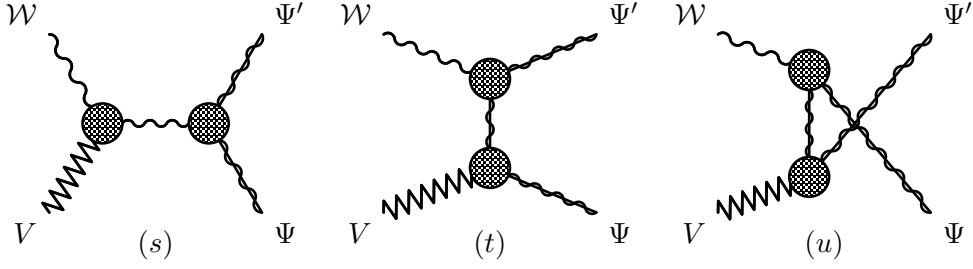
\begin{figure}[h]
\begin{fmffile}{N=4Gravitino2Mixed}
\begin{center}
\begin{tabular}{ c c c c c }
& & & & \\
 \begin{fmfgraph*}(100,67)
   \fmfleft{i1,i2}
   \fmfright{o1,o2}
   \fmf{zigzag}{i1,v1}
   \fmf{zigzag}{i1,v1}
   \fmf{boson}{v1,v2}
   \fmf{boson}{v1,v2}
   \fmf{plain}{v2,o1}
   \fmf{boson}{v2,o1}
   \fmf{boson}{i2,v1}
   \fmf{boson}{i2,v1}
   \fmf{plain}{o2,v2}
   \fmf{boson}{o2,v2} \fmfv{decor.shape=circle,decor.filled=gray50,decor.size=0.15w}{v1,v2}
   \fmflabel{$\mathcal{W}$}{i2}
   \fmflabel{$\Psi'$}{o2}
   \fmflabel{$V$}{i1}
   \fmflabel{$\Psi$}{o1}
 \end{fmfgraph*}
&\,& \begin{fmfgraph*}(100,67)
   \fmfleft{i1,i2}
   \fmfright{o1,o2}
   \fmf{zigzag}{i1,v1}
   \fmf{zigzag}{i1,v1}
   \fmf{plain}{o1,v1,v2,o2}
   \fmf{boson}{o1,v1,v2,o2}
   \fmf{boson}{i2,v2}
   \fmf{boson}{i2,v2} \fmfv{decor.shape=circle,decor.filled=gray50,decor.size=0.15w}{v1,v2}
   \fmflabel{$\mathcal{W}$}{i2}
   \fmflabel{$\Psi'$}{o2}
   \fmflabel{$V$}{i1}
   \fmflabel{$\Psi$}{o1}
 \end{fmfgraph*} &\,& 
 \begin{fmfgraph*}(100,67)
   \fmfleft{i1,i2}
   \fmfright{o1,o2}
   \fmf{zigzag}{i1,v1}
   \fmf{zigzag}{i1,v1}
   \fmf{boson}{i2,v2}
   \fmf{boson}{i2,v2}
   \fmf{plain}{v1,v2}
   \fmf{boson}{v1,v2}
   \fmf{plain,tension=-0.25}{v2,o1}
   \fmf{boson,tension=-0.25}{v2,o1}
   \fmf{plain,tension=-0.25}{v1,o2}
   \fmf{boson,tension=-0.25}{v1,o2}
   \fmf{phantom}{v2,o2}
   \fmf{phantom}{v2,o2}
   \fmf{phantom}{v1,o1}
   \fmf{phantom}{v1,o1} \fmfv{decor.shape=circle,decor.filled=gray50,decor.size=0.15w}{v1,v2}
   \fmflabel{$\mathcal{W}$}{i2}
   \fmflabel{$\Psi'$}{o2}
   \fmflabel{$V$}{i1}
   \fmflabel{$\Psi$}{o1}
 \end{fmfgraph*}\nonumber\\
 $(s)$ &\,& $(t)$ & \, &  $(u)$
\end{tabular}
\end{center}
\end{fmffile}
\caption{Superfactorisation channel for mixed BPS gravitino and vector scattering of different flavours (omitting the $t$-channel).}
\end{figure}
The $t$-channel residue is 
\begin{align}
(s-m_s^2)R_t\sim \frac{-1}{2M_{Pl}^2}\left(\la{\bf{3}}p_1-p_2\rs{\bf{4}}+m_\Psi\ds{\bf{34}}+\left(m_{\Psi'}+2e^{-i\phi}m_\Psi\right)\da{\bf{34}}\right),
\end{align}
and the $u$-channel residue is 
\begin{align}
(t-m_t^2)R_u\sim\frac{-1}{2M_{Pl}^2}\left(-\la{\bf{3}}p_1-p_2\rs{\bf{4}}+(m_\Psi+2e^{i\phi}m_{\Psi'})\ds{\bf{34}}+m_{\Psi'}\da{\bf{34}}\right).
\end{align}
I am assuming that the vector-gravitino couplings have the same crossing signs as the Yukawa couplings in (\ref{3Yukawa}), which were determined by fermionic exchange symmetry (and is justified by the fact that it is required to be true for a correct and sensible result to emerge). If there is a coupling $f_{\mathcal{W}\overline{\mathcal{W}}V}$ permitted between vector multiplets, then there is also a potential $s$-channel contribution:
\begin{align}
(u-m_u^2)R_s\sim \frac{-1}{M_{Pl}}\left(\ds{\bf{34}}+e^{-i\phi}\da{\bf{34}}\right)f_{\mathcal{W}\overline{\mathcal{W}}V}.
\end{align}
Consistency of these residues fixes the vector self-coupling as $f_{\mathcal{W}\overline{\mathcal{W}}V}=-\frac{m_{\Psi}+e^{i\phi}m_{\Psi'}}{M_{Pl}}$.

Repeating a similar calculation with $\overline{\Psi'}$ and $\mathcal{W}'$ instead of $\Psi'$ and $\mathcal{W}$ leads to similar conclusions. Altogether, the vector multiplets have self-coupling constants 
\begin{align}\label{LAN=4SS2}
&f_{\mathcal{W}\overline{\mathcal{W}}V}=-\frac{m_{\Psi}+e^{i\phi}m_{\Psi'}}{M_{Pl}}
&f_{\mathcal{W}'\overline{\mathcal{W}'}V}=-\frac{m_{\Psi}-e^{i\phi}m_{\Psi'}}{M_{Pl}}\nonumber\\
&f_{\mathcal{W}\overline{\mathcal{W}}\overline{V}}=\frac{m_{\Psi}+e^{-i\phi}m_{\Psi'}}{M_{Pl}}
&f_{\mathcal{W}'\overline{\mathcal{W}'}\overline{V}}=\frac{m_{\Psi}-e^{-i\phi}m_{\Psi'}}{M_{Pl}}.
\end{align}
Like the case of the single gravitino pair, further species of vector multiplets cannot be introduced into the theory without introducing higher spin multiplets. This is because gravitationally scattering them off the gravitinos would otherwise enforce the existence of new gravitino-vector $3$-particle superamplitudes in order to explain the cross-channel Mandelstam pole, which is inconsistent with the above analysis. The couplings (\ref{LAN=4SS2}) must be Lie algebra structure constants, but the complete Lie algebra must also include contributions from the graviphotons in the graviton multiplet, as shown in Section \ref{sec:BPSGrav}.

Having now reconstructed the particle spectrum and couplings for the theory when the second gravitino has purely anti-self-dual central charge, I will next focus on the alternative configuration with $\theta\neq 0$ and $\varphi=0$. As explained previously, these central charges in the context of SYM would necessarily be dyonic. However, it was shown in Section \ref{sec:N=4SUGRAsub} that such dyonic configurations can sometimes be reinterpreted as purely electric if additional massless graviphoton multiplets are available to ``gauge'' the otherwise magnetic charges. I will investigate whether this option is also a possibility here. In this case, as explained around (\ref{SuperDegenDyon}) in Section \ref{N=4SuperFact}, the appropriate $3$-particle SUSY delta functions are instead the tiny group charged $\Delta_v\Delta_v$. A sensible superamplitude must then contain further factors of little group spinors to ensure tiny group neutrality. 

In the special case in which one of the particles is massless (as usual, particle $3$), such a tiny group invariant object is easily given by $\Delta_v\Delta_v\mathbf{v}_3^+\mathbf{v}_3^-$. The $3$-particle superamplitudes in Section \ref{sec:SingleGravitino} above can all be converted to this form using the fact that $\Delta_u=\frac{\mathbf{v}_3^+\mathbf{v}_3^-}{m_{\Psi}}\Delta_v$. Similarly, in (\ref{GravitinoGravCoup2Ang}) and (\ref{GravitinoVecCoup2Ang}), using the fact that $\mathbf{u}_3^{(-\theta)+}=m_{\Psi'}e^{-i\theta}/\mathbf{v}_3^{(\theta)-}$, the delta function $\Delta_u^{(-\theta)}=\Delta_v^{(\theta)}\frac{\mathbf{v}_3^+\mathbf{v}_3^-}{m_{\Psi'}e^{-i\theta}}$.  The $\mathbf{v}$ spinors in this latter expression are implicitly defined as being associated with central charge phase angle $\theta$ - I choose to drop the angle superscripts on these little group spinors because this is now the normal choice associated with the unambiguous central charge or mass phase. It is also the same angle as those appearing in the factor of $\Delta_v^{(\theta)}$ already present and accounting for the other pair of supercharges.

As mentioned previously, when all three particles are massive, no SUSY invariant, tiny group neutral Lorentz scalar structure is available. Since the gravitino superfields have spin however, it is possible to construct a tiny group neutral superamplitude in which they couple to massive vector multiplets:
\begin{align}
\begin{split}
\mathcal{A}(\Psi,\Psi',\mathcal{W})&=\gamma e^{-i(\theta+\theta_{W})}\Delta_v\Delta_v\mathbf{v}_1\mathbf{v}_2\nonumber\\
\mathcal{A}(\Psi,\overline{\Psi'},\mathcal{W}')&=-\zeta e^{-i(\theta+\theta_{W'})}\Delta_v\Delta_v\mathbf{v}_1\mathbf{v}_2
\end{split}\qquad\begin{split}
\mathcal{A}(\overline{\Psi},\overline{\Psi'},\overline{\mathcal{W}})&=-\gamma^* e^{-i(2\theta+\theta_{W})}\Delta_v\Delta_v\mathbf{v}_1\mathbf{v}_2\nonumber\\
\mathcal{A}(\overline{\Psi},\Psi',\overline{\mathcal{W}'})&=\zeta^* e^{-i(2\theta+\theta_{W'})}\Delta_v\Delta_v\mathbf{v}_1\mathbf{v}_2.
\end{split}
\end{align}
Here $\theta_{W}$ and $\theta_{W'}$ are the central charge phases of the $\mathcal{W}$ and $\mathcal{W}'$ respectively, which may be easily determined from central charge conservation (along with the masses $m_{W}$ and $m_{W'}$):
\begin{align}
e^{i\theta_{W}}m_{W}=-(m_\Psi+e^{i\theta}m_{\Psi'})\qquad e^{i\theta_{W'}}m_{W'}=-(m_\Psi-e^{i\theta}m_{\Psi'}).
\end{align}
Unlike the $\theta=0$ case above, here there is only a single possible Lorentz structure. For this reason, a unitary rotation can always be performed so that only a single ``colour'' of the vector multiplets is involved in these couplings. Again, as above, $\gamma$ and $\zeta$ are couplings to be determined, but the phase of the vector multiplets can be fixed to set them real and non-negative. 

While natural to continue testing the mixed gravitino superamplitude $\mathcal{A}(\Psi,\overline{\Psi},\Psi',\overline{\Psi'})$ commenced at the beginning of this Section, it is actually easier to instead proceed by testing the Compton scattering superamplitudes of the vector multiplets off the gravitinos. These superampltiudes are also sensitive to the vector multiplet self-couplings:
\begin{align}
\mathcal{A}\left(\mathcal{W},\overline{\mathcal{W}},V\right)&=e^{-2i\theta_W}f_{\mathcal{W}\overline{\mathcal{W}}V}\Delta_v\Delta_v\frac{\mathbf{v}_3^+\mathbf{v}_3^-}{m_{W}}\nonumber\\
\mathcal{A}\left(\mathcal{W},\overline{\mathcal{W}},\overline{V}\right)&=-e^{-2i\theta_W}\left(f_{\mathcal{W}\overline{\mathcal{W}}V}\right)^*\Delta_v\Delta_v\frac{\mathbf{v}_3^+\mathbf{v}_3^-}{m_{W}}
\end{align}
and similarly for $\mathcal{W}'$ instead of $\mathcal{W}$. 

As for the $\theta=0$, $\varphi\neq 0$ case above, additional massless and massive vector multiplets (besides $V$, $\mathcal{W}$, $\mathcal{W}'$ and their conjugates) are forbidden. This is again clear from consistency of their gravitational scattering superamplitudes off the gravitinos. As will now be shown, the graviton exchange residue contains a cross-channel pole and this cannot be canceled by a vector exchange in the same channel. This means that the additional vectors must also couple directly to the gravitinos, in contradiction with the results already established. 

I will begin by testing the residues of the superamplitude $\mathcal{A}\left(\Psi,\overline{\Psi},\mathcal{W},\overline{\mathcal{W}}\right)$. The $t$-channel residue from gravitino exchange is (again dropping the overall SUSY delta functions for brevity)
\begin{align}
R_t&\sim \gamma^2e^{-i(3\theta+\theta_W)}\frac{e^{4i\theta}\mathbf{v}_1\mathbf{v}_2}{(\mathbf{v}^{(13)}_P\cdot\mathbf{v}^{(24)}_P)}\nonumber\\
&=\gamma^2e^{-2i\theta_W}\frac{1}{s}\left(
\frac{1}{2}\left(\la{\bf{1}}p_4-p_3\rs{\bf{2}}+\la{\bf{2}}p_4-p_3\rs{\bf{1}}\right)+m_We^{-i\theta_W}\da{\bf{12}}+m_We^{i\theta_W}\ds{\bf{12}}\right).
\end{align}
I have used (\ref{SpinFact}), (\ref{SpecialSVD2}) and (\ref{CrossContr}) and converted the kinematic structures to an exchange symmetric-looking basis. The $s$-residue receives contributions from both graviton and vector exchanges. The graviton exchange is
\begin{align}
R_s&\sim \frac{1}{M_{Pl}^2}\frac{e^{-2i\theta_W}m_W^2 m_\Psi}{(t-m_t^2)(u-m_u^2)}\left(e^{-i\theta_W}\frac{x_{34}}{x_{12}}\da{\bf{12}}+e^{i\theta_W}\frac{x_{12}}{x_{34}}\ds{\bf{12}}\right)
\left(e^{i\theta_W}\frac{x_{34}}{x_{12}}+e^{-i\theta_W}\frac{x_{12}}{x_{34}}-2\right)\nonumber\\
&=\frac{1}{M_{Pl}^2}\frac{e^{-2i\theta_W}m_W^2 m_\Psi}{(t-m_t^2)(u-m_u^2)}\bigg(\left(\frac{x_{34}}{x_{12}}\right)^2\da{\bf{12}}+\left(\frac{x_{12}}{x_{34}}\right)^2\ds{\bf{12}}+e^{2i\theta_W}\ds{\bf{12}}+e^{-2i\theta_W}\da{\bf{12}}\nonumber\\
&\qquad\qquad\qquad\qquad\qquad\qquad\qquad\qquad\qquad\qquad-2\left(e^{-i\theta_W}\frac{x_{34}}{x_{12}}\da{\bf{12}}+e^{i\theta_W}\frac{x_{12}}{x_{34}}\ds{\bf{12}}\right)\bigg),
\end{align}
while the vector exchange is
\begin{align}
R_s\sim \frac{1}{M_{Pl}}\frac{e^{-2i\theta_W}m_W m_\Psi}{(t-m_t^2)(u-m_u^2)}\left(f_{\mathcal{W}\overline{\mathcal{W}}\overline{V}}\da{\bf{12}}-f_{\mathcal{W}\overline{\mathcal{W}}V}\ds{\bf{12}}\right)
\left(e^{i\theta_W}\frac{x_{34}}{x_{12}}+e^{-i\theta_W}\frac{x_{12}}{x_{34}}-2\right).
\end{align}
There are only three linearly independent and consistent QED residues that the terms containing single powers of $x$-factor ratios can assemble into. These are given by the mixed scalar-fermion photon exchange residue (\ref{sfSQED}) and the scalar QED residue (\ref{sQEDRes}) multiplied by overall spinor factors. Conformity of the combined graviton and vector exchange residue to these structures imposes the constraint $2\frac{m_W}{M_{Pl}}=f_{\mathcal{W}\overline{\mathcal{W}}V}-f_{\mathcal{W}\overline{\mathcal{W}}\overline{V}}$. The mixed scalar-fermion graviton exchange residue (\ref{MixedGravRes}) can also then be recycled to evaluate the complete $s$-channel residue as
\begin{align}
&R_s\sim\frac{1}{M_{Pl}}\frac{e^{-2i\theta_W}m_W m_\Psi}{(t-m_t^2)(u-m_u^2)}\nonumber\\
&\quad\times\bigg(\frac{1}{2}\bigg(-\frac{1}{M_{Pl}}\bigg(\frac{t-m_t^2-(u-m_u^2)+m_t^2-m_u^2}{2m_W m_\Psi}-2\cos\theta_W\bigg)\nonumber\\
&\qquad\qquad\qquad\qquad+i\frac{1}{m_W}\sin\theta_W\left(f_{\mathcal{W}\overline{\mathcal{W}}V}+f_{\mathcal{W}\overline{\mathcal{W}}\overline{V}}\right)\bigg)\left(\la{\bf{1}}p_4-p_3\rs{\bf{2}}+\la{\bf{2}}p_4-p_3\rs{\bf{1}}\right)\nonumber\\
&\qquad+\left(\frac{m_W}{M_{Pl}}\left(e^{2i\theta_W}-1\right)+f_{\mathcal{W}\overline{\mathcal{W}}V}\left(2-\frac{t-m_t^2-(u-m_u^2)+m_t^2-m_u^2}{2m_W m_\Psi}e^{i\theta_W}\right)\right)\ds{\bf{12}}\nonumber\\
&\qquad+\left(\frac{m_W}{M_{Pl}}\left(e^{-2i\theta_W}-1\right)-f_{\mathcal{W}\overline{\mathcal{W}}\overline{V}}\left(2-\frac{t-m_t^2-(u-m_u^2)+m_t^2-m_u^2}{2m_W m_\Psi}e^{-i\theta_W}\right)\right)\da{\bf{12}}\bigg).
\end{align}
The residue consists of terms with either two or one cross-channel Mandelstam pole. However, since this superamplitude has no possible $u$-channel, consistency demands (through (\ref{GenConsFactTriple})) that the two-pole terms must be zero. This implies that, since, by assumption, $\theta_W\neq 0$ (mod $\pi$) (because $\theta\neq 0$ (mod $\pi$) as well), then $(f_{\mathcal{W}\overline{\mathcal{W}}V})^*=f_{\mathcal{W}\overline{\mathcal{W}}V}$ and hence $f_{\mathcal{W}\overline{\mathcal{W}}V}=\frac{m_W}{M_{Pl}}$. Furthermore, equating the single pole terms of the $s$ and $t$-channel residues (as required for (\ref{GenConsFact})) implies that $\gamma=\frac{1}{M_{Pl}}$. 

The analogous calculation with $\Psi'$ instead of $\Psi$ can similarly be conducted. This differs only by the replacement of $\theta_W$ with $\theta_W-\theta$, an overall factor of $e^{-2i\theta}$ and the inclusion of the undetermined couplings of $\Psi'$ to $V$ and $\overline{V}$ in (\ref{GravitinoVecCoup2Ang}) through the replacements $\frac{1}{M_{Pl}}\da{\bf{12}}\mapsto \alpha\ds{\bf{12}}+\beta\da{\bf{12}}$ and $\frac{1}{M_{Pl}}\ds{\bf{12}}\mapsto \alpha^*\da{\bf{12}}+\beta^*\ds{\bf{12}}$ in the $s$-channel vector exchange terms. The consistency constraints in this case reduce to $\alpha+\beta^*=\frac{1}{M_{Pl}}$ and $(\alpha+\beta^*)^*=\alpha+\beta^*$. The testing of the vector-gravitino Compton superamplitudes can then also be repeated with $\mathcal{W}'$ replacing $\mathcal{W}$. This leads to analogous conclusions for this other species of massive vector multiplets, in particular that $f_{\mathcal{W}'\overline{\mathcal{W}'}V}=\frac{m_{W'}}{M_{Pl}}$ and $\zeta=\frac{1}{M_{Pl}}$. 

The remaining undetermined parameters $\alpha$ and $\beta$ can be fixed by the mixed Compton superamplitude $\mathcal{A}\left(\Psi,\Psi',V,\mathcal{W}\right)$. Fixing the couplings in accordance with the results above, the $s$-channel residue is
\begin{align}
R_s&\sim\frac{1}{M_{Pl}^2}\frac{e^{-i(3\theta_W+\theta)}\mathbf{v}_1\mathbf{v}_2\mathbf{v}_3^+\mathbf{v}_3^-}{\left(e^{-2i\theta_W}\mathbf{v}_P^{(12)}\cdot\mathbf{v}_P^{(34)}\right)^2}\nonumber\\
&\sim\frac{1}{M_{Pl}^2}e^{-i(\theta_W+\theta)}\frac{1}{(t-m_t^2)(u-m_u^2)}\nonumber\\
&\qquad\times\Big(-(t-m_t^2)\left(m_We^{i\theta_W}\ds{\bf{12}}+m_We^{-i(\theta_W-\theta)}\da{\bf{12}}\right)\nonumber\\
&\qquad\qquad\qquad+2im_Wm_{\Psi}\sin\theta_W\da{3\bf{1}}\ds{3\bf{2}}+2im_Wm_{\Psi'}e^{i\theta}\sin(\theta_W-\theta)\ds{3\bf{1}}\da{3\bf{2}}\Big).
\end{align}
I have used (\ref{CrossContr}) and (\ref{SpecialSVD2}) and made some simplifications with central charge conservation. The $t$-channel exchange residue is
\begin{align}
R_t&\sim\frac{1}{M_{Pl}^2m_\Psi}\frac{e^{-i(\theta_W+\theta)}\da{\mathbf{1}P^M}\mathbf{v}^{(24)}_{PM}\mathbf{v}_2\mathbf{v}_3^+\mathbf{v}_3^-}{\left(\mathbf{v}_P^{(13)}\cdot\mathbf{v}_P^{(24)}\right)^2}\nonumber\\
&\sim \frac{1}{M_{Pl}^2}e^{-i(\theta_W+\theta)}\frac{1}{(s-m_s^2)(u-m_u^2)}\nonumber\\
&\qquad\times\Big(-(u-m_u^2)\left(\da{3\bf{1}}\ds{3\bf{2}}+m_\Psi\ds{\bf{12}}+m_\Psi e^{i\theta}\da{\bf{12}}\right)\nonumber\\
&\qquad\qquad\qquad+2im_Wm_{\Psi}\sin\theta_W\da{3\bf{1}}\ds{3\bf{2}}+2im_Wm_{\Psi'}e^{i\theta}\sin(\theta_W-\theta)\ds{3\bf{1}}\da{3\bf{2}}\Big),
\end{align}
where $P=P_t$ is the exchanged $t$-channel momentum. As required for self-consistency (\ref{GenConsFactTriple}), the terms containing two Mandelstam poles in $R_t$ and $R_s$ agree. Finally, the $u$-channel residue, which contains the undetermined couplings, is 
\begin{align}
R_u&\sim\frac{1}{M_{Pl}m_{\Psi'}}\frac{e^{-i(\theta_W+3\theta)}\mathbf{v}_1\mathbf{v}_3^+\mathbf{v}_3^-}{\left(e^{-2i\theta}\mathbf{v}_P^{(14)}\cdot\mathbf{v}_P^{(23)}\right)^2}\left(\alpha\ds{\mathbf{2}P^M}-\beta\da{\mathbf{2}P^M}\right)\mathbf{v}^{(14)}_{PM}\nonumber\\
&\sim \frac{1}{M_{Pl}}e^{-i(\theta_W+\theta)}\frac{1}{(s-m_s^2)(t-m_t^2)}\nonumber\\
&\qquad\times\Big(-(s-m_s^2)\Big(-\alpha\da{3\bf{1}}\ds{3\bf{2}}+\beta e^{i\theta}\ds{3\bf{1}}\da{3\bf{2}}\nonumber\\
&\qquad\qquad\qquad\qquad\qquad\qquad+(\alpha+\beta)m_{\Psi'}e^{i\theta}\ds{\bf{12}}+(\alpha+\beta)m_{\Psi'} \da{\bf{12}}\Big)\nonumber\\
&\qquad\qquad\qquad+2im_W\alpha\left(m_{\Psi}\sin\theta_W\da{3\bf{1}}\ds{3\bf{2}}+m_{\Psi'}e^{i\theta}\sin(\theta_W-\theta)\ds{3\bf{1}}\da{3\bf{2}}\right)\nonumber\\
&\qquad\qquad\qquad-2im_W\beta\left(m_{\Psi'}\sin\left(\theta_W-\theta\right)\da{3\bf{1}}\ds{3\bf{2}}+m_{\Psi}e^{i\theta}\sin\theta_W\ds{3\bf{1}}\da{3\bf{2}}\right)\Big),
\end{align}
where here $P=P_u$. It is clear upon applying the $4$-particle test (\ref{GenConsFact}) that $\beta=0$, since the $\ds{3\bf{1}}\da{3\bf{2}}$ structure appears only in the $u$-channel residue. Consistency of the remainder clearly requires $\alpha=\frac{1}{M_{Pl}}$. All of the independent Lorentz structures become consistent with this choice. This result is clearly similar to the $\theta=0$ case, but this time there is no further freedom to allow for a parameter analogous to the phase $\phi$. With the couplings now entirely fixed, it can be verified that the mixed gravitino scattering superamplitude is also consistent and does not provide any further information. 

This completes the reconstruction of the $\mathcal{N}=4$ softly broken SUSY theory with two gravitino pairs. This theory has the required particle content of $\mathcal{N}=8$ SUGRA in the UV (the limit in which the masses are all small compared to the particle energies) and corresponds to the spontaneous breaking $\mathcal{N}=8\rightarrow 4$. Like the theory of a single gravitino pair above, it is rigidly constrained and (almost) without the capacity to incorporate more particles or free parameters while preserving the mass gap with the UV and prohibiting higher spin particles. There are two ``branches'' of solutions. The first admits four independent free parameters: the two gravitino masses and the angles $\varphi$ and $\phi$, all of which parameterise some freedom in the structure of the Lie algebra (the ``gauging''). The second only permits a single angle $\theta$ which allows for comparatively little variation in the gauge structure.

The argument presented here also demonstrates that a third gravitino pair cannot be added to the theory without the introduction of a massive spin-$2$ particle. This is because the above argument could be repeated for any pair of distinct gravitinos, but each such pairing would lead to contradictory demands on the properties of the vector super-Higgs multiplets $V$ and $\overline{V}$. 

The fact that the vector boson couplings conform to the structure of a Lie algebra in the first place can be inferred from properties of the $4$-leg superamplitudes. The residues for the vector boson component amplitudes can be extracted from the residues of the superamplitudes involving the vector, gravitino and graviton multiplets. In general, without SUSY, the vector boson exchange contributions to the vector boson scattering residues scale as $\mathcal{O}(E^3)$ at high energies $E$. The reduction to $\sim E^2$ is only possible if the self-couplings are Lie algebra structure constants or GCS couplings. Graviton and scalar exchange residues scale directly as $\sim E^2$ and do not affect this conclusion. So since the superresidues for all of the $4$-leg superamplitudes in these theories all scale as $\sim E^2$ (this can easily be verified for the cases that have not been explicitly calculated here), it immediately follows that vector bosons contained within the supermultiplets must have $P$-symmetric couplings consistent with Lie algebra structure constants and any $P$-violating couplings must be GCS interactions. The rigidity of these theories largely determines their Lie algebra structure, which can be identified simply by extracting the structure constants from the couplings in the relevant component amplitudes. It can be explicitly verified that they obey the Jacobi identity.

The results of this Section are consistent with the old CSS gaugings of maximal SUGRA, historically obtained from a dimensional reduction from $5d$ to $4d$ \cite{Cremmer:1979uq} and subsequently given a purely $4d$ interpretation \cite{Andrianopoli:2002mf}. These describe the special case with $\theta=\phi=\varphi=0$ (as well as the unbroken $\mathcal{N}=6$ case). The Lie algebra generators associated with the massive vector bosons in either the gravitino multiplets or the vector multiplets are identified with translation-like symmetries that mutually commute with themselves, but which are paired together under the action of a $\mathfrak{u}_1$. Lie algebras with this structure are characteristically non-semisimple.

A broader space of gauged SUGRA theories called $CSO^*$ gaugings were identified and constructed in \cite{Hull:2002cv} and later characterised more extensively in \cite{Catino:2013ppa}. They are predicated upon gauging $\mathfrak{cso}^*_{p,r}$ subalgebras of the $\mathcal{N}=8$ duality symmetries. The $\mathfrak{cso}_{p,q,r}$ algebras are a class of non-semisimple Lie algebras consisting of a $\mathfrak{so}_{p,q}$ subalgebra which is linearly represented upon $r$ copies of $p+q$ dimensional vectors of translation-like generators. There are additionally $\frac{1}{2}r(r-1)$ Abelian central extensions that can be produced by commutators of the translations. The $\mathfrak{cso}^*_{p,r}$ Lie algebras have an analogous structure to $\mathfrak{cso}_{p,0,r}$, except that the rotational subalgebra $\mathfrak{so}_{p}$ is replaced with $\mathfrak{so}^*_{p}$. See \cite{Hull:2002cv} for a more detailed overview of the $CSO(p,q,r)$ and $CSO^*(p,r)$ gaugings and their Lie algebra structure. I will continue to use notation referring to the Lie algebra, rather than the Lie group, since it is only the properties of the algebra that the on-shell arguments are sensitive to, even though characterisations using the group are more common in other works. A limited study of scattering amplitudes in these theories have been previously made in \cite{Chiodaroli:2018dbu} for the purposes of identifying double copies. 

The unbroken $\mathcal{N}=6$ theory above is consistent with the expected structure of the $\mathfrak{cso}^*_{2,6}$ Lie algebra. The vectors in the gravitino multiplets are identified with $6$ pairs of translation-like generators, while the graviphoton gauging the central charge is the $\mathfrak{so}_{1,1}=\mathfrak{u}_1$ generator and the rest are central extensions with purely multipole couplings. In the non-chiral superspace, the $\mathcal{N}=6$ graviton has graviphoton components of the form
\begin{align}
H^+\propto\overline{\eta}_a\gamma^{a+}+\frac{1}{2}\epsilon^{abc}\overline{\eta}_a\overline{\eta}_b\eta^d\gamma^{+}_{c;d}+\left(\prod_a\eta^a\right)\gamma^-+\left(\prod_a\overline{\eta}_a\right)\frac{1}{2}\epsilon_{bcd}\eta^b\eta^c\widetilde{\gamma}^{d+},
\end{align}
where the $R$-index $a\in\{1,2,3\}$. The graviphoton gauging the central charge is the linear combination
\begin{align}
v_c^+=\frac{1}{2}\left(\gamma^++\gamma^{+}_{a;a}\right)
\end{align}
($a$ is summed over). This has archetypical non-semisimple couplings to the massive vectors within the gravitino multiplets. The other massless photons are simply central extensions. Note that a pair of photons coupling to a conjugate pair of massive vectors with equal strength monopoles but each with kinematics structures $\frac{1}{x}\da{\bf{12}}^2$ and $\frac{1}{x}\da{\bf{12}}\ds{\bf{12}}$ can be unitarily rotated into a basis in which one photon manifestly has the archetypical non-semisimple coupling (as explained below (\ref{PartAntiSym})) and the other has only a purely multipolar coupling. 

The gauge structure of the $\mathcal{N}=8\rightarrow 4$ theory is more intricate. The graviphotons in the graviton multiplet have the structure
\begin{align}
H^+\propto\gamma^++\overline{\eta}_a\eta_b\gamma^{+}_{a;b}+\left(\prod_a\overline{\eta}_a\right)\left(\prod_b\eta_b\right)\widetilde{\gamma}^{+}
\end{align}
in the non-chiral superspace. Consider alone the subsector of the theory consisting only of the graviton and vector multiplets. The three vector component amplitudes can be extracted from the superamplitudes in order to identify the Lie algebra. Combinations of the graviphotons $\gamma_{1:2}$ and $\gamma_{2:1}$ are responsible for gauging the central charges (the others don't have $3$-particle couplings with pairs of $W$-bosons, so will be momentarily ignored). As shown in Section \ref{sec:N=4SUGRAsub}, these graviphotons couple to the massive $W$-bosons through Lorentz structures of the form $\frac{1}{x}\da{\bf{12}}\ds{\bf{12}}$ (when they have positive helicity), which, by themselves, characterise non-compact Lie algebras. However, the $W$-bosons also couple minimally to the vector multiplets $V$ and $\overline{V}$. For each pair of $W$-bosons (from the $\mathcal{W}$ and $\mathcal{W}'$ multiplets and their conjugates), a basis for the four active photons can be found with the following properties: one photon sourced by a monopole charge carried by the $W$-bosons, usually also with further multipole couplings, another photon with purely multipolar couplings and the final two photons are fully decoupled. 

Generally, the combinations of photons with these properties is different for the $W$ and $\overline{W}$ pair of bosons and the $W'$ and $\overline{W'}$ pair. Focusing for now on the $\theta=0$ branch of solutions, then by examining the couplings, it can be shown that, with a few exceptional degenerate cases, the Lie algebra always has the identity $\mathfrak{so}^*_4=\mathfrak{su}_2\oplus\mathfrak{su}_{1,1}\rightarrow\mathfrak{u}_1\oplus\mathfrak{u}_1$ (along with an additional unbroken $\mathfrak{u}_1\oplus\mathfrak{u}_1$ subspace that I omit for brevity). Depending upon their central charges, the $W$ and $W'$ boson conjugate pairs correspond to the broken generators of each of the non-Abelian ideals in the direct sum (the pair with the stronger monopole charge is part of the compact subalgebra, the pair with the weaker charge is non-compact), while the unbroken $\mathfrak{u}_1$ residuals are more complicated linear combinations of the four active photons (that mixes states between supermultiplets). 

This $\mathfrak{so}^*_4$ Lie algebra is the expected ``maximal semi-simple'' subalgebra of the $\mathfrak{cso}^*_{4,4}$ gauge algebra identified in \cite{Catino:2013ppa} for a class of maximal SUGRA gaugings with unbroken $\mathcal{N}=4$ SUSY and a Minkowksi vacuum. The remaining (non-trivial) generators are the non-semisimple translation-like generators and are identified with the massive vector bosons within the massive gravitino multiplets. Extracting the relevant component amplitudes from the superamplitudes, it can be verified that their couplings are indeed described by characteristically non-semisimple structure constants. For example, ${f}_{\Psi\Psi'}^{\,\,\,\,\,\,\,\,\,\mathcal{W}}=0$, where $\Psi$ and $\Psi'$ are slang for any of their vector boson members under the condition that they are in conjugate positions in the multiplet, such as $W_1$ and $\widetilde{W'}^1$ (see (\ref{N=2Gravitino})). Likewise for the couplings implicit in superamplitudes involving the gravitinos and the massless vector multiplets under which they carry monopole charges (purely multipolar couplings also arise for some pairings of vectors, reflecting the presence of central extensions of the Lie algebra, although I will refrain from detailing these). This is enough to show non-semisimplicity of the generators corresponding to the vectors in the gravitino multiplets. For non-zero angles $\phi$ and $\varphi$, $P$ and $T$-violating GCS interactions also arise in the couplings between the massive vector multiplets and the vector partners of the gravitinos. 

In the special case that the monopole charges of the vector bosons within the $\mathcal{W}$ and $\mathcal{W}'$ multiplets are equal, then the $\mathfrak{su}_2\oplus\mathfrak{su}_{1,1}$ Lie algebra structure degenerates to the non-semisimple $\mathfrak{e}_2\oplus\mathfrak{e}_{2}$. In this case, the Lie algebra of the full theory instead becomes non-semisimple. 

A more familiar limit is that in which the gravitino pairs degenerate (that is, they have the same central charges, so $m_{\Psi'}=m_{\Psi}$ and $\varphi=0$). In this case, the vectors $\mathcal{W}'$ and $\overline{\mathcal{W}'}$ become massless and they lose their central charges. When $\phi=0$, then the $f_{\mathcal{W}'\overline{\mathcal{W}'}V}$ structure constants vanish as well. This is a special instance of the CSS gaugings. Here, the remaining massive $\mathcal{W}$ pair correspond to the (translation-like) broken parts of a $\mathfrak{e}_2$ algebra while the remaining vectors Abelianise. The unbroken member of the $\mathfrak{e}_2$ is a combination of a graviton-partner photon and a state from the massless super-Higgs multiplets. When $\phi=\pi$ however, then structure constants remain and the (now massless) $\mathcal{W}'$ and $\overline{\mathcal{W}'}$ bosons remain as gluons of an unbroken non-Abelian $\mathfrak{su}_2$ YM Lie algebra. The gravitinos assemble into doublets under this Lie algebra. This is the form of the $\mathfrak{cso}^*_{4,4}$ gauging with an unbroken non-Abelian YM sector that was previously studied in \cite{Chiodaroli:2018dbu}.

The $\theta\neq0$, $\varphi=0$ branch of parameter space resembles the CSS case defined by $\theta=\varphi=\phi=0$. The subsector of vector multiplets and graviphotons is again described by a similar non-semisimple Lie algebra. The angle $\theta$ appears only in the couplings involving vector bosons from the gravitino multiplets and parameterises a mixing between the photon partners of the graviton and the massless photons from the super-Higgs multiplets. The non-compact $\mathfrak{so}_4^*$ region of the gauge sector is inaccessible. 

The argument presented here gives the complete and unique characterisation of the perturbative sector of the gauged maximal SUGRA theories that admit Minkowski vacua with $\mathcal{N}=4$ unbroken SUSY. This is in qualitative agreement with the description of the gauging in \cite{Catino:2013ppa}. Presumably the angles parameterising the couplings can be related to the moduli space constructed in \cite{Catino:2013ppa}, but I leave a detailed off-shell comparison for further work. In any case, the analysis presented here demonstrates the exhaustiveness of the $\mathfrak{cso}^*_{4,4}$ gauging and its geometric parameters in describing the spontaneous breaking of $\mathcal{N}=8\rightarrow 4$ SUSY.

\section{Conclusion}

Consistent complex factorisation (the ``$4$-particle test'') of a perturbative $S$-matrix is a rudimentary yet powerful constraint determining the compatibility of hypothetical Lorentz invariant particle interactions with quantum mechanics and causality. These requirements stringently force massless spinning particles to conform to the structure of famed classes of theories of fundamental physics, in particular, Yang-Mills and general relativity. In this study, I completed the long-standing argument that a massless helicity-$3/2$ particle must be a gravitino in a theory of supergravity. This involved ruling-out some hypothetical non-supersymmetric massless $3$-particle amplitudes using the all-channel pole from \cite{Trott:2026cjj}, but most of the focus was on the construction of massive supermultiplets and the derivation of massive SWIs. Of particular note is the $S$-matrix derivation of the BPS bound and the not-quite universality of the gravitino coupling to the states in the multiplets.

$3$-particle amplitudes of massive particles typically lack the kinematic properties of their massless counterparts that underpin the $4$-particle test. Combining fully massive amplitudes across factorisation channels through unitarity produces residues free of Mandelstam poles. As a result, they do not exhibit any obvious tension with locality. However, when the masses of the particles can be reinterpreted as conserved charges, then the particles' momenta can have special properties analogous to the complex kinematics of $3$-particle massless amplitudes. With extended SUSY, these properties underpin the existence of $3$-particle SUSY invariants that can generate amplitudes involving BPS particles (the special case of amplitudes with two equal mass and one massless particle, even without SUSY, descends from this). The amplitudes involving massive BPS particles can be squeezed into supersymmetric superamplitudes that possess similar inverse kinematic dependence as the massless amplitudes. These superamplitudes are consequently subject to the $4$-particle test. A new feature, special to massive amplitudes, is the possibility of multiple independent Lorentz structures in the numerators. This leads to the potential for multiple consistency constraints within a single amplitude and also raises the issue of basis construction, which is complicated by the fact that only terms in the factorisation residues containing Mandelstam poles are challenged by consistency. In this study, I applied the $4$-particle test to theories of BPS particles with extended SUSY and SUGRA, from which the structure of gauged NL$\Sigma$Ms and gauged SUGRA was elucidated.

In Sections \ref{Sec:LowSpinAmp} and \ref{Sec:VecGrav}, I calculated the $4$-particle tree-level (super)amplitudes for the scattering of massive particles with spin $s\leq 1$ in electrodynamics and gravity and demonstrated how the component amplitudes unify into superamplitudes in the presence of supersymmetry. In \cite{Trott:2026cjj}, I calculated the general tree-level amplitudes induced by the exchange of massive spin $s\leq 1$ particles directly from unitarity in theories without SUSY. This enabled a determination of the conditions on the couplings required for the amplitudes to have suppressed and unitary high-energy dependence, leading to a fully on-shell derivation of the Higgs mechanism. It would be interesting to extend this to supersymmetric theories by performing the analogous calculations with superamplitudes of $\mathcal{N}=1$ and $2$ multiplets. This should give an analogous on-shell demonstration of the supersymmetric Higgs mechanism. Compared to the $\mathcal{N}=0$ case, the decomposition of massive multiplets and rearrangements of states into massless multiplets at high energies should be a new feature. The Higgs boson itself may also be identified as either part of a chiral/hypermultiplet or a massive vector multiplet and the structure of the Higgs potential (being $F$ or $D$-term in origin) should also be identifiable. In Appendix \ref{Sec:3legAmp}, I list all of the $3$-particle superamplitudes necessary for such a calculation, while in Section \ref{sec:SimpleN=1}, I perform these calculations for the simplest case of chiral multiplet scattering with $\mathcal{N}=1$ SUSY (this has also been done previously in \cite{Johansson:2023ymb}). This exposed the additional challenges new to on-shell superamplitudes: the construction of a basis of contact terms and SUSY invariant structures required to be tuned against the leading energy factorisation terms. Such a tuning was required in the calculation of the $4$-leg superamplitude for the WZ model. In this case, some of the challenges were trivialised by the fact that the SUSY-invariant Grassmann structure required for the contact term was unique. However, this need not be the case for supermultiplets with more spin or Grassmann dependence. Even under this simplification, two other challenges were made clear by this example. Firstly, the choice of a little group covariant on-shell superspace ruins manifest parity symmetry. So while linearly independent parity conjugate Lorentz structures without SUSY are easily identified simply as having opposite-shaped spinor brackets, parity conjugate Grassmann structures have completely different appearances. The Grassmann Fourier transform can relate the two, but its application demands algebraic effort and the raw result usually lacks manifest SUSY invariance. The second challenge is that manifest SUSY invariance through use of the building blocks (\ref{SuperBlock}) obscures energy dependence. These blocks each scale as $\sim E$ in the high-energy limit, but higher degree Grassmann structures constructed out of these objects will scale at a weaker power than their degree (since they are Grassmann Fourier transforms of lower degree invariants in a parity-conjugate superspace). This complicates the power counting. 

In Section \ref{sec:ExtendedSSSB}, it was proposed that the theory of $\mathcal{N}=6$ SUGRA with massive gravitinos has the simplest gravitino $S$-matrix, as exemplified by the $4$-leg superamplitude (\ref{4legN=6}). In this theory, the super-Higgs mechanism is automatic. It remains to disassemble (\ref{4legN=6}) into component (super-)amplitudes and demonstrate how they unify in this particularly simple case. Such a unfication involves both scalar and vector super-Higgs particles. Gravitino component amplitudes have been recently calculated on-shell in \cite{Gherghetta:2024tob,Gherghetta:2025tlx,Bellazzini:2025shd}. These studies derive the super-Higgs mechanism for theories in which SUSY is fully broken from $\mathcal{N}=1$ or $2$. However, the amplitudes are not presented in a form with manifest high energy dependence, but instead a form in which the factorisation structure is manifest. It would be interesting to reconcile these results with the beautiful $\mathcal{N}=6$ expression (\ref{4legN=6}), as was done for the analogous vector bosons (super)amplitudes here. This may provide clues toward developing efficient on-shell methods like super-BCFW (which can be presumably applied to the $\mathcal{N}=6$ case here) for computing higher-leg massive amplitudes and reconciling this with the proposal of \cite{Gherghetta:2024tob}.

For theories with $\mathcal{N}\geq 4$ SUGRA, I have been able to fully reconstruct the super-Higgs mechanism purely from the requirements of consistent factorisation of gravitino superamplitudes (or shown that it is impossible). The recent work of \cite{Gherghetta:2024tob,Gherghetta:2025tlx,Bellazzini:2025shd,Desai:2026lyt} has computed gravitino amplitudes with fully broken supersymmetry and derived the super-Higgs mechanism from the requirements of suppressed high-energy dependence and analytic properties. Unlike the extended SUGRA theories considered here, the super-Higgs mechanism in these cases allows for much greater parametric freedom (like the masses of the super-Higgses and their coupling strengths). There still remains to be studied the more complicated cases with more gravitinos flavours or an intermediate amount of unbroken supersymmetry $0<\mathcal{N}<4$. These admit the possibilities of both long and short multiplets. Like the supersymmetric Higgs mechanism mentioned above, a new feature of these theories is the rearrangement of submultiplets between the broken SUSY theories in the IR and the unbroken theory emergent in the UV. The $S$-matrix also provides a framework for the study and construction of theories of gauged SUGRA that cuts through the convoluted geometry associated with off-shell actions. This clarity has been used here to fully characterise the Minkowski vacua preserving $\mathcal{N}=4$ SUSY that can arise from gauging maximal SUGRA. 

Most amplitudes of theories of BPS particles in SYM and SUGRA studied in the past have been equivalent to higher-dimensional massless theories \cite{Osborn:1979tq,Dennen:2009vk,Bern:2010qa}. The special massive kinematics of BPS-like theories established here are a well-known feature of massless amplitudes in $6d$ \cite{Cheung:2009dc,Dennen:2009vk}. I have effectively demonstrated the Scherk-Schwarz mechanism for the $\mathcal{N}=8\rightarrow 6$ supersymmetry breaking and how the massive towers associated with the extra dimensions can be decoupled from a purely $4d$ theory of soft supersymmetry breaking in the IR. This can all presumably be represented as components and subsectors within the massless extra-dimensional theory. Such a connection with the amplitudes of other theories of BPS particles studied here remains to be explicitly made. See \cite{Pokraka:2024fao,Chiodaroli:2023tvo} for some potentially relevant recent work on amplitudes in higher dimensions. 

Relatedly, to my knowledge, amplitudes in the $\mathcal{N}=2$ gauged NL$\Sigma$Ms considered here have never been studied before and are open for further exploration. In Section \ref{sec:N=2}, I construct the $3$-particle superamplitudes of $\mathcal{N}=2$ BPS vector bosons and establish the constraints that SUSY imposes upon the theory (\ref{N=2VacRels}). Subjecting these superamplitudes to the $4$-particle test demanded that their parity-conserving couplings be Lie algebra structure constants and their parity-violating couplings be generalised Chern-Simons terms - the expected features of a gauged NL$\Sigma$M. While I have tested the consistency of the superamplitudes in these theories, I have not explicitly calculated them and it would be interesting to do so and compare them to the general results of \cite{Trott:2026cjj}. Their role in the web of double copies \cite{Bern:2022wqg} also remains to be established (assuming that there is one). Again see \cite{Chiodaroli:2023tvo} for some related recent work in higher dimensions. 

Vacuum energy is a critical part of the off-shell story of supersymmetry breaking, yet the $S$-matrix analysis of this study implicitly assumes a flat spacetime background. It would therefore be of interest to develop and perform an analogous analysis in dS, AdS or cosmology with an appropriate choice of asymptotic observable \cite{Baumann:2022jpr}, especially for processes involving gravitinos. This was recently visited in \cite{Chen:2025foq}.

\acknowledgments 

This work was supported in part by the US Department of Energy under the grant DE-SC0011702.

\appendix

\section{Derivation of Massless Supermultiplets and 3-Particle Gravitino Amplitudes from Consistent Factorisation}\label{sec:MasslessSUGRA}

In this Appendix, I will review the derivation of the gravitino $3$-particle amplitudes from consistent factorisation in massless theories, largely following \cite{McGady:2013sga}. I will assume that the permissible kinematic structures of the $3$-particle amplitudes involving the gravitino have already been established (this is covered in \cite{McGady:2013sga}, along with the two additional exclusions in Section \ref{sec:WrongGravitino}) and that all that remains is a determination of their accompanying coupling constants. The $3$-particle amplitudes determine the grouping of particles into supermultiplets by encoding how they transform under the emission of a gravitino (which effectively induces a half-integer helicity change). These underpin the SWIs. See Section \ref{sec:MasslessSoft} for this to be made explicit. Because this Appendix is largely a review, I may skim over some details. 

The universality of the graviton coupling applies to RS particles as much as anything else. The amplitude 
\begin{align}
A(\widetilde{\psi}^+_a,\widetilde{\psi}^-_b,h^+)=\frac{1}{M_{Pl}}\delta_{ab}\frac{\ds{13}^5}{\ds{12}^2\ds{23}}
\end{align}
(the conjugate helicity amplitude is given by the complex conjugate amplitude) is therefore interpreted as supersymmetrically linking the RS particle to the graviton. The residue for a graviton exchange channel in the amplitude $A(\widetilde{\psi}^+_a,\widetilde{\psi}^+_a,\widetilde{\psi}^-_a,\widetilde{\psi}^-_a)$ contains a single cross-channel pole, which may be interpreted simply as the cross-channel graviton exchange. 

I will begin by reviewing the construction of the graviton multiplets for $\mathcal{N}$ gravitinos. If there are multiple flavours of gravitinos, then each pairing requires a new graviphoton. This can be demonstrated as follows. 
Each such graviphoton $i$ can couple to a pair of gravitino flavours through the amplitude
\begin{align}
A(\widetilde{\psi}^+_a,\widetilde{\psi}^+_b,\gamma^-_i)=C_{ab}\frac{\ds{12}^4}{\ds{13}\ds{23}},
\end{align}
where $C_{ab}=-C_{ba}$ are some coupling constants. If $\gamma_i$ coupled to more than one pair of gravitinos, then they would mediate the amplitude $A(\widetilde{\psi}^+_a,\widetilde{\psi}^+_b,\widetilde{\psi}^-_c,\widetilde{\psi}^-_d)$ in the $s$-channel for some $\{c,d\}\neq\{a,b\}$ (and $a\neq b$ and $c\neq d$). However, the corresponding residue contains a cross-channel pole which cannot be identified with an exchanged boson unless two pairs of indices are equated (in which case, the exchanged particle must be a graviton). Each graviphoton can therefore be labeled by a pair of indices describing the gravitino pair that it couples to: $i=ab$. A basis can always be chosen in which this graviphoton is unique. Consistency of the residues for each channel of $A(\widetilde{\psi}^+_a,\widetilde{\psi}^+_b,\widetilde{\psi}^-_a,\widetilde{\psi}^-_b)$ then requires that the graviphoton coupling to gravitinos be non-zero and gravitational strength (a possible phase can be absorbed into the definition of the graviphoton state), in order to be correctly accounted for by the cross-channel pole induced from the graviton exchange. So $C_{ab}=\frac{1}{M_{Pl}}\epsilon_{ab}$ (where $\epsilon_{ab}$ is just a unit norm antisymmetric tensor in $a$ and $b$) and a unique graviphoton must exist for each pairing of gravitinos (which is the reason that I omitted labeling the coupling with the graviphoton index).


The residue for gravitational scattering of a gravitino off a graviphoton with different flavour labels, that is, the $s$-channel of $A(\widetilde{\psi}^+_a,\gamma^+_{bc},\widetilde{\psi}^-_a,\gamma^-_{bc})$ with $a\neq b,c$, still contains a cross-channel pole. This can be identified with the exchange of a helicity-$1/2$ fermion with interaction of the form 
\begin{align}
A(\chi^-_{i},\gamma^+_{ab},\widetilde{\psi}^+_c)=c_{ab;ci}\frac{\ds{23}^3}{\ds{12}}
\end{align}
(and analogously for its amplitude involving opposite helicity particles), where I momentarily leave the flavour structure of the fermion $\chi$ unspecified. Calling $c_{ab;ci}$ the corresponding coupling constant (the ordering of $a$ and $b$ is meaningless), consistency of the residues implies that $\sum_i|c_{ab;ci}|^2=1/M_{Pl}^2$. Note that this fermion cannot exist if $c\in\{a,b\}$, as the gravitino exchange in the amplitude $a(\widetilde{\psi}^+_a,\gamma^+_{ab},\widetilde{\psi}^-_a,\gamma^-_{ab})$ completely accounts for the cross-channel graviton pole. 

Now, the gravitinos can also mediate the scattering of gravitinos and graviphotons with the flavour configuration $A(\widetilde{\psi}^+_a,\gamma^-_{ac},\widetilde{\psi}^-_b,\gamma^+_{bc})$, but not for any other flavour combination. These residues also contain a cross-channel pole that can be identified by the exchange of a helicity-$1/2$ fermion. For flavour configurations in which there are no possible gravitino or graviton channels, the fermion coupling must be prohibited. This implies that each helicity-$1/2$ fermion can only couple to a gravitino/graviphoton pairing with the same triplet of flavour indices (i.e. if $c_{ab;ci}\neq 0$ for some $\{ab,c\}$ then $c_{de;fi}=0$ for $\{de,f\}\neq \{ab,c\}$). This means that the couplings $\{c_{ab;ci}\}$ are a set of orthonormal vectors, each labeled by the explicit configuration of gravitino flavour indices and with components labeled by their fermion index. Returning to the permitted processes, the gravitino exchange residue of $A(\widetilde{\psi}^+_a,\gamma^-_{ac},\widetilde{\psi}^-_b,\gamma^+_{bc})$ for $a\neq b$ must contain a fermion cross-channel pole of the same form as that described for the gravitational process above. Demanding agreement of the residues requires that $\sum_i(c_{bc;ai})^*c_{ac;bi}=1/M_{Pl}^2$. This implies that $c_{ac;bi}=c_{bc;ai}$. The coupling is therefore uniquely labeled by the unordered triplet of distinct flavour indices $\{a,b,c\}$. Again, since the couplings $c_{ab;ci}$ are orthonormal vectors fully specified by the gravitino flavour triplet, then the associated fermion $i$ is unique and can be replaced by the index $abc$. The triplet therefore uniquely identifies the fermion. The coupling must also be gravitational strength (and an overall phase can be absorbed into the definition of the fermion state). This also demonstrates that the graviphoton cannot otherwise have a matter-like vector coupling to the gravitino and some ``gaugino'' not part of the graviton multiplet. 

This argument can basically be repeated by considering the scattering of a fermion $\chi_{abc}$ off a gravitino $\widetilde{\psi}_d$, with similar conclusions to that described above if $d\neq a,b,c$. A cross-channel pole must exist that can be accounted for by a scalar particle. This is uniquely labeled by the four associated flavour indices $abcd$, again by a similar argument to that stated above for the fermion. If this pole was instead identified as a new vector $v_i$, then the fermion could mediate the $s$-channel of the hypothetical amplitude $A(\widetilde{\psi}^+_b,\gamma^+_{cd},\widetilde{\psi}^-_a,v^-_i)$. However, this residue also contains a cross-channel pole that can only be accounted for by a $u$-channel gravitino exchange. But since the graviphoton $\gamma_{cd}$ cannot couple to any of the pairs of gravitinos involved, this is not possible. The vector $v_i$ therefore cannot exist. 

The argument continues to be repeated by considering the gravitational scattering of the newest particle (now a scalar) off a gravitino and introducing a new particle of helicity $1/2$ unit lower (here, a new fermion) to account for the cross-channel pole. This continues until either the set of distinct flavour indices with which to label the new particles runs out or the new particle has a helicity of size greater than two, which is already inconsistent with these arguments \cite{McGady:2013sga}. 
 
Having derived the graviton multiplets, it remains to construct the matter. I will assume, following \cite{McGady:2013sga}, that there are no massless particles with helicity above $2$, that the graviton is the unique helicity-$2$ particle, that the helicity-$3/2$ particles belong to the graviton supermultiplet (as established above) and that only the unique graviphotons derived above can couple to two gravitinos in a single $3$-particle amplitude. The ``matter'' is therefore allowed to consist only of vectors, fermions and scalars with $3$-particle couplings to the gravitinos of the form $A(\phi,\chi^+,\widetilde{\psi}^+)$, $A(\chi^-,v^+,\widetilde{\psi}^+)$ and their conjugates. 

I will assume, to begin with, that there is only a single RS flavour (so $\mathcal{N}=1$). The next step is to derive the RS particle's coupling to matter. If $n_v$ massless vector bosons $v_i$ are included in the theory, then they can gravitationally scatter off the gravitinos. It is easy to show that the graviton exchange residue for $A(v_i^+,\widetilde{\psi}^+,\tilde{\psi}^-,v_j^-)$ contains a cross-channel pole. However, since the only permitted RS coupling to a vector is given by amplitudes of the form $A(\chi^-_i,v^+_j,\widetilde{\psi}^+)=C^v_{ij}\frac{\ds{12}^3}{\ds{23}}$ and their conjugates (for some $n_\chi$ fermions $\chi_i$), then there is only one other possible channel (given by $\chi_i$ exchange of a fixed helicity). Demanding that the two channels produce consistent residues leads to the constraint $C^v_{ki}(C^v_{kj})^*=\frac{1}{M_{Pl}^2}\delta_{ij}$, demonstrating that the couplings $C^v_{ij}$ are a set of $n_v$ orthonormal vectors of dimension $n_\chi$ (hence $n_v\leq n_\chi$). A basis of states can always be chosen that sets the coupling to $C^v_{ij}=\frac{1}{M_{Pl}}\delta_{ij}$, which is to say that each vector couples to a unique fermion with gravitational strength and phase. This establishes the vector multiplets.



Allowing now for $n_\phi$ scalars to also exist, then their only permitted gravitino coupling is 
\begin{align}\label{SUGRAChiral}
A(\phi_i,\chi^+_j,\widetilde{\psi}^+)=(C^\phi_{ij})^*\frac{\ds{13}^2\ds{12}}{\ds{23}}
\end{align}
(and the parity conjugate amplitude is given by complex conjugation, up to a sign whose computation can be avoided). The scattering of scalars or fermions off the gravitinos can be calculated as for vectors above. Consistency of the residues this time leads to the constraints: $C^\phi_{ik}(C^\phi_{jk})^*+C^\phi_{jk}(C^\phi_{ik})^*=\frac{1}{M_{Pl}^2}\delta_{ij}$ and $C^\phi_{ki}(C^\phi_{kj})^*+C^v_{ik}(C^v_{jk})^*=\frac{1}{M_{Pl}^2}\delta_{ij}$. Given the results from the paragraph above, these imply that the scalars can be grouped into a complex basis $\{\phi,\overline{\phi}\}$ and paired with each fermion not already paired with a vector above, leaving amplitudes $A(\phi,\chi^+,\widetilde{\psi}^+)=\frac{1}{M_{Pl}}\frac{\ds{13}^2\ds{12}}{\ds{23}}$ and $A(\overline{\phi},\chi^-,\widetilde{\psi}^-)=\frac{1}{M_{Pl}}\frac{\da{13}^2\da{12}}{\da{23}}$. 

Having established the supermultiplet structure for $\mathcal{N}=1$ gravitinos, the supersymmetry of the remaining amplitudes, whether $3$-legs or higher, is established from the soft-limits argument given in Section \ref{sec:MasslessSoft}. This is equivalent to determining the constraints on matter interactions from SUSY that could otherwise be inferred by testing case by case for consistent factorisation with gravitino emission. I will not bother to do this here, although I will give an amusing example below. 

It remains to derive the matter multiplet structure (that is, the gravitino-matter couplings) for extended supersymmetry. Since SUSY has been established for a single gravitino, it is easiest to do this using $\mathcal{N}=1$ on-shell superspace, because the analysis is mostly identical to that performed above but with superamplitudes and superfields instead of amplitudes and individual helicity eigenstates. In the case of $\mathcal{N}=2$ (that is, explicit $\mathcal{N}=1$ SUSY with a gravitino multiplet), the derivation of the structure of vector multiplets is close to identical to the derivation of the $\mathcal{N}=1$ multiplets above. The derivation of hypermultiplets is also close to identical, but with the noteworthy feature that the coupling to chiral multiplets, $\mathcal{A}(\Phi^+_i,\Phi^+_j,\widetilde{\Psi}^+)=C^\Phi_{ij}\widetilde{\delta}^{(1)}(Q)\frac{\ds{13}\ds{23}}{\ds{12}}$ (see Appendices \ref{sec:OSsuperfields}, \ref{Sec:3legAmp} and Section \ref{sec:ExtSUGRAder} for notation and review of relevant background theory), requires an antisymmetric coupling constant ($C^\Phi_{ij}=-C^\Phi_{ji}$) in order for it to have the correct identical particle exchange symmetry. This means that the hypermultiplets must be composed of pairs of chiral multiplets - a single ``$1/2$-hypermultiplet'' is not consistent with SUSY. 

I will give an explicit illustration of the mechanics behind the tests of consistent factorisation subsumed into the strategy described in the above paragraph. This will demonstrate part of the emergence of the structure of extended SUGRA from complex factorisation and the all-channel pole. The truncation of the size of a supermultiplet ideally traces back to the fermionic nature of the symmetry, here incarnate as the fermionic exchange symmetry of the gravitinos. A demonstration of this can be made by attempting to construct the amplitude $A(\phi,\widetilde{\psi}^+,\widetilde{\psi}^+,v^+)$ from the all-channel pole, assuming that a helicity-$1/2$ fermion can be exchanged. This has $s$- and $t$-channels related by fermionic exchange symmetry:
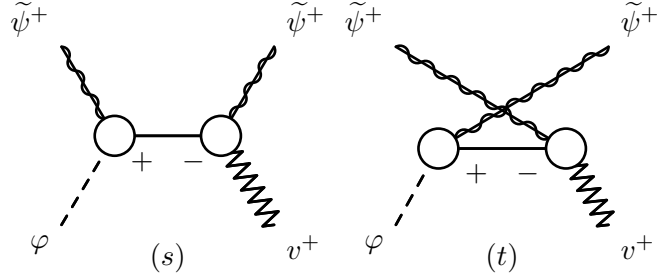
\begin{figure}[h]
\begin{fmffile}{SingleGravitinoCompton}
\begin{center}
\begin{tabular}{ c c c }
& & \\
 \begin{fmfgraph*}(100,67)
   \fmfleft{i1,i2}
   \fmfright{o1,o2}
   \fmf{dashes}{i1,v1}
   \fmf{dashes}{i1,v1}
   \fmf{plain}{i2,v1}
   \fmf{boson}{i2,v1}
   \fmf{zigzag}{v2,o1}
   \fmf{zigzag}{v2,o1}
   \fmf{plain}{v2,o2}
   \fmf{boson}{v2,o2}
   \fmf{plain,label=$+\quad-$}{v1,v2}
   \fmf{plain}{v1,v2} \fmfv{decor.shape=circle,decor.filled=empty,decor.size=0.15w}{v1,v2}
   \fmflabel{$\varphi$}{i1}
   \fmflabel{$\widetilde{\psi}^+$}{i2}
   \fmflabel{$\widetilde{\psi}^+$}{o2}
   \fmflabel{$v^+$}{o1}
 \end{fmfgraph*} & \, &
 \begin{fmfgraph*}(100,67)
   \fmfleft{i1,i2}
   \fmfright{o1,o2}
   \fmf{dashes}{i1,v1}
   \fmf{dashes}{i1,v1}
   \fmf{zigzag}{v2,o1}
   \fmf{zigzag}{v2,o1}
   \fmf{phantom}{v1,i2}
   \fmf{phantom}{v1,i2}
   \fmf{phantom}{v2,o2}
   \fmf{phantom}{v2,o2}
   \fmf{plain,tension=-0.25}{v2,i2}
   \fmf{boson,tension=-0.25}{v2,i2}
   \fmf{plain,tension=-0.25}{v1,o2}
   \fmf{boson,tension=-0.25}{v1,o2}
   \fmf{plain,label=$+\quad-$}{v1,v2}
   \fmf{plain}{v1,v2} \fmfv{decor.shape=circle,decor.filled=empty,decor.size=0.15w}{v1,v2}
   \fmflabel{$\varphi$}{i1}
   \fmflabel{$\widetilde{\psi}^+$}{i2}
   \fmflabel{$\widetilde{\psi}^+$}{o2}
   \fmflabel{$v^+$}{o1}
 \end{fmfgraph*}\nonumber\\
  $(s)$ & \, &  $(t)$ 
\end{tabular}
\end{center}
\end{fmffile}
\caption{On-shell diagrams for $A\left(\phi,\widetilde{\psi}^+,\widetilde{\psi}^+,v^+\right)$.}
\end{figure}
\begin{align}\label{gravitinoTrunc}
A(\phi,\widetilde{\psi}^+,\widetilde{\psi}^+,v^+)&=\frac{\ds{21}\ds{2P_s}^2}{\ds{1P_s}}\frac{1}{s}\frac{\ds{34}^2}{\ds{P_s4}}-\frac{\ds{31}\ds{3P_t}^2}{\ds{1P_t}}\frac{1}{t}\frac{\ds{24}^2}{\ds{P_t4}}\nonumber\\
&=\frac{\da{1q}^2}{\da{2q}\da{3q}}\frac{\ds{12}\ds{34}}{s}\left(\ds{12}\ds{34}+\ds{13}\ds{24}\right),
\end{align}
where I am dropping the overall constants (which are the same in each term) and omitting flavour indices on the scalar and vector legs. This expression is not Lorentz invariant and is therefore only consistent if the accompanying combination of couplings vanish (which has already been established in the analysis above - a fermion belongs either to a vector or a chiral multiplet, not both).

However, if the gravitinos have flavour indices, an alternative presents itself. While $A(\phi,\widetilde{\psi}^+_a,\chi^+)$ and $A(v^+,\widetilde{\psi}^+_a,\chi^-)$ cannot simultaneously exist for a given flavour $a$, they can both exist if the two gravitino flavours are different (i.e. $A(\phi,\widetilde{\psi}^+_a,\chi^+)\neq 0$ and $A(v^+,\widetilde{\psi}^+_b,\chi^-)\neq 0$ with $a\neq b$). More precisely, from the argument described around (\ref{SUGRAChiral}) about the construction of matter multiplets with a single gravitino flavour, the scalar $\phi$ can be paired with a fermion $\chi$ under $\widetilde{\psi}_a$. Likewise, if the scalar $\phi$ is gravitationally scattered off the second gravitino $\widetilde{\psi}_b$, then it must have a $3$-particle coupling involving a fermion in order to explain the cross-channel residue. This could potentially be $\chi$ or some new fermion $\widetilde{\chi}$, so the possible couplings are $A(\phi,\widetilde{\psi}^+_b,\chi^+)$, $A(\overline{\phi},\widetilde{\psi}^+_b,\chi^+)$, $A(\phi,\widetilde{\psi}^+_b,\widetilde{\chi}^+)$ or $A(\overline{\phi},\widetilde{\psi}^+_b,\widetilde{\chi}^+)$. However, the couplings involving $\chi$ give inconsistent contributions to $A(\phi,\widetilde{\psi}^+_b,\widetilde{\psi}^-_a,\overline{\phi})$ and $A(\overline{\phi},\widetilde{\psi}^+_b,\widetilde{\psi}^-_a,\overline{\phi})$, while $A(\overline{\phi},\widetilde{\psi}^+_b,\widetilde{\chi}^+)$ leads to inconsistent $A(\widetilde{\chi}^+,\widetilde{\psi}^+_b,\widetilde{\psi}^-_a,\chi^-)$. This identifies $A(\phi,\widetilde{\psi}^+_b,\widetilde{\chi}^+)$ as the unique possibility and it must have coupling strength $1/M_{Pl}$ (by the same argument as for $A(\phi,\widetilde{\psi}^+_a,\chi^+)$).

Now, under an emission of the gravitino $\widetilde{\psi}_b$, assume that the fermion $\chi$ couples to some massless vector boson. This scenario is described in the paragraph above (\ref{SUGRAChiral}), the conclusion of which is that there must be a vector boson state with a gravitational coupling to both $\widetilde{\psi}_b$ and $\chi$. I will identify this vector as the appropriate analogue of $v$ in the single gravitino flavour amplitude considered above in (\ref{gravitinoTrunc}). Now call $D/M_{Pl}$ the coupling appearing in the possible amplitude $A\left(v^+,\widetilde{\psi}_b^+,\widetilde{\chi}^-\right)$.
Then returning to the amplitude $A(\phi,\widetilde{\psi}^+_a,\widetilde{\psi}^+_b,v^+)$, now with distinct $a\neq b$ gravitino flavours, the analogous $s$ and $t$ channels to (\ref{gravitinoTrunc}) are each given by the exchange of one of the two different fermions:
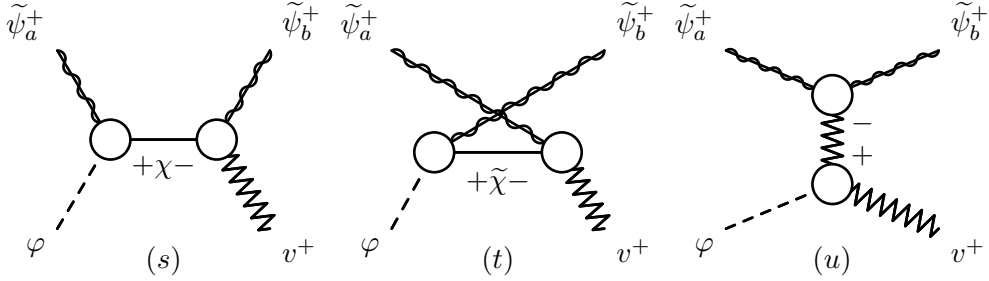
\begin{figure}[h]
\begin{fmffile}{multiGravitinoCompton2}
\begin{center}
\begin{tabular}{ c c c c c}
& & & & \\
 \begin{fmfgraph*}(100,67)
   \fmfleft{i1,i2}
   \fmfright{o1,o2}
   \fmf{dashes}{i1,v1}
   \fmf{dashes}{i1,v1}
   \fmf{plain}{i2,v1}
   \fmf{boson}{i2,v1}
   \fmf{zigzag}{v2,o1}
   \fmf{zigzag}{v2,o1}
   \fmf{plain}{v2,o2}
   \fmf{boson}{v2,o2}
   \fmf{plain,label=$+\chi-$}{v1,v2}
   \fmf{plain}{v1,v2} \fmfv{decor.shape=circle,decor.filled=empty,decor.size=0.15w}{v1,v2}
   \fmflabel{$\varphi$}{i1}
   \fmflabel{$\widetilde{\psi}^+_a$}{i2}
   \fmflabel{$\widetilde{\psi}^+_b$}{o2}
   \fmflabel{$v^+$}{o1}
 \end{fmfgraph*} & \, &
 \begin{fmfgraph*}(100,67)
   \fmfleft{i1,i2}
   \fmfright{o1,o2}
   \fmf{dashes}{i1,v1}
   \fmf{dashes}{i1,v1}
   \fmf{zigzag}{v2,o1}
   \fmf{zigzag}{v2,o1}
   \fmf{phantom}{v1,i2}
   \fmf{phantom}{v1,i2}
   \fmf{phantom}{v2,o2}
   \fmf{phantom}{v2,o2}
   \fmf{plain,tension=-0.25}{v2,i2}
   \fmf{boson,tension=-0.25}{v2,i2}
   \fmf{plain,tension=-0.25}{v1,o2}
   \fmf{boson,tension=-0.25}{v1,o2}
   \fmf{plain,label=$+\widetilde{\chi}-$}{v1,v2}
   \fmf{plain}{v1,v2} \fmfv{decor.shape=circle,decor.filled=empty,decor.size=0.15w}{v1,v2}
   \fmflabel{$\varphi$}{i1}
   \fmflabel{$\widetilde{\psi}^+_a$}{i2}
   \fmflabel{$\widetilde{\psi}^+_b$}{o2}
   \fmflabel{$v^+$}{o1}
 \end{fmfgraph*} &\,& 
  \begin{fmfgraph*}(100,67)
   \fmfleft{i1,i2}
   \fmfright{o1,o2}
   \fmf{dashes}{i1,v1}
   \fmf{dashes}{i1,v1}
   \fmf{plain}{i2,v2}
   \fmf{boson}{i2,v2}
   \fmf{zigzag}{v1,o1}
   \fmf{zigzag}{v1,o1}
   \fmf{plain}{v2,o2}
   \fmf{boson}{v2,o2}
   \fmf{zigzag}{v1,v2}
   \fmf{zigzag}{v1,v2} \fmfv{decor.shape=circle,decor.filled=empty,decor.size=0.15w}{v1,v2}
   \fmflabel{$\varphi$}{i1}
   \fmflabel{$\widetilde{\psi}^+_a$}{i2}
   \fmflabel{$\widetilde{\psi}^+_b$}{o2}
   \fmflabel{$v^+$}{o1}
   \fmfv{label=$-$,label.angle=-45,label.dist=0.1w}{v2}
   \fmfv{label=$+$,label.angle=45,label.dist=0.1w}{v1}
 \end{fmfgraph*}\nonumber\\
  $(s)$ & \, &  $(t)$ & \, & $(u)$
\end{tabular}
\end{center}
\end{fmffile}
\caption{On-shell diagrams for $A\left(\phi,\widetilde{\psi}^+_a,\widetilde{\psi}^+_b,v^+\right)$}
\end{figure}
\begin{align}
&A_{s+t}(\phi,\widetilde{\psi}^+_a,\widetilde{\psi}^+_b,v^+)\nonumber\\
&=\frac{1}{M_{Pl}^2}\frac{\da{1q}^2}{\da{2q}\da{3q}}\frac{\ds{12}\ds{34}}{s}\left(\ds{12}\ds{34}+D\ds{13}\ds{42}\right).
\end{align}
Now, as reviewed above, a graviphoton must exist and couple to the gravitinos with amplitude $A(\widetilde{\psi}^+_a,\widetilde{\psi}^+_b,\gamma^-_{ab})=\epsilon_{ab}\frac{1}{M_{Pl}}\frac{\ds{12}^6}{\ds{13}^2\ds{23}^2}$. 
Introducing a coupling of the graviphoton to the matter $A(\phi,v^+,\gamma^+_{ab})=c_{ab}\ds{23}^2$ (for some coupling constant $c_{ab}=c_{ba}$) then produces a $u$-channel term
\begin{align}
A_{u}(\phi,\widetilde{\psi}^+_a,\widetilde{\psi}^+_b,v^+)&=\frac{c_{ab}}{M_{Pl}}\epsilon_{ab}\frac{\da{1q}^2}{\da{2q}\da{3q}}\frac{\ds{12}\ds{34}}{s}\ds{14}\ds{23}.
\end{align}
If $c_{ab}=1/M_{Pl}$, then this cancels the $s$ and $t$ channels by the Schouten identity to give a consistent amplitude $A(\phi,\widetilde{\psi}^+_a,\widetilde{\psi}^+_b,v^+)=0$, provided that $D=1$. This argument therefore establishes the graviphoton's gravitational strength effective couplings to matter and simultaneously shows how the two distinct gravitino flavours allow for the supermultiplet sizes to be expanded. It also clearly demonstrates that there is substantially more to the argument than originally presented in \cite{GRISARU1977323}.


\section{On-shell Superfields}\label{sec:OSsuperfields}

In what follows, $h^\pm$ will denote a graviton and $\widetilde{\psi}^\pm$ a gravitino. Graviphotons are $v^\pm$ (or sometimes $\gamma^\pm$), while spin-half matter fermions and scalars are respectively $\chi$ and $\phi$. When massive, vector bosons are $W$, while gauginos (massive or massless) are $\lambda$. Superscripts denote helicity, as usual. Tildes and hats may be used to distinguish different species if spin, helicity and $R$-indices do not. See \cite{elvang2015scattering} for a review of massless $\mathcal{N}=8$ superamplitudes.

\subsection{$\mathcal{N}=1$}\label{OSsuperfields}

Chiral and vector multiplets can be found in \cite{Herderschee:2019ofc}. The massive chiral multiplets will be denoted by $\Phi$, the massless ones by $\Phi^\pm$, the massive vectors by $\mathcal{W}^I$ and the massless vectors by $G^\pm$. I will here merely supplement this reference by adding that, choosing a chiral superspace with the highest helicity states as the Clifford vacua, the graviton multiplets are represented as
\begin{align}
H^+&=h^++\eta\widetilde{\psi}^+\\
H^-&=\widetilde{\psi}^-+\eta h^-.
\end{align}

The construction of massive spinning $\mathcal{N}=1$ supermultiplets as fermionic coherent states was explained in \cite{Herderschee:2019ofc}. In general, for spin $s>0$, such a multiplet consists of two spin-$s$ particles (I will denote these by $S_s$ and $S_{\bar{s}}$) and two particles each of spin $s\pm\frac{1}{2}$ (call these $S_{s\pm\frac{1}{2}}$). A coherent state representing the supermultiplet has the structure 
\begin{align}\label{N=1GenSpinSM}
    \mathcal{S}_s^{(I_1\ldots I_{2s})}=S_s^{(I_1\ldots I_{2s})}+\eta_{I_{2s+1}}S_{s+\frac{1}{2}}^{(I_1\ldots I_{2s+1})}+\sqrt{\frac{2s}{2s+1}}\eta^{(I_1}S_{s-\frac{1}{2}}^{I_2\ldots I_{2s})}+\frac{1}{2}\eta_M\eta^MS_{\bar{s}}^{(I_1\ldots I_{2s})}
\end{align}
The first state $S_s$ is the Clifford vacuum chosen so that the multiplet has manifest little group covariance. The superscripts are $SU(2)$ little group indices in symmetric tensor irreps, following \cite{Arkani-Hamed:2017jhn}. The state $S_{s-\frac{1}{2}}$ is accompanied by a non-trivial normalisation factor that must be accounted for when extracting it from the coherent state, as correctly identified by \cite{KNBalasubramanian:2022sae}. The Grassmann integral $\int d^2\eta$ nevertheless correctly implements the state sum, as may be easily verified. The $s=0$ case (a massive chiral multiplet) is special and is given by simply deleting the $S_{s-\frac{1}{2}}$ term in the expression above.

\subsection{$\mathcal{N}=2$}\label{N=2Superspace}

In the standard chiral superspace, massless vector multiplets are represented with two superfields of $CP$-conjugate states
\begin{align}\label{N=2MasslessVec}
G^+&=g^++\eta^a\lambda_a^++\frac{1}{2}\epsilon_{ab}\eta^a\eta^b\phi\\
G^-&=\widetilde{\phi}+\eta^a\lambda_a^-+\frac{1}{2}\epsilon_{ab}\eta^a\eta^bg^-.
\end{align}
Conversion to the non-chiral superspace is usually preferred in theories of BPS particles. This can be performed by applying a Grassmann Fourier transforming in one of the Grassmann variables \cite{Huang:2011um}. Practically, this is tantamount to removing factors of $\eta^a$ (for some specific $a$) from the terms in which they appear in (\ref{N=2MasslessVec}) and placing a conjugate variable $\overline{\eta}_a$ in each of the other terms. In the present example, this corresponds to changing Clifford vacuum to one of the fermions. I will not bother to explicitly do this here, as the primary purpose of this Appendix is to present the multiplet structure and content, while the important features of the transition are largely obvious.

The graviton multiplets are 
\begin{align}
H^+&=h^++\eta^a\widetilde{\psi}^+_a+\frac{1}{2}\epsilon_{ab}\eta^a\eta^b v^+\\
H^-&=v^-+\eta^a\widetilde{\psi}^-_a+\frac{1}{2}\epsilon_{ab}\eta^a\eta^b h^-.
\end{align}

As briefly explained above in Appendix \ref{sec:MasslessSUGRA} and below in Appendix \ref{N=2SUGRAAmps}, the lowest helicity $\mathcal{N}=2$ multiplets must always comes in pairs. Each such complete multiplet representing the SUSY algebra is called a ``$1/2$-hypermultiplet'', while a pair of these containing mutually $CP$-conjugate states is a regular full hypermultiplet. Single ``$1/2$-hypermultiplets'' cannot exist, despite containing enough degrees of freedom to be complete under $CPT$. However, they can still be self-conjugate provided that they pair-up within representations of some other internal symmetry. The massless full hypermultiplet is
\begin{align}
k &= \chi^++\eta^a\phi_a+\frac{1}{2}\epsilon_{ab}\eta^a\eta^b\chi^-\nonumber\\
\overline{k} &= \overline{\chi}^++\eta^a\overline{\phi}_a+\frac{1}{2}\epsilon_{ab}\eta^a\eta^b\overline{\chi}^-
\end{align}
Note that, in this chiral superspace, the hypermultiplet carries helicity that I do not indicate by a superscript. In the non-chiral superspace, they are instead represented by scalar Clifford vacua (which necessarily obscures the possible $SU(2)_R$ $R$-symmetry).

There are two types of massive vector multiplets - long and short. The $\mathcal{N}=2$ long multiplet was constructed in \cite{Herderschee:2019dmc} and is equivalent to the $\mathcal{N}=4$ $\frac{1}{2}$BPS multiplet. The manifest $SU(2)_R$ $R$-symmetry in these expressions corresponds to that of $\mathcal{N}=2$ SUSY. Just as discussed in the context of $\mathcal{N}=4$, the algebra of the supercharges represented on the long multiplets possesses a larger, non-linearly realised $USp(4)$ automorphism symmetry (which is the same as the $R$-symmetry subgroup preserving a single $\frac{1}{2}$BPS vector's central charge matrix in $\mathcal{N}=4$ \cite{Ferrara:1980ra}). The superfield is
\begin{align}\label{LongMultNtwo}
\mathcal{W} =\phi+\eta^a_I\lambda^I_a+\half \eta^a_I \eta^b_J (\epsilon^{IJ} \phi_{(ab)} + \epsilon_{ab} W^{(IJ)})+\frac{1}{2^2}\eta_{bI}\eta_J^b\eta^J_a\widetilde{\lambda}^{aI} +\frac{1}{2^4}\eta_{aI}\eta_{J}^a\eta_{b}^I\eta^{bJ}\widetilde{\phi}.
\end{align}
The $SU(2)_R$ indices can be raised and lowered analogously to the little group indices and are similarly represented in tensor product form rather than as irreps with properly normalised components. 

The $\mathcal{N}=2$ short vector multiplet is equivalent to the $\mathcal{N}=1$ vector multiplet. Just like the massive $\mathcal{N}=1$ vector multiplet, little group covariance prefers that a gaugino be chosen as the Clifford vacuum:
\begin{align}
\mathcal{W}^I = \lambda^I + \frac{1}{\sqrt{2}}\eta^I \phi + \eta_J W^{(IJ)} + \half \eta_J \eta^J \widetilde{\lambda}^I,
\label{vectormult}
\end{align}
The choice of gaugino however manifestly breaks the $SU(2)_R$ $R$-symmetry (as the two form an $SU(2)_R$ doublet), implying that this must be non-linearly realised on the superspace. Only a $U(1)_R$ subgroup is manifest, under which each of the Grassmann variables carries a unit of charge. The same symbols will be used for both $\mathcal{N}=1$ and $\mathcal{N}=2$, which should be clear from context.

Massive hypermultiplets are also necessarily BPS states and contain the same number of degrees of freedom as the massless ones. Because they transform non-trivially under only one of the two supercharges, they are equivalent to massive $\mathcal{N}=1$ chiral multiplets:
\begin{align}\label{N=1Chiral}
K=\phi + \eta_I \chi^I + \half \eta_I \eta^I \widetilde{\phi}.
\end{align}
However, because they are BPS and carry central charge, they must each be paired with separate charge-conjugate multiplets $\overline{K}$ consisting of the antiparticles. As for massive chiral multiplets in $\mathcal{N}=1$ SQCD, these comprise the massive matter of $\mathcal{N}=2$ SQCD. See \cite{Herderschee:2019ofc} for closer examination of the component particles in these multiplets. Just as for the short vector multiplet, the choice of scalar representing the hypermultiplet must break manifest $SU(2)_R$ covariance. See \cite{Abhishek:2022nqv} for some recent analysis making use of these superfields.

\subsection{$\mathcal{N}>2$}\label{N=3graviton}

See \cite{elvang2015scattering} for the well-known massless $\mathcal{N}=4$ vector multiplet. The massive BPS vector multiplet can be found in \cite{Herderschee:2019dmc}, but is equivalent to (\ref{LongMultNtwo}) above. The $\mathcal{N}=4$ graviton multiplets are
\begin{align}
H^+&=h^++\eta^a\widetilde{\psi}^+_a+\eta^a\eta^b v^+_{[ab]}+\frac{1}{3!}\epsilon_{abcd}\eta^a\eta^b\eta^c\chi^{d+}+\frac{1}{4!}\epsilon_{abcd}\eta^a\eta^b\eta^c\eta^d\phi\\
H^-&=\widetilde{\phi}+\eta^a\chi^-_a+\eta^a\eta^bv^-_{[ab]}+\frac{1}{3!}\epsilon_{abcd}\eta^a\eta^b\eta^c\widetilde{\psi}^{d-}+\frac{1}{4!}\epsilon_{abcd}\eta^a\eta^b\eta^c\eta^d h^-.
\end{align}

Massless vectors in $\mathcal{N}=3$ theories have equivalent particle content to $\mathcal{N}=4$, they are just split into separate multiplets of opposite helicity. In contrast, the $\mathcal{N}=3$ graviton multiplets are physically distinct: 
\begin{align}
H^+&=h^++\eta^a\widetilde{\psi}^+_a+\frac{1}{2}\epsilon_{abc}\eta^a\eta^b v^{c+}+\frac{1}{3!}\epsilon_{abc}\eta^a\eta^b\eta^c\chi^{-}\\
H^-&=\chi^-+\eta^av^-_{a}+\frac{1}{2}\epsilon_{abc}\eta^a\eta^b\widetilde{\psi}^{c-}+\frac{1}{3!}\epsilon_{abc}\eta^a\eta^b\eta^ch^-.
\end{align}
Massive vector supermultiplets are $\frac{1}{3}$BPS and are equivalent to $\frac{1}{2}$BPS vectors in $\mathcal{N}=4$ (or long $\mathcal{N}=2$ vectors) (\ref{LongMultNtwo}). 

The graviton multiplets for $\mathcal{N}=6$ are
\begin{align}
H^+&=h^++\eta^a\widetilde{\psi}^+_a+\eta^a\eta^bv_{[ab]}^++\eta^a\eta^b\eta^c\chi^+_{[abc]}+\frac{1}{4!}\epsilon_{abcdef}\eta^a\eta^b\eta^c\eta^d\phi^{[ef]}\nonumber\\
&\qquad\qquad\qquad+\frac{1}{5!}\epsilon_{abcdef}\eta^a\eta^b\eta^c\eta^d\eta^e\chi^{f-}+\frac{1}{6!}\epsilon_{abcdef}\eta^a\eta^b\eta^c\eta^d\eta^e\eta^f v^{-}\nonumber\\
H^-&=v^++\eta^a\chi^+_a+\eta^a\eta^b\phi_{[ab]}+\eta^a\eta^b\eta^c\chi^-_{[abc]}+\frac{1}{4!}\epsilon_{abcdef}\eta^a\eta^b\eta^c\eta^d v^{[ef]-}\nonumber\\
&\qquad\qquad\qquad+\frac{1}{5!}\epsilon_{abcdef}\eta^a\eta^b\eta^c\eta^d\eta^e\widetilde{\psi}^{f-}+\frac{1}{6!}\epsilon_{abcdef}\eta^a\eta^b\eta^c\eta^d\eta^e\eta^f h^{-}
\end{align}
in the usual chiral superspace, although only the non-chiral superspace is ever actually used here. Massive $\frac{1}{2}$BPS multiplets have the same structure as long $\mathcal{N}=3$ massive multiplets (see (\ref{N=3Gravitino}) below for the gravitino).

For $\mathcal{N}=5$ SUGRA, the graviton multiplets are
\begin{align}
H^+&=h^++\eta^a\widetilde{\psi}^+_a+\eta^a\eta^bv_{[ab]}^++\frac{1}{3!}\epsilon_{abcde}\eta^a\eta^b\eta^c\chi^{[de]+}\nonumber\\
&\qquad\qquad\qquad+\frac{1}{4!}\epsilon_{abcde}\eta^a\eta^b\eta^c\eta^d\phi^{e}+\frac{1}{5!}\epsilon_{abcde}\eta^a\eta^b\eta^c\eta^d\eta^e\chi^{-}\nonumber\\
H^-&=\chi^++\eta^a\phi_a+\eta^a\eta^b\chi_{[ab]}^-+\frac{1}{3!}\epsilon_{abcde}\eta^a\eta^b\eta^c v^{[de]-}\nonumber\\
&\qquad\qquad\qquad+\frac{1}{4!}\epsilon_{abcde}\eta^a\eta^b\eta^c\eta^d \widetilde{\psi}^{e-}+\frac{1}{5!}\epsilon_{abcde}\eta^a\eta^b\eta^c\eta^d\eta^e h^-.
\end{align}
These are the only permitted massless multiplets in this theory. There are no vector multiplets. The lowest size massive particle representations are $\frac{2}{5}$BPS. These have the same structure as the $\frac{1}{2}$BPS particles in $\mathcal{N}=6$ SUGRA. The regular three graviton superamplitudes are the only permitted trivalent interactions of massless particles in $\mathcal{N}=5$ and $6$ SUGRA. Both massless theories are completely rigid (in the sense of not permitting ``matter'' sectors with free parameters) without the inclusion of massive particles.

\subsection{Gravitino multiplets}\label{sec:GravitinoSS}


The massive gravitino multiplet with $\mathcal{N}=1$ unbroken supersymmetries is
\begin{align}\label{N=1Gravitino}
\Psi^{(I_1I_2)}=W^{(I_1I_2)}+\eta_J\widetilde{\psi}^{(I_1I_2J)}+\sqrt{\frac{2}{3}}\eta^{(I_1}\chi^{I_2)}+\frac{1}{2}\eta_M\eta^M\widetilde{W}^{(I_1I_2)}.
\end{align}
This is a special instance of the general expression for a massive spinning $\mathcal{N}=1$ multiplet (\ref{N=1GenSpinSM}). The gravitino superfield is evidently represented by one of the two massive vectors that appear in the multiplet. 


Massive gravitino multiplets may be either BPS or non-BPS in the context of extended supersymmetry. The former are the predominant subject of Section \ref{sec:ExtendedSSSB}. The BPS gravitino multiplet with unbroken $\mathcal{N}=2$ SUSY is structurally the same as the $\mathcal{N}=1$ massive gravitino (\ref{N=1Gravitino}). The long $\mathcal{N}=2$ supermultiplet (which is the same as the $\frac{1}{2}$BPS $\mathcal{N}=4$ multiplet) is 
\begin{align}\label{N=2Gravitino}
\Psi^I&=\chi^I+\eta^a_JW^{(IJ)}_a+\frac{1}{\sqrt{2}}\eta^{aI}\phi_a+\frac{1}{2}\epsilon_{ab}\eta^a_J\eta^b_K\widetilde{\psi}^{(IJK)}+\frac{1}{2\sqrt{2}}\epsilon_{ab}\eta^{aI}\eta^b_J\widetilde{\widetilde{\chi}}^J\nonumber\\
&\quad+\frac{1}{2}\epsilon^{JK}\eta^a_J\eta^b_K\chi^I_{(ab)}+\frac{1}{2^2}\eta_{bJ}\eta_{K}^b\eta^{K}_a\widetilde{W}^{a(IJ)}+\frac{1}{2^2\sqrt{2}}\eta_{b}^I\eta_J^b\eta^{J}_a\widetilde{\phi}^a +\frac{1}{2^4}\eta_{aJ}\eta_{K}^a\eta_{b}^J\eta^{bK}\widetilde{\chi}^I.
\end{align}
Here $\widetilde{\widetilde{\chi}}^I$ is another species of fermion (I am running out of symbols). 

The final type of massive gravitino multiplet represents the $\mathcal{N}=3$ SUSY algebra:
\begin{align}\label{N=3Gravitino}
\Psi=&\phi+\eta^a_J\chi^J_a+\frac{1}{2}\epsilon_{abc}\,\eta^a_I\eta^b_JW^{c(IJ)}+\frac{1}{2}\epsilon^{IJ}\eta^a_I\eta^b_J\phi_{(ab)}+\frac{1}{3!}\epsilon_{abc}\,\eta^a_I\eta^b_J\eta^c_K\widetilde{\psi}^{(IJK)}\nonumber\\
&+\frac{1}{2^2}\epsilon_{abd}\,\eta^a_I\eta^b_J\eta^{cJ}\chi^{dI}_c+\frac{1}{2^4}\epsilon_{abe}\,\eta_I^a\eta_{J}^b\epsilon_{cdf}\,\eta^{cI}\eta^{dJ}\phi^{(ef)}
+\frac{1}{2(3!)}\epsilon_{abc}\,\eta_I^a\eta_{J}^b\eta_K^{c}\eta^{dK}\widetilde{W}^{(IJ)}_d \nonumber\\
&+\frac{1}{2^33!}\epsilon_{abc}\,\eta^a_J\eta^b_K\eta^c_I\epsilon_{def}\,\eta^{dJ}\eta^{eK}\widetilde{\chi}^{fI}+\frac{1}{2^3(3!)^2}\epsilon_{abc}\,\eta^a_I\eta^b_J\eta^c_K\epsilon_{def}\,\eta^{dI}\eta^{eJ}\eta^{fK}\widetilde{\phi}.
\end{align}
This is the minimum amount of SUSY required for such a multiplet to be self-conjugate, although they can still appear in pairs with opposite charges. Multiplets of this structure also describe $\frac{1}{2}$BPS gravitinos with unbroken $\mathcal{N}=6$ SUSY.

Massive gravitino multiplets in all higher SUSY theories are shortened to BPS representations equivalent to the three types just stated. Their superampltiudes are studied in Section \ref{sec:ExtendedSSSB}.

\section{Elementary 3-Particle Superamplitudes}\label{Sec:3legAmp}

This Appendix classifies and finishes constructing all $3$-particle superamplitudes in rigid SUSY and SUGRA theories involving matter particles of spin $\leq 1$. This includes the elementary amplitudes that fully generate the minimal, perturbative $S$-matrices of theories such as regular YM and Einsteinian gravity from their factorisation properties. 
The intention is to simply provide the kinematical parts these amplitudes. For this reason, I will consistently omit the inclusion of coupling constants and internal indices unless there is some point of significance to be made. The couplings are, in any case, subject to the same constraints (such as unitarity) required by their $\mathcal{N}=0$ component amplitudes. See \cite{elvang2015scattering} for review of basics of superamplitudes and also \cite{KNBalasubramanian:2022sae} for analysis of higher-spin multiplets.

\subsection{Rigid $\mathcal{N}=1$}\label{N=1Rigid}
All rigid $\mathcal{N}=1$ $3$-particle superamplitudes of vector and chiral multiplets were constructed in \cite{Herderschee:2019ofc} and, while I will not reproduce those results here, I will make an exception to provide an alternative presentation of the $3$ vector superamplitude in order to connect it to the general non-supersymmetric result (\ref{3legVec}), as well as the vector-Higgs coupling (\ref{2vec1scalar}).

The superamplitude between two massive vectors and one chiral multiplet is 
\begin{align}
A\left(\mathcal{W}_A,\Phi_i,\mathcal{W}_B\right)=\delta^{(2)}(Q^\dagger)\left(c^i_{AB}\da{\bf{13}}+\lambda^i_{AB}\ds{\bf{13}}+\left(\bar{c}^i_{AB}\ds{\bf{13}}+\bar{\lambda}^i_{AB}\da{\bf{13}}\right)\frac{m_1m_3}{2m_2^2}F_{2J}F_{2}^J\right).
\end{align}
There are two Lorentz structures at each Grassmann order. The terms with couplings $c$ and $\bar{c}$ are supersymmetrisations of the higher-dimensional interactions of the form $\phi F^2$ from (\ref{2vec1scalar}) (where $F$ is the Yang-Mills curvature, $\phi$ is a scalar field). The partner interactions to these include mixed matter-gaugino dipole interactions, as well as lower-dimensional scalar cubic and Yukawa couplings. The terms proportional to the couplings $\lambda$ and $\bar{\lambda}$ are supersymmetrisations of Higgs couplings to two vectors. They include purely chiral couplings between the matter fermion and the gaugino to a vector. The Grassmann structure of the superamplitude reflects the symmetry of the natural Grassmann coherent state representations that have been made here (see Appendix \ref{sec:OSsuperfields}). If the multiplets are self-conjugate, then $\bar{c}=c^*$ and $\bar{\lambda}=\lambda^*$.

Next I will examine the three vector superamplitude. I will divide these into two sets of terms. The first is a single independent SUSY structure given by
\begin{align}\label{N=1selfHiggs}
A_h\left(\mathcal{W}_A,\mathcal{W}_B,\mathcal{W}_C\right)=\delta^{(2)}(Q^\dagger)\frac{\lambda_{ABC}}{4m_1^2}\left(\left(m_3^2-m_2^2\right)\ds{\bf{23}}+2m_3\da{\bf{31}}\ds{\bf{21}}+2m_2\da{\bf{21}}\ds{\bf{31}}\right)\cdot\bf{F}_{1}
\end{align}
where $\bf{F}_1$ is just $F_1^I$ with the spin index suppressed (just as for the massive spinors) and the dot contraction is intended to indicate the partial spin index contraction between implicitly symmetrised $SU(2)$ tensors. For example, $\da{\mathbf{13}}\ds{\mathbf{12}}\cdot \mathbf{F}_{1} =\da{1^{(I_1}\mathbf{3}}\ds{1^{I_2)}\mathbf{2}}F_{1I_2}$ and $\ds{\mathbf{23}}\cdot \mathbf{F}_{1}=\ds{\mathbf{23}}F_1^{I_1}$.
This superamplitude (\ref{N=1selfHiggs}) corresponds to the supersymmetrisation of a coupling in which the scalar particles in the vector multiplet have a Higgs-like interaction with two vector bosons. This arises when the Higgs boson is part of a massive vector multiplet and, in the high energy limit, is identified as a scalar partner to a Goldstone boson in a massless chiral multiplet. The superamplitude (\ref{N=1selfHiggs}) does not contain a three vector component amplitude.

The remaining set of terms are supersymmetrisations of the Lorentz structures in the general three vector amplitude (\ref{3legVec}), with the exception of the terms proportional to the couplings $g_{ABC}$ (off-shell, operators of the form $F^3$), which are prohibited by the supersymmetry. The novel complication here is that different choices of representation of the Grassmann polynomial correspond to different choices of fixing the redundancy in (\ref{3legVec}) provided by (\ref{3PMassRed}). Choosing a 
representation of the superamplitude that breaks the particle permutation symmetry, as was done in Section \ref{sec:SimpleN=1} for the purposes of simplifying the state sum in the factorisation of the $4$-leg superamplitude, also breaks the permutation symmetry in the representation of the implicit three vector component amplitude (\ref{3legVec}). Proceeding nevertheless with $\bf{F}_1$ as the Grassmann structure, the possible terms in the superamplitude are
\begin{align}\label{N=13vec}
A_v\left(\mathcal{W}_A,\mathcal{W}_B,\mathcal{W}_C\right)&=\delta^{(2)}(Q^\dagger)(a\da{\bf{21}}\da{\bf{31}}+b\ds{\bf{23}}+c\ds{\bf{21}}\ds{\bf{31}}+d\da{\bf{23}}\nonumber\\
&\qquad\qquad\quad+e\left(\da{\bf{31}}\ds{\bf{21}}+\da{\bf{21}}\ds{\bf{31}}\right)+f\left(\da{\bf{31}}\ds{\bf{21}}-\da{\bf{21}}\ds{\bf{31}}\right))\cdot\mathbf{F}_{1}.
\end{align}
Matching the three vector component amplitude onto (\ref{3legVec}) and restoring the colour indices, the couplings are identified as
\begin{align}
a&=\frac{i}{2m_1^2}\left(({f}_{AB}^{\,\,\,\,\,\,\,\,\,C})^*-({f}_{CA}^{\,\,\,\,\,\,\,\,\,B})^*\right)\nonumber\\
b&=\frac{-i}{4m_1^2m_2m_3}\big(im_1^2\left(m_3^2\left(\Im{f}_{BC}^{\,\,\,\,\,\,\,\,\,A}-\Im{f}_{AB}^{\,\,\,\,\,\,\,\,\,C}\right)+m_2^2\left(\Im{f}_{BC}^{\,\,\,\,\,\,\,\,\,A}-\Im{f}_{CA}^{\,\,\,\,\,\,\,\,\,B}\right)\right)\nonumber\\
&\qquad\qquad\qquad\qquad+2m_1^2(m_2^2+m_3^3)\Re {f}_{BC}^{\,\,\,\,\,\,\,\,\,A}-m_1^2\left(m_3^2\,\Re{f}_{AB}^{\,\,\,\,\,\,\,\,\,C}+m_2^2\,\Re{f}_{CA}^{\,\,\,\,\,\,\,\,\,B}\right)\nonumber\\
&\qquad\qquad\qquad\qquad\qquad\qquad\qquad\qquad\qquad+(m_2^2-m_3^2)\left(m_3^2\,\Re{f}_{AB}^{\,\,\,\,\,\,\,\,\,C}-m_2^2\,\Re{f}_{CA}^{\,\,\,\,\,\,\,\,\,B}\right)\big)\nonumber\\
c&=\frac{-i}{2m_1^2m_2m_3}\left(m_3^2\left({f}_{AB}^{\,\,\,\,\,\,\,\,\,C}-{f}_{BC}^{\,\,\,\,\,\,\,\,\,A}\right)-m_2^2\left({f}_{CA}^{\,\,\,\,\,\,\,\,\,B}-{f}_{BC}^{\,\,\,\,\,\,\,\,\,A}\right)\right)\nonumber\\
d&=\frac{-i}{4m_1}\left(\left(1+\frac{m_3^2-m_2^2}{m_1^2}\right)({f}_{AB}^{\,\,\,\,\,\,\,\,\,C})^*+\left(1-\frac{m_3^2-m_2^2}{m_1^2}\right)({f}_{CA}^{\,\,\,\,\,\,\,\,\,B})^*\right)\nonumber\\
e&=\frac{-i}{4m_1^3m_2m_3}\big((m_2^2-m_3^2)\left(m_3\Re {f}_{AB}^{\,\,\,\,\,\,\,\,\,C}+m_2\Re {f}_{CA}^{\,\,\,\,\,\,\,\,\,B}\right)\nonumber\\
&\qquad\qquad\qquad\qquad-m_2m_3(m_2+m_3)i\left(\Im {f}_{AB}^{\,\,\,\,\,\,\,\,\,C}-\Im {f}_{CA}^{\,\,\,\,\,\,\,\,\,B}\right)-m_1^2(m_2-m_3)\Re {f}_{BC}^{\,\,\,\,\,\,\,\,\,A}\big)\nonumber\\
f&=\frac{-i}{4m_1^3m_2m_3}\big((m_2^2-m_3^2)\left(m_3\Re {f}_{AB}^{\,\,\,\,\,\,\,\,\,C}-m_2\Re {f}_{CA}^{\,\,\,\,\,\,\,\,\,B}\right)\nonumber\\
&\qquad\qquad\qquad\qquad-m_2m_3(m_2-m_3)i\left(\Im {f}_{AB}^{\,\,\,\,\,\,\,\,\,C}-\Im {f}_{CA}^{\,\,\,\,\,\,\,\,\,B}\right)+m_1^2(m_2+m_3)\Re {f}_{BC}^{\,\,\,\,\,\,\,\,\,A}\big).
\end{align}
The representation of the three vector component amplitude (\ref{3legVec}) implicit in (\ref{N=13vec}) corresponds to the elimination of the $\da{\bf{23}}\ds{\bf{12}}\ds{\bf{13}}$ term.

Notably, the partner amplitudes contained in this expression include operator dimension $5$ effective interactions of the form $\sim \phi F^2$ (and its $P$-violating counterpart) and anomalous gaugino dipoles. In the limit in which the couplings are restricted to being that of standard Yang-Mills however, these also vanish. 

I will leave a computation of the four vector superamplitude, as well as a complete demonstration of the Higgs mechanism with $\mathcal{N}=1$ SUSY, for future work. See \cite{Argyres:2001eva} for the expected off-shell story.

\subsection{$\mathcal{N}=1$ SUGRA}\label{N=1SUGRA}

Extending the rigid $3$-particle superamplitudes classified in \cite{Herderschee:2019ofc} to include gravitons yields the following possibilities uniquely determined by symmetries, the mass shell condition and ``locality'' in the sense of requiring (at least) a non-negative mass scaling in kinematic variables. The requirement of consistent factorisation of component amplitudes \cite{McGady:2013sga} must be invoked to additionally rule-out a few further cases. The same rules about helicity counting and mass dimension used in \cite{Herderschee:2019ofc} may be used to derive these results. 

The SUSY structure of the massless $3$-particle superamplitudes is determined entirely by the SUSY delta functions:
\begin{align}
\delta^{(2)}(Q^\dagger)&=\eta_1\da{12}\eta_2+\eta_2\da{23}\eta_3+\eta_3\da{31}\eta_1\nonumber\\
\widetilde{\delta}^{(1)}\left(Q\right)&=\ds{12}\eta_3+\ds{23}\eta_1+\ds{31}\eta_2.
\end{align}
The latter invariant is the Grassmann Fourier transform of the delta function of the left-handed supercharge $Q$ when represented in the superspace of opposite chirality. The superscripts denote Grassmann degree. 

The resulting possible graviton superamplitudes with massless particles are, in the usual superspace, 
\begin{equation}
\begin{split}
\mathcal{A}(\Phi^+,H^-,\Phi^-)&=\delta^{(2)}(Q^\dagger)\frac{\da{12}\da{23}^2}{\da{13}^2}\\
\mathcal{A}(G^+,H^-,G^-)&=\delta^{(2)}(Q^\dagger)\frac{\da{23}^3}{\da{13}^2}\\
\mathcal{A}(H^-,H^+,H^-)&=\delta^{(2)}(Q^\dagger)\frac{\da{13}^5}{\da{12}^2\da{23}^2}\\
\mathcal{A}(\Phi^-,H^-,H^-)&=\delta^{(2)}(Q^\dagger)\da{23}^3
\end{split}
\qquad\begin{split}
\mathcal{A}(\Phi^+,H^+,\Phi^-)&=\widetilde{\delta}^{(1)}(Q)\frac{\ds{12}^2\ds{23}}{\ds{13}^2}\\
\mathcal{A}(G^+,H^+,G^-)&=\widetilde{\delta}^{(1)}(Q)\frac{\ds{12}^3}{\ds{13}^2}\\
\mathcal{A}(H^+,H^-,H^+)&=\widetilde{\delta}^{(1)}(Q)\frac{\ds{13}^5}{\ds{12}^2\ds{23}^2}\\
\mathcal{A}(\Phi^+,H^+,H^+)&=\widetilde{\delta}^{(1)}(Q)\ds{23}^3.
\end{split}
\end{equation}
Overall constants have been omitted. These are the supersymmetrisations of the graviton self-coupling, matter couplings to gravity and a higher-dimensional $\phi R^2$ operator gravitational coupling (for some quadratic combination of curvature tensors $R$). 

Just as for vectors, supersymmetry rules-out the three graviton superamplitude if all gravitons have the same helicity. Non-universal field theoretic effective operators $\sim R^3$ are not consistent with SUSY, similarly to $\sim F^3$ operators for massless vector fields. However, also notable is that there are no possible amplitudes between two vectors and one graviton arising with higher kinematical mass dimension beyond that of minimal coupling. This is expected from the locality and unitarity arguments in \cite{McGady:2013sga} and SUSY rules-out the remaining case where the vectors are non-self-interacting photons. The $R_{\mu\nu\rho\sigma}F^{\mu\nu}F^{\rho\sigma}$ effective operator (the only operator of the form $RF^2$ that contributes to an on-shell $3$-particle amplitude) is therefore incompatible with supersymmetry, at least when the vectors are massless and originate from vector multiplets. These constraints will continue to hold with extended SUSY below.


With massive chiral multiplets, the possible superamplitudes are
\begin{equation}
\begin{split}
\mathcal{A}(\Phi,H^+,\Phi)&=\delta^{(2)}(Q^\dagger)\frac{1}{x^2}\\
\mathcal{A}(\Phi,H^+,H^+)&=\delta^{(2)}(Q^\dagger)\ds{23}^4
\end{split}
\qquad\begin{split}
\mathcal{A}(\Phi,H^-,\Phi)&=\delta^{(2)}(Q^\dagger)x^2F_2\\
\mathcal{A}(\Phi,H^-,H^-)&=\delta^{(2)}(Q^\dagger)\da{23}^3.
\end{split}
\end{equation}
With massive vectors, they are
\begin{align}
\mathcal{A}(\mathcal{W},H^+,\mathcal{W})&=\delta^{(2)}(Q^\dagger)\frac{1}{mx^2}\da{\bf{13}}\label{N=1VecMatter}\\
\mathcal{A}(\mathcal{W},H^-,\mathcal{W})&=\delta^{(2)}(Q^\dagger)\frac{x^2}{m}\ds{\bf{13}}F_2.
\end{align}
In all cases the massive states must have equal mass $m$. These amplitudes are identical to the analogous superamplitudes with minimally coupled massless vectors instead of gravitons, but with an extra factor of $x$ to provide the additional unit of little group scaling. The anomalous gravitational dipole is prohibited by self-consistency \cite{Chung:2018kqs}, so has been omitted in the expressions above (but is otherwise consistent with the symmetries). This also  rules-out mixed-spin dipole couplings $\mathcal{A}(\mathcal{W},H^\pm,\Phi)$ which are otherwise consistent with SUSY and Lorentz invariance. Since the anomalous gravitational quadrupole moment of the vectors is determined through SUSY by the gravitational dipole moments, these are also prohibited. Minimal gravitational coupling of the vector multiplets is therefore fixed by self-consistency and SUSY.

\subsection{$\mathcal{N}=2$}\label{sec:N=2App}

\subsubsection{Rigid SUSY}\label{N=2Rigid}
All possible massless $3$-particle superamplitudes that are consistent with symmetries and locality are constructed here. In the chiral superspace, these may be determined to be
\begin{equation}\label{N=2List}
\begin{split}
\mathcal{A}(k,G^+,k)&=\widetilde{\delta}^{(2)}(Q)\frac{1}{\ds{13}}\\
\mathcal{A}(G^+,G^-,G^+)&=\widetilde{\delta}^{(2)}(Q)\frac{\ds{13}}{\ds{12}\ds{23}}\\
\mathcal{A}(G^+,G^+,G^+)&=\widetilde{\delta}^{(2)}(Q)\\
\end{split}
\qquad\begin{split}
\mathcal{A}(k,G^-,k)&=\delta^{(4)}(Q^\dagger)\frac{1}{\da{13}}\\
\mathcal{A}(G^-,G^+,G^-)&=\delta^{(4)}(Q^\dagger)\frac{\da{13}}{\da{12}\da{23}}\\
\mathcal{A}(G^-,G^-,G^-)&=\delta^{(4)}(Q^\dagger).
\end{split}
\end{equation}
Notably, the hypermultiplets cannot interact with themselves trivalently (so there is no $\mathcal{N}=2$ analogue of a cubic Wess-Zumino theory). Unlike for the $\mathcal{N}=1$ case, three vector multiplets of the same helicity do have an admissible superamplitude. However, these do not contain the three vector component amplitudes $A(V^+,V^+,V^+)$ and instead represent dim-$5$ couplings of the form $\phi F^2$ and anomalous dipoles for the gauginos. 

The gauge couplings of the hypermultiples are remarkable because, unlike for the matter-gauge interactions in the $\mathcal{N}=0$ and $\mathcal{N}=1$ cases, the kinematical factor of the superamplitude already has the required exchange symmetry of the matter particles. The coupling constant must therefore be symmetric under exchange of the internal state
labels of the matter. The scalar-vector component amplitude can be extracted and it has an antisymmetric kinematic factor under scalar exchanges. Bosonic exchange symmetry is accounted for by the $R$-indices that must label the scalar states: 
\begin{align}
\mathcal{A}(k_i,G^+,k_j)\ni A(\phi_{ia},g^+,\phi_{jb})=\epsilon_{ab}S_{ij}\frac{\ds{23}\ds{12}}{\ds{13}},
\end{align}
where $S_{ij}=S_{ji}$ are some coupling constants. 

I will consider separately $\mathcal{N}=2$ theories of long and short massive vector multiplets. In field-theoretic SYM, the long multiplets are expected to be produced in theories with ``Higgs branches'' of vacua in which scalars from hypermultiplets acquire VEVs. These produce massive vectors by augmenting massless vectors multiplets with ``longitudinal'' hypermultiplets and without generating central charges. Short multiplets arise in theories with ``Coulomb branches'' of vacua in which scalars from vector multiplets acquire VEVs, producing short massive vector multiplets with central charges (provided by the couplings to the unbroken Abelian generators of the multiplets with the VEVs) saturating the BPS bound. The massive short vectors are produced by vector multiplets Higgsing themselves and do not required the identification of new multiplets emerging in the high energy limit \cite{Fayet:1978ig}. 
For all intents and purposes, I will consider two general cases of theories, the first of which consists of long massive multiplets without central charges and the second of which will consist of only BPS massive vector and hypermultiplets (in addition to massless particles). Theories of short multiplets are the subject of Section \ref{sec:N=2}, so superamplitudes of long muliplets will be the predominant subject of this Appendix. The case of long multiplets will be studied in the standard chiral superspace used to describe the massless superamplitudes above and to present the superfields in Appendix \ref{sec:OSsuperfields}. Just as for $\mathcal{N}=4$ however, an analogous non-chiral superspace is more suitable for the theory of BPS multiplets \cite{Huang:2011um}. To convert the massless superamplitudes above to the non-chiral superspace, the delta functions are simply replaced as 
\begin{align}
\delta^{(4)}(Q^\dagger)\rightarrow \delta^{(2)}(Q^{\dagger 1})\widetilde{\delta}^{(1)}(Q^{\dagger2})\\
\widetilde{\delta}^{(2)}(Q)\rightarrow \delta^{(2)}(Q_2)\widetilde{\delta}^{(1)}(Q_1),
\end{align}
where
\begin{align}
\delta^{(2)}(Q_2)&=\overline{\eta}_1\ds{12}\overline{\eta}_2+\overline{\eta}_2\ds{23}\overline{\eta}_3+\overline{\eta}_3\ds{31}\overline{\eta}_1\nonumber\\
\widetilde{\delta}^{(1)}(Q^{\dagger 2})&=\da{12}\overline{\eta}_3+\da{31}\overline{\eta}_2+\da{23}\overline{\eta}_1
\end{align}
One of the $R$-components of the doublet of Grassmann variables $\eta_i^a$ is replaced (through Grassmann Fourier transform) by the Grassmann variables of the conjugate chirality superspace $\overline{\eta}_{ia}$ (here this is chosen to be $a=2$, I omit the $R$-labels on the Grassmann variables in the expression above). This effectively lowers the helicity of the $\mathcal{N}=2$ superfields from Appendix \ref{sec:OSsuperfields} by $1/2$. See \cite{ArkaniHamed:2008gz} for further details. More generally, mixed branches can also exist involving non-BPS particles with central charges, but I will omit these for simplicity. See \cite{Chen:2021hjl,Chen:2021huj} for a recent study involving amplitudes in this general case. 

%
%

I will next classify the $3$-particle superamplitudes of long vector multiplets. Firstly, the superamplitudes involving two massless vectors may be determined from symmetries to be
\begin{align}\label{N=2Higgs2massless}
\begin{split}
\mathcal{A}(\mathcal{W},G^+,G^+)&=\delta^{(4)}(Q^\dagger)\ds{23}^2
\end{split}\qquad
\begin{split}
\mathcal{A}(\mathcal{W},G^-,G^-)&=\delta^{(4)}(Q^\dagger).
\end{split}
\end{align}
These are effective operators that combine only fermion dipole moments, $\phi F^2$ couplings and Yukawa interactions among various components. There is no violation of the Landau-Yang theorem because there is no component amplitude involving only vector particles. 

With two or three massive long vector multiplets, the permitted superamplitudes are
\begin{align}
\mathcal{A}(\mathcal{W},G^+,\mathcal{W})&=\delta^{(4)}(Q^\dagger)\left(\frac{a^+}{m}\frac{1}{x}+c^+\frac{1}{2}\epsilon_{ab}{F}_2^a{F}_2^b\right)
\\
\mathcal{A}(\mathcal{W},G^-,\mathcal{W})&=\delta^{(4)}(Q^\dagger)\left(c^-+\frac{a^-}{m}x\frac{1}{2}\epsilon_{ab}{F}_2^a{F}_2^b\right)
\\
\mathcal{A}(\mathcal{W},\mathcal{W},\mathcal{W})&=\delta^{(4)}(Q^\dagger)\Big(c+\lambda_{ab}\frac{m_1m_3}{2m_2^2}F_{2J}^{(a}F_{2}^{b)J}+\bar{c}\prod_a \frac{m_1m_3}{2m_2^2}F_{2J}^aF_{2}^{aJ}\nonumber\\
&\qquad\qquad\qquad\qquad\qquad\qquad-\frac{a}{2m_3^3}\epsilon_{ab}F_{2J}^a\la{2^J}p_1\rs{2^Q}F_{2Q}^b\Big)\label{N=2WWW}
\end{align}
In the superamplitudes with only two massive vectors, the terms proportional to $a^\pm$ can only exist if the masses are equal ($m_1=m_3=m$). These terms correspond simply to the standard minimal coupling. Restoring colour indices $A,B,C$ to particles $1,2,3$ respectively, these couplings match onto (\ref{2Mass1MasslessVec}) as $a^\pm=-if_{ABC}$. The terms proportional to $c^\pm$ contain the usual dimension-$5$ effective $3$-particle interactions (gaugino dipole interactions and scalar-vector axion/modulus-like couplings). 

In the superamplitude of three massive vectors, the SUSY structure in the second line of (\ref{N=2WWW}) represents a SYM-like coupling. Since this is the only SUSY structure that contributes to the three vector component amplitude, then just as for $\mathcal{N}=4$, the coupling must be fully exchange antisymmetric when the colour indices are restored. Matching onto (\ref{3legVec}) with (\ref{StandardLA}) gives $a=-if_{ABC}$. However, unlike for $\mathcal{N}=4$, these presumably need only obey the Jacobi identity if the four vector superamplitude is required to have suppressed high energy dependence. This is not a necessary consequence of consistency with locality and SUSY. In any case however, once the Jacobi identity has been established, the standard properties of the YM Lie algebra are automatic (semi-simplicity, compactness, homogeneous generator normalisations). I have not bothered to present (\ref{N=2WWW}) in a form with manifest particle exchange symmetries, but the SUSY structure of this expression is clear. 

Turning to the first line of the superamplitude (\ref{N=2WWW}),
there are five independent SUSY structures labeled by couplings $c$, $\bar{c}$ and $\lambda_{ab}=\lambda_{ba}$ (the latter of which explicitly break $SU(2)_R$), each of which corresponds to one of the five scalars in the long vector multiplets (this is manifest in (\ref{N=2WWW}) for multiplet $2$). The $c$ and $\bar{c}$ terms describe the same type of dimension-$5$ effective interactions as their counterparts $c^\pm$ in the previous superamplitudes with a massless vector. The $\lambda_{ab}$ terms are supersymmetrisations of Higgs-like interactions between the scalars in (\ref{LongMultNtwo}) with the same $R$-index configuration and two vectors. See (\ref{2vec1scalar}) for comparison.

Note that even the pure SYM sector interactions in both $\mathcal{N}=4$ and $\mathcal{N}=2$ differ as a consequence of the central charges despite the multiplets having the same composition. While the three vector component amplitude is the same, the partner interactions are not. The obvious example is Higgs couplings, which are fixed by SUSY in $\mathcal{N}=4$, but are parameterically free in $\mathcal{N}=2$. In the latter case, this allows for them to combine, in the high-energy limit, with other Goldstone bosons or matter states into different representations of the YM Lie algebra besides the adjoint. 


The possible vector couplings to hypermultiplets (which, by assumption of the absence of central charges, are massless here) are
\begin{align}
\mathcal{A}(\mathcal{W},k,k)&=\delta^{(4)}(Q^\dagger)\ds{23}\\
\mathcal{A}(\mathcal{W},k,\mathcal{W})&=\delta^{(4)}(Q^\dagger)\lambda^a{F}_2^a.\label{WhW}
\end{align}
The first superamplitude is simply the minimal coupling of massless matter to a massive vector multiplet. The second (\ref{WhW}) explicitly breaks the $SU(2)_R$ through the coupling constants $\lambda^a$ that have been explicitly included. This contains Higgs-like interactions between the vectors and the corresponding combination of $R$-indexed scalars in the hypermultiplet determined by $\lambda^a$. 

Just as for $\mathcal{N}=1$, there are therefore two possible implementations of the Higgs mechanism from the low energy perspective that, in the high energy limit, are qualitatively the same. From state counting in the high-energy limit, a long vector multiplet is expected to decompose into a massless vector and (full) hypermultiplet. These ``longitudinal'' Goldstone hypermultiplets should then combine, possibly with other matter hypermultiplets, into a complete linear representation of the broken Lie algebra. The $\mathcal{N}=2$ super-Goldstone and Higgs bosons are therefore expected to belong to hypermultiplets in the high energy limit. If the Higgs and Goldstone bosons belong to separate hypermultiplets in the high energy limit, then the Higgs remains at low energies as part of a massless hypermultiplet with coupling (\ref{WhW}). If they are parts of the same hypermultiplets, then, in the low-energy massive theory, the Higgs bosons are identified with scalars in the massive vector multiplets and the Higgs couplings are given by the $\lambda_{ab}$ terms in (\ref{N=2WWW}). Of course, combinations of both of these scenarios are also possible. In either case, it is clear that the Higgs mechanism necessarily breaks the otherwise emergent $SU(2)_R$ $R$-symmetry. The scalars in the massless hypermultiplets comprise fundamental $SU(2)_R$ representations, yet they must be necessarily split-up when identified with either the longitudinal modes of the vectors or the Higgs bosons. Since the massive multiplets and the YM coupling preserve this symmetry, the interaction responsible for driving this rearrangement of states must come from the Higgs coupling. 
I will leave a direct demonstration of this through the computation of the $4$-particle superamplitude for future work. See \cite{Fayet:1978ig} for an account of the expected off-shell story of the supersymmetric Higgs mechanism.

Despite the breaking of $SU(2)_R$, a separate $U(1)_R$ symmetry remains in the SYM and Higgs superamplitudes. In the choices of superspace used here, the Clifford vacua of the superfields have $U(1)_R$ charges $[k]_R=1$, $[G^+]_R=0$, $[G^-]_R=2$ and $[\mathcal{W}]_R=2$, while the Grassmann variables carry charge $[\eta]_R=1$. This is potentially broken by the higher-dimension effective interactions that are otherwise permitted by SUSY.

\subsubsection{SUGRA}\label{N=2SUGRAAmps}

The possible superamplitudes of gravitons and massless matter are
\begin{equation}\label{N=2SUGRA}
\begin{split}
\mathcal{A}(k,H^-,k)&=\delta^{(4)}(Q^\dagger)\frac{\da{12}\da{23}}{\da{13}^2}\\
\mathcal{A}(G^+,H^+,G^-)&=\widetilde{\delta}^{(2)}(Q)\frac{\ds{12}^2}{\ds{13}^2}\\
\mathcal{A}(G^+,H^+,H^+)&=\widetilde{\delta}^{(2)}(Q)\ds{23}^2\\
\mathcal{A}(H^-,H^+,H^-)&=\delta^{(4)}(Q^\dagger)\frac{\da{13}^4}{\da{12}^2\da{23}^2}
\end{split}\qquad
\begin{split}
\mathcal{A}(k,H^+,k)&=\widetilde{\delta}^{(2)}(Q)\frac{\ds{12}\ds{23}}{\ds{13}^2}\\
\mathcal{A}(G^+,H^-,G^-)&=\delta^{(4)}(Q^\dagger)\frac{\da{23}^2}{\da{13}^2}\\
\mathcal{A}(G^-,H^-,H^-)&=\delta^{(4)}(Q^\dagger)\da{23}^2\\
\mathcal{A}(H^+,H^-,H^+)&=\widetilde{\delta}^{(2)}(Q)\frac{\ds{13}^4}{\ds{12}^2\ds{23}^2}.
\end{split}
\end{equation}
Note that for $\mathcal{N}=2$ SUSY, the counting rules described in \cite{Herderschee:2019ofc} must be modified because both delta functions have even helicity weight.

The three graviton superamplitudes $\mathcal{A}(H^+,H^+,H^+)=\widetilde{\delta}^{(2)}(Q)\ds{12}\ds{23}\ds{31}$ (and its parity conjugate) are consistent with SUSY, contrary to $\mathcal{N}=1$. These do not contain three graviton component amplitudes, so parametrically free $R^3$ operators are still forbidden, as is well-known. However, this superamplitude has the wrong particle exchange symmetry, so would require further internal quantum numbers, violating uniqueness of the graviton \cite{weinberg1964photons,McGady:2013sga}. Also of note in the catalogue above is that the gravitinos couple to the graviphoton by a helicity flip (that is, an incoming gravitino's helicity flips upon emission), which is in contrast to a minimally coupled, massless spin-$1/2$ fermion. This must be accompanied by a change in the $SU(2)_R$ component of the gravitino. 

The hypermultiplet's coupling to the graviton in (\ref{N=2SUGRA}) contains the wrong exchange symmetry, necessitating the existence of internal quantum number labels. Calling $A_{ij}=-A_{ji}$ some coupling constants, the scalar component amplitude can be extracted as
\begin{align}
\mathcal{A}(k_i,H^+,k_j)=A_{ij}\widetilde{\delta}^{(2)}(Q)\frac{\ds{12}\ds{23}}{\ds{13}^2}\ni A(\phi_{ia},h^+,\phi_{jb})=\epsilon_{ab}A_{ij}\frac{\ds{23}^2\ds{12}^2}{\ds{13}^2},
\end{align}
which is now symmetric under scalar exchange because of the $R$-index Levi-Civita symbol. This is another illustration of $S$-matrix inconsistency of $1/2$-hypermultiplets without internal representations. The most obvious option is for two $1/2$-hypermultiplets to be paired with each other in a $U(1)$ representation, making the fermions Dirac. This pair is typically what is referred to as a ``hypermultiplet'', although non-trivial pseudo-real representations of $1/2$-hypermultiplets are also possible, as is well-known.

The superamplitudes coupling the graviton to long massive vectors are 
\begin{align}
\begin{split}
\mathcal{A}(\mathcal{W},H^+,\overline{\mathcal{W}})&=\delta^{(4)}(Q^\dagger)\frac{1}{x^2}\\
\mathcal{A}(\mathcal{W},H^+,H^+)&=\delta^{(4)}(Q^\dagger)\ds{23}^4
\end{split}\qquad
\begin{split}
\mathcal{A}(\mathcal{W},H^-,\overline{\mathcal{W}})&=\delta^{(4)}(Q^\dagger)x^2\frac{1}{2}\epsilon_{ab}{F}_2^a{F}_2^b\\
\mathcal{A}(\mathcal{W},H^-,H^-)&=\delta^{(4)}(Q^\dagger)\da{23}^2.
\end{split}
\end{align}
The standard minimally coupled graviton-matter superamplitudes are given in the first line and are identical to their electromagnetic counterparts, but with some extra factors to account for the different little group scaling. An anomalous gravitational dipole term is consistent with SUSY, but has been omitted here because of inconsistency with locality \cite{Chung:2018kqs}. It is assumed here that the long vectors do no carry central charge. Consequently, they do not couple to the graviphoton. The second pair of interactions are supersymmetrisations of $\phi R^2$-type couplings (which were also possible for massless vectors).

The minimal coupling superamplitudes presented here preserve the full $\mathcal{N}=2$ $U(2)_R$ $R$-symmetry, assigning $U(1)_R$ charges to the gravitons as $[H^+]_R=0$ and $[H^-]_R=2$. The high-dimensional effective interactions break $U(1)_R$, but preserve $SU(2)_R$.

The gravitational superamplitudes of short multiplets have been presented in previous Sections: Hypermultiplet couplings are presented in Section \ref{sec:LowSpinGrav}, while vectors are explained in Section \ref{sec:BPSGrav}.

\subsection{$\mathcal{N}=4$}

The premitted graviton superamplitudes with massless matter (in the usual chiral superspace) are:
\begin{equation}
\begin{split}
\mathcal{A}(G,H^+,G)&=\widetilde{\delta}^{(4)}(Q)\frac{1}{\ds{13}^2}\\
\mathcal{A}(H^+,H^-,H^+)&=\widetilde{\delta}^{(4)}(Q)\frac{\ds{13}^2}{\ds{12}^2\ds{23}^2}\\
\mathcal{A}(H^+,H^+,H^+)&=\widetilde{\delta}^{(4)}(Q)
\end{split}\qquad
\begin{split}
\mathcal{A}(G,H^-,G)&=\delta^{(8)}(Q^\dagger)\frac{1}{\da{13}^2}\\
\mathcal{A}(H^-,H^+,H^-)&=\delta^{(8)}(Q^\dagger)\frac{\da{13}^2}{\da{12}^2\da{23}^2}\\
\mathcal{A}(H^-,H^-,H^-)&=\delta^{(8)}(Q^\dagger).
\end{split}
\end{equation}
The superamplitudes of all same-sign helicity graviton multiplets combine $\phi R^2$-type couplings, dipole moments for the gravitinos, gravitational quadrupole moments mixing gravitinos with helicity-$1/2$ fermions, and interactions of the off-shell form $RF^2$. The latter is possible because the graviphotons are not the highest weight states in the supermultiplet, in contrast to vectors originating from vector multiplets, so there is no tension with the SWIs. The massive vector self-couplings were given in Section \ref{Sec:N=4SYM} and couplings to gravitons in Section \ref{sec:BPSGrav}.




\subsection{$\mathcal{N}=3$}

The massless $\mathcal{N}=3$ SUGRA-matter superamplitudes are similar to the $\mathcal{N}=4$ ones listed above and I will not bother to reproduce them here. The only significant difference is that $\mathcal{A}(H^+,H^+,H^+)$ and its parity conjugate are not permissible, failing to be consistent with both $\mathcal{N}=3$ SUSY and Lorentz invariance. Minimal gravity and Yang-Mills are therefore the only possible massless trivalent interactions.

\section{Comments on Dyonic Superamplitudes}\label{Dyons}

An $S$-matrix for dyons has been recently proposed in \cite{Csaki:2020inw}, based on \cite{Zwanziger:1972sx}, in which all pairs of external scattering states are classified by their tiny group representations, analogous to Wigner's classification of single particle states into little group representations. If any two scattering states have relative dyonic charges, then their $U(1)$ tiny group charge determines the angular momentum stored in their electromagnetic fields at large separation.  Kinematic variables with tiny group charges were constructed from which dyonic $S$-matrix entries could be built, along with the usual spinors describing external polarisations. This led to the identification of angular momentum selection rules for the scattering of dyons. 

The appearance of tiny group charges in the discussion of $3$-particle kinematics in Section \ref{3PSK} suggests that the spinor variables described there are natural ingredients for a dyonic $S$-matrix. In this Appendix, I briefly sketch some simple ideas for $3$-particle superamplitudes of BPS dyons constructed out of SUSY delta functions and little group spinors. As dyonic scattering does not admit a perturbative expansion in a small coupling constant like electron scattering in QED, it is not clear how these superamplitudes are to be interpreted, especially since they rely on complexified (non-Minkowskian) kinematics in order to exist. 
Amplitudes with incoming particles are also not equivalent to amplitudes with outgoing particles under analytic continuation. The proposals in this Appendix are purely speculative and merely represent a shallow pursuit of the suggestion of \cite{Csaki:2020inw} and (\ref{SpecialSVD}). 

In the following discussion, the tiny group fixing described around equation (\ref{Trimmed3PSK}) will be temporarily retracted. The incoming and outgoing states are characterised by tiny group representations of each pairing of the particles that constitute them. An amplitude of three particles must therefore be either a $1\rightarrow 2$ or $2\rightarrow 1$ process and only the two particle state can have a non-trivial tiny group representation. The partitioning of states into incoming or outgoing breaks the equivalence of each pairing's tiny groups. However, it is appropriate (as far as I can see) to identify the tiny group charges of $\mathbf{v}_{j|i}^I$ and $\mathbf{v}_{i|j}^J$ as opposites, which further allows for the fixing $\beta_{ij}=-\alpha_{ij}$. It is then possible to construct scalars $\alpha_{ik}\alpha_{kj}/\alpha_{ji}$ that carry two spinor's worth of the $U(1)_{i,j}$ tiny group charge, while the spinors $\mathbf{u}_{i-1|i}$ and $\mathbf{v}_{i+1|i}$ can be regarded as redundant, just as for the purely electric case. 

Consider three $\frac{1}{2}$BPS particles with the dyonic central charge configuration (\ref{CentralChargesMagN=4}). The corresponding representations of the supercharges are given in (\ref{Scharge1/2dyon}). As mentioned previously, because the phases in (\ref{Scharge1/2dyon}) now appear with the same sign in both pairs of supercharges, $\ds{v_{i|j}Q_{a+2}}=\da{v_{i|j}Q^{\dagger a}}$ now holds for both $a=1,2$ (where $\rs{v_{i|j}}=\mathbf{v}_{i|j}^J\rs{j_J}$ etc). As a result, the $v$-type spinors appear in both pairs of SUSY delta functions in (\ref{SuperBPSDelta}). For a superamplitude $\mathcal{A}(\mathcal{W}_{\{q_1+q_2,p_2\}}\rightarrow\mathcal{W}_{\{q_1,0\}},\mathcal{W}_{\{q_2,p_2\}})$ (where the subscripts on the particles denote electric and magnetic charges respectively), it would be natural to select $\mathbf{v}_{i|j}=\mathbf{v}_{2|3}$ in the definition of the SUSY delta functions. Factors of $\alpha_{21}\alpha_{13}/\alpha_{32}$ can then dress the delta functions to give the superamplitude the required tiny group charge, which here is given by $q_1p_2$ \cite{Csaki:2020inw}. If the external states are spinning, then further factors of the little group spinors $\mathbf{v}_{i|j}$ and $\mathbf{u}_{i|j}$ should be included, again possibly balanced by factors of $\alpha_{ij}$ to give the required tiny group charges. Alternatively, it could be that the dyonic charges involved are not the central charges of the SUSY algebra. In this case, the usual SUSY delta functions defined in (\ref{SuperBPSDelta}) are still applicable (with the spinors appearing within them, $\rs{v_{i|j}}$ and $\rs{u_{i|j}}$, chosen to have opposite tiny group charges). Again, the full dyonic superamplitude would be given by dressing this with terms constructed out of factors of $\mathbf{u}_{i|j}$, $\mathbf{v}_{i|j}$ and $\alpha_{ij}$ to provide the necessary little and tiny group representations. Note that while the particles are implicitly all outgoing in the representation of the supercharges in (\ref{Scharge1/2dyon}), crossing the incoming leg simply introduces some relative signs that do not affect any of the critical ingredients in the construction (and likewise for the relations in (\ref{3PSKMassive})). For this reason I will not bother to state the effects here.

Another example, alluded to in Section \ref{sec:1/4BPS}, is the hypothetical superamplitude describing the decay of a $\frac{1}{4}$BPS dyon into a pair of $\frac{1}{2}$BPS states (one carrying the magnetic charge and the other the electric charge). The SUSY invariant for this process would be given again by the hybrid $\Delta_v\delta^{(2)}(Q^{\dagger 2})\delta^{(2)}(Q_{4})$, where this time the supercharges $Q^{\dagger 2}$ and $Q_4$ are given by combining together (\ref{supercharges}), (\ref{Scharge1/2dyon}) and (\ref{SUSYRepsLongGen}). Some phases may need to be manually introduced into these expressions depending upon whether a particular leg is to be interpreted as incoming or outgoing, but the general structure of the SUSY invariant should not be affected by this. Further kinematical factors of little group spinors and spacetime spinor bilinears would presumably also be required to ensure that the superamplitude has the required spin and tiny group properties.

\section{Amplitudes with Gravitational Quadrupoles}\label{Quadrupoles}

In this Appendix, I compile the possible contributions from independent gravitational quadrupole moments to the scattering amplitudes of massive vector bosons computed in Section \ref{sec:pureGrav}. The general amplitudes for a massive vector boson coupled to a graviton are given by 
\begin{align}
A(W,\overline{W},h^+)&=\frac{1}{M_{Pl}}\left(\frac{1}{x^2}\da{\bf{12}}^2+\frac{Q}{\Lambda^2}\ds{3\bf{1}}^2\ds{3\bf{2}}^2\right)\nonumber\\
A(W,\overline{W},h^-)&=\frac{1}{M_{Pl}}\left(x^2\ds{\bf{12}}^2+\frac{Q^*}{\Lambda^2}\da{3\bf{1}}^2\da{3\bf{2}}^2\right),
\end{align}
where $\Lambda$ is some higher energy scale and $Q$ is the ``anomalous'' gravitational quadrupole moment (defined here precisely as the departure of the quadrupole from minimal coupling). Off-shell, the quadrupole terms correspond to operators of the form $RF^2$, which is clear from taking the high energy limit. With any amount of SUSY, the independent quadrupoles for the vector bosons are prohibited, since they become constrained by the illegal dipole moments. 

The quadrupole contribution to the vector-scalar amplitude (\ref{vsgrav}) can be easily computed to give
\begin{align}
&A(W,\overline{W},\varphi,\overline{\varphi})
\nonumber\\
&=-\frac{1}{s}\frac{1}{\left(2M_{Pl}\Lambda\right)^2}\Big(Q\left((u-t)\ds{\bf{12}}+M\left(\la{\bf{1}}p_3-p_4\rs{\bf{2}}+\ls{\bf{1}}p_3-p_4\ra{\bf{2}}\right)\right)^2+\text{Parity conj.}\Big).
\end{align}
The parity conjugate terms are given, as usual, by swapping the bracket shapes and complex conjugating the coefficients. In the high energy limit, this amplitude does not match onto a contact interaction, so cannot be canceled by one. However, there are contact terms of the same (or lower) order in energy dependence that must presumably be included as part of a consistent EFT (see e.g. \cite{Henriksson:2022oeu}). The quadrupole contribution to the vector-fermion amplitude (\ref{vfgrav}) is similarly given by
\begin{align}
&A(W,\overline{W},\psi,\overline{\psi})\nonumber\\
&=-\frac{1}{s}\frac{1}{\left(2M_{Pl}\Lambda\right)^2M}\Big(Q\ds{\bf{34}}\left((u-t)\ds{\bf{12}}+M\left(\la{\bf{1}}p_3-p_4\rs{\bf{2}}+\ls{\bf{1}}p_3-p_4\ra{\bf{2}}\right)\right)^2\nonumber\\
&\qquad\qquad\qquad\qquad\qquad\qquad\qquad\qquad\qquad\qquad\qquad\qquad\qquad\qquad+\text{Parity conj.}\Big).
\end{align}

For vector bosons scattering off other vector bosons (\ref{vvgrav}), there can be single or double quadrupole insertions. The single insertion terms are simply
\begin{align}
&A_Q(W,\overline{W},W',\overline{W'})\nonumber\\
&=-\frac{1}{s}\frac{1}{\left(2M_{Pl}\Lambda\right)^2}\Big(Q\ds{\bf{34}}^2\left((u-t)\ds{\bf{12}}+M\left(\la{\bf{1}}p_3-p_4\rs{\bf{2}}+\ls{\bf{1}}p_3-p_4\ra{\bf{2}}\right)\right)^2\nonumber\\
&\qquad\qquad\qquad\qquad+Q'\ds{\bf{12}}^2\left((u-t)\ds{\bf{34}}+m\left(\la{\bf{3}}p_1-p_2\rs{\bf{4}}+\ls{\bf{3}}p_1-p_2\ra{\bf{4}}\right)\right)^2\nonumber\\
&\qquad\qquad\qquad\qquad\qquad\qquad\qquad\qquad\qquad\qquad\qquad\qquad\qquad\qquad\quad+\text{Parity conj.}\Big),
\end{align}
where $Q$ and $Q'$ are the quadrupoles of $W$ and $W'$ respectively. The double insertion terms are 
\begin{align}
&A_{QQ}(W,\overline{W},W',\overline{W'})\nonumber\\
&=-\frac{1}{s}\frac{1}{\left(M_{Pl}\Lambda^2\right)^2}\bigg(Q(Q')^*\bigg(u\ds{\bf{12}}\da{\bf{34}}-M^2\ds{\bf{12}}\ds{\bf{34}}-m^2\da{\bf{12}}\da{\bf{34}}+2mM\mathbf{\Pi}_u\bigg)\nonumber\\
&\qquad\qquad\qquad\qquad\quad\times\bigg(-t\ds{\bf{12}}\da{\bf{34}}+M^2\ds{\bf{12}}\ds{\bf{34}}+m^2\da{\bf{12}}\da{\bf{34}}+2mM\mathbf{\Pi}_t\bigg)\nonumber\\
&\qquad\qquad\qquad\qquad\qquad\qquad\qquad\qquad\qquad\qquad\qquad\qquad\qquad\qquad\qquad+\text{Parity conj.}\bigg),
\end{align}
Again, the leading high energy divergences cannot be completely tuned against possible contact terms.

\bibliography{superspace}{}
\bibliographystyle{JHEP}
\end{document}